\newif\ifjoesubmission
\joesubmissionfalse
\documentclass[11pt]{article}
\usepackage{style}
\ifjoesubmission
\usepackage[mathlines]{lineno}
\linenumbers
\newcommand{\tablesize}{\footnotesize}
\newcommand{\affilsize}{\small}
\else
\newcommand{\tablesize}{\small}
\newcommand{\affilsize}{}
\fi

\title{\papertitle\thanks{We thank St\'ephane Bonhomme, Azeem Shaikh, Vira Semenova, Kirill Ponomarev, John Pepper, Ahnaf Rafi, Rahul Singh and Michal Koles\'ar for helpful suggestions. All remaining errors are our own.}}

\author{
\begin{tabular}[t]{c}
Geonwoo Kim\\
{\affilsize Department of Economics}\\
{\affilsize University of Chicago}\\
{\affilsize\texttt{kimgeonwoo@uchicago.edu}}
\end{tabular}
\quad
\begin{tabular}[t]{c}
Suyong Song\\
{\affilsize Department of Economics and Finance}\\
{\affilsize Tippie College of Business, University of Iowa}\\
{\affilsize\texttt{suyong-song@uiowa.edu}}
\end{tabular}
}
\date{November 2026}

\begin{document}

\maketitle

\begin{abstract}
We develop inference on a treatment coefficient in a partially linear model whose nuisance functions are approximately linear in high-dimensional controls observed only through \(K\ge2\) repeated, error-prone measurements. Averaging the measurements leaves a nonvanishing error that can shift the double machine learning (DML) moment by a bilinear form in the nuisance coefficients and the error covariance, a bias that better first stages cannot remove. Within-unit contrasts identify this covariance, including correlations across controls. \CORE{} recalibrates each cross-fitted score with the unit's own covariance contribution. The score averages residual products from
distinct measurements and carries covariance-estimation uncertainty into the standard error. It is centered and Neyman orthogonal for affine nuisances. Under approximate sparsity and generic rate conditions on linear first stages, verified for a corrected Lasso, the estimator is uniformly asymptotically linear and its confidence intervals are uniformly valid. Simulations and repeated lottery valuations from the Caltech Cohort Study illustrate the method.

\vspace{\baselineskip}


\end{abstract}

\section{Introduction}\label{sec:intro}

Controls used to adjust for confounding are often measured with error. Linked survey and administrative records report different earnings for the same worker~\citep{boundkrueger1991extent,abowdstinson2013estimating}, experiments elicit the same preference twice~\citep{gillensnowbergyariv2019}, and epidemiological studies replicate laboratory and monitor measurements on the same individuals~\citep[Secs.~4.2 and~4.3]{keogh2020stratos}. Controlling for a noisy proxy only partly controls for the variable it measures, so part of the confounding survives the adjustment~\citep{pei2019poorly}, and with many controls the adjustment must itself be regularized. This paper develops inference on a scalar treatment coefficient when the nuisance functions are approximately linear in high-dimensional controls that are observed only through \(K\ge2\) repeated, error-prone measurements.

Double/debiased machine learning (DML) delivers such inference when the controls are observed: orthogonal scores make the estimator insensitive, to first order, to errors in regularized first stages, and cross-fitting removes their own-observation bias~\citep{chernozhukov2018double}. With mismeasured controls the score is not centered at the truth. In a partially linear model with outcome \(Y_i\), treatment \(D_i\), latent controls \(X_i\) and linear nuisance functions \(X_i'\betao\) and \(X_i'\gammao\), replacing \(X_i\) by a proxy \(Z_i=X_i+\xi_i\), such as the average of the measurements, gives, when the errors have conditional mean zero,
\begin{equation}
\E\bigl[\{Y_i-\thetao D_i-Z_i'\betao\}\{D_i-Z_i'\gammao\}\bigr]
=\betao'\Omegao\gammao,\qquad \Omegao:=\E[\xi_i\xi_i'],
\label{eq:intro-naive-shift}
\end{equation}
at the true treatment effect \(\thetao\) and the true coefficients. The shift is a population quantity, so neither more data nor better first stages remove it. It involves every pair of controls on which the outcome and treatment equations load: when the two equations load on different controls whose errors are correlated, a correction that uses only the error variances removes none of it.

Each unit's repeated measurements give an unbiased estimate of \(\Omegao\), because the latent controls cancel in the differences among a unit's measurements and the errors of distinct measurements are uncorrelated. We propose \CORE{}, a covariance-recalibrated DML estimator that subtracts from each cross-fitted score the bilinear form \(\beta'\Omega_i\gamma\) evaluated at the unit's own covariance contribution \(\Omega_i\). An exact algebraic identity shows that the recalibrated score is an average, over ordered pairs of distinct measurements, of products of residuals in which one measurement residualizes the outcome and another residualizes the treatment. With affine nuisance functions, the corrected score is centered at the treatment coefficient and is Neyman orthogonal to the nuisance coefficients. Under approximate sparsity, it has the same mean as the latent linear score. Our rate condition makes the remaining approximation and first-stage terms negligible for inference. The unit-level score has no covariance nuisance, and its standard error accounts for covariance estimation. With a plug-in score, covariance-estimation error enters the influence function at first order, as~\citet*{belloni2017confidence} show for a diagonal covariance.  Conditional on the training sample, each held-out score has the population mean at the fitted coefficients. Thus only the prediction rates of the fitted affine first stages, including their intercepts, matter.

The paper makes three contributions. First, it shows that \CORE{} is uniformly asymptotically linear, that its variance estimator is consistent and that its confidence intervals are uniformly valid, over classes of data-generating processes with many controls and approximately sparse nuisance functions. The conditions are rates for generic cross-fitted linear first stages, and the product-rate condition is that of DML without measurement error. Estimating the covariance on the full sample instead, and plugging it into every held-out score, gives an estimator with the same influence function when, in addition, the \(\ell_1\) errors of the first stages are small enough that the interaction of the covariance estimate with the first stages is negligible. Second, it verifies the rate conditions for a corrected Lasso, the Convex Conditioned Lasso (CoCoLasso) of~\citet{datta2017cocolasso}, computed with the replicate-based covariance of the training folds. Our sufficient sparsity budget is \(s^2\log\max\{p,n\}=o(\sqrt n)\) when the coefficients are bounded only in \(\ell_2\) norm, because the projection that makes CoCoLasso convex perturbs its score by up to the \(\ell_1\) norm of the target times the error of the corrected Gram matrix. The familiar \(s\log\max\{p,n\}=o(\sqrt n)\) suffices under bounded \(\ell_1\) norms or when \(p\) is a small fraction of \(n\), and, up to a factor \((\log\log n)^2\), when the penalties are selected by any training-sample rule, cross-validation included, from a grid whose largest value eventually shrinks with the training-sample size (\Cref{rem:thy-cv-not-covered}), on a subclass with bounded \(\ell_1\) norms in which the outcome and the treatment each have a covariance bounded away from zero with at least one control. Third, it evaluates the method in simulated and real data with repeated measurements.

\begin{figure}[t]
\centering
\includegraphics[width=\linewidth]{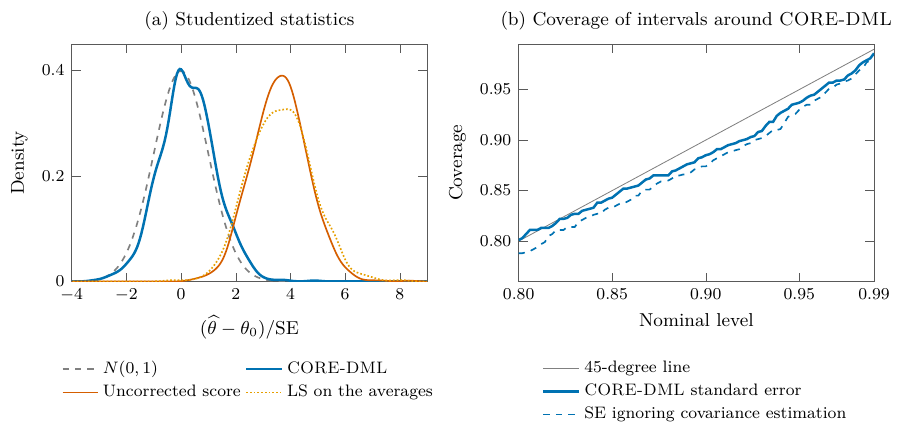}
\caption{Measurement error in the controls and the \CORE{} correction in the canonical simulation design of~\Cref{sec:simulations} with \((n,p)=(1000,200)\), \(K=2\) measurements with single-measurement reliability 0.8 and AR\((0.5)\) errors, and 1{,}000 replications. Panel~(a): densities of the studentized statistic for least squares on the averaged measurements, for the uncorrected DML score on the corrected first stages of \CORE{}, and for \CORE{}, against the standard normal density. Panel~(b): coverage, at nominal levels from 80\% to 99\%, of intervals centered at the \CORE{} estimates, with the \CORE{} standard error and with the standard error that ignores covariance estimation, which replaces each \(\Omega_i\) by \(\Pn\Omega_i\) (\Cref{subsec:estimator-variance}); neither uses the true covariance. At the 95\% level they cover 0.937 and 0.930; the \CORE{} standard error is the larger in every replication, so its interval contains the other and covers in 7 replications where the other does not.}
\label{fig:sim-illustrative}
\end{figure}

\Cref{fig:sim-illustrative} shows the main features in a simulated design. The two procedures that leave the score uncorrected are centered more than three standard errors from the truth, whereas the bias of \CORE{} is 0.22 of its standard deviation. The standard error of \CORE{} includes the contribution of the estimated covariance, and intervals around the same estimates with a standard error that ignores this contribution cover less often. In this design, knowing the covariance would improve precision slightly: on the same draws, the standard deviation of \CORE{} is 1.021 times that of the infeasible known-covariance estimator (\Cref{subsec:sim-canonical}). In designs with fewer controls than observations, sparse or polynomially decaying coefficients, correlated errors, unequal error covariances across measurements, heteroskedasticity and heavy tails, \CORE{} removes the bias of procedures that ignore the measurement error, and the coverage of its intervals ranges from 0.918 to 0.946; in the canonical sequence of~\Cref{sec:simulations} the remaining shortfall is the second-order bias of the first stages, which we compute exactly from the first-stage fits and the true parameters of the design. When the supports grow with \(p\), coverage deteriorates, even for DML with the latent controls, whose bias is smaller. In the Caltech Cohort Study of~\citet{gillensnowbergyariv2019}, students valued the same lotteries at two stakes, and we use the paired valuations as repeated measurements of their attitudes toward risk and ambiguity. The male--female gap in choosing tournament pay is 13.41 percentage points after the correction, against 13.29 without it. The correction is small because the valuations barely predict gender once performance and stated rank, which we treat as measured without error, are controlled for; two observable implications of the measurement model are not rejected. In an online labor market with hundreds of text-based controls and simulated measurement error, the correction removes most of the displacement that the error causes.

\paragraph{Related literature.} The paper builds on the errors-in-variables literature that uses additional measurements to identify latent moments and correct regression bias~\citep{fuller1980properties,fuller1987measurement,hausmanneweyichimurapowell1991,carroll2006measurement,schennach2016recent,schennach2022measurement}. Under independence conditions, two measurements of a scalar variable also identify its distribution through Kotlarski's lemma~\citep{livuong1998multiple} and nonseparable and quantile regression models through Kotlarski-type deconvolution~\citep{songschennachwhite2015nonseparable,firpogalvaosong2017quantile}. Corrections based on a repeated measurement can change economic conclusions, as in the S-shaped cross-sectional investment--\(q\) relation that~\citet{songwee2026investment} recover with an analyst-based measurement of Tobin's~\(q\) and the eigendecomposition argument of~\citet{huschennach2008instrumental}. Here the mismeasured variables are many controls that enter approximately linearly, so we restrict only conditional means and cross-measurement covariances, which identify the error covariance but not the distribution of the latent controls. For treatment effects with mismeasured covariates, \citet{battistinchesher2014treatment} approximate the bias of estimators that ignore the error under small-error asymptotics, \citet{nagasawa2024noisy} constructs control variables from a noisy proxy and an excluded variable, \citet{agarwalsingh2026noisy} exploit a low-rank structure in a single noisy measurement of many covariates, and~\citet{xuetal2024pollutants} combine DML with a regression-calibration correction, estimated in an external validation study, when both the exposure and the confounders are mismeasured. More generally,~\citet{arganarazescanciano2025dml} characterize orthogonal moments for functionals of unobserved heterogeneity, with a factor model for two measurements among their applications. In our setting a fixed number of repeated measurements of the whole high-dimensional control vector identifies a possibly nondiagonal error covariance from second moments, without validation data or a low-rank restriction on the controls, and the errors need not be small.

High-dimensional regression with mismeasured regressors is studied by~\citet{loh2012high}, \citet{rosenbaumtsybakov2010sparse,rosenbaumtsybakov2013improved}, \citet{datta2017cocolasso}, \citet*{bellonirosenbaumtsybakov2017conic} and~\citet{rudelson2017errors}, among others. Except for the matrix uncertainty selector of~\citet{rosenbaumtsybakov2010sparse}, which bounds the effect of a small deterministic error, these estimators subtract a known error covariance or an estimate of it and deliver estimation and selection bounds. The estimate is obtained from replicated measurements or independent samples of the noise, as~\citet[Sec.~3.2.1]{loh2012high} and~\citet[Sec.~4.1]{datta2017cocolasso} point out, or, in~\citet{rudelson2017errors}, from the observed design when the trace of the covariate covariance is known; the bounds of~\citet{loh2012high} remain valid with a covariance estimated from an independent sample of the noise, and~\citet{rosenbaumtsybakov2013improved} also give conservative finite-sample confidence intervals that require an upper bound on the sparsity. \citet{luolipengyue2025care} study estimation and false-discovery control in high-dimensional partially linear models whose linearly entering covariates are mismeasured. For inference, \citet{lilima2021measurement} construct a corrected decorrelated score test for the coefficients of a fixed number of mismeasured covariates when the other high-dimensional covariates are measured without error and the error moments are known, and~\citet{cuiguoyangzhang2025def} construct a doubly robust corrected score for the coefficient of a single mismeasured regressor with a known error variance.

The analysis closest to ours is that of~\citet*{belloni2017confidence}, who construct uniformly valid confidence intervals and simultaneous confidence bands for the coefficients of a high-dimensional linear model whose regressors are observed with additive errors. Their orthogonal score decorrelates a bias-corrected least-squares moment through an errors-in-variables regression of the target regressor on the others and, for the coefficient of an error-free regressor, is the corrected score~\eqref{eq:score-corrected}, which does not require a diagonal covariance; their conditions take the error covariance and its estimate to be diagonal (their Conditions~A(v) and~D). When the covariance is replaced by a diagonal estimate whose entries converge at the rate \(\{\log(np)/n\}^{1/2}\), they show that the score is not orthogonal to the covariance, so that the error of the estimate enters at first order (their Theorem~3); when the estimate is asymptotically linear, they derive the resulting influence-function term (their Corollary~3) and note that this representation can be used to justify a multiplier bootstrap that accounts for it, and in a missing-data example they verify these conditions and show that replacing the influence function of the covariance estimate by an estimate has a negligible effect on the bootstrap (their Section~4.2). Their formal coverage results (their Corollary~2 and Theorem~2) are stated for a known covariance. Our setting and results differ from theirs in five respects. First, the error covariance is that of the averaged measurements; it is identified by within-unit contrasts of the repeated measurements and may be nondiagonal, so the correction also removes the part of the shift that comes from correlations of the errors across coordinates. Second, the cross-measurement score evaluates the correction at each unit's own covariance contribution, so the nondiagonal analog of the term of their Corollary~3, \(-\betao'(\Omega_i-\Omegao)\gammao\), is part of the score and need not be estimated separately. Third, the parameter is a scalar treatment coefficient in a partially linear model whose nuisance functions are approximately rather than exactly sparse. Fourth, the nuisance coefficients are learned by cross-fitting under prediction-rate conditions, whereas their first steps are computed on the full sample and must produce sparse estimates. Fifth, our variance estimator~\eqref{eq:est-variance} contains the covariance-estimation term, and we prove that the resulting intervals are uniformly valid. Their simultaneous inference on many coefficients, including those of mismeasured regressors, is outside our scope. With coefficients bounded in \(\ell_2\) norm their growth condition is \(s\log(np)=o(\sqrt n)\), which is less restrictive than our sufficient budget for CoCoLasso unless \(p\) is a small fraction of \(n\) (\Cref{coro:cocolasso-core}); their condition takes the rates of their first steps as given, whereas our budget includes the cost of the projection that makes CoCoLasso convex.

In the cross-measurement form, a single ordering of the measurements is a just-identified high-dimensional instrumental-variables problem~\citep{bellonihansennewey2022many,canerkock2019gmm} and the linear case of the two-proxy structure of proxy-control and proximal causal inference~\citep{griliches1977estimating,miaogengtchetgen2018proxy,deaner2021many,tchetgentchetgenyingcuishimiao2024proximal,cuipushimiaotchetgen2024semiparametric}. \Cref{subsec:score-cross} explains what the common scale of repeated measurements adds. Our orthogonality and cross-fitting arguments follow~\citet{chernozhukov2018double}, building on~\citet{robinson1988rootn}, \citet{newey1994asymptotic} and~\citet{belloni2014inference} for orthogonality and on~\citet{neweyrobins2017crossfitting} for cross-fitting.

The framework covers nuisance functions that are approximately linear in the latent controls (\Cref{subsec:model-sparsity}). It requires errors that are on the scale of the latent controls, have conditional mean zero and are uncorrelated across measurements, together with a fixed number of measurements; it allows the errors to be heteroskedastic, correlated across controls and of different quality across measurements. \Cref{sec:model} presents the model and the measurement system, \Cref{sec:score} the corrected and cross-measurement scores, \Cref{sec:estimator} the estimator and \Cref{sec:theory} the asymptotic theory and the corrected-Lasso first stage. \Cref{sec:simulations,sec:application} report the simulations and the applications, and \Cref{sec:conclusion} concludes. The appendices contain the proofs, further results on the measurement system and the first stages, and the details of the simulations and the applications.
\section{Model and Measurement System}\label{sec:model}

\paragraph{Notation.}
Let \(n\) denote the sample size and \(p_n\) the dimension of the latent controls, which may grow with \(n\). We set \(\bar p_n=\max\{p_n,n\}\) and write \(p\) and \(\bar p\) for \(p_n\) and \(\bar p_n\) when the dependence on \(n\) need not be shown. Measurements are indexed by \(k,h\in\{1,\ldots,K\}\), with \(K\ge2\) fixed, and cross-fitting folds by \(\ell\in\{1,\ldots,L\}\), with \(L\) fixed. A subscript \(0\) marks the true value of a parameter, and estimators carry a hat. Write \(\Pn A_i=n^{-1}\sum_{i=1}^n A_i\), and let \(\E_P\) and \(\Pr_P\) denote expectation and probability under \(P\); the subscript is dropped when no confusion can arise. For a sequence of classes \(\mathcal P_n\) of distributions and a deterministic sequence \(d_n\), the statement \(A_n=O_P(d_n)\) uniformly over \(\mathcal P_n\) means
\[
\lim_{M\to\infty}\limsup_{n\to\infty}
\sup_{P\in\mathcal P_n}
\Pr_P\{|A_n|>Md_n\}=0,
\]
and uniform \(o_P(d_n)\) is defined analogously. Unless stated otherwise, all stochastic order statements are uniform over \(\mathcal P_n\). For vectors, \(\norm{\cdot}_q\) denotes the \(\ell_q\) norm and \(\norm{b}_0\) the number of nonzero entries of \(b\). For matrices, \(\norm{\cdot}_{\mathrm{op}}\) and \(\norm{\cdot}_{\max}\) denote the operator norm and the largest absolute entry, \(\lambda_{\min}\) and \(\lambda_{\max}\) the smallest and largest eigenvalues of a symmetric matrix, and \(A\succeq B\) means that \(A-B\) is positive semidefinite. For a scalar random variable \(G\) and \(q\ge1\), let \(\norm{G}_{P,q}=(\E_P|G|^q)^{1/q}\); when \(G\) depends on fitted quantities, the norm is computed with those quantities held fixed, over an independent draw from \(P\). For \(q\in\{1,2\}\), \(\norm{G}_{\psi_q}=\inf\{t>0:\E_P[\exp(|G|^q/t^q)]\le2\}\) is the Orlicz norm, and for a random vector it is the supremum of this norm over linear combinations with Euclidean-unit coefficients. Throughout, \(c,C>0\) denote constants that do not depend on \(n\) or \(P\); they may change from line to line, except where they are fixed in the definition of \(\mathcal P_n\) in~\Cref{def:dgpclass}. For positive sequences, \(a_n\lesssim b_n\) means \(a_n\le Cb_n\), \(a_n\asymp b_n\) means \(a_n\lesssim b_n\lesssim a_n\), and \(a_n\ll b_n\) means \(a_n/b_n\to0\). The indicator function is \(\ind\{\cdot\}\).

\subsection{Partially linear model}\label{subsec:model-plm}

For \(i=1,\ldots,n\), the outcome \(Y_i\in\R\), the treatment \(D_i\in\R\) and the latent controls \(X_i\in\R^{p}\) follow the partially linear model~\citep{robinson1988rootn}
\begin{align}
Y_i &= \thetao D_i+f_0(X_i)+U_i,
\label{eq:model-plr-y}\\
D_i &= g_0(X_i)+V_i,
\label{eq:model-plr-d}
\end{align}
where \(\thetao\in\R\) is the parameter of interest and \(f_0\) and \(g_0\) are unknown nuisance functions. The disturbances satisfy
\begin{gather}
\E[U_i\mid D_i,X_i]=0,
\qquad
\E[V_i\mid X_i]=0,
\label{eq:model-moment-restrictions}\\
0<c\le J_0:=\E[V_i^2]\le C<\infty.
\label{eq:model-j0}
\end{gather}
The first restriction is conditional mean exogeneity given the latent controls, the second makes \(g_0(X_i)\) the conditional mean of the treatment, and~\eqref{eq:model-j0} requires residual variation in the treatment. If \(X_i\) were observed, inference could be based on the score
\begin{equation}
\psi_i^0(\theta,f,g)
=\{Y_i-\theta D_i-f(X_i)\}\{D_i-g(X_i)\},
\label{eq:model-latent-score}
\end{equation}
whose mean vanishes at \((\thetao,f_0,g_0)\), whose derivative in \(\theta\) is \(-J_0\), and which is Neyman orthogonal with respect to \((f,g)\)~\citep{bickel1993efficient,chernozhukov2018double}. When the effect of \(D_i\) varies with \(X_i\), the latent-score moment \(\E[\psi_i^0(\theta,f_0,g_0)]=0\) identifies a weighted average of the conditional effects, with weights proportional to the conditional variance of the treatment; the functional of~\Cref{prop:observed-identification}, which \CORE{} targets, has this interpretation when \(g_0\) is affine and otherwise contains additional projection terms (\Cref{rem:heterogeneous}). When \(D_i\) is not randomly assigned given \(X_i\), as in the application of~\Cref{subsec:app-caltech}, \(\thetao\) is an adjusted association.

\subsection{Repeated measurements}\label{subsec:model-design}

Repeated measurements are a standard source of identification in measurement-error models~\citep{hausmanneweyichimurapowell1991,livuong1998multiple,li2002robust,schennach2004estimation,schennach2020mismeasured,schennach2022measurement}. For each unit we observe \(K\ge2\) measurements of the latent controls, \(W_{ik}=X_i+\nu_{ik}\) for \(k=1,\ldots,K\), where \(\nu_{ik}\in\R^{p}\) is the error of measurement \(k\). Let \(\mathcal F_i:=\sigma(X_i,U_i,V_i)\).

\begin{assumption}[Repeated measurements]\label{def:model-repeated-measurement}
For all \(k,h\in\{1,\ldots,K\}\),
\begin{align}
\E[\nu_{ik}\mid \mathcal F_i]&=0,
\label{eq:model-repeat-mean-zero}\\
\E[\nu_{ik}\nu_{ih}']&=0\quad(k\ne h).
\label{eq:model-repeat-cross-orthogonality}
\end{align}
We write \(\Sigma_{\nu,k}:=\E[\nu_{ik}\nu_{ik}']\) for the error covariance of measurement \(k\).
\end{assumption}

\Cref{def:model-repeated-measurement} leaves the conditional covariance \(\E[\nu_{ik}\nu_{ik}'\mid\mathcal F_i]\) unrestricted: the errors may be heteroskedastic, correlated across the coordinates of a measurement and distributed differently across measurements. Errors that are mean zero and independent of \(\mathcal F_i\) are what~\citet{schennach2016recent} calls strongly classical; \Cref{def:model-repeated-measurement} weakens this to conditional mean zero and also requires the errors of distinct measurements to be uncorrelated.

The estimators use an average of the measurements. For deterministic weights \(\alpha_1,\ldots,\alpha_K\) that sum to one, define the aggregate proxy, its error and the proxy-error covariance
\begin{gather}
Z_i:=\textstyle\sum_{k=1}^K \alpha_k W_{ik}=X_i+\xi_i,
\qquad
\xi_i:=\textstyle\sum_{k=1}^K \alpha_k\nu_{ik},
\label{eq:model-aggregate-proxy}\\
\Omegao:=\E[\xi_i\xi_i']=\textstyle\sum_{k=1}^K \alpha_k^2\Sigma_{\nu,k},
\label{eq:model-omega}
\end{gather}
where the last equality uses~\eqref{eq:model-repeat-cross-orthogonality}. We call the bilinear adjustment \(\beta'\Omega\gamma\) of the score the \emph{covariance correction}. The results of~\Cref{subsec:score-shift,subsec:score-corrected-score} use only~\eqref{eq:model-repeat-mean-zero}; the covariance contributions of~\Cref{subsec:score-cross}, and hence the estimator of~\Cref{sec:estimator}, also use~\eqref{eq:model-repeat-cross-orthogonality}. If some coordinates of \(X_i\) are measured exactly, \(\Omegao\) vanishes on their rows and columns, and nothing else changes (\Cref{rem:mixed-controls}).

\paragraph{What the measurement system requires.} The measurements of a control share the scale of the latent control and refer to the same period, and none of them is treated as error free; proxies with their own scales or loadings are the domain of proxy-control methods~\citep{deaner2021many} and of measurement systems with factor loadings~\citep{cunhaheckmanschennach2010}. Examples are linked survey and administrative records of the same earnings~\citep{abowdstinson2013estimating}, replicate laboratory or monitor measurements on the same individuals~\citep[Secs.~4.2, 4.3.4 and~4.3.5]{keogh2020stratos}, panel waves of a time-invariant characteristic and duplicate elicitations in experiments, such as lottery valuations put on a common prize scale~\citep{gillensnowbergyariv2019}. Restriction~\eqref{eq:model-repeat-mean-zero} is the weakly classical condition of~\citet{schennach2016recent}, with the conditioning set enlarged to the structural disturbances; because \((X_i,U_i,V_i)\) and \((X_i,D_i,Y_i)\) generate the same \(\sigma\)-field, it states that each measurement is conditionally unbiased and nondifferential in mean, \(\E[W_{ik}\mid X_i,D_i,Y_i]=X_i\). It excludes errors correlated with the true value, errors shared with a self-reported outcome and misclassified discrete controls (\Cref{app:measurement}). Restriction~\eqref{eq:model-repeat-cross-orthogonality} is natural for distinct sources or independent elicitations and fails when a component, such as a person-specific reporting bias, enters every measurement; \Cref{app:sim-crossdep} gives the resulting bias and a sensitivity analysis, which~\Cref{subsec:app-caltech} applies. Correlation across the coordinates of a measurement, as when related items are misreported together, is allowed subject to \(\|\Sigma_{\nu,k}\|_{\mathrm{op}}\le C\) (\Cref{def:dgpclass}), and \CORE{} estimates the resulting nondiagonal \(\Omegao\). Measurements recorded on scales that differ by known affine maps can be aligned first, and additive measurement-specific intercepts are accommodated by centering the within-unit contrasts (\Cref{rem:source-intercepts}).

\paragraph{A fixed number of measurements.} Two measurements suffice to identify each \(\Sigma_{\nu,k}\), \(\Omegao\) and, with affine nuisance functions, \(\thetao\) from second moments (\Cref{prop:observed-identification,rem:cov-weights-identification}). With \(K\) fixed, averaging can leave measurement error in the proxy. Both the score shift derived in~\Cref{sec:score} and the contribution of estimating \(\Omegao\) to the asymptotic variance can persist as \(n\) grows. Unlike small-error approximations~\citep{battistinchesher2014treatment,evdokimovzeleneev2025simple}, the theory does not require either to be small.

\subsection{Covariance correction and identification}\label{subsec:model-omega}

The estimator of~\Cref{sec:estimator} is invariant to the location of the controls, so we normalize \(\E[X_i]=0\). Unpenalized intercepts absorb \(\E[f_0(X_i)]\) and \(\E[g_0(X_i)]\), and all population displays refer to centered variables. With \(\Sigma_X:=\E[X_iX_i']\), the correction identity
\begin{equation}
\E[Z_iZ_i']-\Omegao=\Sigma_X
\label{eq:model-covariance-correction}
\end{equation}
underlies both the corrected regressions and the corrected score. Because \(\E[\xi_i\mid X_i]=0\), the identity implies \(\E[(R_i-Z_i'b)^2-b'\Omegao b]=\E[(R_i-X_i'b)^2]\) for every \(b\in\R^{p}\) and every scalar \(R_i\) with \(\E[\xi_iR_i]=0\), such as \(Y_i\) and \(D_i\), which are functions of \((X_i,U_i,V_i)\). Corrected least squares on the proxy therefore has the population objective of least squares on the latent controls. With affine nuisance functions, the measurements also identify \(\thetao\) without reference to \(\Omegao\).

\begin{proposition}[Identification from the observed distribution]\label{prop:observed-identification}
Suppose that~\eqref{eq:model-plr-y}--\eqref{eq:model-j0} hold with \(f_0(x)=\mu_f+x'\betao\) and \(g_0(x)=\mu_g+x'\gammao\) for constants \(\mu_f\) and \(\mu_g\), that~\Cref{def:model-repeated-measurement} holds, and that all variables are square integrable. Let \(\pio:=\betao+\thetao\gammao\), \(\Sigma_{12}:=\operatorname{Cov}(W_{i1},W_{i2})\), \(c_{DZ}:=\operatorname{Cov}(Z_i,D_i)\) and \(c_{YZ}:=\operatorname{Cov}(Z_i,Y_i)\), where \(Z_i\) is the aggregate proxy~\eqref{eq:model-aggregate-proxy} for any deterministic weights that sum to one. Then \(\Sigma_{12}=\operatorname{Var}(X_i)\), \(c_{DZ}=\Sigma_{12}\gammao\), \(c_{YZ}=\Sigma_{12}\pio\), and
\begin{equation*}
\thetao=\frac{\operatorname{Cov}(Y_i,D_i)-c_{DZ}'\Sigma_{12}^{\dagger}c_{YZ}}{\operatorname{Var}(D_i)-c_{DZ}'\Sigma_{12}^{\dagger}c_{DZ}},
\qquad
\operatorname{Var}(D_i)-c_{DZ}'\Sigma_{12}^{\dagger}c_{DZ}=J_0,
\end{equation*}
where \(\Sigma_{12}^\dagger\) is the Moore--Penrose inverse. Hence \(\thetao\) is identified from the distribution of the observed variables \((Y_i,D_i,W_{i1},\ldots,W_{iK})\); the value of the ratio does not depend on the aggregation weights, and the ratio involves neither \(\Omegao\) nor \(\E[X_i]\).
\end{proposition}

The proof (\Cref{app:model-identification}) is the standard argument in which one measurement instruments another~\citep{schennach2016recent} and repeated measurements identify the error covariance~\citep[Case~(iii)]{fuller1980properties}. The ratio requires a generalized inverse of the high-dimensional matrix \(\Sigma_{12}\); \CORE{} does not compute it and recenters a score instead.

\subsection{Approximate sparsity and scope}\label{subsec:model-sparsity}

Following~\citet{belloni2012sparse}, we approximate \(f_0\) and \(g_0\) by their best sparse linear mean-square approximations, with sparsity levels \(s_{f,n}\) and \(s_{g,n}\) and approximation-error bounds \(a_{f,n}\) and \(a_{g,n}\):
\begin{align}
&f_0(X_i)-\E[f_0(X_i)]=X_i'\betao+r_{f,i},
\label{eq:model-f-approx}\\
&g_0(X_i)-\E[g_0(X_i)]=X_i'\gammao+r_{g,i},
\label{eq:model-g-approx}\displaybreak[1]\\
&\betao\in\argmin\nolimits_{b\in\R^{p}:\,\|b\|_0\le s_{f,n}}\E\bigl[\{f_0(X_i)-X_i'b\}^2\bigr],
\label{eq:model-best-s-beta}\\
&\gammao\in\argmin\nolimits_{b\in\R^{p}:\,\|b\|_0\le s_{g,n}}\E\bigl[\{g_0(X_i)-X_i'b\}^2\bigr],
\label{eq:model-best-s-gamma}\\
&\{\E[r_{f,i}^2]\}^{1/2}\le a_{f,n},
\qquad
\{\E[r_{g,i}^2]\}^{1/2}\le a_{g,n}.
\label{eq:model-approx-errors}
\end{align}

\begin{definition}[Approximate sparsity]\label{def:approx-sparsity}
The pair \((f_0,g_0)\) is approximately sparse in the dictionary \(X_i\) if there exist sequences \((s_{f,n},s_{g,n},a_{f,n},a_{g,n})\) and coefficients \((\betao,\gammao)\) that satisfy~\eqref{eq:model-f-approx}--\eqref{eq:model-approx-errors} with \(a_{f,n}\to0\) and \(a_{g,n}\to0\).
\end{definition}
Because \(\E[X_i]=0\), the approximation errors have mean zero; exact sparsity is the case \(r_{f,i}=r_{g,i}=0\). The sparsity levels are restricted only through the rate conditions of~\Cref{sec:theory}.

The framework therefore concerns nuisance functions that are approximately linear in the latent controls, with many controls and regularized first stages. Unlike DML with accurately measured controls, it cannot enlarge the dictionary with powers, interactions or other transformations of the measurements: the error of a transformed measurement is in general neither conditionally mean zero nor uncorrelated across measurements, so a richer basis requires measurement restrictions of its own (\Cref{rem:linear-scope}). Transformations of exactly measured controls carry no measurement error and may be added, provided the enlarged dictionary satisfies the conditions of~\Cref{def:dgpclass}, including the sub-Gaussian bound~\eqref{eq:thy-vector-subgaussian}; their interactions with mismeasured controls are transformations of the measurements and are not covered. When \(f_0\) or \(g_0\) is not affine, the functional of~\Cref{prop:observed-identification} identifies a projection parameter \(\theta^\ast\) with \(|\theta^\ast-\thetao|\le a_{f,n}a_{g,n}/J_0\), which is \(o(n^{-1/2})\) under the rate conditions of~\Cref{sec:theory}, and \(\theta^\ast=\thetao\) whenever either nuisance is affine (\Cref{rem:projection-target}). High dimensionality here means many controls relative to the information in the sample, as with correlated controls or approximate sparsity; \(p_n>n\) is allowed but not required.
\section{The Score}\label{sec:score}

\subsection{The naive proxy moment}\label{subsec:score-shift}

Replacing \(X_i\) by the proxy \(Z_i\) in the linear version of the score~\eqref{eq:model-latent-score} gives the naive proxy moment
\begin{equation*}
\psi_i^{\mathrm{naive}}(\theta,\beta,\gamma)
=
\{Y_i-\theta D_i-Z_i'\beta\}\{D_i-Z_i'\gamma\}.
\end{equation*}
Let \(\Delta_{fg,n}:=\E[r_{f,i}r_{g,i}]\), where \(r_{f,i}\) and \(r_{g,i}\) are the approximation errors in~\eqref{eq:model-f-approx}--\eqref{eq:model-g-approx}; by the Cauchy--Schwarz inequality, \(\abs{\Delta_{fg,n}}\le a_{f,n}a_{g,n}\).

\begin{proposition}[Shift of the naive moment]\label{prop:naive-shift}
Suppose that~\eqref{eq:model-plr-y}--\eqref{eq:model-moment-restrictions} and~\eqref{eq:model-repeat-mean-zero} hold and all variables are square integrable, let \(\betao\) and \(\gammao\) satisfy~\eqref{eq:model-f-approx}--\eqref{eq:model-g-approx}, and let \(Z_i\) and \(\Omegao\) be as in~\eqref{eq:model-aggregate-proxy}--\eqref{eq:model-omega}. Then
\begin{equation}
\E\bigl[\psi_i^{\mathrm{naive}}(\thetao,\betao,\gammao)\bigr]
=
\betao'\Omegao\gammao+\Delta_{fg,n}.
\label{eq:score-naive-shift}
\end{equation}
\end{proposition}

At the true coefficients, the two proxy residuals are \(U_i+r_{f,i}-\xi_i'\betao\) and \(V_i+r_{g,i}-\xi_i'\gammao\). Conditional mean zero removes every product of an error index \(\xi_i'b\) with a latent residual, but the product of the two error indices has mean \(\betao'\Omegao\gammao\) (\Cref{app:score-proofs}). Because the shift arises at the true coefficients, neither better first stages nor orthogonalizing the score in the nuisance coefficients can remove it. With a single control it produces the familiar errors-in-variables bias, whose sign is that of \(\betao\gammao\) (\Cref{rem:score-scalar}). The shift involves every pair of coordinates on which \(\betao\) and \(\gammao\) load: when the outcome and treatment equations load on different controls whose errors are correlated, \(\betao'\diag(\Omegao)\gammao=0\) and a diagonal correction removes none of it. The approximation term \(\Delta_{fg,n}\) is present even without measurement error, and the rate condition~\eqref{eq:thy-product-rate} makes it negligible.

\subsection{The corrected score}\label{subsec:score-corrected-score}

Subtracting the measurement-error part of the shift gives the corrected score
\begin{equation}
\psi_i(\theta,\beta,\gamma,\Omega)
=
\{Y_i-\theta D_i-Z_i'\beta\}\{D_i-Z_i'\gamma\}-\beta'\Omega\gamma,
\label{eq:score-corrected}
\end{equation}
a moment version of the corrected score of~\citet{nakamura1990corrected}. It is the orthogonal errors-in-variables score of~\citet[eq.~(2.4)]{belloni2017confidence} for the coefficient of an error-free regressor, here the treatment; they take the error covariance to be diagonal, which this score does not require. At \(\Omega=\Omegao\) its mean equals that of the latent linear score \(\{Y_i-\theta D_i-X_i'\beta\}\{D_i-X_i'\gamma\}\) for every \((\theta,\beta,\gamma)\) (\Cref{app:score-proofs}). Write the population moment as
\begin{equation}
M(\theta,\beta,\gamma,\Omega)
:=\E\bigl[\psi_i(\theta,\beta,\gamma,\Omega)\bigr].
\label{eq:score-population-map}
\end{equation}

\begin{lemma}[Centering]\label{lemm:centering}
Under the conditions of~\Cref{prop:naive-shift}, \(M(\thetao,\betao,\gammao,\Omegao)=\Delta_{fg,n}\). In particular, \(M(\thetao,\betao,\gammao,\Omegao)=0\) whenever \(\Delta_{fg,n}=0\), as under exact sparsity.
\end{lemma}

\begin{lemma}[Neyman orthogonality in the coefficients]\label{lemm:orthogonality}
Suppose that the conditions of~\Cref{prop:naive-shift} hold and that the sparse approximations are exact. Then, for all fixed directions \(h_\beta,h_\gamma\in\R^{p}\),
\begin{align}
\partial_\beta M(\thetao,\betao,\gammao,\Omegao)[h_\beta]&=0,
\label{eq:score-beta-orthogonality}\\
\partial_\gamma M(\thetao,\betao,\gammao,\Omegao)[h_\gamma]&=0.
\label{eq:score-gamma-orthogonality}
\end{align}
Moreover, \(\partial_\theta M(\thetao,\betao,\gammao,\Omegao)=-J_0\), and the derivative with respect to \(\Omega\) in a symmetric direction \(A\) is
\begin{equation}
\partial_\Omega M(\thetao,\betao,\gammao,\Omegao)[A]
=-\betao'A\gammao.
\label{eq:score-omega-derivative}
\end{equation}
\end{lemma}

The correction removes the measurement-error part of the shift and of the derivatives with respect to the nuisance coefficients, but it does not make the score insensitive to \(\Omega\): the derivative~\eqref{eq:score-omega-derivative} vanishes in every direction only if \(\betao=0\) or \(\gammao=0\). A plug-in estimate \(\widehat\Omega\) therefore moves the score at first order, through \(-\betao'(\widehat\Omega-\Omegao)\gammao\), and its estimation error enters the influence function through the usual first-step adjustment~\citep{newey1994asymptotic,chernozhukovescancianoichimuranewey2022locally}. For a diagonal covariance estimated by a diagonal estimator, \citet[Thm.~3 and Cor.~3]{belloni2017confidence} derive this term. Under approximate sparsity the derivatives contain approximation terms that involve only the latent design (\Cref{app:score-proofs}), and orthogonality carries over to the reduced-form parameterization \((\pi,\gamma)\), \(\pi=\beta+\theta\gamma\), used in estimation (\Cref{rem:score-reduced-form}).

\subsection{Observation-level covariance contributions and the cross-measurement score}\label{subsec:score-cross}

With equal weights the proxy is the mean of the measurements, \(Z_i=\bar W_i:=K^{-1}\sum_{k=1}^KW_{ik}\), with error \(\xi_i=\bar\nu_i\) and covariance \(\Omegao=K^{-2}\sum_{k=1}^K\Sigma_{\nu,k}\), and the observation-level covariance contribution and its centered version are
\begin{equation}
\Omega_i
:=
\frac{1}{K(K-1)}
\sum_{k=1}^K
(W_{ik}-\bar W_i)(W_{ik}-\bar W_i)',
\qquad
H_i:=\Omega_i-\Omegao.
\label{eq:cov-bessel-row}
\end{equation}
Within-unit centering removes the latent controls, \(W_{ik}-\bar W_i=\nu_{ik}-\bar\nu_i\), and~\eqref{eq:model-repeat-cross-orthogonality} then makes each \(\Omega_i\) unbiased for \(\Omegao\), \(\E[H_i]=0\) (\Cref{app:covariance-estimator}). With two measurements, \(\Omega_i=\tfrac14(W_{i1}-W_{i2})(W_{i1}-W_{i2})'\) is unbiased for \(\Omegao=(\Sigma_{\nu,1}+\Sigma_{\nu,2})/4\) whether or not \(\Sigma_{\nu,1}=\Sigma_{\nu,2}\). Its average \(\Pn\Omega_i\) is the classical within-unit estimator of the error covariance (\citealp[Case~(iii)]{fuller1980properties}; \citealp[Sec.~5]{lianghardlecarroll1999}) divided by \(K\), because it targets the error covariance of the averaged measurement.

Evaluated at the unit's own contribution \(\Omega_i\), the corrected score becomes a moment in products of distinct measurements.

\begin{proposition}[Cross-measurement representation]\label{prop:cov-cross-replicate-score}
With the equal-weight proxy \(Z_i=\bar W_i\) and the contributions~\eqref{eq:cov-bessel-row}, for every \((\theta,\beta,\gamma)\) and all scalars \(a_Y,a_D\),
\begin{multline}
\{Y_i-a_Y-\theta D_i-Z_i'\beta\}\{D_i-a_D-Z_i'\gamma\}-\beta'\Omega_i\gamma\\
=\frac{1}{K(K-1)}\sum_{k\ne h}
\{Y_i-a_Y-\theta D_i-W_{ik}'\beta\}
\{D_i-a_D-W_{ih}'\gamma\}.
\label{eq:cov-cross-replicate-score}
\end{multline}
The identity holds realization by realization. Under~\Cref{def:model-repeated-measurement}, if all variables are square integrable, the right side has mean \(M(\theta,\beta,\gamma,\Omegao)\) at \(a_Y=a_D=0\) for every \((\theta,\beta,\gamma)\); and, on every realization,
\begin{equation}
\psi_i(\theta,\beta,\gamma,\Omega_i)=\psi_i(\theta,\beta,\gamma,\Omegao)-\beta'H_i\gamma .
\label{eq:cov-score-decomposition}
\end{equation}
\end{proposition}

With \(K=2\), the cross-measurement score residualizes the outcome with one measurement and the treatment with the other, and averages the two orderings:
\[
\tfrac12\{Y_i-\theta D_i-W_{i1}'\beta\}\{D_i-W_{i2}'\gamma\}+\tfrac12\{Y_i-\theta D_i-W_{i2}'\beta\}\{D_i-W_{i1}'\gamma\}.
\]
Three properties follow from~\Cref{prop:cov-cross-replicate-score}. First, the score contains no covariance estimate, so there is no covariance nuisance to which it could fail to be orthogonal. Second, because its mean is \(M(\theta,\beta,\gamma,\Omegao)\), it inherits the centering of~\Cref{lemm:centering} and the orthogonality of~\Cref{lemm:orthogonality} in \((\beta,\gamma)\). Third, by~\eqref{eq:cov-score-decomposition} its value at the true coefficients is the corrected score at \(\Omegao\) minus \(\betao'H_i\gammao\), which is exactly the first-order effect of estimating \(\Omegao\) by \(\Pn\Omega_i\): the uncertainty from covariance estimation is part of the score rather than an adjustment to it.

The representation also connects the correction to instrumental variables and proxy controls. Stacking the \(K(K-1)\) ordered pairs, with regressors \((1,D_i,W_{ih}')\), instruments \((1,D_i,W_{ik}')\) and coefficients common to all pairs, gives a just-identified system whose estimator is full-sample corrected least squares and whose standard error clustered by unit includes the covariance-estimation term (\Cref{rem:stacked-iv}). For a scalar regressor of interest, the obviously related instrumental variables (ORIV) estimator of~\citet{gillensnowbergyariv2019} instruments each replicate with the others, with replicate-specific first stages and intercepts. A single ordering is also the linear case of the two-proxy structure of proxy-control and proximal causal inference~\citep{miaogengtchetgen2018proxy,deaner2021many,deaner2023proxy,tchetgentchetgenyingcuishimiao2024proximal,cuipushimiaotchetgen2024semiparametric}, with one measurement as the treatment-inducing proxy and another as the outcome-inducing proxy: under~\Cref{def:model-repeated-measurement} with affine nuisances, two distinct measurements satisfy the zero-correlation exclusion restrictions of~\citet[Assumption~1.1]{deaner2021many} in either role, and the single-ordering moment \(\E[(Y_i-\mu-\theta D_i-W_{ih}'\beta)(1,D_i,W_{ik}')']=0\) is his instrumental-variables moment~\citep[eq.~(8)]{deaner2021many}, which goes back to~\citet{griliches1977estimating}. The proxy framework does not require the proxies to share the scale of the latent variable and, in its nonparametric form, identifies bridge functions under completeness conditions. The common scale of repeated measurements buys something different: the measurements can be averaged, the covariance of the averaged error is identified by within-unit contrasts and estimated as a positive-semidefinite matrix, and the shift of the DML moment on the averaged measurements, at the true coefficients, is the finite-dimensional term \(\betao'\Omegao\gammao\), which a recentered orthogonal score removes. The price is the same-scale error structure of~\Cref{def:model-repeated-measurement}.
\section{The CORE-DML Procedure}\label{sec:estimator}

\subsection{Cross-fitting and first stages}\label{subsec:est-crossfit}

As in cross-fitted semiparametric estimation and DML~\citep{neweyrobins2017crossfitting,chernozhukov2018double}, the nuisance coefficients used for the observations of a fold are estimated from the other folds. The folds \(I_1,\ldots,I_L\), with \(L\) fixed, partition \(\{1,\ldots,n\}\) independently of the data, with sizes that differ by at most one. We write \(I_\ell^c\) for the complement of \(I_\ell\), \(\ell(i)\) for the fold of observation \(i\), \(n_\ell:=|I_\ell|\), \(\mathbb P_{n,\ell}A_i:=n_\ell^{-1}\sum_{i\in I_\ell}A_i\) for held-out-fold averages and \(\mathbb P_{n,-\ell}A_i:=|I_\ell^c|^{-1}\sum_{i\in I_\ell^c}A_i\) for training-sample averages.

A first-stage learner is any rule that maps the training observations of fold \(\ell\) into estimates \(\widehat\pi_{-\ell}\) and \(\widehat\gamma_{-\ell}\) of \(\pio=\betao+\thetao\gammao\) and \(\gammao\), the coefficients of the sparse approximations to the regressions of \(Y_i\) and \(D_i\) on the latent controls. The rule may use the training-sample covariance estimate \(\widehat\Omega_{-\ell}:=\mathbb P_{n,-\ell}\Omega_i\) and auxiliary randomness independent of the data, such as the inner folds of cross-validation. The intercepts are not penalized and come from the training means, \(\widehat\mu_{Y,-\ell}=\bar Y_{-\ell}-\bar Z_{-\ell}'\widehat\pi_{-\ell}\) and \(\widehat\mu_{D,-\ell}=\bar D_{-\ell}-\bar Z_{-\ell}'\widehat\gamma_{-\ell}\). The baseline learner is the corrected Lasso of~\Cref{subsec:est-nuisance}.

\subsection{The estimator}\label{subsec:score-ratio}

For \(i\in I_\ell\), let \(\widetilde Y_i:=Y_i-\widehat\mu_{Y,-\ell}-Z_i'\widehat\pi_{-\ell}\) and \(\widetilde D_i:=D_i-\widehat\mu_{D,-\ell}-Z_i'\widehat\gamma_{-\ell}\) be the cross-fitted residuals, and evaluate the corrected score at the unit's own covariance contribution:
\begin{equation}
N_i:=\widetilde Y_i\widetilde D_i-\widehat\pi_{-\ell(i)}'\Omega_i\widehat\gamma_{-\ell(i)},
\qquad
Q_i:=\widetilde D_i^2-\widehat\gamma_{-\ell(i)}'\Omega_i\widehat\gamma_{-\ell(i)}.
\label{eq:est-NQ}
\end{equation}
Because \(Y_i-\theta D_i-Z_i'\beta=(Y_i-Z_i'\pi)-\theta(D_i-Z_i'\gamma)\) when \(\beta=\pi-\theta\gamma\), the cross-fitted score at \(\theta\) is \(\widehat\psi_i(\theta):=N_i-\theta Q_i\), and by~\Cref{prop:cov-cross-replicate-score}
\[
\widehat\psi_i(\theta)=\frac{1}{K(K-1)}\sum_{k\ne h}\bigl\{Y_i-\widehat\mu_{Y,-\ell(i)}-\theta(D_i-\widehat\mu_{D,-\ell(i)})-W_{ik}'\widehat\beta_{\ell(i)}(\theta)\bigr\}\bigl\{D_i-\widehat\mu_{D,-\ell(i)}-W_{ih}'\widehat\gamma_{-\ell(i)}\bigr\},
\]
with \(\widehat\beta_\ell(\theta):=\widehat\pi_{-\ell}-\theta\widehat\gamma_{-\ell}\). The root of \(\Pn\widehat\psi_i(\theta)=0\) is
\begin{equation}
\widehat\theta
=
\widehat N/\widehat J,
\qquad
\widehat N:=\Pn N_i,
\qquad
\widehat J:=\Pn Q_i .
\label{eq:est-ratio}
\end{equation}
This is the DML2 estimator of~\citet[Def.~3.2]{chernozhukov2018double} for the cross-measurement score. The numerator corrects the residual covariance of outcome and treatment, and the denominator \(\widehat J\), an estimate of the corrected Jacobian \(J_0\), corrects the residual variance of the treatment.

\subsection{Standard error and confidence interval}\label{subsec:estimator-variance}

With \(\widehat\psi_i:=\widehat\psi_i(\widehat\theta)=N_i-\widehat\theta Q_i\), the variance estimator and the confidence interval are
\begin{gather}
\widehat\sigma^2
:=
\widehat J^{-2}\Pn\widehat\psi_i^2,
\label{eq:est-variance}\\
\widehat C_{1-\alpha}:=\bigl[
\widehat\theta-z_{1-\alpha/2}\,\widehat\sigma/\sqrt n,\
\widehat\theta+z_{1-\alpha/2}\,\widehat\sigma/\sqrt n
\bigr],
\label{eq:est-ci}
\end{gather}
where \(\alpha\in(0,1)\) and \(z_{1-\alpha/2}\) is the standard normal \((1-\alpha/2)\)-quantile; \(\Pn\widehat\psi_i=0\), so~\eqref{eq:est-variance} is also the centered second moment. No separate covariance-estimation term is needed: by~\eqref{eq:cov-score-decomposition}, at the true coefficients \(\widehat\psi_i\) is the corrected score at \(\Omegao\) minus \(\betao'H_i\gammao\), so \(\Pn\widehat\psi_i^2=\widehat J^{2}\widehat\sigma^2\) estimates
\begin{equation}
J_0^2\sigma^2
=\operatorname{Var}\{\psi_i(\thetao,\betao,\gammao,\Omegao)\}+\operatorname{Var}(\betao'H_i\gammao)-2\operatorname{Cov}\{\psi_i(\thetao,\betao,\gammao,\Omegao),\betao'H_i\gammao\},
\label{eq:est-variance-decomposition}
\end{equation}
where \(\sigma^2\) is the asymptotic variance of \(\sqrt n(\widehat\theta-\thetao)\) (\Cref{theo:main}). Replacing \(\Omega_i\) by \(\Pn\Omega_i\) in \(\widehat\psi_i\), with \(\widehat J\) unchanged and the centered second moment as for the variants of~\Cref{subsec:est-variants}, gives the standard error that ignores covariance estimation. It omits the last two terms of~\eqref{eq:est-variance-decomposition}; when the fold terms \(\widehat\beta_\ell(\widehat\theta)'(\Pn\Omega_i-\Omegao)\widehat\gamma_{-\ell}\) are \(o_P(1)\), as they are under the \(\ell_1\)-rate condition~\eqref{eq:thy-covariance-rate} of~\Cref{prop:plugin-inference}, it therefore estimates the asymptotic standard deviation of the infeasible known-covariance estimator, which uses \(\Omegao\) in place of \(\Omega_i\) (\Cref{coro:known-covariance}), and it is generally inconsistent for that of \CORE{}. The last term of~\eqref{eq:est-variance-decomposition} is zero when the measurement errors are i.i.d.\ Gaussian across measurements, with common covariance \(\Sigma_\nu\), and independent of \(\mathcal F_i\), and the omission then understates the variance whenever \(\betao'\Sigma_\nu\betao\,\gammao'\Sigma_\nu\gammao>0\); with unequal source covariances, heteroskedasticity or high-kurtosis measurement errors, the net term can have either sign (\Cref{rem:est-unaugmented}).

\begin{definition}[\CORE{}]\label{def:core-generic}
Given fold-specific first-stage estimates \(\widehat\pi_{-\ell}\) and \(\widehat\gamma_{-\ell}\), \(\ell=1,\ldots,L\), with training-mean intercepts, \CORE{} is the triple \((\widehat\theta,\widehat\sigma,\widehat C_{1-\alpha})\) of~\eqref{eq:est-ratio}--\eqref{eq:est-ci}, with \(\widehat\theta:=0\) and \(\widehat\sigma:=0\) on the event \(\{\widehat J=0\}\).
\end{definition}

The name refers to the covariance recalibration of the DML score: each held-out score is recalibrated by its own repeated-measurement covariance contribution. \Cref{alg:coredml} in~\Cref{app:sim-numerics} lists the steps. The estimate depends on the random partition into folds; as~\citet[Sec.~3.4]{chernozhukov2018double} recommend, the application of~\Cref{subsec:app-caltech} reports the median over repeated partitions with the median-adjusted standard error.

\subsection{Two variants}\label{subsec:est-variants}

The covariance input of the held-out score can instead be estimated. The \emph{plug-in} variant replaces \(\Omega_i\) in~\eqref{eq:est-NQ} by the full-sample estimate \(\widehat\Omega^{\mathrm{full}}:=\Pn\Omega_i\), and the \emph{training-fold} variant by \(\widehat\Omega_{-\ell(i)}\). Under the conditions of~\Cref{prop:plugin-inference}, their influence functions coincide with that of \CORE{}, but their standard errors must add the covariance-estimation term explicitly, as in~\citet[Cor.~3]{belloni2017confidence}. For a variant with estimate \(\widehat\theta^{v}\), let \(\widehat J_v\) be the average of \(Q_i\) with the variant's covariance input in place of \(\Omega_i\). The standard error of the variant is \(\widehat\sigma_v/\sqrt n\), with \(\widehat\sigma_v^2:=\widehat J_v^{-2}\,\Pn\{\widehat\psi_i(\widehat\theta^{v})-\Pn\widehat\psi_i(\widehat\theta^{v})\}^2\), where \(\widehat\psi_i(\theta)=N_i-\theta Q_i\) are the contributions of the cross-measurement score of~\Cref{subsec:score-ratio}, which use the unit's own \(\Omega_i\). For the plug-in variant \(\widehat\theta^{\mathrm{pl}}\) these are the plug-in contributions augmented by the covariance-estimation term,
\[
\widehat\psi_i(\widehat\theta^{\mathrm{pl}})
=(\widetilde Y_i-\widehat\theta^{\mathrm{pl}}\widetilde D_i)\widetilde D_i-\widehat\beta_{\ell(i)}(\widehat\theta^{\mathrm{pl}})'\widehat\Omega^{\mathrm{full}}\widehat\gamma_{-\ell(i)}-\widehat\beta_{\ell(i)}(\widehat\theta^{\mathrm{pl}})'(\Omega_i-\widehat\Omega^{\mathrm{full}})\widehat\gamma_{-\ell(i)},
\]
and the training-fold contributions are augmented in the same way, with \(\widehat\Omega_{-\ell(i)}\) in place of \(\widehat\Omega^{\mathrm{full}}\). \CORE{} and the plug-in estimator are related exactly by
\begin{equation}
\widehat\theta-\widehat\theta^{\mathrm{pl}}
=-\widehat J^{-1}\sum_{\ell=1}^L\frac{n_\ell}{n}\,\widehat\beta_\ell(\widehat\theta^{\mathrm{pl}})'\bigl(\bar\Omega_{(\ell)}-\widehat\Omega^{\mathrm{full}}\bigr)\widehat\gamma_{-\ell},
\qquad
\bar\Omega_{(\ell)}:=\mathbb P_{n,\ell}\Omega_i,
\label{eq:est-cm-plugin-identity}
\end{equation}
so \CORE{} is the plug-in estimator with the covariance input of fold~\(\ell\) replaced by \(\bar\Omega_{(\ell)}\), the average over that fold's held-out observations, which are independent of the fold-\(\ell\) coefficients. Because \(\widehat\Omega^{\mathrm{full}}\) and \(\widehat\Omega_{-\ell}\) average training observations from which the fold-\(\ell\) coefficients are learned, the bilinear terms in~\eqref{eq:est-cm-plugin-identity} cannot be controlled by conditioning on the training sample (see equation~\eqref{eq:app-variant-identity} for the training-fold variant); \Cref{prop:plugin-inference} controls them for the variants under additional \(\ell_1\)-rate conditions on the first stages, which \CORE{} does not need. The two variants and \CORE{} are numerically close in the simulations, as are the plug-in variant and \CORE{} in both applications.
\section{Asymptotic Theory}\label{sec:theory}

The main theorem covers \CORE{} computed from any cross-fitted linear first stages that attain the prediction rates of~\Cref{assu:first-stage-rates}. \Cref{subsec:fs-levels} gives conditions under which a corrected Lasso attains them. In rate expressions, \(\log p\) stands for \(\log\max\{p,2\}\).

\subsection{Data-generating processes and first-stage rates}\label{subsec:theory-dgp}

Let \(\rho_{\pi,n}\) and \(\rho_{g,n}\) be deterministic sequences that converge to zero and bound the prediction errors of the two first stages. They enter the definition of the class of data-generating processes.

\begin{assumption}[Data-generating processes]\label{def:dgpclass}
Fix constants \(0<c<C<\infty\), the number of folds \(L\ge2\) and the number of measurements \(K\ge2\). For each \(n\), \(\Po\) is a set of distributions \(P\) of \(n\) i.i.d.\ vectors \((Y_i,D_i,X_i,U_i,V_i,W_{i1},\ldots,W_{iK})\), of which only \((Y_i,D_i,W_{i1},\ldots,W_{iK})\) are observed, with \(p=p_n\), such that every \(P\in\Po\) satisfies the following conditions.
\begin{enumerate}[label=\textup{(\arabic*)}]
\item \emph{Structural model.} The equations~\eqref{eq:model-plr-y}--\eqref{eq:model-plr-d} and the restrictions~\eqref{eq:model-moment-restrictions} hold, and
\begin{equation*}
|\thetao|+\|\betao\|_2+\|\gammao\|_2+\|\pio\|_2\le C,
\qquad
c\le J_0(P):=\E_P[V_i^2]\le C .
\end{equation*}

\item \emph{Approximate sparsity.} For deterministic sequences \(s_{f,n},s_{g,n},s_{\pi,n},a_{f,n},a_{g,n},a_{\pi,n}\), the representations~\eqref{eq:model-f-approx}--\eqref{eq:model-g-approx} hold with
\[
\|\betao\|_0\le s_{f,n},\qquad \|\gammao\|_0\le s_{g,n},\qquad \|r_f\|_{P,2}\le a_{f,n},\qquad \|r_g\|_{P,2}\le a_{g,n}.
\]
The reduced-form coefficient \(\pio=\betao+\thetao\gammao\) and approximation error \(r_\pi=r_f+\thetao r_g\) satisfy
\begin{equation*}
\|\pio\|_0\le s_{\pi,n},\qquad
\|r_\pi\|_{P,2}\le a_{\pi,n},
\qquad
\E_P[r_{\pi,i}^4]+\E_P[r_{g,i}^4]\le C,
\end{equation*}
where \(s_{\pi,n}\le s_{f,n}+s_{g,n}\), \(a_{\pi,n}\le a_{f,n}+Ca_{g,n}\) and \(a_{f,n},a_{g,n},a_{\pi,n}\to0\). Any \((\betao,\gammao)\) with these properties may be used, in particular the best sparse coefficients of~\eqref{eq:model-best-s-beta}--\eqref{eq:model-best-s-gamma}.

\item \emph{Measurements.} \Cref{def:model-repeated-measurement} holds with \(\|\Sigma_{\nu,k}\|_{\mathrm{op}}\le C\) for every \(k\), and the proxy uses equal weights, \(Z_i=K^{-1}\sum_{k=1}^K W_{ik}\), so that \(\Omegao=K^{-2}\sum_k\Sigma_{\nu,k}\) and \(H_i=\Omega_i-\Omegao\), with \(\Omega_i\) as in~\eqref{eq:cov-bessel-row}.

\item \emph{Tails.} The latent controls and the measurement errors satisfy the sub-Gaussian bounds
\begin{equation}
\sup_{\|u\|_2=1}\|u'X_i\|_{\psi_2}
+
\max_{k\le K}\sup_{\|u\|_2=1}\|u'\nu_{ik}\|_{\psi_2}
\le C ,
\label{eq:thy-vector-subgaussian}
\end{equation}
the disturbances satisfy \(\|U_i\|_{\psi_2}+\|V_i\|_{\psi_2}\le C\), and the observed variables satisfy
\begin{equation*}
\max_{j\le p}\|Z_{ij}Y_i\|_{\psi_1}
+
\max_{j\le p}\|Z_{ij}D_i\|_{\psi_1}
+
\max_{j,m\le p}\|Z_{ij}Z_{im}-\E_P[Z_{ij}Z_{im}]\|_{\psi_1}
\le C .
\end{equation*}

\item \emph{Design.} For a fixed constant \(c_X>0\),
\begin{equation}
c_X\le\lambda_{\min}(\Sigma_X)\le\lambda_{\max}(\Sigma_X)\le C,
\qquad
\|\Omegao\|_{\mathrm{op}}\le C.
\label{eq:thy-design-eigenvalues}
\end{equation}

\item \emph{Influence function.} Let \(\Delta_{fg,n}(P):=\E_P[r_{f,i}r_{g,i}]\) and
\begin{equation}
\varphi_i(P)
:=
J_0(P)^{-1}
\left\{
\psii(\thetao,\betao,\gammao,\Omegao)
-\E_P[\psii(\thetao,\betao,\gammao,\Omegao)]
-\betao'H_i\gammao
\right\},
\label{eq:thy-dgp-if}
\end{equation}
where, as in all population displays, \(\psii\) is evaluated at the centered variables \(Y_i-\E_PY_i\) and \(D_i-\E_PD_i\), and \(\E_P[\varphi_i(P)]=0\) because \(\E_P[H_i]=0\). For constants \(c_\sigma\), \(C_\sigma\) and \(\varpi>0\) that do not depend on \(n\) or \(P\),
\begin{equation*}
0<c_\sigma\le
\sigma^2(P):=\E_P[\varphi_i(P)^2]\le C_\sigma<\infty,
\qquad
\E_P\left|\frac{\varphi_i(P)}{\sigma(P)}\right|^{2+\varpi}\le C .
\end{equation*}

\item \emph{Rates.} With \(\rho_{\beta,n}:=\rho_{\pi,n}+|\thetao|\rho_{g,n}\), we require \(a_{\pi,n}\le C\rho_{\pi,n}\), \(a_{g,n}\le C\rho_{g,n}\) and
\begin{equation}
\sqrt n\Bigl\{
|\Delta_{fg,n}|
+a_{g,n}\rho_{\beta,n}
+a_{f,n}\rho_{g,n}
+\rho_{\beta,n}\rho_{g,n}
\Bigr\}
\longrightarrow0,
\label{eq:thy-product-rate}
\end{equation}
where, because \(\thetao\) and \(\Delta_{fg,n}\) depend on \(P\), the left-hand side is bounded by a deterministic sequence that converges to zero and, like \(c\) and \(C\), is part of the specification of \(\Po\).
\end{enumerate}
\end{assumption}

The four terms in~\eqref{eq:thy-product-rate} bound, in turn, the centering error from approximate sparsity, the two interactions between approximation and estimation errors, and the product of the two first-stage errors. Without measurement error, \eqref{eq:thy-product-rate} is the familiar product-rate condition of DML; with measurement error it is unchanged, because the cross-measurement score involves the measurement errors only through terms that average out conditionally on the training sample.

\begin{assumption}[First-stage prediction rates]\label{assu:first-stage-rates}
Uniformly over folds and \(P\in\Po\),
\begin{align}
\|\widehat\mu_{Y,-\ell}-\mu_{Y0}+X_i'(\widehat\pi_{-\ell}-\pio)\|_{P,2}
&=O_P(\rho_{\pi,n}),
\label{eq:thy-highlevel-pi}\\
\|\widehat\mu_{D,-\ell}-\mu_{D0}+X_i'(\widehat\gamma_{-\ell}-\gammao)\|_{P,2}
&=O_P(\rho_{g,n}),
\label{eq:thy-highlevel-gamma}
\end{align}
where \(\widehat\mu_{Y,-\ell}\) and \(\widehat\mu_{D,-\ell}\) are the training-mean intercepts of~\Cref{subsec:est-crossfit} and \(\mu_{Y0}\) and \(\mu_{D0}\) the population intercepts after the normalization of~\Cref{subsec:model-omega}. The fitted coefficients and intercepts of fold \(\ell\) are measurable functions of the training observations \(\{i\notin I_\ell\}\) and of auxiliary randomness independent of the data.
\end{assumption}

\subsection{Uniform inference}\label{subsec:theory-inference}\label{subsec:theory-covariance}

\begin{theorem}[Uniform inference for \CORE{}]\label{theo:main}
Suppose that~\Cref{def:dgpclass} holds, and let \CORE{} be computed as in~\Cref{def:core-generic} from cross-fitted first stages that satisfy~\Cref{assu:first-stage-rates}. Then the following hold uniformly over \(P\in\Po\).
\begin{enumerate}[label=\textup{(\roman*)}]
\item \emph{Asymptotic linearity.} With \(\varphi_i(P)\) the influence function~\eqref{eq:thy-dgp-if},
\begin{equation}
\sqrt n(\widehat\theta-\thetao)
=
\frac{1}{\sqrt n}\sum_{i=1}^n\varphi_i(P)+r_n(P),
\qquad
r_n(P)=o_P(1).
\label{eq:thy-uniform-expansion}
\end{equation}
\item \emph{Variance consistency.} With \(\sigma^2(P)=\E_P[\varphi_i(P)^2]\),
\(
|\widehat\sigma^2-\sigma^2(P)|
=o_P(1).
\)
\item \emph{Coverage.} For every fixed \(\alpha\in(0,1)\), the confidence interval~\eqref{eq:est-ci} satisfies
\begin{equation}
\lim_{n\to\infty}
\sup_{P\in\Po}
\left|
\Pr_P\{\thetao\in\widehat C_{1-\alpha}\}-(1-\alpha)
\right|=0.
\label{eq:thy-ci-validity}
\end{equation}
\end{enumerate}
\end{theorem}

The proof (\Cref{app:main-proof}) follows the argument of~\citet{chernozhukov2018double} for the cross-measurement score, and the measurement structure enters at one point. Conditionally on the training sample of fold \(\ell\), the fitted coefficients are fixed and each held-out \(\Omega_i\) is a function of unit \(i\) alone, so conditional mean zero and uncorrelatedness across measurements make the held-out score exactly centered at the population moment \(M\) of~\eqref{eq:score-population-map} evaluated at the fitted slopes and the true covariance, plus the product of the two intercept errors (\Cref{lemm:cm-centering}). Orthogonality then leaves only the second-order terms controlled by~\eqref{eq:thy-product-rate}, and~\eqref{eq:cov-score-decomposition} places \(-\betao'H_i\gammao\) in the influence function. The first stages may use the training-fold covariance estimate in any way. Only their affine prediction rates~\eqref{eq:thy-highlevel-pi}--\eqref{eq:thy-highlevel-gamma}, which include the errors of the training-mean intercepts, matter; the proof uses no \(\ell_1\) rate, sparsity of the coefficient errors or restriction on \(\log p\). When \(p/n\to\infty\), the slope prediction rates alone need not bound the intercept errors, which contain \(\bar Z_{-\ell}'(\widehat\gamma_{-\ell}-\gammao)\) and its analog for \(\widehat\pi_{-\ell}\); for the corrected Lasso, \Cref{lem:affine-intercept} bounds this term by \(\|\bar Z_{-\ell}\|_\infty\|\widehat\gamma_{-\ell}-\gammao\|_1\), which requires a weaker \(\ell_1\) rate than condition~\eqref{eq:thy-covariance-rate} imposes for the variants of~\Cref{prop:plugin-inference}.

Comparing~\eqref{eq:thy-dgp-if} with the influence function \(\varphi_i^{\Omegao}(P)=J_0^{-1}\{\psi_i(\thetao,\betao,\gammao,\Omegao)-\E_P\psi_i\}\) of the infeasible known-covariance estimator (\Cref{coro:known-covariance}) isolates the effect of estimating the covariance: the asymptotic variances differ by \(J_0^{-2}\{\operatorname{Var}(\betao'H_i\gammao)-2\operatorname{Cov}(\psi_i,\betao'H_i\gammao)\}\), the last two terms of~\eqref{eq:est-variance-decomposition}. With \(K\) fixed, this difference can persist as \(n\) grows and can have either sign (\Cref{rem:est-unaugmented}); it is positive in the canonical simulation design (\Cref{subsec:sim-canonical}).

The two variants of~\Cref{subsec:est-variants} have the same influence function under stronger first-stage conditions.

\begin{proposition}[Plug-in and training-fold variants]\label{prop:plugin-inference}
Suppose that the conditions of~\Cref{theo:main} hold, that \(\log p=o(n)\), and that, uniformly over folds and \(P\in\Po\), \(\|\widehat\pi_{-\ell}-\pio\|_1=O_P(\zeta_{\pi,n})\) and \(\|\widehat\gamma_{-\ell}-\gammao\|_1=O_P(\zeta_{g,n})\) for deterministic sequences with
\begin{equation}
\sqrt{\log p}\,\{\zeta_{\beta,n}+\zeta_{g,n}\}
\longrightarrow0 ,
\qquad
\zeta_{\beta,n}:=\zeta_{\pi,n}+C\zeta_{g,n},
\label{eq:thy-covariance-rate}
\end{equation}
where \(C\) is the constant of~\Cref{def:dgpclass}, which bounds \(|\thetao|\) and makes \(\zeta_{\beta,n}\) deterministic, so that \(\|\widehat\beta_\ell(\thetao)-\betao\|_1=O_P(\zeta_{\beta,n})\). Then the plug-in and training-fold estimators differ from \(\widehat\theta\) by \(o_P(n^{-1/2})\), their standard errors with the covariance-estimation term agree with \(\widehat\sigma/\sqrt n\) to first order, and the conclusions of~\Cref{theo:main} hold for them.
\end{proposition}

Condition~\eqref{eq:thy-covariance-rate} bounds the interaction of the covariance estimate with the first-stage errors in~\eqref{eq:est-cm-plugin-identity}. It can fail while~\eqref{eq:thy-product-rate} holds, for example for first stages with dense coefficient errors, whose \(\ell_1\) errors are large relative to their prediction errors. With a known covariance, the covariance-estimation term disappears: with \(\Omegao\) in place of \(\Omega_i\) in~\eqref{eq:est-NQ}, first stages that may use \(\Omegao\) or the training-fold estimates, and condition~(6) of~\Cref{def:dgpclass} imposed on \(\varphi_i^{\Omegao}\) in place of \(\varphi_i\), the conclusions of~\Cref{theo:main} hold with influence function \(\varphi_i^{\Omegao}\) (\Cref{coro:known-covariance}). Additive measurement-specific intercepts, as in~\Cref{rem:source-intercepts}, are handled by centering each unit's within-unit deviations at their held-out-fold means; the influence function is then~\eqref{eq:thy-dgp-if} for the measurements net of their intercepts (\Cref{coro:source-intercepts}).

\subsection{A corrected-Lasso first stage}\label{subsec:est-nuisance}\label{subsec:theory-nuisance}\label{subsec:fs-levels}

The baseline first stage is the Convex Conditioned Lasso (CoCoLasso) of~\citet{datta2017cocolasso}, an \(\ell_1\)-penalized corrected least-squares regression that is the sample analog of the identity of~\Cref{subsec:model-omega}. For fold \(\ell\) and \(R_i\in\{Y_i,D_i\}\), let
\begin{equation*}
\widehat\Sigma_{Z,-\ell}
:=\mathbb P_{n,-\ell}\{(Z_i-\bar Z_{-\ell})(Z_i-\bar Z_{-\ell})'\}-\widehat\Omega_{-\ell},
\qquad
\widehat c_{RZ,-\ell}
:=\mathbb P_{n,-\ell}\{(Z_i-\bar Z_{-\ell})(R_i-\bar R_{-\ell})\}.
\end{equation*}
By~\eqref{eq:cov-distinct-products}, up to the centering terms, \(\widehat\Sigma_{Z,-\ell}\) is the average of the cross-products \(W_{ik}W_{ih}'\) of distinct measurements, so CoCoLasso is a Lasso on the cross-measurement Gram matrix, a regularized counterpart of the stacked instrumental-variables system of~\Cref{subsec:score-cross}. The corrected Gram matrix need not be positive semidefinite. Following~\citet[Sec.~2 and App.~A]{datta2017cocolasso}, we replace it by a nearest matrix, in the maximum norm, among the matrices with eigenvalues of at least a fixed floor \(\lambda_{\mathrm{fl}}\in(0,c_X)\):
\begin{equation}
\widehat\Sigma^{+}_{Z,-\ell}
\in
\argmin_{\Sigma\succeq\lambda_{\mathrm{fl}} I}
\norm{\Sigma-\widehat\Sigma_{Z,-\ell}}_{\max},
\label{eq:est-psd-projection}
\end{equation}
or by a near-minimizer whose distance is certified to be at most \(\kappa\ge1\) times the minimum plus a tolerance \(\bar\eta\ge0\) (\Cref{lemm:psd-projection-cost,rem:projection-certificate}). The fitted slopes are
\begin{equation}
\widehat b_{R,-\ell}(\lambda_R)
\in
\argmin_{b\in\R^{p}}
\left\{
\frac{1}{2}b'\widehat\Sigma^{+}_{Z,-\ell}b
-
\widehat c_{RZ,-\ell}'b
+\lambda_R\norm{b}_1
\right\},
\label{eq:est-cocolasso-objective}
\end{equation}
with \(\widehat\pi_{-\ell}:=\widehat b_{Y,-\ell}(\lambda_Y)\) and \(\widehat\gamma_{-\ell}:=\widehat b_{D,-\ell}(\lambda_D)\) used without post-selection refitting, because the score is orthogonal and consistent model selection is not needed. Problem~\eqref{eq:est-cocolasso-objective} is the additive-error CoCoLasso of~\citet[Sec.~4.1]{datta2017cocolasso}. Their error bounds take the error covariance as known, and they note that with replicated measurements an estimate of it can be subtracted instead (p.~2408), as do~\citet[Sec.~3.2.1]{loh2012high} for their corrected Lasso; here the subtracted matrix is estimated from the training-fold measurements and the proxy is their average. The penalties are chosen by a version of the calibrated cross-validation of~\citet[Sec.~5]{datta2017cocolasso}, with projected validation Gram matrices and four inner folds (\Cref{app:sim-numerics}).

\begin{definition}[Feasible \CORE{}]\label{def:core-dml}
The feasible \CORE{} estimator is the estimator of~\Cref{def:core-generic} computed from the CoCoLasso first stages~\eqref{eq:est-cocolasso-objective}, with each penalty selected by calibrated cross-validation over the shrinking grid of~\Cref{rem:thy-cv-not-covered}.
\end{definition}

\Cref{tab:first-stage-routes} summarizes sufficient conditions under which CoCoLasso first stages satisfy~\Cref{assu:first-stage-rates} and~\eqref{eq:thy-product-rate}. They rest on the score and restricted-eigenvalue conditions of~\Cref{prop:corrected-lasso-specialization}. A certified projection moves the corrected Gram matrix, in the maximum norm, by at most \(\kappa\) times the estimation error of that matrix plus \(\bar\eta\), and that error is of order \(\sqrt{\log p/n}\) (\Cref{prop:sparse-bilinear}); with \(\bar\eta\le\sqrt{\log\bar p/n}\), the gradient of the projected objective at a target \(b\) therefore differs from that of the unprojected one by at most \(\|b\|_1\) times a term of order \(\sqrt{\log\bar p/n}\) (\Cref{lemm:psd-projection-cost}). When the targets are bounded only in \(\ell_2\) norm, \(\|b\|_1\le\sqrt s\|b\|_2\), so our sufficient penalty carries a factor \(\sqrt s\) and the budget becomes \(s_g(s_\pi+s_g)\log\bar p=o(\sqrt n)\), that is, \(s^2\log\bar p=o(\sqrt n)\) when \(s_\pi\) and \(s_g\) are of order \(s\); \citet[Thm.~1]{datta2017cocolasso} and~\citet[Lem.~1]{lilima2021measurement} obtain conditions of the same type. The familiar budget \(s\log\bar p=o(\sqrt n)\) is recovered under bounded target \(\ell_1\) norms (\Cref{coro:cocolasso-core}(b)) or when \(p\) is a small fraction of \(n\) (\Cref{coro:cocolasso-core}(c)); in the second case the corrected Gram matrix satisfies the floor with probability approaching one, and the budget holds for projections that then leave it unchanged or are certified with \(\bar\eta\le\sqrt{\log\bar p/(np)}\). Whether the factor \(\sqrt s\) is necessary for the exact projection is not settled here; for some certified near-minimizers with \(\kappa>1\) the projection increment at the target is of the exact order \(d\|b\|_1\), where \(d\) is the max-norm distance of the corrected Gram matrix to the floored set (\Cref{rem:active-projection-tight}), so the factor \(\sqrt s\) is attained whenever \(d\) is of order \(\sqrt{\log\bar p/n}\) and \(\|b\|_1\) of order \(\sqrt s\).

\begin{table}[!htbp]\centering\tablesize
\caption{First-stage rate conditions for \CORE{}: the generic requirement and sufficient conditions for CoCoLasso first stages under exact sparsity.}\label{tab:first-stage-routes}
\setlength{\tabcolsep}{3pt}
\begin{tabular}{@{}>{\raggedright\arraybackslash}p{4.1cm}c>{\centering\arraybackslash}p{2.3cm}>{\raggedright\arraybackslash}p{3.8cm}>{\raggedright\arraybackslash}p{2.5cm}@{}}
\toprule
First stage & Penalty & Prediction rate & Sufficient budget & Result \\
\midrule
Any linear learner & --- & \(\rho_{\pi,n},\rho_{g,n}\) & \eqref{eq:thy-product-rate} & \Cref{theo:main} \\
Lasso, error-free controls &\(\sqrt{\log\bar p/n}\) & \(\sqrt{s\log\bar p/n}\) & \(s\log\bar p=o(\sqrt n)\) & BCH \\
\addlinespace
\multicolumn{5}{@{}l@{}}{\emph{CoCoLasso, deterministic penalties}}\\
\(\|b\|_2\le C\) & \(\sqrt{s\log\bar p/n}\) & \(s\sqrt{\log\bar p/n}\) & \(s_g(s_\pi+s_g)\log\bar p=o(\sqrt n)\) & \Cref{coro:cocolasso-core}(a) \\
\(\|b\|_1\le C_1\) & \(\sqrt{\log\bar p/n}\) & \(\sqrt{s\log\bar p/n}\) & \(s\log\bar p=o(\sqrt n)\) & \Cref{coro:cocolasso-core}(b) \\
\(\|b\|_2\le C\), \(p\le c_\ast n\) & \(\sqrt{\log n/n}\) & \(\sqrt{s\log n/n}\) & \(s\log n=o(\sqrt n)\) & \Cref{coro:cocolasso-core}(c) \\
\addlinespace
\multicolumn{5}{@{}l@{}}{\emph{CoCoLasso, penalties selected by any training-sample rule from the shrinking grid}}\\
\(\|b\|_1\le C_1\), \(\|\operatorname{Cov}_P(Z_i,R_i)\|_\infty\ge c_{ZR}\) & CV grid & \(\upsilon_n\sqrt{s\log\bar p/n}\) & \(s\upsilon_n^2\log\bar p=o(\sqrt n)\) & \Cref{coro:cocolasso-core}(d) \\
\bottomrule
\end{tabular}
\par\smallskip\parbox{\linewidth}{\footnotesize Rates and budgets for targets \(b\in\{\pio,\gammao\}\) with \(s_\pi\) and \(s_g\) nonzero coefficients, where \(\bar p=\max\{p,n\}\). In the columns Penalty and Prediction rate, \(s\) is \(s_\pi\) or \(s_g\), according to the target; in the column Sufficient budget, \(s=s_\pi+s_g\). The constant \(c_\ast\) is that of~\Cref{lemm:inactive-projection}, \(\upsilon_n=\max\{1,\log\log n\}\) and the CV (cross-validation) grid is the shrinking grid of~\Cref{rem:thy-cv-not-covered}. The row \(p\le c_\ast n\) also assumes projections with \(\bar\eta\le\sqrt{\log\bar p/(np)}\) or projections that return the corrected Gram matrix unchanged when it satisfies the floor. The results for the plug-in and training-fold variants also use~\eqref{eq:thy-covariance-rate}, which adds \((s_\pi^{3/2}+s_g^{3/2})\log\bar p=o(\sqrt n)\) in the row of~\Cref{coro:cocolasso-core}(a). BCH:~\citet{belloni2014inference}.}
\end{table}

\begin{corollary}[CoCoLasso first stages for \CORE{}]\label{coro:cocolasso-core}
Let conditions~\textup{(1)}--\textup{(6)} of~\Cref{def:dgpclass} hold with \(c_X>\lambda_{\mathrm{fl}}>0\) and sparsity sequences \(s_{\pi,n},s_{g,n}\ge1\), and let the first stages be the corrected Lasso~\eqref{eq:est-cocolasso-objective} with projections certified as in~\eqref{eq:thy-certified-projection} with a fixed \(\kappa\ge1\) and \(\bar\eta_n\le\sqrt{\log\bar p_n/n}\). In each case below, if~\eqref{eq:thy-product-rate} holds for the stated rates with a deterministic bound, the conclusions of~\Cref{theo:main} hold for \CORE{} uniformly over the stated class.
\begin{enumerate}[label=\textup{(\alph*)}]
\item Penalties \(\lambda_Y=C_\lambda\sqrt{s_{\pi,n}\log\bar p_n/n}\) and \(\lambda_D=C_\lambda\sqrt{s_{g,n}\log\bar p_n/n}\), with \(C_\lambda\) the constant of~\Cref{lemm:penalty-level}, and rates \(\rho_{\pi,n}=s_{\pi,n}\sqrt{\log\bar p_n/n}+a_{\pi,n}\) and \(\rho_{g,n}=s_{g,n}\sqrt{\log\bar p_n/n}+a_{g,n}\). Under exact sparsity it suffices that \(s_{g,n}(s_{\pi,n}+s_{g,n})\log\bar p_n=o(\sqrt n)\).
\item On the subclass with \(\max\{\|\pio\|_1,\|\gammao\|_1\}\le C_1\), penalties \(\lambda_Y=\lambda_D=C_\lambda^{(1)}\sqrt{\log\bar p_n/n}\), with \(C_\lambda^{(1)}\) the constant of~\Cref{lemm:penalty-level-l1}, and rates \(\rho=\sqrt{s\log\bar p_n/n}+a\), evaluated at \((s,a)=(s_{\pi,n},a_{\pi,n})\) and \((s_{g,n},a_{g,n})\). Under exact sparsity it suffices that \((s_{\pi,n}+s_{g,n})\log\bar p_n=o(\sqrt n)\).
\item If \(p_n\le c_\ast n\), penalties \(\lambda_Y=\lambda_D=C_\lambda^{(2)}\sqrt{\log\bar p_n/n}\) and the rates of~(b), with no bound on \(\|\pio\|_1\) or \(\|\gammao\|_1\), for projections that, in addition, satisfy \(\bar\eta_n\le\sqrt{\log\bar p_n/(np_n)}\) or return \(\widehat\Sigma_{Z,-\ell}\) unchanged whenever \(\widehat\Sigma_{Z,-\ell}\succeq\lambda_{\mathrm{fl}}I\), as the implementation of Appendix~\ref{app:sim-numerics} does up to rounding; \(c_\ast\), \(C_\lambda^{(2)}\) and this condition are those of~\Cref{lemm:inactive-projection}. Under exact sparsity it suffices that \((s_{\pi,n}+s_{g,n})\log n=o(\sqrt n)\).
\item On the subclass of~(b) with \(\|\operatorname{Cov}_P(Z_i,Y_i)\|_\infty\ge c_{ZR}\) and \(\|\operatorname{Cov}_P(Z_i,D_i)\|_\infty\ge c_{ZR}\) for a fixed \(c_{ZR}>0\), penalties selected from the shrinking grid of~\Cref{rem:thy-cv-not-covered} by any rule that depends on the training observations and an auxiliary randomization only, and rates \(\rho=\upsilon_n\sqrt{s\log\bar p_n/n}+a\), with \(\upsilon_n=\max\{1,\log\log n\}\). Under exact sparsity it suffices that \(\upsilon_n^2(s_{\pi,n}+s_{g,n})\log\bar p_n=o(\sqrt n)\).
\end{enumerate}
\end{corollary}

The proof (Appendix~\ref{app:cocolasso-core-proof}) combines the slope bounds of~\Cref{prop:corrected-lasso-specialization} with the training-mean intercept bound of~\Cref{lem:affine-intercept}; in every case the intercept error is of the order of the prediction rate, so no condition on the \(\ell_1\) rates beyond those implied by the corrected-Lasso analysis is needed. \Cref{exam:fs-nonempty} gives Gaussian designs that satisfy every condition of each of the cases~(a)--(d), so none of the four classes is empty. Case~(b) covers linear nuisances with polynomially decaying coefficients, \(|b_j|\asymp j^{-\varsigma}\), for every \(\varsigma>1\) when \(\log\bar p=O(\log n)\) and \(n^{1/(4\varsigma-2)}=o(p)\), with deterministic penalties and suitable sparsity levels (\Cref{coro:l1-decay}). Under exact sparsity, the restricted-eigenvalue constants can be taken free of the projection floor \(\lambda_{\mathrm{fl}}\) (\Cref{lemm:re-floor-free}).

\begin{remark}[Cross-validated penalties]\label{rem:thy-cv-not-covered}
The feasible estimator selects each penalty by calibrated cross-validation over thirty log-spaced values from \(\lambda_{\mathrm{sh}}:=\lambda_{\mathrm{top}}\min\{1,100\,\upsilon_mt_m\}\) down to \(\lambda_{\mathrm{sh}}/100\), where \(\lambda_{\mathrm{top}}=\|\widehat c_{RZ,-\ell}\|_\infty\), \(t_m=\sqrt{\log\max(p,m)/m}\), \(\upsilon_m=\max\{1,\log\log m\}\) and \(m\) is the training-sample size. Case~(d) of~\Cref{coro:cocolasso-core} covers this rule, as it covers any selection from this grid based on the training sample: its proof never uses the cross-validation criterion. The factor \(\min\{1,100\,\upsilon_mt_m\}\) equals one, whatever \(p\), unless the training sample has more than about \(9.46\times10^5\) observations, so at every sample size in~\Cref{sec:simulations,sec:application} the grid is the fixed-ratio grid from \(\lambda_{\mathrm{top}}\) down to \(\lambda_{\mathrm{top}}/100\), and the shrinking on which case~(d) rests does not occur. The factor matters for the asymptotic theory: over the fixed-ratio grid, the first stages are inconsistent when \(\|\operatorname{Cov}_P(Z_i,R_i)\|_\infty\) is bounded away from zero (\Cref{prop:grid-floor}). The penalties selected by \CORE{} in the simulations are interior to the grid, and in the approximately sparse and growing-support designs their medians are 0.74--3.89 times \(\sqrt{\log\bar p/n}\) (\Cref{app:sim-partc}); cross-validation is the rule that~\citet[Sec.~5]{datta2017cocolasso} propose for CoCoLasso, and~\citet{chetverikov2021cross} analyze it for the ordinary Lasso.
\end{remark}
\section{Monte Carlo Evidence}\label{sec:simulations}

\subsection{Designs and procedures}\label{subsec:sim-design}

All designs have many controls but fewer controls than observations, with \(p/n\) between \(0.1\) and \(0.25\), so that regularization is needed for precision but corrected least squares remains well defined. They differ in the support of the nuisance coefficients and in the measurement errors.

The \emph{canonical sequence} takes its latent design, \(\betao\) and outcome noise from the additive-error design of~\citet[Sec.~6.1]{datta2017cocolasso}. Its observations are i.i.d.\ with \(X_i\sim N(0,\Sigma_X)\), \((\Sigma_X)_{jm}=0.5^{|j-m|}\), \(D_i=X_i'\gammao+V_i\) and \(Y_i=\thetao D_i+X_i'\betao+U_i\), where \(\thetao=1\), and \(V_i\sim N(0,1)\) and \(U_i\sim N(0,9)\) are independent of each other and of \(X_i\). The coefficient \(\betao\) has entries \((3,1.5,2)\) on coordinates \(\{1,2,5\}\) and \(\gammao\) has entries \((0.8,1,0.6)\) on \(\{3,4,6\}\), so the two supports are disjoint but adjacent. Each observation has \(K=2\) measurements whose errors \(\nu_{ik}\sim N(0,\Sigma_\nu)\), \(\Sigma_\nu=\tfrac14\Sigma_{0.5}\) with \(\Sigma_{0.5}\) the AR\((0.5)\) correlation matrix, are independent of each other and of \((X_i,U_i,V_i)\); a single measurement has reliability \(\operatorname{Var}(X_{ij})/\operatorname{Var}(W_{ik,j})=0.8\). The sequence is \((n,p)\in\{(500,100),(1000,200),(2000,500),(4000,1000)\}\), with 1{,}000 replications per design. The shift~\eqref{eq:intro-naive-shift} is \(\betao'\Omegao\gammao=0.508\), and because the supports are disjoint it comes entirely from the correlation of the errors across coordinates: \(\betao'\diag(\Omegao)\gammao=0\). Sixteen further designs keep the canonical distributions and change the coefficients: twelve have polynomially decaying coefficients outside the canonical supports, with exponents \(\varsigma\in\{1.75,2,2.5\}\), and four are exactly sparse on supports \(\{1,\ldots,\lceil3\log p\rceil\}\) that grow with \(p\) (\Cref{app:sim-partc}). Each has 500 replications.

The \emph{heterogeneous-error designs} ask whether the conclusions depend on the knife-edge structure of the canonical errors. They have \((n,p)=(1000,100)\) and 1{,}000 replications, the canonical latent design and disturbances, and overlapping supports: \(\betao\) has entries \((3,1.5,2)\) and \(\gammao\) entries \((0.8,1,0.6)\), both on \(\{1,3,5\}\), so the shift has nonzero diagonal and off-diagonal parts. Design~1 keeps Gaussian errors with equal source covariances. Design~2 doubles the error covariance of the second measurement, so the single-measurement reliabilities are \(0.8\) and \(2/3\). Design~3 adds conditional heteroskedasticity, with errors scaled by a factor that increases with \(|X_{i1}|\), and draws standardized \(t_5\) errors. Design~4 has \(K=3\) measurements with covariances in the ratios \(1:1.5:2.5\) and the errors of design~3. Design~5 adds the nonlinear term \(0.3\,\Phi(X_{i1})X_{i2}\) to \(f_0\), with \(g_0\) linear, so the projection target equals \(\thetao\) (\Cref{rem:projection-target}). Designs~6 and~7 are designs~1 and~3 at \((n,p)=(2000,200)\), with 500 replications. \Cref{app:sim-hetero} gives the exact construction. 
In designs~3, 4 and~7, the \(t_5\) errors violate the sub-Gaussian condition~(4) of~\Cref{def:dgpclass}. The common sample normalization also makes observations dependent. Design~5 lies outside condition~(2) because the error of the best linear approximation to \(f_0\) does not vanish with \(n\). These four designs examine the procedure beyond the conditions of~\Cref{sec:theory}.

\phantomsection\label{subsec:sim-estimators}\CORE{} denotes the feasible estimator of~\Cref{def:core-dml} with five outer folds; the same estimate with the standard error that ignores covariance estimation, which replaces each \(\Omega_i\) by \(\Pn\Omega_i\) in the score contributions (\Cref{subsec:estimator-variance}), and the plug-in variant (\Cref{subsec:est-variants}) use the same fits. The infeasible \emph{known-covariance estimator} uses \(\Omegao\) in place of every covariance estimate, in the first stages, the scores and the variance (\Cref{coro:known-covariance}), with the folds, penalty grid and tuning rule of \CORE{}; it is computed in two canonical designs and in designs~1--4. The \emph{uncorrected score} combines the CoCoLasso fits of \CORE{} with the uncorrected DML score, so comparing it with \CORE{} isolates the score correction. \emph{Least squares on the averaged measurements} applies the uncorrected score to full-sample least-squares fits on the proxy. \emph{DML on the averaged measurements} (\NAIVEDML{}) sets the covariance to zero in the first stages and the score, and the \emph{diagonal correction} keeps only the diagonal of the covariance estimate in both; the two share the folds, penalty grid and tuning rule of \CORE{}. \OracleZ{} evaluates the corrected score at the true nuisance coefficients and covariance in the full sample. We report the bias, the standard deviation (SD), the root mean squared error (RMSE), the ratio SE/SD of the mean standard error to the SD, the coverage of nominal \(95\%\) intervals and their mean length. The Monte Carlo standard error of a coverage rate near \(0.95\) is about \(0.007\) with 1{,}000 replications and \(0.010\) with 500. Every CoCoLasso fit selects its penalties by cross-validation, and none of the 180{,}000 penalties selected by \CORE{} in the canonical, approximately sparse, growing-support and heterogeneous-error designs lies at an end of the grid; no replication failed the numerical checks of~\Cref{app:sim-numerics}.

\subsection{The canonical sequence}\label{subsec:sim-canonical}

\begin{table}[t]\centering\tablesize
\caption{Canonical sequence: \(K=2\), single-measurement reliability \(0.8\), AR\((0.5)\) measurement errors, disjoint supports.}\label{tab:sim-canonical}
{\setlength{\tabcolsep}{3pt}\begin{tabular}{llrrrrrr}
\toprule
$(n,p)$ & Procedure & Bias & SD & RMSE & SE/SD & Coverage & Length \\
\midrule
$(500,100)$ & Least squares on the averages & 0.322 & 0.142 & 0.352 & 0.89 & 0.301 & 0.499 \\
 & Uncorrected score & 0.320 & 0.129 & 0.345 & 1.00 & 0.308 & 0.503 \\
 & \CORE{} & 0.036 & 0.183 & 0.187 & 0.96 & 0.928 & 0.693 \\
 & \quad SE ignoring covariance estimation &  &  &  & 0.94 & 0.921 & 0.673 \\
 & \OracleZ{} & \ensuremath{-}0.009 & 0.180 & 0.180 & 1.01 & 0.954 & 0.709 \\
\addlinespace
$(1000,200)$ & Least squares on the averages & 0.330 & 0.101 & 0.345 & 0.90 & 0.059 & 0.356 \\
 & Uncorrected score & 0.329 & 0.089 & 0.341 & 1.02 & 0.042 & 0.357 \\
 & \CORE{} & 0.028 & 0.129 & 0.132 & 0.99 & 0.937 & 0.500 \\
 & \quad SE ignoring covariance estimation &  &  &  & 0.96 & 0.930 & 0.485 \\
 & \OracleZ{} & \ensuremath{-}0.002 & 0.129 & 0.129 & 1.00 & 0.953 & 0.506 \\
\addlinespace
$(2000,500)$ & Least squares on the averages & 0.331 & 0.074 & 0.340 & 0.87 & 0.002 & 0.252 \\
 & Uncorrected score & 0.333 & 0.064 & 0.339 & 1.01 & 0.000 & 0.252 \\
 & \CORE{} & 0.019 & 0.093 & 0.095 & 0.98 & 0.937 & 0.357 \\
 & \quad SE ignoring covariance estimation &  &  &  & 0.95 & 0.931 & 0.346 \\
 & \OracleZ{} & \ensuremath{-}0.001 & 0.090 & 0.090 & 1.01 & 0.958 & 0.358 \\
\addlinespace
$(4000,1000)$ & Least squares on the averages & 0.334 & 0.052 & 0.338 & 0.87 & 0.000 & 0.178 \\
 & Uncorrected score & 0.339 & 0.044 & 0.342 & 1.03 & 0.000 & 0.178 \\
 & \CORE{} & 0.015 & 0.066 & 0.067 & 0.99 & 0.936 & 0.255 \\
 & \quad SE ignoring covariance estimation &  &  &  & 0.96 & 0.929 & 0.246 \\
 & \OracleZ{} & 0.002 & 0.064 & 0.064 & 1.01 & 0.958 & 0.253 \\
\bottomrule
\end{tabular}}
\par\smallskip\parbox{0.97\linewidth}{\footnotesize Procedures and measures as in~\Cref{subsec:sim-estimators}; bias, SD, RMSE and length (mean interval length) in units of \(\theta\) (\(\thetao=1\)); 1{,}000 replications per design, the same draws for every procedure.}
\end{table}

\begin{table}[t]\centering\tablesize
\caption{Estimated versus known covariance on the same draws and folds: (1)~\CORE{}; (2)~the same estimates with the standard error that ignores covariance estimation; (3)~the infeasible known-covariance estimator.}\label{tab:sim-known-cov}
{\setlength{\tabcolsep}{3pt}\begin{tabular}{lrrrcrrrrrr}
\toprule
 & \multicolumn{4}{c}{Standard deviation} & \multicolumn{3}{c}{Coverage} & \multicolumn{3}{c}{Length} \\
\cmidrule(lr){2-5}\cmidrule(lr){6-8}\cmidrule(lr){9-11}
Design & (1) & (3) & (1)/(3) & 95\% interval & (1) & (2) & (3) & (1) & (2) & (3) \\
\midrule
Canonical $(500,100)$ & 0.183 & 0.177 & 1.036 & [1.018,\,1.056] & 0.928 & 0.921 & 0.932 & 0.693 & 0.673 & 0.673 \\
Canonical $(1000,200)$ & 0.129 & 0.126 & 1.021 & [1.002,\,1.039] & 0.937 & 0.930 & 0.938 & 0.500 & 0.485 & 0.485 \\
\addlinespace
Design 1 & 0.134 & 0.129 & 1.033 & [1.015,\,1.052] & 0.934 & 0.922 & 0.935 & 0.497 & 0.478 & 0.477 \\
Design 2 & 0.155 & 0.147 & 1.054 & [1.029,\,1.081] & 0.920 & 0.907 & 0.916 & 0.563 & 0.534 & 0.533 \\
Design 3 & 0.159 & 0.162 & 0.981 & [0.950,\,1.011] & 0.918 & 0.917 & 0.915 & 0.591 & 0.584 & 0.583 \\
Design 4 & 0.141 & 0.142 & 0.998 & [0.970,\,1.023] & 0.918 & 0.920 & 0.921 & 0.515 & 0.517 & 0.516 \\
\bottomrule
\end{tabular}}
\par\smallskip\parbox{0.97\linewidth}{\footnotesize (1)/(3): ratio of the standard deviations, with a paired bootstrap 95\% interval over replications (2{,}000 resamples). Coverage and mean length of nominal \(95\%\) intervals. Designs as in~\Cref{tab:sim-canonical,tab:sim-hetero}; 1{,}000 replications each.}
\end{table}

\begin{figure}[t]
\centering
\includegraphics[width=\linewidth]{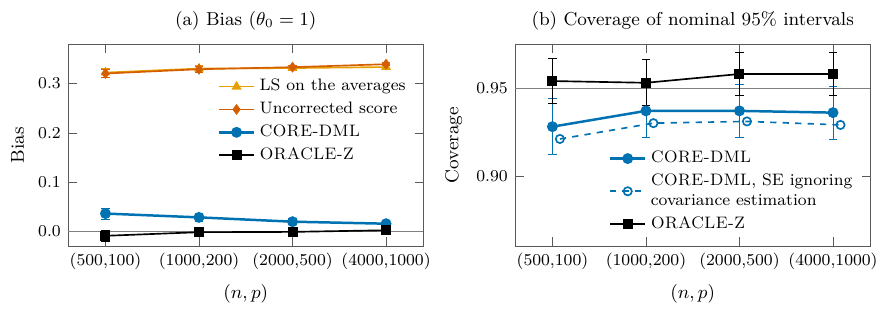}
\caption{Canonical sequence of~\Cref{tab:sim-canonical}: (a)~bias and (b)~coverage of nominal \(95\%\) intervals, with \(\pm1.96\) Monte Carlo standard-error bars (1{,}000 replications). The dashed line in~(b) uses the \CORE{} estimates with the standard error that ignores covariance estimation; its difference from \CORE{} is measured on the same replications (paired Monte Carlo standard error at most 0.003) and is drawn without bars. The two uncorrected procedures lie below the axis of~(b), with coverage 0.301 (least squares) and 0.308 (uncorrected score) at \((500,100)\) and 0.000 at \((4000,1000)\). \OracleZ{} is infeasible.}
\label{fig:sim-canonical}
\end{figure}

The canonical sequence asks whether the corrected score removes the shift~\eqref{eq:intro-naive-shift} when the nuisance functions are learned from the measurements, and how close the intervals come to nominal coverage as \(n\) and \(p\) grow together. The two procedures with the uncorrected score are biased by about a third of \(\thetao\) in every design (\Cref{tab:sim-canonical,fig:sim-canonical}), and because their standard deviation shrinks, their coverage falls from about \(0.30\) to zero. The uncorrected score uses the same fits as \CORE{}, so correcting the first stages alone leaves the population shift in place; correcting the score removes it.

\CORE{} has bias 0.015--0.036, SE/SD 0.96--0.99 and coverage 0.928--0.937; the infeasible \OracleZ{} covers at 0.953--0.958. The remaining bias is the second-order term of the first stages: in every design it is within two Monte Carlo standard errors of the mean bias \(b_{\mathrm{Rem}}\) implied by the product of the first-stage errors, the term controlled by the rate condition~\eqref{eq:thy-product-rate} and isolated exactly by~\Cref{lemm:cm-centering}, and infeasible intervals recentered by \(b_{\mathrm{Rem}}\) cover at 0.946--0.955 (\Cref{tab:app-remainder}). Because \(p\) grows with \(n\), this term shrinks slowly along the sequence: \(\sqrt n\) times the bias is 0.80--0.97.

The covariance-estimation term multiplies the standard error that ignores covariance estimation by 1.029--1.033, against \(f_H=1.039\) at the true parameters of this design (\Cref{rem:est-unaugmented}), and raises coverage by 0.6--0.7 percentage points. The two intervals share the point estimate, and the \CORE{} standard error is the larger in every replication from \((1000,200)\) on; the \CORE{} interval covers in 27 replications of the four designs in which the other does not and never conversely, and the paired Monte Carlo standard error of the difference is at most 0.003. The effect of estimating the covariance on precision is measured instead by the known-covariance estimator on the same draws and folds (\Cref{tab:sim-known-cov}): at \((1000,200)\) and \((500,100)\) the standard deviation of \CORE{} is 1.021 and 1.036 times that of this estimator (paired bootstrap 95\% intervals 1.002--1.039 and 1.018--1.056), and the known-covariance estimator, with its own standard error, covers about as often as \CORE{} with shorter intervals. To first order, the standard error that ignores covariance estimation is the standard deviation of the known-covariance estimator; intervals around the \CORE{} estimates with this standard error are therefore about as short as those of that estimator and cover less often. The plug-in variant differs from \CORE{} by at most 0.022 in any replication, 0.13 Monte Carlo standard deviations, and covers at 0.928--0.940 (\Cref{tab:app-variants}). Regularization pays even with \(p<n\): classical corrected least squares on all measurements, the pooled cross-measurement instrumental-variables estimator (\Cref{rem:stacked-iv}), has a standard deviation 1.61--1.92 times and an RMSE 2.09--5.46 times those of \CORE{} (\Cref{app:sim-canonical-extra}).

\subsection{Approximately sparse and growing-support designs}\label{subsec:sim-partc}

\begin{table}[t]\centering\tablesize
\caption{Approximately sparse and growing-support designs.}\label{tab:sim-partc}
{\setlength{\tabcolsep}{5pt}\begin{tabular}{lrrrrr}
\toprule
 & \multicolumn{4}{c}{\CORE{}} & Uncorrected \\
\cmidrule(lr){2-5}\cmidrule(lr){6-6}
$(n,p)$ & Bias/SD & $b_{\mathrm{Rem}}$/SD & SE/SD & Coverage & Coverage \\
\midrule
\multicolumn{6}{l}{\textit{Canonical exact sparsity, $s_\pi=6$, $s_g=3$}}\\
$(500,100)$ & 0.20 & 0.26 & 0.96 & 0.928 & 0.308 \\
$(1000,200)$ & 0.22 & 0.24 & 0.99 & 0.937 & 0.042 \\
$(2000,500)$ & 0.21 & 0.22 & 0.98 & 0.937 & 0.000 \\
$(4000,1000)$ & 0.23 & 0.20 & 0.99 & 0.936 & 0.000 \\
\addlinespace
\multicolumn{6}{l}{\textit{Polynomial decay, $\varsigma=1.75$}}\\
$(500,100)$ & 0.04 & 0.03 & 0.97 & 0.942 & 0.460 \\
$(1000,200)$ & \ensuremath{-}0.04 & 0.00 & 0.99 & 0.940 & 0.166 \\
$(2000,500)$ & \ensuremath{-}0.04 & \ensuremath{-}0.03 & 0.95 & 0.940 & 0.010 \\
$(4000,1000)$ & \ensuremath{-}0.08 & \ensuremath{-}0.05 & 0.96 & 0.938 & 0.000 \\
\addlinespace
\multicolumn{6}{l}{\textit{Polynomial decay, $\varsigma=2$}}\\
$(500,100)$ & 0.01 & 0.02 & 0.92 & 0.926 & 0.446 \\
$(1000,200)$ & \ensuremath{-}0.06 & \ensuremath{-}0.01 & 0.93 & 0.930 & 0.190 \\
$(2000,500)$ & \ensuremath{-}0.01 & \ensuremath{-}0.04 & 0.97 & 0.944 & 0.020 \\
$(4000,1000)$ & \ensuremath{-}0.03 & \ensuremath{-}0.05 & 0.96 & 0.926 & 0.000 \\
\addlinespace
\multicolumn{6}{l}{\textit{Polynomial decay, $\varsigma=2.5$}}\\
$(500,100)$ & 0.03 & 0.02 & 0.94 & 0.928 & 0.454 \\
$(1000,200)$ & \ensuremath{-}0.03 & \ensuremath{-}0.03 & 0.97 & 0.940 & 0.154 \\
$(2000,500)$ & \ensuremath{-}0.02 & \ensuremath{-}0.03 & 1.00 & 0.946 & 0.012 \\
$(4000,1000)$ & \ensuremath{-}0.01 & \ensuremath{-}0.04 & 0.92 & 0.924 & 0.000 \\
\addlinespace
\multicolumn{6}{l}{\textit{Growing exact sparsity, $s_\pi=s_g=\lceil 3\log p\rceil$, fixed-size coefficients}}\\
$(500,100)$ & 0.50 & 0.51 & 0.91 & 0.880 & 0.208 \\
$(1000,200)$ & 0.90 & 0.91 & 1.02 & 0.856 & 0.008 \\
$(2000,500)$ & 1.37 & 1.42 & 0.94 & 0.682 & 0.000 \\
$(4000,1000)$ & 1.82 & 1.80 & 1.03 & 0.554 & 0.000 \\
\bottomrule
\end{tabular}}
\par\smallskip\parbox{0.97\linewidth}{\footnotesize Bias/SD: bias over the Monte Carlo standard deviation; \(b_{\mathrm{Rem}}\)/SD: mean bias implied by the first-stage remainder of~\Cref{lemm:cm-centering}, over the same SD; Uncorrected: the uncorrected score on the fits of \CORE{}. Canonical rows: 1{,}000 replications; other designs: 500 replications each.}
\end{table}

\begin{figure}[t]
\centering
\includegraphics[width=\linewidth]{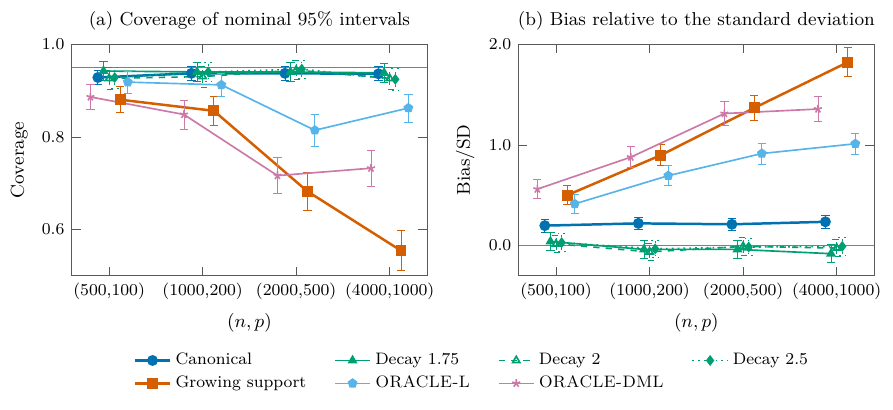}
\caption{\CORE{} in the designs of~\Cref{tab:sim-partc}: (a)~coverage of nominal \(95\%\) intervals and (b)~bias/SD, with \(\pm1.96\) Monte Carlo standard-error bars. The growing-support designs add, on the same draws, two infeasible references that share first-stage slopes learned by the Lasso on the latent controls: \mbox{ORACLE-L} (the plug-in variant of \CORE{} with these slopes) and \mbox{ORACLE-DML} (DML on the latent controls).}
\label{fig:sim-sparsity}
\end{figure}

In the decay designs, whose exponents lie above the threshold \(3/2\) of~\Cref{rem:polynomial-decay}, \CORE{} covers as in the canonical sequence, with less bias (\Cref{tab:sim-partc,fig:sim-sparsity}): its coverage is 0.924--0.946, its absolute bias at most 0.083 of its standard deviation and SE/SD 0.92--1.00. The growing-support designs add \(\lceil3\log p\rceil-3\) coefficients of fixed size and alternating sign to each equation and violate condition~(1) of~\Cref{def:dgpclass} as \(p\to\infty\); their first-stage errors barely fall: \(\sqrt n\) times the product of the root mean squared first-stage prediction errors rises from 9.6 to 19.0 along the sequence, whereas it falls from 4.1 to 2.2 in the canonical sequence (\Cref{app:sim-partc}). Coverage falls from 0.880 to 0.554 and bias/SD rises from 0.50 to 1.82. The bias, 0.093--0.124, is again the first-stage remainder, 0.095--0.128 on average, and infeasible intervals recentered by it cover at 0.924--0.958 (\Cref{tab:app-remainder}). The shortfall is not specific to measurement error, but measurement error enlarges it: the references of~\Cref{fig:sim-sparsity}, whose first stages are learned from the latent controls, are biased by 0.060--0.090 and cover at 0.716--0.918, and the bias of the plug-in variant of \CORE{} is 1.19--1.82 times that of \mbox{ORACLE-L}, which differs from it only in its first stages (\Cref{tab:app-refset-grow}).

\subsection{Heterogeneous measurement errors}\label{subsec:sim-hetero}

\begin{table}[tp]\centering\tablesize
\caption{Heterogeneous-error designs: overlapping supports, \((n,p)=(1000,100)\) unless stated otherwise.}\label{tab:sim-hetero}
{\setlength{\tabcolsep}{3.2pt}\begin{tabular*}{\linewidth}{@{\extracolsep{\fill}}llrrrrr@{}}
\toprule
 & Procedure & Bias & SD & SE/SD & Coverage & Length \\
\midrule
\multicolumn{7}{@{}l}{\rlap{\textit{Design 1: Overlapping supports, equal sources, Gaussian errors; $\betao'\Omegao\gammao=0.886$, diagonal part $0.638$}}}\\
 & \NAIVEDML{} & 0.614 & 0.093 & 0.97 & 0.000 & 0.353 \\
 & Diagonal correction & 0.228 & 0.117 & 0.94 & 0.440 & 0.433 \\
 & \CORE{} & 0.028 & 0.134 & 0.95 & 0.934 & 0.497 \\
 & \quad SE ignoring covariance estimation &  &  & 0.91 & 0.922 & 0.478 \\
 & \OracleZ{} & \ensuremath{-}0.002 & 0.130 & 0.98 & 0.942 & 0.498 \\
\addlinespace
\multicolumn{7}{@{}l}{\rlap{\textit{Design 2: Unequal source covariances; $\betao'\Omegao\gammao=1.329$, diagonal part $0.956$}}}\\
 & \NAIVEDML{} & 0.792 & 0.091 & 0.96 & 0.000 & 0.345 \\
 & Diagonal correction & 0.278 & 0.136 & 0.88 & 0.371 & 0.472 \\
 & \CORE{} & 0.037 & 0.155 & 0.93 & 0.920 & 0.563 \\
 & \quad SE ignoring covariance estimation &  &  & 0.88 & 0.907 & 0.534 \\
 & \OracleZ{} & \ensuremath{-}0.003 & 0.148 & 0.98 & 0.953 & 0.569 \\
\addlinespace
\multicolumn{7}{@{}l}{\rlap{\textit{Design 3: Unequal sources, heteroskedastic $t_5$ errors; $\betao'\Omegao\gammao=1.329$, diagonal part $0.956$}}}\\
 & \NAIVEDML{} & 0.791 & 0.093 & 0.99 & 0.000 & 0.360 \\
 & Diagonal correction & 0.283 & 0.144 & 0.86 & 0.373 & 0.487 \\
 & \CORE{} & 0.043 & 0.159 & 0.95 & 0.918 & 0.591 \\
 & \quad SE ignoring covariance estimation &  &  & 0.94 & 0.917 & 0.584 \\
 & \OracleZ{} & \ensuremath{-}0.004 & 0.165 & 0.98 & 0.949 & 0.634 \\
\addlinespace
\multicolumn{7}{@{}l}{\rlap{\textit{Design 4: Three sources, heteroskedastic $t_5$ errors; $\betao'\Omegao\gammao=0.984$, diagonal part $0.708$}}}\\
 & \NAIVEDML{} & 0.663 & 0.094 & 0.98 & 0.000 & 0.362 \\
 & Diagonal correction & 0.251 & 0.125 & 0.92 & 0.400 & 0.448 \\
 & \CORE{} & 0.037 & 0.141 & 0.93 & 0.918 & 0.515 \\
 & \quad SE ignoring covariance estimation &  &  & 0.93 & 0.920 & 0.517 \\
 & \OracleZ{} & 0.001 & 0.143 & 0.98 & 0.947 & 0.548 \\
\addlinespace
\multicolumn{7}{@{}l}{\rlap{\textit{Design 5: Nonlinear outcome equation; $\betao'\Omegao\gammao=0.886$, diagonal part $0.638$}}}\\
 & \NAIVEDML{} & 0.632 & 0.092 & 0.99 & 0.000 & 0.355 \\
 & Diagonal correction & 0.243 & 0.114 & 0.97 & 0.409 & 0.435 \\
 & \CORE{} & 0.034 & 0.133 & 0.96 & 0.935 & 0.501 \\
 & \quad SE ignoring covariance estimation &  &  & 0.92 & 0.927 & 0.482 \\
 & \OracleZ{} & 0.002 & 0.126 & 1.02 & 0.956 & 0.501 \\
\addlinespace
\multicolumn{7}{@{}l}{\rlap{\textit{Design 6: Design 1 at $(n,p)=(2000,200)$; $\betao'\Omegao\gammao=0.886$, diagonal part $0.638$}}}\\
 & \NAIVEDML{} & 0.607 & 0.066 & 0.96 & 0.000 & 0.251 \\
 & Diagonal correction & 0.215 & 0.085 & 0.93 & 0.274 & 0.309 \\
 & \CORE{} & 0.017 & 0.094 & 0.96 & 0.932 & 0.355 \\
 & \quad SE ignoring covariance estimation &  &  & 0.93 & 0.916 & 0.341 \\
 & \OracleZ{} & \ensuremath{-}0.005 & 0.094 & 0.96 & 0.948 & 0.353 \\
\addlinespace
\multicolumn{7}{@{}l}{\rlap{\textit{Design 7: Design 3 at $(n,p)=(2000,200)$; $\betao'\Omegao\gammao=1.329$, diagonal part $0.956$}}}\\
 & \NAIVEDML{} & 0.789 & 0.062 & 1.04 & 0.000 & 0.254 \\
 & Diagonal correction & 0.247 & 0.101 & 0.90 & 0.246 & 0.356 \\
 & \CORE{} & 0.027 & 0.113 & 0.96 & 0.930 & 0.426 \\
 & \quad SE ignoring covariance estimation &  &  & 0.94 & 0.932 & 0.419 \\
 & \OracleZ{} & 0.002 & 0.106 & 1.07 & 0.964 & 0.445 \\
\bottomrule
\end{tabular*}}
\par\smallskip\parbox{0.97\linewidth}{\footnotesize Designs, procedures and measures as in~\Cref{subsec:sim-design}, except that RMSE is not reported; 1{,}000 replications in designs~1--5 and 500 in designs~6--7; \(\thetao=1\). In design~5, \OracleZ{} and the value of \(\betao'\Omegao\gammao\) in the header use the coefficients of the linear part of \(f_0\); at the projection coefficient \(\betao+0.15e_2\) of \(f_0\), where \(e_2\) is the second unit vector, the shift is 0.904, with the same diagonal part.}
\end{table}

When the outcome and treatment equations load on the same controls, part of the shift is diagonal, so a diagonal correction removes part of it, and its comparison with \CORE{} isolates the value of the off-diagonal covariances (\Cref{tab:sim-hetero}). The diagonal correction imposes a diagonal error covariance, which these designs do not have, and it is not the procedure of~\citet{belloni2017confidence}. In designs~1--4 the diagonal correction is biased by 0.228--0.283, of the order of the off-diagonal part of the shift (0.25--0.37), and covers at 0.371--0.440; \NAIVEDML{} never covers. \CORE{} covers at 0.918--0.934, with SE/SD 0.93--0.95, against 0.942--0.953 for \OracleZ{}. Unequal source covariances, heteroskedasticity and heavy-tailed errors leave the correction centered, as~\Cref{lemm:centering,prop:cov-cross-replicate-score} predict: \OracleZ{}, which evaluates the corrected score at the true coefficients and covariance, has bias at most 0.006 in absolute value in every design, and because each \(\Omega_i\) remains unbiased for \(\Omegao\), the remaining bias of \CORE{} comes from the estimated first stages (\Cref{lemm:cm-centering}, whose centering argument uses only second moments). In the Gaussian designs intervals with the standard error that ignores covariance estimation cover less often than those of \CORE{}, and in designs~1 and~2 the known-covariance estimator is more precise: the standard deviation of \CORE{} is 1.033 and 1.054 times that of this estimator (\Cref{tab:sim-known-cov}). With heteroskedastic \(t_5\) errors the covariance-estimation term contributes little and its net effect on the standard error is close to zero or slightly negative, which~\Cref{rem:est-unaugmented} shows can happen; in designs~3 and~4 the ratio of the standard deviations is 0.981 and 0.998, and both 95\% intervals contain one. In all seven designs the denominator correction is substantial, \(\Pn\widehat\gamma'\Omega_i\widehat\gamma\) being 0.29--0.45 of \(\widehat J\) at the median, yet the denominator is never below 0.78.

The conclusions do not change in designs~5--7. With the nonlinear outcome equation of design~5, \CORE{} covers at 0.935. Designs~6 and~7 double \(n\) and \(p\): the bias of \CORE{} falls from 0.028 and 0.043 in designs~1 and~3 to 0.017 and 0.027, with coverage 0.932 and 0.930, whereas the coverage of the diagonal correction falls from 0.440 and 0.373 to 0.274 and 0.246: its bias, of the order of the off-diagonal part of the shift, remains of the same size while its standard deviation falls.

\subsection{First-stage learners and other comparisons}\label{subsec:sim-firststage}

Correcting the first stages matters with the corrected score as well. Along the canonical sequence, the corrected score on ordinary Lasso fits to the proxy covers less often than \CORE{}, by 0.083 at \((500,100)\) and 0.384 at \((4000,1000)\) (first 250 replications, on which \CORE{} covers at 0.912), whereas the corrected Lasso of~\citet{loh2012high}, which at a given penalty differs from CoCoLasso only when its \(\ell_1\) constraint binds or the projection of the training Gram matrix is active, gives the same estimates in at least 97.2\% of the replications and coverage 0.912--0.937; \mbox{ORACLE-L} with the score of \CORE{} covers more often than \CORE{} by 0.010--0.020 (\Cref{fig:app-firststage-canonical}). The single orderings of the cross-measurement moment, which residualize the outcome with one measurement and the treatment with the other, cover at 0.932--0.947 on the fits of \CORE{}, with standard deviations 1.15--1.23 times and intervals 1.18--1.19 times those of the symmetrized \CORE{} (\Cref{app:sim-comparison-suite}). \Cref{subsec:app-mturk} compares the procedure of~\citet{belloni2017confidence} with \CORE{} in the designs where the diagonal-covariance assumption of that procedure holds.
\section{Empirical Applications}\label{sec:application}

The two applications are complementary. The first uses measurements that were actually repeated, but only three controls are measured with error. The second has hundreds of controls, but the repeated measurements are generated by adding errors to observed controls.

\subsection{Tournament entry and repeated lottery valuations}\label{subsec:app-caltech}

How much of the gender gap in tournament entry remains after accounting for performance, beliefs about rank and attitudes toward risk and ambiguity, when the attitudes are measured with error? We use the Caltech Cohort Study of~\citet{gillensnowbergyariv2019} (GSY), in which the same students valued lotteries twice in fall 2014 and chose how to be paid for a task in spring 2015~\citep{gillensnowbergyariv2019data}.

\paragraph{Design and estimand.} In the spring task, students first had three minutes to solve as many sums of five two-digit numbers as they could under a winner-take-all tournament (40 tokens per correct sum for the best score in a random group of four), and then guessed their rank in that group. Before a second three-minute round they chose between a piece rate of 10 tokens per correct sum and a tournament paying 40 tokens per correct sum if their second-round score exceeded the first-round scores of three other students. The outcome \(Y_i\) is the choice of the tournament and \(D_i\) the archive's male indicator. The recorded controls, which we treat as measured without error, are the number of correct first-round sums and indicators for a stated rank of 2, 3 and 4; second-round performance follows the choice and is excluded. The controls measured with error are the student's valuations of three lotteries: a risky urn with equal numbers of red and black balls, a compound urn whose number of red balls was drawn uniformly, and an ambiguous urn whose composition was chosen by the dean and not disclosed. Each was valued twice, with a 20-ball urn paying 100 tokens and a 30-ball urn paying 150 tokens, in multiple price lists with 10-token steps; the three lottery blocks were spread across the survey, and the two stakes of a block were presented one after the other in random order. We code each switching value at the midpoint of its 10-token interval and, as GSY do, divide the 30-ball value by 1.5 to put the two stakes on a common scale under constant relative risk aversion (their footnote~22): \(W_{i1,j}=S_{i,20,j}-5\) and \(W_{i2,j}=(S_{i,30,j}-5)/1.5\) for lottery \(j\), where \(S_{i,20,j}\) and \(S_{i,30,j}\) are the recorded switching values. The estimand is the coefficient on \(D_i\) in the partially linear model with these latent fall valuations as controls. Gender is not assigned, so \(\thetao\) is an adjusted difference in entry probabilities, not a causal effect. Of the 783 students in the archive, 774 have complete fall valuations; 189 of 465 men (40.65\%) and 66 of 309 women (21.36\%) chose the tournament.

\begin{table}[t]\centering\tablesize
\caption{Male--female gap in choosing tournament pay (percentage points).}\label{tab:gsymain}
\begin{tabular*}{\linewidth}{@{\extracolsep{\fill}}lrrr@{}}
\toprule
Specification & Gap & SE & 95\% interval \\
\midrule
Unadjusted difference in entry rates & 19.29 & 3.26 & [12.89, 25.68] \\
First-round score and stated rank only & 13.24 & 3.04 & [7.27, 19.20] \\
\addlinespace
Averaged valuations, \NAIVEDML{} & 13.29 & 3.00 & [7.40, 19.17] \\
Diagonal correction & 13.47 & 3.00 & [7.60, 19.34] \\
\CORE{} & \textbf{13.41} & \textbf{3.00} & \textbf{[7.54, 19.28]} \\
Plug-in variant & 13.41 & 2.99 & [7.54, 19.28] \\
Corrected least squares & 13.25 & 3.02 & [7.33, 19.16] \\
\bottomrule
\end{tabular*}
\par\smallskip\parbox{\linewidth}{\footnotesize All rows use the same 774 students; rows 2--7 include first-round score and stated rank, and rows 3--7 also include the three lottery valuations. Rows 1--2 and 7 are full-sample least-squares regressions with heteroskedasticity-robust standard errors (row~7 clusters the stacked cross-measurement system by student, \Cref{rem:stacked-iv}). Rows 3--6 are medians over 100 random partitions into five folds, with the median-adjusted standard error of~\citet[Sec.~3.4]{chernozhukov2018double}. Intervals use normal critical values.}
\end{table}

\paragraph{Results.} \Cref{tab:gsymain} reports the estimates. The raw gap of 19.29 points falls to 13.24 points once first-round performance and stated rank enter, and adding the averaged valuations without a correction (\NAIVEDML{}, \Cref{subsec:sim-estimators}) leaves it at 13.29 points. \CORE{} estimates a gap of 13.41 points (SE 3.00); the plug-in variant and corrected least squares give 13.41 and 13.25. The correction changes the estimate by 0.12 points at the median, between \ensuremath{-}0.35 and 0.69 points across partitions. Almost all of this change comes from the corrected first stages, whose cross-validated penalties and fitted coefficients differ from those of \NAIVEDML{}: with the same first stages the score correction moves the estimate by 0.005 points at the median (between \ensuremath{-}0.028 and 0.027), and in least squares adding the averaged valuations moves the gap from 13.24 to 13.23 points and correcting their measurement error moves it to 13.25. The correction is small for a reason that the data reveal directly: given the recorded controls, the valuations barely predict gender or entry, with partial correlations of at most 0.024 in absolute value (unconditional correlations at most 0.026 with gender and 0.027 with entry), so the fitted slopes of gender on the valuations are near zero and the fold-averaged shift \(\sum_\ell(n_\ell/n)\widehat\beta_\ell(\widehat\theta^{\mathrm{pl}})'\widehat\Omega^{\mathrm{full}}\widehat\gamma_{-\ell}\) is \ensuremath{-}0.000013 in score units, against a corrected Jacobian of 0.241 (medians over the 100 partitions). This agrees with GSY's finding that the risky-urn valuations, unlike their other risk measures, are uncorrelated with gender \citep[Sec.~IV.F]{gillensnowbergyariv2019}; here the compound- and ambiguous-urn valuations are too. The estimate varies more with the random partition into folds, from 12.51 to 14.46 across the 100 partitions, than with the correction, which is why the table reports medians.

\begin{table}[t]\centering\tablesize
\caption{What the paired urn valuations reveal about measurement error.}\label{tab:gsymeas}
\begin{tabular*}{\linewidth}{@{\extracolsep{\fill}}lrrrrr@{}}
\toprule
Lottery & 20-ball mean & 30-ball mean & Difference (SE) & Repeat correlation & Reliability of average \\
\midrule
Risky & 54.01 & 53.65 & \ensuremath{-}0.35 (0.37) & 0.748 & 0.855 \\
Ambiguous & 51.52 & 51.27 & \ensuremath{-}0.25 (0.36) & 0.789 & 0.881 \\
Compound & 51.07 & 50.85 & \ensuremath{-}0.22 (0.43) & 0.747 & 0.855 \\
\bottomrule
\end{tabular*}
\par\smallskip\parbox{\linewidth}{\footnotesize Values in tokens, coded at the midpoints of the switching intervals, with the 30-ball values divided by 1.5. Difference: 30-ball minus 20-ball mean; repeat correlation: correlation between the 20-ball and 30-ball values. Reliability is \(1-\widehat\Omega_{jj}/\widehat{\operatorname{Var}}(Z_{ij})\) under the maintained measurement model, with \(\widehat\Omega\) the sample covariance matrix of the stake contrasts \((W_{i2}-W_{i1})/2\). Sample as in~\Cref{tab:gsymain}.}
\end{table}

\paragraph{The measurements.} \Cref{tab:gsymeas} shows that the two stakes measure the valuations with substantial but not overwhelming noise: the repeat correlations are 0.747--0.789 and the reliability of the average is 0.855--0.881. The errors are correlated across lotteries, with correlations of the stake differences of 0.147 (risky and ambiguous), 0.242 (risky and compound) and 0.172 (ambiguous and compound), so the estimated covariance is not diagonal; a diagonal correction gives 13.47 points. Part of the error is rounding to the price-list grid, which is a function of the latent valuation, so conditional mean zero holds only approximately; because the two grids differ, the rounding errors differ across stakes, which~\Cref{def:model-repeated-measurement} allows.

\paragraph{When is the correction credible?} Identification requires the two stakes to measure the same valuation on a common scale, errors with conditional mean zero given the latent controls, gender and entry, and errors uncorrelated across stakes. A student-specific response to stake size, a departure from the rescaling that constant relative risk aversion justifies, would violate the first condition. Two observable implications are not rejected: the mean valuations are equal across stakes (\(\chi^2(3)=1.27\), \(p=0.736\)), and the centered stake differences are uncorrelated with gender and with entry, in a student-level robust test of six restrictions (\(\chi^2(6)=0.67\), \(p=0.995\)). A reporting error shared by the two stakes cancels in their difference and escapes both tests, and it is plausible here, because the two stakes of a lottery were valued one after the other; GSY recommend placing duplicate items in different parts of a survey for this reason. A shared error makes the correction too small. Multiplying the covariance estimate by 1.22 or 1.67 in the first stages, the score and the variance corresponds to a shared error component with covariance \(\bar\chi(\Sigma_{\nu,1}+\Sigma_{\nu,2})/2\), \(\bar\chi=0.10\) or \(0.25\), and hence to cross-stake error correlations of at least 0.10 and 0.25 (\Cref{app:cal-sensitivity}); the gap becomes 13.46 and 13.65 points. At the point estimates, values of \(\bar\chi\) above 0.40, and hence, when the two stakes have equal error covariances, cross-stake correlations above 0.40, would make the implied covariance matrix of the latent controls indefinite; the bound is itself estimated, and a bootstrap over students gives a 95\% interval of 0.26 to 0.49 for it.

\begin{table}[t]\centering\tablesize
\caption{Alternative repeated controls and their observable restrictions.}\label{tab:gsyrobu}
{\setlength{\tabcolsep}{3pt}\begin{tabular*}{\linewidth}{@{\extracolsep{\fill}}lrrrrr@{}}
\toprule
 & & & \CORE{} & \multicolumn{2}{c@{}}{$p$-value of the test} \\
\cmidrule(l){5-6}
Repeated controls & Students & \NAIVEDML{} & [95\% interval] & Uncorrelated & Equal means \\
\midrule
Fall urn valuations & 774 & 13.29 & 13.41 [7.54, 19.28] & 0.995 & 0.736 \\
Spring confidence ratings & 783 & 12.19 & 12.11 [6.07, 18.16] & 0.138 & $<0.001$ \\
Fall--spring risk measures & 775 & 9.18 & 6.42 [\ensuremath{-}0.38, 13.22] & $<0.001$ & $<0.001$ \\
\bottomrule
\end{tabular*}}
\par\smallskip\parbox{\linewidth}{\footnotesize Gaps in percentage points, medians over 100 partitions; the first row is the main design of~\Cref{tab:gsymain}; every specification includes first-round score and stated rank, and each uses its own complete cases. For the two alternative designs, whose measurement means differ, \CORE{} uses the centered contributions of~\Cref{rem:source-intercepts}. Uncorrelated: test that the centered within-student contrasts are uncorrelated with gender and entry (six moments for two measurements of three controls, four for three measurements of one control). Equal means: test of equal measurement means.}
\end{table}

\paragraph{Alternative repeated controls.} \Cref{tab:gsyrobu} reports two other designs. In the spring survey, students guessed the number of jelly beans in each of three pictured jars and rated their confidence in each guess on a six-point scale. We treat these ratings as three measurements of one latent control, which assumes that the ratings of the three jars share its scale (\Cref{rem:source-intercepts}). The correlation restriction is not rejected (\(p=0.138\)), but the three ratings have different means (\(p<0.001\)); centering the within-student contrasts, which removes differences in means but not in scales, leaves the gap at 12.11 points, against 12.19 without correction. A common scale and errors uncorrelated across ratings make the covariances of the three pairs of ratings equal (\Cref{rem:source-intercepts,rem:cov-weights-identification}); equality is not rejected (\(\chi^2(2)=0.28\), \(p=0.871\); \Cref{app:cal-tests}), and the implied scales of the second and third ratings relative to the first are 1.02 and 1.00, each with a standard error of 0.05. The second design pairs the same risk items across the fall and spring waves: the second risky-project allocation, the average of the two fair-urn valuations as shares of their prizes and self-rated willingness to take risks. It produces the largest change under correction, from 9.18 to 6.42 points, but its restrictions are rejected (\(\chi^2(6)=31.75\), \(p=1.8\times10^{-5}\)). Preferences that change between waves or differential reporting would produce this pattern, and GSY's corrected correlation of only 0.42 between the fall and spring urn valuations \citep[p.~1855]{gillensnowbergyariv2019} points the same way; the smaller estimate therefore has little support as a measurement-error adjustment.

GSY's own instrumental-variables specification, which instruments spring measures of overconfidence and risk attitudes with their duplicates and conditions on second-round performance, estimates a male coefficient of 0.63 points (SE 5.4; their Table~3, column~8). It conditions on different student characteristics and corrects the error in different measures, so its estimate is not comparable with those in~\Cref{tab:gsymain}. In our specification, first-round performance and stated rank reduce the raw gap by about a third (from 19.29 to 13.24 points); adding the fall valuations changes it little and the correction for their measurement error is small.

\subsection{Simulated measurement error in an online labor market}\label{subsec:app-mturk}\label{subsec:app-design}\label{subsec:app-results}

The second exercise asks how well \CORE{} recovers a known estimate when many controls are mismeasured. We take the outcome, treatment and controls from the public replication data of~\citet{dubejacobsnaidusuri2020,dubejacobsnaidusuri2020data}, who study the labor supply facing requesters on Amazon Mechanical Turk, add simulated measurement error to most controls and compare each procedure with the estimate computed from the original controls. The outcome \(Y_i\) is the log duration of a task and the treatment \(D_i\) its log posted reward. Because tasks posted by the same requester are dependent, we keep one randomly drawn task per requester, which leaves \(n=24{,}923\) of 258{,}352 tasks. The controls are \(p=302\) standardized features available when a task is posted: hashed text features, topic shares and category scores, and 16 exactly measured controls (\Cref{app:emp-sample}). We add Gaussian errors to the 286 text, topic and category coordinates, independent of the data, of each other across measurements and, except in one design, across coordinates. A single measurement has reliability \(\varrho=1/\{1+\operatorname{Var}(\nu_{ik,j})\}\). The five primary designs use \(K=2\) and \(\varrho\in\{0.90,0.80,0.60,0.40,0.25\}\); two further designs use \(K=3\) with \(\varrho=0.80\), and \(K=2\) with \(\varrho=0.80\) and AR\((0.5)\) correlation of the errors across coordinates. Each design has 200 measurement-error draws; 10 of the 1{,}400 draws stopped with a numerical error and are excluded (\Cref{app:emp-complete}).

\begin{table}[t]\centering\tablesize
\caption{Recovery of the clean-data estimate under simulated measurement error.}\label{tab:app-main}
{\setlength{\tabcolsep}{3.5pt}\begin{tabular}{@{}llrrrrrr@{}}
\toprule
 & & Mean & Mean abs. & RMS & & Mean & \\
Design & Procedure & displ. & displ. & displ. & SD & SE & Incl. \\
\midrule
$\varrho=0.90$, $K=2$ & \NAIVEDML{} & 0.20 & 0.21 & 0.22 & 0.09 & 1.00 & 1.000 \\
 & \CORE{} & 0.03 & 0.08 & 0.10 & 0.10 & 1.00 & 1.000 \\
\addlinespace
$\varrho=0.80$, $K=2$ & \NAIVEDML{} & 0.39 & 0.39 & 0.41 & 0.11 & 0.99 & 1.000 \\
 & \CORE{} & 0.07 & 0.12 & 0.16 & 0.14 & 1.01 & 1.000 \\
\addlinespace
$\varrho=0.60$, $K=2$ & \NAIVEDML{} & 0.81 & 0.81 & 0.82 & 0.17 & 0.99 & 1.000 \\
 & \CORE{} & 0.20 & 0.25 & 0.31 & 0.24 & 1.02 & 1.000 \\
\addlinespace
$\varrho=0.40$, $K=2$ & \NAIVEDML{} & 1.34 & 1.34 & 1.36 & 0.21 & 0.98 & 0.995 \\
 & \CORE{} & 0.38 & 0.44 & 0.53 & 0.36 & 1.04 & 1.000 \\
\addlinespace
$\varrho=0.25$, $K=2$ & \NAIVEDML{} & 1.99 & 1.99 & 2.00 & 0.23 & 0.96 & 0.330 \\
 & \CORE{} & 0.82 & 0.83 & 0.96 & 0.50 & 1.08 & 0.990 \\
\addlinespace
$\varrho=0.80$, $K=3$ & \NAIVEDML{} & 0.28 & 0.28 & 0.30 & 0.11 & 1.00 & 1.000 \\
 & \CORE{} & 0.04 & 0.11 & 0.14 & 0.13 & 1.00 & 1.000 \\
\addlinespace
$\varrho=0.80$, $K=2$, AR$(0.5)$ errors & \NAIVEDML{} & 0.38 & 0.38 & 0.40 & 0.11 & 0.99 & 1.000 \\
 & \CORE{} & 0.04 & 0.13 & 0.16 & 0.16 & 1.01 & 1.000 \\
\bottomrule
\end{tabular}}
\par\smallskip\parbox{0.97\linewidth}{\footnotesize \(\varrho\): reliability of a single measurement; \(K\): number of measurements. Entries other than Incl.\ are in units of the standard error of the clean-data estimate. Mean displ.: mean over draws of the signed displacement \(\widehat\theta-\widehat\theta_{\mathrm{clean}}\), positive when toward zero because \(\widehat\theta_{\mathrm{clean}}<0\); abs.: absolute; RMS: root mean squared; SD: standard deviation across draws; Mean SE: average standard error, which measures sampling uncertainty and is not comparable with the SD; Incl.: share of nominal 95\% intervals that contain the clean-data estimate, a stability diagnostic rather than coverage. Draws per design: 195--200.}
\end{table}

The benchmark is DML with the same sample, dictionary, folds and penalty rule applied to the original controls, \(\widehat\theta_{\mathrm{clean}}=\ensuremath{-}0.393\) with conventional DML standard error 0.013. For every draw we compute \CORE{} and \NAIVEDML{}, which sets the covariance to zero in the first stages and the score (\Cref{subsec:sim-estimators}). Variation across draws measures how well each procedure recovers the fixed clean-data estimate, not sampling uncertainty. \NAIVEDML{} is displaced toward zero in every design, by 0.20 benchmark standard errors at \(\varrho=0.90\) and 1.99 at \(\varrho=0.25\) (\Cref{tab:app-main}). \CORE{} removes 59--88 percent of this mean displacement, including in the design whose error covariance is not diagonal, and its remaining displacement is 0.03--0.07 benchmark standard errors in the designs with \(\varrho\ge0.80\). As with other measurement-error corrections, the gain in centering comes with more dispersion: the standard deviation across draws of \CORE{} is 1.10--2.19 times that of \NAIVEDML{}. Corrected least squares on the 302 controls fails our positive-definiteness safeguard in every draw with \(\varrho\le0.60\): its corrected Gram matrix then has an eigenvalue below \(10^{-4}\) (\Cref{app:emp-complete}). In the six designs whose error covariance is diagonal, \Cref{app:emp-bck} also computes the procedure of~\citet{belloni2017confidence} with the tuning of their simulations, with and without our refit of its first steps: measured from each procedure's own clean-data value, both versions are less displaced than \CORE{} at \(\varrho\le0.40\) and more displaced at \(\varrho\ge0.80\), and the refit fails in 17 of the 200 draws at \(\varrho=0.25\).
\section{Conclusion}\label{sec:conclusion}

When high-dimensional controls are observed only through repeated noisy measurements, the DML score is shifted by a bilinear form in the nuisance coefficients and the covariance of the measurement error, a shift that better first stages cannot remove. \CORE{} recalibrates each cross-fitted score with the unit's own repeated-measurement covariance contribution. The score averages
residual products from distinct measurements and carries covariance-estimation uncertainty into the standard error. With affine nuisances, it is centered and Neyman orthogonal. Under approximate sparsity, our rate conditions make the remaining drift negligible for inference. A corrected Lasso attains the required rates under sufficient sparsity, yielding uniformly valid intervals.

The identifying restriction that the errors of distinct measurements are uncorrelated must be argued from how the measurements are produced; \Cref{app:sim-crossdep} gives the bias when it fails and a sensitivity analysis, which~\Cref{subsec:app-caltech} applies. Implications of conditional mean zero, such as equal measurement means and within-unit contrasts uncorrelated with the treatment and the outcome, can be tested, as in that application; with three or more measurements, so can an implication of a common scale and uncorrelated errors: equal covariances across pairs of measurements (\mbox{\Cref{rem:cov-weights-identification}}). Four extensions are natural. The first is to use information in the measurements beyond second moments, which may require stronger distributional restrictions and raises the question of the efficiency bound~\citep[cf.][]{tsiatis2004locally,ma2006locally}. The second is to allow nonlinear nuisance functions: with four or more measurements whose errors are mutually independent conditionally on the latent controls and the structural disturbances, products of up to four distinct measurements are conditionally unbiased for the corresponding products of the latent controls, which suggests a corrected dictionary of interactions. The third is to let the number of measurements grow with the sample size. The fourth is to develop inference that remains valid when the rate condition~\eqref{eq:thy-product-rate} fails, for example by bounding or estimating the first-stage remainder of~\Cref{lemm:cm-centering}.

\section*{Data and Code Availability}
The Caltech application uses the replication archive of~\citet{gillensnowbergyariv2019data}, supplementary material to the Journal of Political Economy article (doi:\doi{10.1086/701681}). The online labor market application uses the task sample labeled \texttt{ipeirotis} in the replication archive of~\citet{dubejacobsnaidusuri2020data} (openICPSR project 231430, version V1, doi:\doi{10.3886/E231430V1}); its measurement errors are simulated. The replication package which is available upon request contains the estimator, the programs that generate the Monte Carlo draws and estimates, the per-replication records and the programs that produce every table and figure from them.

\clearpage
\bibliographystyle{elsarticle-harv}
\bibliography{references}

\clearpage
\appendix
\crefalias{section}{appendix}\crefalias{subsection}{subappendix}\crefalias{subsubsection}{subsubappendix}
\counterwithin{table}{section}
\counterwithin{figure}{section}

\begingroup\emergencystretch=1.5em
\section{Proofs and Remarks for Sections 2--4}\label{app:model-proofs}
\subsection{Identification and the scope of the model}\label{app:model-identification}

\begin{proof}[Proof of~\Cref{prop:observed-identification}]
Conditional mean zero makes \(\nu_{ik}\) uncorrelated with every square-integrable function of \((X_i,U_i,V_i)\), and~\eqref{eq:model-repeat-cross-orthogonality} gives \(\operatorname{Cov}(\nu_{i1},\nu_{i2})=0\); hence \(\Sigma_{12}=\operatorname{Var}(X_i)\).  Because \(D_i\) and \(Y_i\) are functions of \((X_i,U_i,V_i)\) and \(\E[V_i\mid X_i]=\E[U_i\mid D_i,X_i]=0\), we have \(c_{DZ}=\Sigma_{12}\gammao\), \(c_{YZ}=\thetao c_{DZ}+\Sigma_{12}\betao=\Sigma_{12}\pio\), \(\operatorname{Var}(D_i)=\gammao'\Sigma_{12}\gammao+J_0\) and \(\operatorname{Cov}(Y_i,D_i)=\thetao\operatorname{Var}(D_i)+\betao'\Sigma_{12}\gammao\).  The identity \(\Sigma_{12}\Sigma_{12}^\dagger\Sigma_{12}=\Sigma_{12}\) gives \(c_{DZ}'\Sigma_{12}^\dagger c_{YZ}=\gammao'\Sigma_{12}\betao+\thetao\gammao'\Sigma_{12}\gammao\) and \(c_{DZ}'\Sigma_{12}^\dagger c_{DZ}=\gammao'\Sigma_{12}\gammao\), so the numerator equals \(\thetao J_0\) and the denominator equals \(J_0>0\).
\end{proof}

The slopes \(\betao\) and \(\gammao\) of the affine nuisances of~\Cref{prop:observed-identification} coincide with the sparse approximation coefficients of~\Cref{subsec:model-sparsity} whenever \(s_{f,n}\ge\|\betao\|_0\) and \(s_{g,n}\ge\|\gammao\|_0\), because \(\Sigma_X\) is nonsingular under the conditions of~\Cref{sec:theory}.

\begin{remark}[Population target without approximate sparsity]\label{rem:projection-target}
Let \(r^\ast_f\) and \(r^\ast_g\) be the residuals of the population linear projections, with intercept, of \(f_0(X_i)\) and \(g_0(X_i)\) on \(X_i\).  By the corrected least-squares identity of~\Cref{subsec:model-omega}, applied with an intercept, the population corrected least-squares coefficients of \(Y_i\) and \(D_i\) on \(Z_i\) are those of the linear projections, with intercept, of \(Y_i\) and \(D_i\) on \(X_i\).  Because the covariance terms cancel exactly in population, the corrected ratio functional evaluated at these coefficients equals
\begin{equation*}
\theta^\ast=\thetao+\frac{\E[r^\ast_{f,i}r^\ast_{g,i}]}{\E[(V_i+r^\ast_{g,i})^2]},
\end{equation*}
which is also the value of the observed-distribution functional of~\Cref{prop:observed-identification} when the nuisances are not affine.  Thus \(\theta^\ast=\thetao\) exactly when \(\E[r^\ast_{f,i}r^\ast_{g,i}]=0\), in particular when either nuisance is affine in the dictionary; nonlinearities in different coordinates can still displace \(\theta^\ast\) when the controls are correlated.  Because the projections minimize mean-square error over all intercepts and coefficient vectors, \(\E[(r^\ast_{f,i})^2]\le a_{f,n}^2\) and \(\E[(r^\ast_{g,i})^2]\le a_{g,n}^2\) for the best sparse approximation errors of~\Cref{subsec:model-sparsity}; these bounds and \(\E[(V_i+r^\ast_{g,i})^2]=J_0+\E[(r^\ast_{g,i})^2]\ge J_0\) give \(|\theta^\ast-\thetao|\le a_{f,n}a_{g,n}/J_0\).  Under~\Cref{def:dgpclass}, \(J_0\ge c\) and \(a_{g,n}\le C\rho_{g,n}\) (conditions~(1) and~(7)), so the term \(\sqrt n\,a_{f,n}\rho_{g,n}\) in~\eqref{eq:thy-product-rate} gives \(\sqrt n\,|\theta^\ast-\thetao|\to0\): the discrepancy is negligible for root-\(n\) inference on \(\thetao\).

The limit of the estimator depends on its first stages: if they converge to affine limits with slopes \((\bar\pi,\bar\gamma)\) and \(\widehat\Omega\to_P\Omegao\), the population corrected ratio at these limits is \(\bar\theta=\thetao+{\E[\bar r_{f,i}\bar r_{g,i}]/\E[(V_i+\bar r_{g,i})^2]}\), where \(\bar r_{g,i}=g_0(X_i)-\E g_0(X_i)-X_i'\bar\gamma\) and \(\bar r_{f,i}=f_0(X_i)-\E f_0(X_i)-X_i'(\bar\pi-\thetao\bar\gamma)\). By the same identity, the ratio involves the latent residuals \(\thetao(V_i+\bar r_{g,i})+\bar r_{f,i}+U_i\) of \(Y_i\) and \(V_i+\bar r_{g,i}\) of \(D_i\), and \(\E[U_i(V_i+\bar r_{g,i})]=\E[\bar r_{f,i}V_i]=0\).  Hence \(\bar\theta=\theta^\ast\) when \(\bar\gamma\) is the projection coefficient of \(g_0(X_i)\) on \(X_i\), whatever \(\bar\pi\), because \(r^\ast_{g,i}\) is uncorrelated with \((1,X_i')\); otherwise \(\bar\theta\) depends on the learner.
\end{remark}

\begin{remark}[Heterogeneous effects]\label{rem:heterogeneous}
Suppose instead that \(Y_i=\vartheta(X_i)D_i+f_0(X_i)+U_i\) with \(\E[U_i\mid D_i,X_i]=0\), that~\eqref{eq:model-plr-d} holds with \(\E[V_i\mid X_i]=0\) and \(J_0>0\), that all variables are square integrable and that the measurements satisfy the conditions of~\Cref{prop:observed-identification}.  The proof of that proposition, in which \(D_i\) and \(Y_i\) are still functions of \((X_i,U_i,V_i)\), gives \(\Sigma_{12}=\operatorname{Var}(X_i)\), \(c_{DZ}=\operatorname{Cov}(X_i,D_i)\) and \(c_{YZ}=\operatorname{Cov}(X_i,Y_i)\), so by the Frisch--Waugh--Lovell theorem the observed-distribution functional is the coefficient of \(D_i\) in the population linear regression of \(Y_i\) on \((1,D_i,X_i)\), \(\E[Y_iD_i^\perp]/\E[(D_i^\perp)^2]\) with \(D_i^\perp=V_i+r^\ast_{g,i}\) and \(r^\ast_{g,i}\) as in~\Cref{rem:projection-target}.  When \(g_0\) is affine, the two moment restrictions reduce it to \(\E[\vartheta(X_i)\sigma_V^2(X_i)]/\E[\sigma_V^2(X_i)]\), with \(\sigma_V^2(X_i):=\E[V_i^2\mid X_i]\), which equals the average effect \(\E[\vartheta(X_i)]\) only when \(\vartheta(X_i)\) and \(\sigma_V^2(X_i)\) are uncorrelated; when \(g_0\) is not affine, the same calculation adds \(\E[\{\vartheta(X_i)g_0(X_i)+f_0(X_i)\}r^\ast_{g,i}]\) to the numerator and \(\E[(r^\ast_{g,i})^2]\) to the denominator.
\end{remark}

\begin{remark}[Mixed exact and mismeasured controls]\label{rem:mixed-controls}
When some coordinates of \(X_i\) are observed exactly, the aggregate proxy has no error in them, so \(\Omegao\) vanishes on their rows and columns, and \(\betao'\Omegao\gammao\) involves only the mismeasured coordinates of the structural and treatment nuisance vectors.  The corrected score, the variance estimator and the contributions \(\Omega_i\) and \(H_i\) apply unchanged.
\end{remark}

\begin{remark}[Measurement-specific intercepts and scales]\label{rem:source-intercepts}
Suppose \(W_{ik}=\mu_k+X_i+\nu_{ik}\), where the fixed vectors \(\mu_k\) are offsets specific to measurement \(k\), as when two elicitation formats produce systematically different reports after a common scale has been imposed.  A common shift of all \(\mu_k\) is absorbed by the intercepts, so take \(\sum_k\mu_k=0\); the equal-weight proxy is then unchanged.  The within-unit deviations become \(W_{ik}-\bar W_i=\mu_k+\nu_{ik}-\bar\nu_i\), so the contributions~\eqref{eq:cov-bessel-row} have mean \(\Omegao+\{K(K-1)\}^{-1}\sum_k\mu_k\mu_k'\), and the uncorrected offsets bias the correction.  Centering each deviation at its average over the held-out fold, with the divisor \(n_\ell-1\), and, in the first stages, at its training-sample average, with the divisor \(|I_\ell^c|-1\), removes the bias exactly in the averages that enter the estimator and the first stages, and leaves the influence function unchanged up to replacing \(\Omega_i\) by its version at the population offsets (\Cref{coro:source-intercepts}).  The offsets are identified because the latent controls cancel in the deviations.  When the measurements also have measurement-specific scales, \(W_{ik}=\mu_k+A_kX_i+\nu_{ik}\) with \(\nu_{ik}\) as in~\Cref{def:model-repeated-measurement} and the normalization \(A_1=I_p\), which fixes the units of \(X_i\), the scales \(A_k\) are identified from second moments in some designs.  When \(K\ge3\) and \(\Sigma_X\) is nonsingular, \(\operatorname{Cov}(W_{ik},W_{ih})=A_k\Sigma_XA_h'\) for \(k\ne h\) gives \(A_k=\operatorname{Cov}(W_{ik},W_{ih})\{\operatorname{Cov}(W_{i1},W_{ih})\}^{-1}\), \(k\ge2\), for every \(h\notin\{1,k\}\) with \(A_h\) nonsingular.  When \(K=2\) and \(A_2\) is diagonal, \(\operatorname{Cov}(W_{i2},D_i)=A_2\operatorname{Cov}(W_{i1},D_i)\) identifies the \(j\)th diagonal entry of \(A_2\) whenever \(\operatorname{Cov}(X_{ij},D_i)\ne0\), and likewise with \(Y_i\); with \(K=2\) and an unrestricted \(A_2\), these moments do not identify the scales in general.  \CORE{} requires a common scale: known nonsingular scales are removed by using \(A_k^{-1}(W_{ik}-\mu_k)\) as the \(k\)th measurement, and the theory does not cover estimated scales.
\end{remark}

\begin{remark}[Nuisances linear in the latent controls]\label{rem:linear-scope}
The dictionary in~\Cref{def:approx-sparsity} is fixed by the measurement system: it is the latent vector \(X_i\) whose measurements satisfy~\Cref{def:model-repeated-measurement}, up to the linear reparameterizations of~\Cref{rem:sparsity-representation}.  With accurately measured controls, DML can enlarge the dictionary with powers, interactions, splines or other nonlinear transformations of the controls. Such transformations of mismeasured controls are not admissible regressors, because the errors of transformed measurements are in general neither conditionally mean zero nor orthogonal across measurements; for example, \(\E[W_{ik,j}^2\mid\mathcal F_i]=X_{ij}^2+\E[\nu_{ik,j}^2\mid\mathcal F_i]\).  Correcting them would require higher-order restrictions on the measurement errors or deconvolution, which we do not use.  Transformations of exactly measured controls carry no measurement error and may be added to the dictionary (\Cref{rem:mixed-controls}), provided the enlarged dictionary satisfies the conditions of~\Cref{def:dgpclass}, in particular the tail and design conditions~(4) and~(5); the square of a Gaussian control, for example, violates~\eqref{eq:thy-vector-subgaussian}.  Their interactions with mismeasured controls are transformations of the measurements and are not covered.  The theory therefore covers nuisance functions that are approximately linear in \(X_i\), with approximation errors small enough for the rate condition~\eqref{eq:thy-product-rate}; otherwise~\Cref{rem:projection-target} describes the population target and the estimator's limit.
\end{remark}

\begin{remark}[Dictionary dependence and invariance under linear reparameterization]\label{rem:sparsity-representation}
Approximate sparsity is defined relative to the displayed coordinates of \(X_i\), and an invertible linear reparameterization preserves the control span but can change \(\ell_0\) and \(\ell_1\) complexity, and hence both a sparsity-based estimator and the plausibility of a given sparsity budget~\citep{kolesarmullerroelsgaard2026fragility}.  By contrast, the measurement-error correction is invariant.  For nonsingular \(A\in\R^{p_n\times p_n}\), let \(\widetilde X_i=AX_i\), \(\widetilde Z_i=AZ_i\), \(\widetilde\Omega_0=A\Omega_0A'\), \(\widetilde\beta=(A')^{-1}\beta\) and \(\widetilde\gamma=(A')^{-1}\gamma\).  Then \(\widetilde Z_i'\widetilde\beta=Z_i'\beta\), \(\widetilde Z_i'\widetilde\gamma=Z_i'\gamma\) and \(\widetilde\beta'\widetilde\Omega_0\widetilde\gamma=\beta'\Omega_0\gamma\), so the population corrected score is numerically unchanged when the nuisance coefficients and the covariance are transformed accordingly; the sparse approximation class and the feasible \(\ell_1\) learner are not invariant.
\end{remark}

\subsection{Approximate sparsity: formulations}\label{app:model-sparsity-details}

We follow the approximate-sparsity framework of~\citet{belloni2012sparse}, used in~\citet{belloni2014inference} and~\citet{chernozhukov2018double}; \citet{bradic2025minimax} study a related minimax class.  The sequences \((s_{f,n},s_{g,n},a_{f,n},a_{g,n})\) index the triangular-array approximation that the DML expansion requires, with exact sparsity as the zero-residual special case.  The coefficients in~\eqref{eq:model-best-s-beta}--\eqref{eq:model-best-s-gamma} are best sparse approximation coefficients and may differ from the coefficients of the population linear projection on all \(p_n\) controls.

Weak-\(\ell_q\) coefficient decay gives a primitive sufficient condition for the \emph{approximation error} in this formulation.  Suppose, for example, that \(f_0\) has a linear expansion in the coordinates of \(X_i\) with coefficients whose ordered magnitudes obey \(|b|_{(j)}\le A_f j^{-\varsigma_f}\) for some \(\varsigma_f>1/2\) (a weak-\(\ell_q\) condition with \(q=1/\varsigma_f<2\)), and that \(\lambda_{\max}(\Sigma_X)\le C\), so that the part of the expansion outside its \(s\) largest coefficients, with coefficient vector \(b_{\mathrm{tail}}\), satisfies \(\E[(X_i'b_{\mathrm{tail}})^2]\le C\norm{b_{\mathrm{tail}}}_2^2\).  Then the best \(s\)-term truncation has root mean-square approximation error at most of order \(s^{1/2-\varsigma_f}\), up to the design constant; the same statement applies to \(g_0\).  The all-coordinate score event, condition~(i) of~\Cref{prop:corrected-lasso-specialization}, which is stated for the positive-semidefinite projection used by \mbox{CoCoLasso}, holds at a penalty of order \(\sqrt{s\log\bar p_n/n}\) (\Cref{lemm:penalty-level}).  Best-sparse \(L^2\) approximation, and decay with \(1/2<\varsigma_f\le1\), do not imply bounded \(\ell_1\) norms of the best-\(s\) vectors, which the proof of~\Cref{lemm:penalty-level} bounds by \(C\sqrt s\); with \(\varsigma_f>1\) these norms are bounded (\Cref{lemm:l1-decay}), and~\Cref{lemm:penalty-level-l1} gives the event at a penalty of order \(\sqrt{\log\bar p_n/n}\), which~\Cref{coro:l1-bounded} uses.

For a fixed effective sparsity level \(s\) and a square-integrable nuisance function \(h_0\), define a best sparse approximation coefficient by
\[
b_s^\star\in\mathop{\arg\min}_{b\in\mathbb R^{p_n}:\ \|b\|_0\le s}\E\{h_0(X_i)-X_i'b\}^2.
\]  Because \(b_s^\star\) minimizes over all \(s\)-sparse vectors, an \(s\)-sparse \(\widetilde b_s\) with \(\|h_0(X_i)-X_i'\widetilde b_s\|_{P,2}\le \varepsilon_s\) exists if and only if \(\|h_0(X_i)-X_i'b_s^\star\|_{P,2}\le \varepsilon_s\), so the best-sparse and existential formulations impose the same approximation-rate requirement.  Best-sparse approximation need not make the residual orthogonal to every coordinate of \(X_i\): an \(s\)-sparse least-squares approximation satisfies normal equations only along directions that preserve its support, so the specialization in Appendix~\ref{app:firststage-cl} keeps the approximation residual explicit.

The feasible algorithm learns \(\gammao\) and the reduced-form coefficient \(\pio=\betao+\thetao\gammao\), which approximates \(f_0(X_i)+\thetao g_0(X_i)\) with error \(r_{\pi,i}:=r_{f,i}+\thetao r_{g,i}\); its sequences satisfy \(\norm{\pio}_0\le s_{\pi,n}\le s_{f,n}+s_{g,n}\) and \(\{\E r_{\pi,i}^2\}^{1/2}\le a_{\pi,n}\le a_{f,n}+Ca_{g,n}\), so support overlap is allowed.  The first-stage rates of~\Cref{sec:theory} refer to these two regressions: the product-rate condition~\eqref{eq:thy-product-rate} involves \(a_{f,n}\) and \(a_{g,n}\), and \(a_{\pi,n}\) enters through the reduced-form rate condition (\(a_{\pi,n}\le C\rho_{\pi,n}\)) and~\Cref{prop:corrected-lasso-specialization}.

\subsection{The corrected score: remarks and proofs}\label{app:score-proofs}

\begin{remark}[Scalar case: the covariance correction]\label{rem:score-scalar}
When \(p_n=1\), the bilinear shift becomes the familiar scalar errors-in-variables covariance correction.  Suppose \(X_i\in\R\), the outcome equation is \(Y_i=\thetao D_i+\betao X_i+U_i\), the treatment equation is \(D_i=\gammao X_i+V_i\), and the proxy is \(Z_i=X_i+\xi_i\) with \(\E[\xi_i^2]=\sigma_\xi^2\).  Direct substitution into the naive proxy moment gives \(\E[(Y_i-\thetao D_i-Z_i\betao)(D_i-Z_i\gammao)]=\betao\gammao\sigma_\xi^2\), so subtracting \(\betao\gammao\sigma_\xi^2\) recenters the score~\citep{fuller1987measurement}.  The bias removed by this correction need not be toward zero: its direction depends on the sign of \(\betao\gammao\).  In high dimensions, \(\betao'\Omegao\gammao\) is the corresponding bilinear covariance form.
\end{remark}

The corrected score~\eqref{eq:score-corrected} at \(\Omega=\Omegao\) and the latent linear score \(\{Y_i-\theta D_i-X_i'\beta\}\{D_i-X_i'\gamma\}\) agree in unconditional mean for every \((\theta,\beta,\gamma)\), and in conditional mean given \(\mathcal F_i\) only when \(\E[\xi_i\xi_i'\mid\mathcal F_i]=\Omegao\): by~\eqref{eq:model-repeat-mean-zero}, the conditional mean of their difference is \(\beta'\{\E[\xi_i\xi_i'\mid\mathcal F_i]-\Omegao\}\gamma\).

\begin{proof}[Proof of~\Cref{prop:naive-shift}]
By~\eqref{eq:model-f-approx}--\eqref{eq:model-g-approx} and \(Z_i=X_i+\xi_i\), \(Y_i-\thetao D_i-Z_i'\betao=U_i+r_{f,i}-\xi_i'\betao\) and \(D_i-Z_i'\gammao=V_i+r_{g,i}-\xi_i'\gammao\), so
\begin{align*}
\psi_i^{\mathrm{naive}}(\thetao,\betao,\gammao)
&=(U_i+r_{f,i})(V_i+r_{g,i})
-(U_i+r_{f,i})\xi_i'\gammao
-(V_i+r_{g,i})\xi_i'\betao
+(\xi_i'\betao)(\xi_i'\gammao).
\end{align*}
Because \(r_{f,i}\) and \(r_{g,i}\) are \(\sigma(X_i)\)-measurable, \eqref{eq:model-moment-restrictions} gives \(\E[U_i r_{g,i}]=0\) and \(\E[V_i r_{f,i}]=0\).  Moreover, \(V_i=D_i-g_0(X_i)\) is \(\sigma(D_i,X_i)\)-measurable, so \(\E[U_iV_i]=0\).  The aggregate proxy error inherits \(\E[\xi_i\mid X_i,U_i,V_i]=0\), and \(U_i+r_{f,i}\) and \(V_i+r_{g,i}\) are functions of \((X_i,U_i,V_i)\), so \(\E[(U_i+r_{f,i})\xi_i'\gammao]=0\) and \(\E[(V_i+r_{g,i})\xi_i'\betao]=0\), while \(\E[(\xi_i'\betao)(\xi_i'\gammao)]=\betao'\E[\xi_i\xi_i']\gammao=\betao'\Omegao\gammao\).  Together with \({\E[(U_i+r_{f,i})(V_i+r_{g,i})]}=\E[r_{f,i}r_{g,i}]\), this yields~\eqref{eq:score-naive-shift}.
\end{proof}

\begin{proof}[Proof of~\Cref{lemm:centering}]
The corrected score subtracts \(\betao'\Omegao\gammao\) from the naive proxy moment, so~\Cref{prop:naive-shift} gives \(M(\thetao,\betao,\gammao,\Omegao)=\E[\psi_i^{\mathrm{naive}}(\thetao,\betao,\gammao)]-\betao'\Omegao\gammao=\Delta_{fg,n}\), which vanishes when \(\Delta_{fg,n}=0\).
\end{proof}

\begin{proof}[Proof of~\Cref{lemm:orthogonality}]
Under exact sparsity, \(Y_i-\thetao D_i-Z_i'\betao=U_i-\xi_i'\betao\) and \(D_i-Z_i'\gammao=V_i-\xi_i'\gammao\).  Differentiating~\eqref{eq:score-population-map} in a direction \(h_\beta\) gives \(\partial_\beta M(\thetao,\betao,\gammao,\Omegao)[h_\beta]=-\E[Z_i'h_\beta\{D_i-Z_i'\gammao\}]-h_\beta'\Omegao\gammao=-h_\beta'\E[Z_i(V_i-\xi_i'\gammao)]-h_\beta'\Omegao\gammao\), and \(\E[Z_i(V_i-\xi_i'\gammao)]=-\Omegao\gammao\) because \(Z_i=X_i+\xi_i\), \(\E[X_iV_i]=0\), \(\E[\xi_iV_i]=0\), \(\E[X_i\xi_i']=0\) and \(\E[\xi_i\xi_i']=\Omegao\); this is~\eqref{eq:score-beta-orthogonality}.  In a direction \(h_\gamma\), \(\partial_\gamma M(\thetao,\betao,\gammao,\Omegao)[h_\gamma]=-\E[\{Y_i-\thetao D_i-Z_i'\betao\}Z_i']h_\gamma-\betao'\Omegao h_\gamma=-\E[(U_i-\xi_i'\betao)Z_i']h_\gamma-\betao'\Omegao h_\gamma\).  The moment restrictions and~\eqref{eq:model-repeat-mean-zero} give \(\E[U_iX_i']=0\) and \(\E[U_i\xi_i']=0\), while \(\E[(\xi_i'\betao)Z_i']=\betao'\Omegao\), so \(\E[(U_i-\xi_i'\betao)Z_i']=-\betao'\Omegao\), and~\eqref{eq:score-gamma-orthogonality} follows.  Finally, \(\partial_\theta M(\thetao,\betao,\gammao,\Omegao)=-\E[D_i\{D_i-Z_i'\gammao\}]=-\E[D_iV_i]=-\E[V_i^2]=-J_0\), because the proxy-error term has mean zero conditional on \((X_i,U_i,V_i)\), while~\eqref{eq:score-omega-derivative} is the derivative of \(-\beta'\Omega\gamma\) at \((\betao,\gammao)\).
\end{proof}

Under approximate sparsity, the derivative calculations of~\Cref{lemm:orthogonality} contain terms in the approximation residuals; at \((\thetao,\betao,\gammao,\Omegao)\),
\begin{align}
M(\thetao,\betao,\gammao,\Omegao)&=\Delta_{fg,n},\notag\\
\partial_\beta M(\thetao,\betao,\gammao,\Omegao)[h_\beta]&=-h_\beta'\E[X_i r_{g,i}],
\label{eq:score-approx-beta-derivative}\\
\partial_\gamma M(\thetao,\betao,\gammao,\Omegao)[h_\gamma]&=-\E[r_{f,i}X_i']h_\gamma,
\label{eq:score-approx-gamma-derivative}\\
\partial^2_{\beta\gamma}M(\thetao,\betao,\gammao,\Omegao)[h_\beta,h_\gamma]
&= h_\beta'\Sigma_X h_\gamma.
\label{eq:score-second-cross-derivative}
\end{align}
These are the drift, approximation-derivative and quadratic terms that the product-rate condition~\eqref{eq:thy-product-rate} controls (\Cref{rem:thy-rate-conditions}); by~\eqref{eq:model-covariance-correction}, the cross-derivative is the prediction-norm inner product behind the product-rate remainder.  Because \(r_{f,i}\) and \(r_{g,i}\) are functions of \(X_i\) and \(\E[\xi_i\mid X_i]=0\), the derivative drifts involve only the latent design, so measurement error adds no first-order approximation drift beyond the centering shift already removed.

\begin{remark}[Orthogonality under the reduced-form parameterization]\label{rem:score-reduced-form}
The feasible learner estimates \((\pi,\gamma)\) and uses \(\beta(\theta)=\pi-\theta\gamma\).  At \(\theta=\thetao\) and under the conditions of~\Cref{lemm:orthogonality}, the chain rule gives \(\partial_\pi M=\partial_\beta M=0\) and \(\partial_\gamma M|_{(\pi,\gamma)}=\partial_\gamma M|_{(\beta,\gamma)}-\thetao\partial_\beta M=0\), so the orthogonality carries over to the learned parameterization, and the expansion behind~\Cref{theo:main}(i) uses the structural coefficient error \(\delta_\beta=\delta_\pi-\thetao\delta_\gamma\), with \(\delta_\pi=\widehat\pi_{-\ell}-\pio\) and \(\delta_\gamma=\widehat\gamma_{-\ell}-\gammao\), stacked with the intercept errors into the affine vectors \(\delta_{\beta,\ell}\) and \(\delta_{\gamma,\ell}\) of~\Cref{app:nuisance-and-inference-proofs}.
\end{remark}

\subsection{The repeated-measurement covariance estimator}\label{app:covariance-estimator}

\paragraph{Unbiasedness and pooled pairs.} Let \(S_\nu:=\sum_{k=1}^K\Sigma_{\nu,k}\).  Under~\Cref{def:model-repeated-measurement}, \eqref{eq:model-repeat-cross-orthogonality} gives \(\E[\sum_{k=1}^K(\nu_{ik}-\bar\nu_i)(\nu_{ik}-\bar\nu_i)']=S_\nu-K\E[\bar\nu_i\bar\nu_i']=S_\nu-S_\nu/K=(K-1)S_\nu/K\), and dividing by \(K(K-1)\) yields \(\E[\Omega_i]=S_\nu/K^2=\Omegao\).

For vectors \(w_1,\ldots,w_K\in\R^{p_n}\) with mean \(\bar w=K^{-1}\sum_kw_k\), we have \(\sum_{k\ne h}w_kw_h'=K^2\bar w\bar w'-\sum_k w_kw_k'\) and \(\sum_kw_kw_k'=\sum_k(w_k-\bar w)(w_k-\bar w)'+K\bar w\bar w'\).  Expanding the pair sum and substituting these two identities gives
\begin{equation}
\sum_{k<h}(w_k-w_h)(w_k-w_h)'
=(K-1)\sum_k w_kw_k'-\sum_{k\ne h}w_kw_h'
=K\sum_k(w_k-\bar w)(w_k-\bar w)',
\label{eq:cov-pair-centered-identity}
\end{equation}
so the pooled-pairs form \(\{K^2(K-1)\}^{-1}\sum_{k<h}(W_{ik}-W_{ih})(W_{ik}-W_{ih})'\) equals \(\Omega_i\) for every observation.  With \(w_k=W_{ik}\), the same two identities give
\begin{equation}
\frac{1}{K(K-1)}\sum_{k\ne h}W_{ik}W_{ih}'=Z_iZ_i'-\Omega_i,
\qquad
\frac{1}{K(K-1)}\sum_{k\ne h}W_{ih}=Z_i,
\label{eq:cov-distinct-products}
\end{equation}
so the corrected score at \(\Omega_i\) is a score in products of distinct measurements, and the uncentered corrected Gram matrix \(\Pn(Z_iZ_i'-\Omega_i)\) is the average cross-product of distinct measurements.

\begin{proof}[Proof of~\Cref{prop:cov-cross-replicate-score}]
Expanding the right side of~\eqref{eq:cov-cross-replicate-score} and substituting~\eqref{eq:cov-distinct-products}, with \(Y_i-a_Y-\theta D_i\) and \(D_i-a_D\) in place of \(Y_i-\theta D_i\) and \(D_i\), gives \({(Y_i-a_Y-\theta D_i)(D_i-a_D)}-{(Y_i-a_Y-\theta D_i)Z_i'\gamma}-{(D_i-a_D)Z_i'\beta}+{\beta'(Z_iZ_i'-\Omega_i)\gamma}\), which is the left side.  Because \(\psi_i\) is linear in its covariance argument, \(\psi_i(\theta,\beta,\gamma,\Omega_i)=\psi_i(\theta,\beta,\gamma,\Omegao)-\beta'(\Omega_i-\Omegao)\gamma\), which is~\eqref{eq:cov-score-decomposition}; taking expectations and using \(\E[\Omega_i]=\Omegao\), established above, gives the mean statement.
\end{proof}

\begin{remark}[Stacked instrumental variables based on distinct measurements]\label{rem:stacked-iv}
For every unit, stack the \(K(K-1)\) ordered pairs \((k,h)\), \(k\ne h\), with regressors \((1,D_i,W_{ih}')\), instruments \((1,D_i,W_{ik}')\) and coefficients common to all pairs.  By~\eqref{eq:cov-distinct-products}, the unit average of the stacked moment is the corrected normal-equation row of \(Y_i\) on \((1,D_i,Z_i)\) with correction \(\Omega_i\).  Hence, whenever the corrected Gram matrix of \((1,D_i,Z_i)\) is nonsingular, the just-identified stacked estimator is full-sample corrected least squares with \(\widehat\Omega^{\mathrm{full}}=\Pn\Omega_i\), and, by~\Cref{prop:cov-cross-replicate-score}, its standard error clustered by unit (without small-sample correction) is the standard error of corrected least squares that includes the covariance-estimation term, the full-sample analog of~\eqref{eq:est-variance}.  Both identities are algebraic: they hold in any dimension, for raw measurements or measurements shifted by a common vector, and without linear nuisances.  In population the stacked moment requires only equal measurement means, measurement errors uncorrelated across measurements and with \((X_i,U_i,V_i)\) (mean independent of \(X_i\) if the nuisances are not affine); it does not involve the error covariances, which may be heteroskedastic or specific to a source.  Its Neyman-orthogonal score for \(\theta\) is proportional to the cross-measurement score~\eqref{eq:cov-cross-replicate-score} at the coefficients of the population linear projections of \(Y_i\) on \((1,D_i,X_i)\) and of \(D_i\) on \((1,X_i)\), which are the true coefficients when the nuisances are affine (by the Frisch--Waugh--Lovell theorem, the \(\theta\)-row of the inverse population Jacobian \(-\E[(1,D_i,X_i')'(1,D_i,X_i')]\), applied to the unit average of the stacked moment, is \(-\{\E[(V_i+r^\ast_{g,i})^2]\}^{-1}\) times that score).  Because \(\psi_i(\theta,\beta,\gamma,\Omegao)-\beta'H_i\gamma=\psi_i(\theta,\beta,\gamma,\Omega_i)\), this score differs from the corrected score at \(\Omegao\) by \(-\beta'H_i\gamma\) at the same coefficients, which is \(-\betao'H_i\gammao\) when the nuisances are affine.  \CORE{} is therefore the cross-fitted DML estimator of~\citet[Def.~3.2]{chernozhukov2018double} for this linear score (\Cref{theo:main}), whose nuisances \((\pi,\gamma)\) are the coefficients of high-dimensional instrumental-variables regressions of \(Y_i\) and \(D_i\) on one measurement instrumented by another (\Cref{subsec:est-nuisance}); the plug-in variant, which evaluates the held-out score at \(\widehat\Omega^{\mathrm{full}}\), is not of this form (\Cref{subsec:est-variants}).  With clustering by unit, the stacked regression is a pooled variant of the obviously related instrumental variables (ORIV) estimator of~\citet{gillensnowbergyariv2019}, which instruments each replicate of a scalar regressor of interest with the others, using replicate-specific first stages and intercepts.
\end{remark}

\begin{remark}[Weights and identification]\label{rem:cov-weights-identification}
Equal weighting is not necessary for identification from the full observed distribution.  Condition~\eqref{eq:model-repeat-cross-orthogonality} gives \(\E[(W_{ik}-\bar W_i)(W_{ik}-\bar W_i)']=(1-2/K)\Sigma_{\nu,k}+S_\nu/K^2\), so at \(K=2\) these within-unit contrasts identify only \(S_\nu\).  The maintained restrictions, however, also imply \(\E[W_{i1}W_{i2}']=\E[X_iX_i']\) and \(\Sigma_{\nu,k}=\E[W_{ik}W_{ik}']-\E[W_{i1}W_{i2}']\), \(k=1,2\).  Thus \(\alpha_1^2\Sigma_{\nu,1}+\alpha_2^2\Sigma_{\nu,2}\) is identified for any fixed weights summing to one, even with unequal source covariances, and the same identities identify every \(\Sigma_{\nu,k}\) when \(K\ge3\).  These raw-moment identities do not require a zero latent mean, and sample cross moments may be symmetrized.  Equal weighting is retained because contrasts remove the latent vector observation by observation and yield a positive-semidefinite covariance estimate.  Unequal or observation-specific weights and varying numbers of measurements would need their own score and variance procedures; \citet{lubotsky2006interpretation} study how to combine several proxies of a single mismeasured regressor.  When the measurements differ in quality, the symmetric average over ordered pairs is in general not efficient, and the stacked system of~\Cref{rem:stacked-iv} suggests an overidentified estimator that weights the ordered pairs optimally; we do not pursue it.  With \(K\ge3\), equality of \(\E[W_{ik}W_{ih}']\) across distinct pairs is a testable implication of~\eqref{eq:model-repeat-cross-orthogonality}, and the direction that matters for the estimator is low dimensional, \(\betao'(\E[W_{ik_1}W_{ih_1}']-\E[W_{ik_2}W_{ih_2}'])\gammao\) for two such pairs \((k_1,h_1)\ne(k_2,h_2)\); with the offsets and scales of~\Cref{rem:source-intercepts}, the covariances \(\operatorname{Cov}(W_{ik},W_{ih})=A_k\Sigma_XA_h'\) replace these moments, and their equality also requires a common scale.  \Cref{subsec:app-caltech} tests low-dimensional implications of~\eqref{eq:model-repeat-mean-zero} (equal measurement means and contrasts uncorrelated with \(D_i\) and \(Y_i\)), tests the equality of the covariances for its three confidence ratings and, for the two stakes of its main design, assesses~\eqref{eq:model-repeat-cross-orthogonality} by the sensitivity analysis of~\Cref{app:sim-crossdep}.
\end{remark}

\subsection{The variance of the feasible estimator}\label{app:estimator-details}

\begin{remark}[The covariance-estimation term in the variance]\label{rem:est-unaugmented}
The asymptotic variance of \CORE{} makes the role of covariance estimation explicit:
\begin{multline*}
J_0(P)^2\sigma^2(P)
=\operatorname{Var}_P\{\psi_i(\thetao,\betao,\gammao,\Omegao)\}+\operatorname{Var}_P(\betao'H_i\gammao)\\
-2\operatorname{Cov}_P\{\psi_i(\thetao,\betao,\gammao,\Omegao),\betao'H_i\gammao\}.
\end{multline*}
The covariance term can have either sign, so estimating \(\Omegao\) need not increase the standard error.  If the measurement errors \(\nu_{i1},\ldots,\nu_{iK}\) are i.i.d.\ \(N(0,\Sigma_\nu)\) and independent of \(\mathcal F_i\), their mean \(\xi_i=\bar\nu_i\) is independent of their deviations \(\nu_{ik}-\bar\nu_i\). The score contribution depends on the errors only through \(\xi_i\), whereas \(\betao'H_i\gammao\) depends on them only through the deviations, so independence from \(\mathcal F_i\) gives \(\operatorname{Cov}_P\{\psi_i(\thetao,\betao,\gammao,\Omegao),\betao'H_i\gammao\}=0\), and the variance of the covariance-estimation term adds exactly to the score variance.  More generally, let \(K=2\) and suppose that \(\nu_{ik}=\tau_i\epsilon_{ik}\), where the scale \(\tau_i\ge0\) is \(\mathcal F_i\)-measurable with \(\E_P[\tau_i^2]=1\) and \(\epsilon_{i1},\epsilon_{i2}\) are i.i.d.\ \(N(0,\Sigma_\nu)\) and independent of \(\mathcal F_i\); this special case of~\Cref{def:model-repeated-measurement} confines heteroskedasticity to a common scale.  Because \(\epsilon_{i1}+\epsilon_{i2}\) and \(\epsilon_{i1}-\epsilon_{i2}\) are then independent of each other and of \(\mathcal F_i\), a direct calculation with Isserlis' formula gives the net covariance-estimation term
\begin{align}
&\operatorname{Var}_P(\betao'H_i\gammao)-2\operatorname{Cov}_P\{\psi_i(\thetao,\betao,\gammao,\Omegao),\betao'H_i\gammao\}
\notag\\
&\qquad=\tfrac14\bigl\{\E_P[\tau_i^4]\,\sigma_\beta^2\sigma_\gamma^2+\sigma_{\beta\gamma}^2\bigr\}
-\sigma_{\beta\gamma}\,\E_P\bigl[(U_i+r_{f,i})(V_i+r_{g,i})(\tau_i^2-1)\bigr],
\label{eq:est-net-augmentation}
\end{align}
where \(\sigma_\beta^2:=\betao'\Sigma_\nu\betao\), \(\sigma_\gamma^2:=\gammao'\Sigma_\nu\gammao\) and \(\sigma_{\beta\gamma}:=\betao'\Sigma_\nu\gammao\).  The last expectation vanishes when \(\tau_i\equiv1\).  When \(r_{f,i}=r_{g,i}=0\) it equals \(\E_P[U_iV_i\tau_i^2]\), which also vanishes when the scale does not depend on \(U_i\) (because \(\E_P[U_i\mid X_i,V_i]=0\) by~\eqref{eq:model-moment-restrictions}) and when \(U_i\) is symmetric conditional on \((X_i,V_i)\) and the scale depends on \(U_i\) only through \(|U_i|\).  In these cases the net term is nonnegative, and it is positive unless \(\sigma_\beta\sigma_\gamma=0\).  \Cref{sec:simulations} reports the contribution of the covariance-estimation term in the canonical and heterogeneous-error designs.

Even within~\Cref{def:dgpclass}, the net term can be negative when the scale covaries with \((U_i+r_{f,i})(V_i+r_{g,i})\), including through \(r_{f,i}r_{g,i}\) when the scale depends on \(X_i\); when the two source covariances are sufficiently unequal; and, even with \(\tau_i\equiv1\), when the measurement errors are non-Gaussian with large kurtosis, bounded errors included.  For example, with \(p_n=1\) and measurement errors \(\nu_{i1},\nu_{i2}\) that are i.i.d.\ and independent of \(\mathcal F_i\), it equals \(\betao^2\gammao^2\{7(\E_P\nu_{ik}^2)^2-\E_P\nu_{ik}^4\}/8\), which is negative when the kurtosis of \(\nu_{ik}\) exceeds~\(7\); with independent Gaussian measurement errors of unequal variances instead, it is negative when the ratio of the two variances exceeds \((1+\sqrt2)^2\approx5.8\).

In the Gaussian case of the first paragraph, with equal weights, \(\betao'\Omega_i\gammao=S_{\beta\gamma,i}/K\), where \(S_{\beta\gamma,i}\) is the sample covariance of \(\betao'\nu_{ik}\) and \(\gammao'\nu_{ik}\) over \(k=1,\dots,K\), so the Gaussian sample-covariance formula gives
\begin{equation*}
V_H:=\operatorname{Var}_P(\betao'H_i\gammao)=\frac{\sigma_\beta^2\sigma_\gamma^2+\sigma_{\beta\gamma}^2}{K^2(K-1)}.
\end{equation*}
This component falls at the rate \(1/\{K^2(K-1)\}\) as independent measurements are added, which does not imply that coverage is monotone in \(K\).  Its share grows as reliability falls: if \(\Sigma_\nu=\omega\Lambda\) for a scalar \(\omega>0\) and a fixed matrix \(\Lambda\), \(r_{f,i}=r_{g,i}=0\) and \(U_i\) is independent of \((X_i,V_i)\), as in the canonical sequence, then the corrected-score variance \(V_\psi:=\operatorname{Var}_P\{\psi_i(\thetao,\betao,\gammao,\Omegao)\}\) equals \(a_0+a_1\omega+a_2\omega^2\) with \(a_0=\operatorname{Var}(U_i)J_0>0\), \(a_1=\{\operatorname{Var}(U_i)\gammao'\Lambda\gammao+J_0\betao'\Lambda\betao\}/K\ge0\) and \(a_2=\{\betao'\Lambda\betao\,\gammao'\Lambda\gammao+(\betao'\Lambda\gammao)^2\}/K^2\ge0\), and \(V_H=c_H\omega^2\) with \(c_H=a_2/(K-1)\), so that \(\mathrm d(V_H/V_\psi)/\mathrm d\omega=c_H\omega(2a_0+a_1\omega)/(a_0+a_1\omega+a_2\omega^2)^2\) is positive whenever \(c_H>0\).  The factor \(f_H=\{1+V_H/V_\psi\}^{1/2}\) by which the covariance-estimation term scales the standard error at the true parameters therefore increases strictly as reliability falls, and the standard error that ignores covariance estimation understates the dispersion of \CORE{} by this factor.  At the coefficients of the canonical sequence, with Gaussian disturbances, \(f_H=1.039\) in the canonical design and \(f_H=1.002\) in the same design with \(K=5\).  Over the single-measurement reliabilities \(0.5,0.6,\ldots,0.9\) (AR\((0.5)\) errors) and \(\operatorname{Var}(U_i)\in\{1,4,9,16\}\), the maximum of \(f_H\) is \(1.271\) with \(K=2\) and \(1.045\) with \(K=5\), both attained at reliability \(0.5\) and \(\operatorname{Var}(U_i)=1\); these values follow from the closed forms above.  In the Gaussian case with equal weights, if \(\Sigma_\nu\) is also diagonal, as~\citet{belloni2017confidence} assume, replacing each \(\Omega_i\) by its diagonal leaves the score centered and replaces \(V_H\) by \(2\sum_j\beta_{0j}^2\gamma_{0j}^2(\Sigma_\nu)_{jj}^2/\{K^2(K-1)\}\), which is smaller by \(\sum_{j\ne m}(\Sigma_\nu)_{jj}(\Sigma_\nu)_{mm}(\beta_{0j}\gamma_{0m}+\beta_{0m}\gamma_{0j})^2/\{2K^2(K-1)\}\), where \(\beta_{0j}\) and \(\gamma_{0j}\) are the \(j\)th entries of \(\betao\) and \(\gammao\): a correct diagonal restriction is at least as precise, and an incorrect one biases the estimator (\Cref{subsec:sim-hetero}).
\end{remark}
\section{The Measurement System}\label{app:measurement}

Restriction~\eqref{eq:model-repeat-mean-zero} is the weakly classical mean-independence condition.  It excludes nonclassical errors: biases specific to one measurement other than the constant offsets that~\Cref{rem:source-intercepts} accommodates, errors shared with a self-reported outcome, misclassified discrete controls~\citep[Sec.~6.1]{schennach2016recent} and reports whose errors are correlated with the true value, such as optimal predictions~\citep{hyslop2001bias} or the mean-reverting survey earnings errors found in validation studies~\citep{boundkrueger1991extent,pischke1995measurement}, which partly reflect treating the administrative record as error free~\citep{kapteynypma2007measurement}.  Panel waves are repeated measurements when the latent control is constant across waves and the errors are serially uncorrelated.

Repeated measurements also arise as repeated survey reports, as the same question put to different respondents~\citep[Sec.~4.1]{schennach2016recent} and as replicate laboratory or monitor measurements~\citep{keogh2020stratos}.  \Cref{def:model-repeated-measurement} leaves each \(\Sigma_{\nu,k}\) unrestricted, so, unlike under the classical assumption as summarized by~\citet{bound2001measurement}, the errors in different coordinates of a measurement may be correlated.  Exposure measurements with errors correlated across pollutants~\citep{xuetal2024pollutants} illustrate this correlation within a measurement, which the assumption allows; twenty-four-hour dietary recalls, whose errors are correlated with those of food-frequency questionnaires~\citep[Sec.~4.3.1]{keogh2020stratos}, illustrate correlation across measurements, which it excludes.

\Cref{def:model-repeated-measurement} restricts only the conditional mean of the errors and their unconditional covariances across measurements, so the conditional covariance \(\E[\nu_{ik}\nu_{ik}'\mid\mathcal F_i]\) may depend on \((X_i,U_i,V_i)\) in any way that the tail bounds of~\Cref{def:dgpclass} allow. Illustrative calculations such as~\eqref{eq:est-net-augmentation} use the special case \(\nu_{ik}=\tau_i\epsilon_{ik}\) of~\Cref{rem:est-unaugmented}, with a scalar \(\mathcal F_i\)-measurable scale and innovations independent of \(\mathcal F_i\), which is compatible with those bounds when \(\epsilon_{ik}\) is sub-Gaussian and \(\tau_i\) is bounded. The higher moments and distributions of \(X_i\) and of the errors, which~\Cref{def:model-repeated-measurement} leaves unidentified, would require further restrictions, such as the independence conditions behind Kotlarski's lemma~\citep[Sec.~4.1]{schennach2016recent}.

\subsection{Dependence of errors across measurements}\label{app:sim-crossdep}

Without orthogonality of the errors across measurements, let \(C_{kh}:=\E[\nu_{ik}\nu_{ih}']\) for \(k\neq h\) and \(\bar C:=\{K(K-1)\}^{-1}\sum_{k\neq h}C_{kh}\); because \(C_{hk}=C_{kh}'\), \(\bar C\) is symmetric and depends only on the symmetric parts of the \(C_{kh}\).  For the equal-weight proxy and the contributions~\eqref{eq:cov-bessel-row}, the unbiasedness calculation of~\Cref{app:covariance-estimator} gives \(\Omegao=K^{-2}(\sum_k\Sigma_{\nu,k}+\sum_{k\neq h}C_{kh})\) and \(\E[\Omega_i]=\Omegao-\bar C\), so that \(\E[Z_iZ_i']-\E[\Omega_i]=\Sigma_X+\bar C\), whereas \(\operatorname{Cov}(Z_i,R_i)=\operatorname{Cov}(X_i,R_i)\) for \(R\in\{D,Y\}\) under conditional mean zero.  At the true nuisances, the corrected moment with \(\E[\Omega_i]\) in place of \(\Omegao\) is therefore displaced by \(\betao'\bar C\gammao\).  With affine nuisances and \(\Sigma_X\) and \(\Sigma_X+\bar C\) nonsingular, the population corrected normal equations are solved by \((\Sigma_X+\bar C)^{-1}\Sigma_X\gammao\) and \((\Sigma_X+\bar C)^{-1}\Sigma_X\pio\), and, when its denominator is nonzero, the corrected ratio evaluated at these solutions is
\begin{equation*}
\theta_{\bar C}
=\thetao+\frac{\gammao'\mathcal M_{\bar C}\betao}{J_0+\gammao'\mathcal M_{\bar C}\gammao},
\qquad
\mathcal M_{\bar C}:=\Sigma_X-\Sigma_X(\Sigma_X+\bar C)^{-1}\Sigma_X=\Sigma_X(\Sigma_X+\bar C)^{-1}\bar C.
\end{equation*}
The denominator is at least \(J_0\) when \(\bar C\succeq0\), and \(\mathcal M_{\bar C}=\bar C-\bar C(\Sigma_X+\bar C)^{-1}\bar C\) gives \(\theta_{\bar C}-\thetao=\betao'\bar C\gammao/J_0+O(\|\bar C\|^2)\).  Thus \(\theta_{\bar C}=\thetao\) whenever \(\bar C=0\), a condition weaker than pairwise orthogonality and unaffected by antisymmetric dependence across measurements; conversely, \(\theta_{\bar C}=\thetao\) for all \((\betao,\gammao)\) only if \(\mathcal M_{\bar C}=0\), that is, only if \(\bar C=0\).

A shared systematic error, such as a person-specific reporting bias, a common recall shock, coding error or imputation rule, or a copied value, enters all measurements.  In the simplest case, \(\nu_{ik}=c_i+e_{ik}\), where \(c_i\) is mean zero and uncorrelated with \((X_i,U_i,V_i)\) and with every \(e_{ik}\), and the measurement-specific errors \(e_{ik}\) satisfy~\Cref{def:model-repeated-measurement}.  Then \(\bar C=\operatorname{Var}(c_i)\) and \(\operatorname{Var}(X_i+c_i)=\Sigma_X+\bar C\), so \(\theta_{\bar C}\) is the coefficient of \(D_i\) in the population linear projection of \(Y_i\) on \(D_i\) and the noisier controls \(X_i+c_i\), a coefficient that can retain part of the confounding bias~\citep{pei2019poorly}.  Because \(\Omegao=\E[\Omega_i]+\bar C\), a sensitivity analysis can add hypothesized values of \(\bar C\) to the covariance estimates, in the first stages and in the score; first and second moments of the data, which are all that \CORE{} uses, do not identify \(\bar C\), although the restriction \(\Sigma_X\succeq0\) bounds it (\Cref{app:cal-sensitivity}).

For example, if each measurement has error covariance $\Sigma_\nu$ and each distinct pair has cross covariance $\chi_\nu\Sigma_\nu$, then
\begin{equation*}
\Omega_0=\frac{1+(K-1)\chi_\nu}{K}\Sigma_\nu,
\qquad
\E[\Omega_i]=\frac{1-\chi_\nu}{K}\Sigma_\nu,
\end{equation*}
so the within-unit correction leaves the fraction $K\chi_\nu/\{1+(K-1)\chi_\nu\}$ of the shift $\betao'\Omegao\gammao$ uncorrected, $0.40$ for $\chi_\nu=0.25$ and $2/3$ for $\chi_\nu=0.5$ when $K=2$.  This fraction describes the moment at the true nuisances; the estimator's own bias ratio also reflects first stages that use the same biased covariance.  Dependence of the errors across measurements thus recreates the centering failure that the correction removes, whatever the calibration of the standard error.  In this family the sensitivity analysis multiplies the covariance estimates by \(\{1+(K-1)\chi_\nu\}/(1-\chi_\nu)\).

\subsection{Single ordered-pair moment versus its symmetrized counterpart}\label{app:measurement-one-ordering}\label{app:sim-population-displays}

With \(K=2\), the influence function of \CORE{} averages, over both ordered pairs of measurements, the just-identified moment that instruments one measurement by the other (\Cref{prop:cov-cross-replicate-score}).  At the true parameters of the canonical sequence its asymptotic variance is 0.69 times that of one ordering (Gaussian measurements and disturbances).  One ordering with nuisance coefficients specific to it is never more efficient than the symmetric average when the two measurements have equal error covariances and the errors and disturbances are mutually independent, and in the canonical design it is strictly less efficient.  When the measurements differ in quality, an ordering can be more efficient: with \(p=1\), mutually independent disturbances and errors, and error variances \(s_1\) and \(s_2\), the asymptotic variance of the ordering that residualizes the outcome with measurement \(h\) and the treatment with measurement \(k\) exceeds that of the symmetric average by \(J_0^{-2}\{\operatorname{Var}(U_i)\gammao^2(3s_k-s_h)+J_0\betao^2(3s_h-s_k)\}/4\), which is negative, for example, when \(s_h\) is close to zero and \(J_0\betao^2>3\operatorname{Var}(U_i)\gammao^2\) (cf.~\Cref{rem:cov-weights-identification}).  One ordering requires less of the measurements.  A measurement \(a+AX_i+\nu_{ik}\) with \(A\) nonsingular, used either as the instrument or as the instrumented measurement, leaves \(\thetao\) identified in that ordering, whereas such a measurement shifts the population limit of \CORE{}, which pools the coefficients of the two orderings.  In the ordering that instruments \(W_{ih}\) by \(W_{ik}\), the orthogonal score residualizes the outcome with \(W_{ih}\) and the treatment with \(W_{ik}\), and its mean at the true parameters is
\[
\E[(U_i-\nu_{ih}'\betao)(V_i-\nu_{ik}'\gammao)]=-\gammao'\E[\nu_{ik}U_i]-\betao'\E[\nu_{ih}V_i]+\betao'\E[\nu_{ih}\nu_{ik}']\gammao.
\]
One ordering therefore also tolerates errors of the instrumenting measurement correlated with the treatment disturbance and of the instrumented measurement correlated with the outcome disturbance, whereas the symmetric average needs both measurements free of such correlation, as~\Cref{def:model-repeated-measurement} requires.  This is the asymmetric exclusion structure of the proxy-control framework, in which the treatment-inducing and outcome-inducing proxies face different restrictions~\citep[Assumption~1.1]{deaner2021many}; when one source is more plausibly nondifferential, for example an administrative record, the ordering that uses it as the instrument may be preferable.  The single orderings reported in~\Cref{app:sim-comparison-suite} use the pooled fits of \CORE{}, so they share its population limit and do not inherit this robustness; an ordering-specific estimator would fit its own high-dimensional instrumental-variables first stages.
\section{Covariance Concentration}\label{app:covariance-proofs}

Each result in this appendix that invokes conditions of~\Cref{def:dgpclass} holds for every distribution \(P\) that satisfies those conditions with the constants of that assumption, and its stochastic orders are uniform over the class of all such distributions, which contains \(\Po\). Expectations and probabilities are under such a \(P\), \(C\) is the constant of~\Cref{def:dgpclass}, and all other constants depend only on the constants of that assumption.

\Needspace{4\baselineskip}\begin{lemma}[Sub-exponential Bernstein inequality]\label{lemm:app-cov-bernstein}
Let \(G_{1,n},\ldots,G_{n,n}\) be independent and mean zero with \(\max_i\|G_{i,n}\|_{\psi_1}\le B\).  There exists \(c_B>0\), depending only on \(B\), such that for every \(t>0\),
\[
\Pr\!\left(\left|\mathbb P_n G_{i,n}\right|>t\right) \le 2\exp\{-c_B n\min(t^2,t)\}.
\]
The bound holds uniformly over any family of distributions sharing the bound \(B\) on the \(\psi_1\)-norms.
\end{lemma}

\begin{proof}
This is Bernstein's inequality for averages of independent sub-exponential variables~\citep[Cor.~2.8.3]{vershynin2018high}.
\end{proof}

\begin{lemma}[Products of sub-Gaussian variables; centering]\label{lemm:app-cov-product}
If scalar random variables \(G_1,G_2\) satisfy \(\|G_1\|_{\psi_2}\le C_1\) and \(\|G_2\|_{\psi_2}\le C_2\), then \(\|G_1G_2\|_{\psi_1}\le C_{\psi}C_1C_2\) with \(C_{\psi}=1\), and fixed finite sums of such products are sub-exponential with a \(\psi_1\)-norm bound that depends only on the number of summands and the bounds on the \(\psi_2\)-norms.  Moreover, \(\|G-\E G\|_{\psi_1}\le2\|G\|_{\psi_1}\) for every sub-exponential \(G\).
\end{lemma}

\begin{proof}
The product bound is Lemma~2.7.7 of~\citet{vershynin2018high}; it follows from Young's inequality \(|G_1G_2|/(C_1C_2)\le G_1^2/(2C_1^2)+G_2^2/(2C_2^2)\) and \(e^{(s+t)/2}\le(e^s+e^t)/2\), which give \(\E\exp\{|G_1G_2|/(C_1C_2)\}\le2\).  The finite-sum statement follows from the triangle inequality for the \(\psi_1\)-norm.  For the centering bound, let \(c=\|G\|_{\psi_1}\).  Jensen's inequality gives \(\exp(\E|G|/c)\le\E\exp(|G|/c)\le2\), so \(|\E G|\le c\log2\), while a constant \(a\) has \(\|a\|_{\psi_1}=|a|/\log2\); hence \(\|\E G\|_{\psi_1}\le c\), and the triangle inequality gives the bound (cf.~\citealp[Exercise~2.7.10]{vershynin2018high}).
\end{proof}

\begin{lemma}[Moment bounds for outcome and treatment variables]\label{lemm:app-cov-means}
Under conditions~\textup{(1)}--\textup{(5)} of~\Cref{def:dgpclass}, there is a constant \(C_R\), depending only on the constants of~\Cref{def:dgpclass}, such that, for \(R\in\{Y,D\}\),
\[
\operatorname{Var}_P(R_i)+|\E_PR_i|+\max_{j\le p_n}\|Z_{ij}(R_i-\E_PR_i)\|_{\psi_1}\le C_R.
\]
\end{lemma}

\begin{proof}
With intercepts \(\mu_{D0}\) and \(\mu_{Y0}\), the reduced forms are \(D_i=\mu_{D0}+X_i'\gammao+r_{g,i}+V_i\) and \(Y_i=\mu_{Y0}+X_i'\pio+r_{\pi,i}+U_i+\thetao V_i\), so conditions~(1), (2) and~(4), with \(\E_PG^2\le\|G\|_{\psi_2}^2\), bound \(\operatorname{Var}_P(R_i)\) by a constant \(V_R\).  Write \(\bar\mu_R:=\E_PR_i\) and \(\tilde R_i:=R_i-\bar\mu_R\).  By~\eqref{eq:thy-vector-subgaussian}, \(\kappa_Z:=\|Z_{ij}\|_{\psi_2}\le2C\), so \(\E_PZ_{ij}^2\le\kappa_Z^2\) and, because \(e^x\ge x^2/2\), \(\E_PZ_{ij}^4\le4\kappa_Z^4\).  By~\eqref{eq:model-covariance-correction} and condition~(5), \(\E_PZ_{ij}^2=(\Sigma_X)_{jj}+(\Omegao)_{jj}\ge c_X\).  H\"older's inequality \(\E Z^2\le(\E|Z|)^{2/3}(\E Z^4)^{1/3}\) gives \(\E_P|Z_{ij}|\ge c_X^{3/2}/(2\kappa_Z^2)\ge c_X^{3/2}/(8C^2)\), and Jensen's inequality gives \(\E|G|\le\|G\|_{\psi_1}\) for every \(G\).  Hence, by condition~(4) and the Cauchy--Schwarz inequality,
\[
|\bar\mu_R|\,c_X^{3/2}/(8C^2)\le\E_P|\bar\mu_RZ_{ij}|\le\E_P|Z_{ij}R_i|+\E_P|Z_{ij}\tilde R_i|\le C+2C\,V_R^{1/2},
\]
which bounds \(|\bar\mu_R|\).  Finally, \(\|Z_{ij}\tilde R_i\|_{\psi_1}\le\|Z_{ij}R_i\|_{\psi_1}+|\bar\mu_R|\,\|Z_{ij}\|_{\psi_1}\), and \(\|Z_{ij}\|_{\psi_1}\le C_{\psi}\kappa_Z/\sqrt{\log2}\) by~\Cref{lemm:app-cov-product} with \(G_2\equiv1\).
\end{proof}

\begin{lemma}[Observation-level covariance contributions]\label{theo:cov-interface}
Under conditions \textup{(3)} and \textup{(4)} of~\Cref{def:dgpclass}, the contributions~\eqref{eq:cov-bessel-row} and their full-sample average \(\widehat\Omega^{\mathrm{full}}=\Pn\Omega_i\) satisfy
\begin{equation*}
\widehat\Omega^{\mathrm{full}}-\Omegao=\Pn H_i,
\qquad
H_i=\Omega_i-\Omegao,
\qquad
\E_P[H_i]=0,
\end{equation*}
for every distribution \(P\) that satisfies these conditions, and, for all deterministic \(u,v\in\R^{p_n}\),
\[
\|u'\Omega_iv\|_{\psi_1}\le C_{\psi}C^2\|u\|_2\|v\|_2, \quad \|u'H_iv\|_{\psi_1}\le2C_{\psi}C^2\|u\|_2\|v\|_2, \quad \E_P(u'H_iv)^2\le16C_{\psi}^2C^4\|u\|_2^2\|v\|_2^2,
\]
with \(C_{\psi}\) the constant of~\Cref{lemm:app-cov-product}; these bounds do not depend on \(K\ge2\), and they hold conditionally on any \(\sigma\)-field that is independent of unit \(i\), for \(u,v\) measurable with respect to it.  The pooled-pairs form has the same contributions \(\Omega_i\) by~\eqref{eq:cov-pair-centered-identity}.
\end{lemma}

\begin{proof}[Proof of~\Cref{theo:cov-interface}]
The identity \(\widehat\Omega^{\mathrm{full}}-\Omegao=\Pn H_i\) holds on every realization by the definition of \(H_i\), and Appendix~\ref{app:covariance-estimator} establishes \(\E_P[\Omega_i]=\Omegao\) from the orthogonality~\eqref{eq:model-repeat-cross-orthogonality} of the errors across measurements, hence \(\E_P[H_i]=0\).  Since \(W_{ik}-\bar W_i=\nu_{ik}-\bar\nu_i\), we have \(\Omega_i={\{K(K-1)\}^{-1}\sum_{k=1}^K(\nu_{ik}-\bar\nu_i)(\nu_{ik}-\bar\nu_i)'}\).  For deterministic \(u\), \(u'(\nu_{ik}-\bar\nu_i)=\sum_{h=1}^K(\ind\{h=k\}-K^{-1})u'\nu_{ih}\) has \(\psi_2\)-norm at most \(2(K-1)K^{-1}C\|u\|_2\) by~\eqref{eq:thy-vector-subgaussian}.  By~\Cref{lemm:app-cov-product}, each product \(\{u'(\nu_{ik}-\bar\nu_i)\}\{v'(\nu_{ik}-\bar\nu_i)\}\) has \(\psi_1\)-norm at most \(4(K-1)^2K^{-2}C_{\psi}C^2\|u\|_2\|v\|_2\), so \(u'\Omega_iv\) has \(\psi_1\)-norm at most \(4(K-1)K^{-2}C_{\psi}C^2\|u\|_2\|v\|_2\le C_{\psi}C^2\|u\|_2\|v\|_2\), because \(4(K-1)\le K^2\).  The centering bound of~\Cref{lemm:app-cov-product} gives the \(\psi_1\) bound for \(u'H_iv\), and \(e^x\ge x^2/2\) gives \(\E G^2\le4\|G\|_{\psi_1}^2\) for every \(G\), hence the second-moment bound.  Conditionally on a \(\sigma\)-field independent of unit \(i\), vectors measurable with respect to it are fixed and unit \(i\) keeps its distribution, so the same argument gives the conditional bounds.
\end{proof}

\Needspace{4\baselineskip}\begin{proposition}[Covariance concentration and fixed-vector contraction]\label{prop:sparse-bilinear}
Suppose that conditions~\textup{(1)}--\textup{(4)} of~\Cref{def:dgpclass} hold and \(\log p_n=o(n)\).  Let \(S\) range over the \(2L+1\) index sets \(\{1,\ldots,n\},I_1,\ldots,I_L,\allowbreak I_1^c,\ldots,I_L^c\), which do not depend on the data, write \(\mathbb P_SA_i:=|S|^{-1}\sum_{i\in S}A_i\), \(\bar Z_S:=\mathbb P_SZ_i\) and \(\bar R_S:=\mathbb P_SR_i\) for \(R\in\{Y,D\}\), and let
\[
\widehat\Omega_S:=\mathbb P_S\Omega_i,
\qquad
\widehat\Sigma_{Z,S}:=\mathbb P_S\{(Z_i-\bar Z_S)(Z_i-\bar Z_S)'\}-\widehat\Omega_S,
\qquad
\widehat c_{RZ,S}:=\mathbb P_S\{(Z_i-\bar Z_S)(R_i-\bar R_S)\}.
\]
Thus \(\widehat\Omega_S\) is \(\widehat\Omega^{\mathrm{full}}\) for \(S=\{1,\ldots,n\}\), the held-out-fold average \(\bar\Omega_{(\ell)}\) of~\eqref{eq:est-cm-plugin-identity} for \(S=I_\ell\) and the training-fold average \(\widehat\Omega_{-\ell}\) for \(S=I_\ell^c\); for \(S=I_\ell^c\), \(\widehat\Sigma_{Z,S}\) and \(\widehat c_{RZ,S}\) are the training-centered moments \(\widehat\Sigma_{Z,-\ell}\) and \(\widehat c_{RZ,-\ell}\) of~\Cref{subsec:est-nuisance}.  Uniformly over the distributions that satisfy these conditions and over \(S\),
\begin{align}
\|\widehat\Omega_S-\Omegao\|_{\max}
&=
\Op\bigl(\sqrt{\log p_n/n}\bigr),
\label{eq:thy-cov-max}\\
\|\widehat\Sigma_{Z,S}-\Sigma_X\|_{\max}
&=
\Op\bigl(\sqrt{\log p_n/n}\bigr).
\label{eq:thy-gram-max}
\end{align}
For every deterministic \(b\), which may depend on \(P\),
\begin{equation}
\|(\widehat\Omega_S-\Omegao)b\|_\infty
=
\Op\bigl(\|b\|_2\sqrt{\log p_n/n}\bigr),
\label{eq:thy-fixed-vector-contraction}
\end{equation}
the same holds for \(b'(\widehat\Omega_S-\Omegao)\), and, for \(R\in\{Y,D\}\),
\begin{equation}
\begin{aligned}
\|(\widehat\Sigma_{Z,S}-\Sigma_X)b\|_\infty
&=
\Op\bigl(\|b\|_2\sqrt{\log p_n/n}\bigr),\\
\|\widehat c_{RZ,S}-\E_P[Z_iR_i]\|_\infty
&=
\Op\bigl(\sqrt{\log p_n/n}\bigr),
\end{aligned}
\label{eq:thy-gram-fixed-vector}
\end{equation}
where the bound on the cross-moments also requires condition~\textup{(5)}.  Finally, for every \(S\) and all vectors \(u,v\), deterministic or random and selected by the same data,
\begin{equation}
|u'(\widehat\Omega_S-\Omegao)v|
\le
\|u\|_1\|v\|_1\|\widehat\Omega_S-\Omegao\|_{\max}
\label{eq:thy-random-sparse-bilinear}
\end{equation}
on every realization, so that combining it with~\eqref{eq:thy-cov-max} requires no union bound over data-selected supports.
\end{proposition}

\begin{proof}[Proof of~\Cref{prop:sparse-bilinear}]
Consider first \(S=\{1,\ldots,n\}\) and the uncentered averages.  For every \((j,m)\), the entry \((H_i)_{jm}\) is centered, independent across observations and uniformly sub-exponential by~\Cref{theo:cov-interface}.  \Cref{lemm:app-cov-bernstein} and a union bound over \(p_n^2\) entries give~\eqref{eq:thy-cov-max} for this \(S\), uniformly because the bound is common; \(\log p_n=o(n)\) ensures that \(t\asymp\sqrt{\log p_n/n}\) lies in the quadratic branch of the Bernstein bound.  Next, \(\Pn(Z_iZ_i')-\widehat\Omega^{\mathrm{full}}-\Sigma_X=\{\Pn(Z_iZ_i')-\E_P[Z_iZ_i']\}-\{\widehat\Omega^{\mathrm{full}}-\Omegao\}\) by~\eqref{eq:model-covariance-correction}, and the first bracket has the same max-norm rate by the \(\psi_1\) bound for \(Z_{ij}Z_{im}-\E_P[Z_{ij}Z_{im}]\) in \dgpc{4},~\Cref{lemm:app-cov-bernstein} and another union bound.

For~\eqref{eq:thy-fixed-vector-contraction} with this \(S\), \(e_j'H_ib\) has \(\psi_1\)-norm at most \(2C_{\psi}C^2\|b\|_2\) by~\Cref{theo:cov-interface}; \Cref{lemm:app-cov-bernstein} applied to \(e_j'H_ib/\|b\|_2\) and a union bound over \(j\le p_n\) give the rate, and symmetry gives the row-vector version.  Next, \(\|Z_{ij}\|_{\psi_2}\le2C\) and \(\|Z_i'b\|_{\psi_2}\le2C\|b\|_2\) by~\eqref{eq:thy-vector-subgaussian}, so \(Z_{ij}Z_i'b-\E_P[Z_{ij}Z_i'b]\) has \(\psi_1\)-norm at most \(8C_{\psi}C^2\|b\|_2\) by~\Cref{lemm:app-cov-product}; \Cref{lemm:app-cov-bernstein}, applied to this variable divided by \(\|b\|_2\), and a union bound over \(j\le p_n\) give \(\|\{\Pn(Z_iZ_i')-\E_P[Z_iZ_i']\}b\|_\infty=\Op(\|b\|_2\sqrt{\log p_n/n})\).

Each index set \(S\) is chosen independently of the data and has \(|S|\ge\lfloor n/L\rfloor\ge n/(2L)\) for \(n\ge L\); \Cref{lemm:app-cov-bernstein} applied with \(|S|\) in place of \(n\) changes only constants, and a union bound over the \(2L+1\) sets extends these rates to the uncentered averages over every \(S\).  This proves~\eqref{eq:thy-cov-max} and~\eqref{eq:thy-fixed-vector-contraction}, and it gives the rates of~\eqref{eq:thy-gram-max} and of the first bound in~\eqref{eq:thy-gram-fixed-vector} for the uncentered matrix \(\mathbb P_S(Z_iZ_i')-\widehat\Omega_S\) in place of \(\widehat\Sigma_{Z,S}\).  For the \(S\)-centered versions, write \(\tilde R_i:=R_i-\E_PR_i\).  Because \(\E_PZ_i=0\) (\Cref{subsec:model-omega}), replacing \(R_i\) by \(\tilde R_i\) changes neither the \(S\)-centered cross-moment nor \(\E_P[Z_iR_i]\). By~\Cref{lemm:app-cov-means}, \(\|Z_{ij}\tilde R_i\|_{\psi_1}\le C_R\), so \(Z_{ij}\tilde R_i-\E_P[Z_{ij}R_i]\) has \(\psi_1\)-norm at most \(2C_R\), and~\Cref{lemm:app-cov-bernstein} and a union bound over \(j\le p_n\) give the rate \(\sqrt{\log p_n/n}\) for the \(S\)-average of \(Z_i\tilde R_i\).  Centering then subtracts \(\bar Z_S\bar Z_S'\), \(\bar Z_S(\bar Z_S'b)\) and \(\bar Z_S\bar{\tilde R}_S\), with \(\bar{\tilde R}_S\) the \(S\)-average of \(\tilde R_i\).  A union bound, \(\E_PZ_i=0\) and~\eqref{eq:thy-vector-subgaussian} give \(\|\bar Z_S\|_\infty=\Op(\sqrt{\log p_n/n})\); by Chebyshev's inequality, \(|\bar Z_S'b|=\Op(\|b\|_2n^{-1/2})\) and \(|\bar{\tilde R}_S|=\Op(n^{-1/2})\), since \(Z_i'b\) and \(\tilde R_i\) are centered with variances at most \(4C^2\|b\|_2^2\) and \(C_R\).  The three terms are therefore \(\Op(\log p_n/n)\), \(\Op(\|b\|_2\sqrt{\log p_n}/n)\) and \(\Op(\sqrt{\log p_n}/n)\) in the max and sup norms, each of smaller order than the corresponding rate.  This proves~\eqref{eq:thy-gram-max} and~\eqref{eq:thy-gram-fixed-vector}.

Finally,~\eqref{eq:thy-random-sparse-bilinear} is the pathwise inequality \(|u'Av|\le\|u\|_1\|v\|_1\|A\|_{\max}\), valid for every matrix \(A\) and all vectors \(u,v\).
\end{proof}
\section{Proofs of the Inference Results}\label{app:nuisance-and-inference-proofs}

Throughout this appendix, expectations and probabilities are under an arbitrary \(P\in\Po\), and all stochastic orders are uniform over \(\Po\) and over the fixed number of folds. Of the restrictions on the measurement errors, the proofs use~\eqref{eq:model-repeat-mean-zero}--\eqref{eq:model-repeat-cross-orthogonality} and the tail bound~\eqref{eq:thy-vector-subgaussian} of condition~(4), which implies \(\|\Sigma_{\nu,k}\|_{\mathrm{op}}\le C^2\) and enters through~\Cref{lemm:cm-perturbation}, the fourth moments in the proof of~\Cref{theo:main}, the proof of~\Cref{coro:source-intercepts} and the results of~\Cref{app:covariance-proofs}; no restriction on the conditional covariance of the errors is needed. Under the domination condition in~(7) of~\Cref{def:dgpclass}, \(|\Delta_{fg,n}|\le a_{f,n}a_{g,n}\le Ca_{f,n}\rho_{g,n}\), so the projection parameter \(\theta^\ast\) of~\Cref{rem:projection-target} is root-\(n\) close to \(\thetao\).

\paragraph{Affine notation.} Let \(T_i=(1,X_i')'\) and \(T_{ik}=(1,W_{ik}')'\). Write \(g_0(X_i)=T_i'\check\gamma_0+r_{g,i}\) and \(f_0(X_i)=T_i'\check\beta_0+r_{f,i}\), and set \(\check\pi_0=\check\beta_0+\thetao\check\gamma_0\); these vectors stack a population intercept and the slopes \(\gammao,\betao,\pio\). The latent coordinates are centered in population without loss of generality, so \(\E[T_iT_i']=\diag(1,\Sigma_X)\), whose smallest eigenvalue is at least \(\min\{1,c_X\}\). For fold \(\ell\), let \(\check\pi_\ell\) and \(\check\gamma_\ell\) stack the training-mean intercepts and the slopes \(\widehat\pi_{-\ell}\) and \(\widehat\gamma_{-\ell}\), and let \(\check\beta_\ell(\theta)=\check\pi_\ell-\theta\check\gamma_\ell\). By~\Cref{prop:cov-cross-replicate-score}, applied with the intercepts as the scalars \(a_Y\) and \(a_D\), the terms of~\eqref{eq:est-NQ} are, for \(i\in I_\ell\),
\begin{equation}
\begin{aligned}
N_i&=\frac{1}{K(K-1)}\sum_{k\ne h}(Y_i-T_{ik}'\check\pi_\ell)(D_i-T_{ih}'\check\gamma_\ell),\\
Q_i&=\frac{1}{K(K-1)}\sum_{k\ne h}(D_i-T_{ik}'\check\gamma_\ell)(D_i-T_{ih}'\check\gamma_\ell),
\end{aligned}
\label{eq:app-NQ-cross}
\end{equation}
and \(\widehat\psi_i(\theta)=N_i-\theta Q_i\). Let \(\delta_{\beta,\ell}:=\check\beta_\ell(\thetao)-\check\beta_0\) and \(\delta_{\gamma,\ell}:=\check\gamma_\ell-\check\gamma_0\). By~\Cref{assu:first-stage-rates}, \(\|T_i'\delta_{\gamma,\ell}\|_{P,2}=O_P(\rho_{g,n})\) and \(\|T_i'\delta_{\beta,\ell}\|_{P,2}\le\|T_i'(\check\pi_\ell-\check\pi_0)\|_{P,2}+|\thetao|\,\|T_i'\delta_{\gamma,\ell}\|_{P,2}=O_P(\rho_{\beta,n})\), and the eigenvalue bound gives \(\|\delta\|_2\le\max\{1,c_X^{-1/2}\}\|T_i'\delta\|_{P,2}\) for every affine coefficient vector \(\delta\). Finally, let \(\psi^\times_{i,0}:=\psii(\thetao,\betao,\gammao,\Omega_i)\) be the cross-measurement score at the population affine coefficients and \(Q^\times_{i,0}\) the corresponding Jacobian term; by~\eqref{eq:cov-score-decomposition} and~\Cref{lemm:centering}, \(\psi^\times_{i,0}=\psii(\thetao,\betao,\gammao,\Omegao)-\betao'H_i\gammao\) and
\begin{equation}
\psi^\times_{i,0}-\Delta_{fg,n}=J_0(P)\varphi_i(P).
\label{eq:app-score-if}
\end{equation}

\begin{remark}[What the rate conditions control]\label{rem:thy-rate-conditions}
The terms of~\eqref{eq:thy-product-rate} control the drift \(\Delta_{fg,n}=M(\thetao,\betao,\gammao,\Omegao)\) and the terms of~\eqref{eq:score-approx-beta-derivative}--\eqref{eq:score-second-cross-derivative}: the two approximation derivatives in the directions of the first-stage errors, bounded by \(a_{g,n}\rho_{\beta,n}\) and \(a_{f,n}\rho_{g,n}\), and the quadratic term \(\E[(T_i'\delta_{\beta,\ell})(T_i'\delta_{\gamma,\ell})]=\delta_{\beta,\ell}'\diag(1,\Sigma_X)\delta_{\gamma,\ell}\), the cross-derivative~\eqref{eq:score-second-cross-derivative} in the slope errors plus the product of the intercept errors. The second bound involves \(a_{f,n}\) because \(r_\pi=r_f+\thetao r_g\) can be small through cancellation even when \(r_f\) is not. \Cref{lemm:cm-centering} shows that these are the only terms through which the first stages affect the conditional mean of the held-out score.
\end{remark}

\subsection{\texorpdfstring{Proof of~\Cref*{theo:main}}{Proof of Theorem 5.3}}\label{app:main-proof}

\begin{lemma}[Exact conditional centering]\label{lemm:cm-centering}
Under conditions~\textup{(1)}--\textup{(4)} of~\Cref{def:dgpclass}, for every fold \(\ell\) and \(i\in I_\ell\), conditionally on the training observations of fold \(\ell\) and the auxiliary randomization,
\begin{align*}
\E[\widehat\psi_i(\thetao)\mid I_\ell^c]
&=\Delta_{fg,n}-\E[(T_i'\delta_{\beta,\ell})r_{g,i}]-\E[r_{f,i}(T_i'\delta_{\gamma,\ell})]+\E[(T_i'\delta_{\beta,\ell})(T_i'\delta_{\gamma,\ell})],\\
\E[Q_i\mid I_\ell^c]
&=J_0+\E[r_{g,i}^2]-2\E[(T_i'\delta_{\gamma,\ell})r_{g,i}]+\E[(T_i'\delta_{\gamma,\ell})^2],
\end{align*}
where the expectations on the right are over an independent draw of \((X_i,U_i,V_i)\) with the fitted coefficients held fixed.
\end{lemma}

\begin{proof}
Conditionally on \(I_\ell^c\), the fitted affine vectors are fixed, and the held-out unit is independent of them. Write \(\nu^\ast_{ik}:=(0,\nu_{ik}')'\), so that \(T_{ik}=T_i+\nu^\ast_{ik}\). For \(k\ne h\),
\[
Y_i-\thetao D_i-T_{ik}'\check\beta_\ell(\thetao)=U_i+r_{f,i}-T_i'\delta_{\beta,\ell}-\nu^{\ast\prime}_{ik}\check\beta_\ell(\thetao),
\qquad
D_i-T_{ih}'\check\gamma_\ell=V_i+r_{g,i}-T_i'\delta_{\gamma,\ell}-\nu^{\ast\prime}_{ih}\check\gamma_\ell.
\]
By~\eqref{eq:model-repeat-mean-zero}, each product of an error index with a function of \((X_i,U_i,V_i)\) has mean zero, and by~\eqref{eq:model-repeat-cross-orthogonality} the product of the two error indices has mean \(\check\beta_\ell(\thetao)'\E[\nu^\ast_{ik}\nu^{\ast\prime}_{ih}]\check\gamma_\ell=0\). The conditional mean of the product is therefore that of \((U_i+r_{f,i}-T_i'\delta_{\beta,\ell})(V_i+r_{g,i}-T_i'\delta_{\gamma,\ell})\). Because \(\E[U_i\mid D_i,X_i]=0\) and \(V_i\) is a function of \((D_i,X_i)\), \(\E[U_iV_i]=\E[U_ir_{g,i}]=\E[U_iT_i]=0\); because \(\E[V_i\mid X_i]=0\), \(\E[V_ir_{f,i}]=\E[V_iT_i]=0\); and \(\E[r_{f,i}r_{g,i}]=\Delta_{fg,n}\). Since \(\widehat\psi_i(\thetao)=N_i-\thetao Q_i\) is the average of these products over the ordered pairs, the first display follows. The second display follows in the same way from \((V_i+r_{g,i}-T_i'\delta_{\gamma,\ell})^2\) and \(\E[V_i^2]=J_0\).
\end{proof}

\begin{lemma}[Cross-fitted perturbation]\label{lemm:cm-perturbation}
Under conditions~\textup{(1)}--\textup{(5)} of~\Cref{def:dgpclass} and~\Cref{assu:first-stage-rates}, for every fold \(\ell\) and \(i\in I_\ell\),
\[
\bigl\|\widehat\psi_i(\thetao)-\psi^\times_{i,0}\bigr\|_{P,2}=O_P(\rho_{\beta,n}+\rho_{g,n}+\rho_{\beta,n}\rho_{g,n}),
\qquad
\bigl\|Q_i-Q^\times_{i,0}\bigr\|_{P,2}=O_P(\rho_{g,n}+\rho_{g,n}^2),
\]
with norms conditional on the training observations of fold \(\ell\) and the auxiliary randomization.
\end{lemma}

\begin{proof}
With \(\mathcal U_{ik}:=U_i+r_{f,i}-\nu_{ik}'\betao\) and \(\mathcal G_{ih}:=V_i+r_{g,i}-\nu_{ih}'\gammao\), the difference \(\widehat\psi_i(\thetao)-\psi^\times_{i,0}\) is the average over \(k\ne h\) of \(-(T_{ik}'\delta_{\beta,\ell})\mathcal G_{ih}-\mathcal U_{ik}(T_{ih}'\delta_{\gamma,\ell})+(T_{ik}'\delta_{\beta,\ell})(T_{ih}'\delta_{\gamma,\ell})\). By~\eqref{eq:thy-vector-subgaussian}, every linear combination of \(T_{ik}\) is sub-Gaussian, so \(\|T_{ik}'\delta\|_{P,4}\le C\|\delta\|_2\le C'\|T_i'\delta\|_{P,2}\) for fixed \(\delta\); the fourth moments of \(\mathcal U_{ik}\) and \(\mathcal G_{ih}\) are bounded by conditions~(2) and~(4), with \(\E r_{f,i}^4\le C\) through \(r_f=r_\pi-\thetao r_g\). H\"older's inequality gives the first bound; the second follows in the same way.
\end{proof}

\begin{proof}[Proof of~\Cref{theo:main}(i)]
\emph{Numerator.} Write \(w_\ell:=n_\ell/n\). For each fold,
\[
\mathbb P_{n,\ell}\widehat\psi_i(\thetao)=\mathbb P_{n,\ell}\psi^\times_{i,0}+\bigl\{\E[\widehat\psi_i(\thetao)\mid I_\ell^c]-\Delta_{fg,n}\bigr\}+(\mathbb P_{n,\ell}-\E[\,\cdot\mid I_\ell^c])\{\widehat\psi_i(\thetao)-\psi^\times_{i,0}\},
\]
because \(\E[\psi^\times_{i,0}]=\Delta_{fg,n}\). By~\Cref{lemm:cm-centering} and the Cauchy--Schwarz inequality, the braces are bounded by
\[
a_{g,n}\|T_i'\delta_{\beta,\ell}\|_{P,2}+a_{f,n}\|T_i'\delta_{\gamma,\ell}\|_{P,2}+\|T_i'\delta_{\beta,\ell}\|_{P,2}\|T_i'\delta_{\gamma,\ell}\|_{P,2}=O_P(a_{g,n}\rho_{\beta,n}+a_{f,n}\rho_{g,n}+\rho_{\beta,n}\rho_{g,n}),
\]
which is \(o_P(n^{-1/2})\) by~\eqref{eq:thy-product-rate}. Conditionally on \(I_\ell^c\), the last term is an average of \(n_\ell\) independent centered variables with conditional variance at most \(n_\ell^{-1}\|\widehat\psi_i(\thetao)-\psi^\times_{i,0}\|_{P,2}^2\), which is \(o_P(n^{-1})\) by~\Cref{lemm:cm-perturbation}; by the conditional Chebyshev inequality, applied on training-measurable events on which the rates hold with a slowly diverging factor, and \(n_\ell\asymp n\), the last term is \(o_P(n^{-1/2})\). Summing over the \(L\) folds with weights \(w_\ell\), which requires no independence between their overlapping training samples, and using~\eqref{eq:app-score-if},
\begin{equation}
\widehat M_n:=\Pn\widehat\psi_i(\thetao)=J_0(P)\,\Pn\varphi_i(P)+\Delta_{fg,n}+o_P(n^{-1/2}).
\label{eq:app-cm-numerator}
\end{equation}
\emph{Denominator.} By~\Cref{lemm:cm-centering}, \(\E[Q_i\mid I_\ell^c]=J_0+O_P(a_{g,n}^2+a_{g,n}\rho_{g,n}+\rho_{g,n}^2)=J_0+o_P(1)\); by~\Cref{lemm:cm-perturbation} and conditional Chebyshev, \(\mathbb P_{n,\ell}(Q_i-Q^\times_{i,0})-\E[Q_i-Q^\times_{i,0}\mid I_\ell^c]=o_P(1)\); and \(\Pn Q^\times_{i,0}-\E Q^\times_{i,0}=O_P(n^{-1/2})\) by Chebyshev's inequality and the fourth moments of conditions~(2) and~(4). Hence \(\widehat J-J_0(P)=o_P(1)\) and, since \(J_0(P)\ge c\), \(\sup_{P\in\Po}\Pr_P(\widehat J<c/2)\to0\).

\emph{Ratio.} Because \(\widehat\psi_i\) is affine in \(\theta\), on \(\{\widehat J\ne0\}\) we have \(\sqrt n(\widehat\theta-\thetao)=\widehat J^{-1}\sqrt n\,\widehat M_n\). Chebyshev's inequality and condition~(6) give \(n^{-1/2}\sum_i\varphi_i(P)=O_P(1)\), and \(\sqrt n|\Delta_{fg,n}|\to0\) by~\eqref{eq:thy-product-rate}, so \(\sqrt n\,\widehat M_n=O_P(1)\). On \(\{\widehat J\ge c/2\}\),
\[
\sqrt n(\widehat\theta-\thetao)-\frac1{\sqrt n}\sum_{i=1}^n\varphi_i(P)
=\bigl(\widehat J^{-1}-J_0(P)^{-1}\bigr)\sqrt n\,\widehat M_n+J_0(P)^{-1}\Bigl\{\sqrt n\,\widehat M_n-\frac{J_0(P)}{\sqrt n}\sum_{i=1}^n\varphi_i(P)\Bigr\},
\]
where \(|\widehat J^{-1}-J_0(P)^{-1}|\le2c^{-2}|\widehat J-J_0(P)|=o_P(1)\) and the braces equal \(\sqrt n\Delta_{fg,n}+o_P(1)=o_P(1)\) by~\eqref{eq:app-cm-numerator}. Both terms are \(o_P(1)\), which proves~\eqref{eq:thy-uniform-expansion}; the event \(\{\widehat J<c/2\}\), on which the convention of~\Cref{def:core-generic} may apply, has probability tending to zero uniformly.
\end{proof}

\begin{proof}[Proof of~\Cref{theo:main}(ii)]
Write \(\widehat\psi_i=\widehat\psi_i(\widehat\theta)\). For \(i\in I_\ell\), \(\widehat\psi_i-J_0(P)\varphi_i(P)=\{\widehat\psi_i(\thetao)-\psi^\times_{i,0}\}+\Delta_{fg,n}-(\widehat\theta-\thetao)Q_i\) by~\eqref{eq:app-score-if}. The conditional Markov inequality, applied to the squared first term on the same training-measurable events, and~\Cref{lemm:cm-perturbation} give \(\Pn\{\widehat\psi_i(\thetao)-\psi^\times_{i,0}\}^2=o_P(1)\); \(|\Delta_{fg,n}|\le a_{f,n}a_{g,n}=o(1)\); and \(\Pn Q_i^2=O_P(1)\), by the same arguments and the fourth moments, while \(\widehat\theta-\thetao=O_P(n^{-1/2})\) by part~(i). Hence \(\Pn\{\widehat\psi_i-J_0(P)\varphi_i(P)\}^2=o_P(1)\) and, by Cauchy--Schwarz, \(\Pn\widehat\psi_i^2-J_0(P)^2\,\Pn\varphi_i(P)^2=o_P(1)\).

It remains to show \(\Pn\varphi_i(P)^2-\sigma^2(P)=o_P(1)\) uniformly. Let \(A_i:=\varphi_i(P)^2-\sigma^2(P)\) and \(\bar q:=\min\{1+\varpi/2,2\}\). By condition~(6), \(\E_P|A_i|^{\bar q}\le2^{\bar q}\E_P|\varphi_i(P)|^{2\bar q}\le C'\) uniformly, with \(C'\) depending only on \(C\), \(C_\sigma\) and \(\varpi\). For \(M>0\) put \(A_i^{M}:=A_i\ind\{|A_i|\le M\}-\E_P[A_i\ind\{|A_i|\le M\}]\). Then \(\Var_P(\Pn A_i^{M})\le M^2/n\) and, since \(\E_PA_i=0\), \(\E_P|\Pn(A_i-A_i^{M})|\le2\E_P[|A_i|\ind\{|A_i|>M\}]\le2C'M^{1-\bar q}\), so that \(\sup_{P\in\Po}\Pr_P(|\Pn A_i|>\varepsilon)\le4M^2/(n\varepsilon^2)+4C'M^{1-\bar q}/\varepsilon\) for every \(\varepsilon>0\). Taking \(M=n^{1/4}\) gives the claim. Finally, \(\widehat J^{-2}=J_0(P)^{-2}+o_P(1)\) by the proof of part~(i), which proves \(|\widehat\sigma^2-\sigma^2(P)|=o_P(1)\).
\end{proof}

\begin{lemma}[Uniform normal approximation]\label{lemm:uniform-clt}
Let \(G_{i,n}(P)\) be row-wise i.i.d.\ triangular arrays satisfying \(\mathbb E_PG_{i,n}=0\), \(\mathbb E_PG_{i,n}^2=1\), and \(\sup_{P\in\Po}\mathbb E_P|G_{i,n}|^{2+\varpi}\le C\) for some \(\varpi>0\). Then
\[
\lim_{n\to\infty}\sup_{P\in\Po}\sup_{x\in\mathbb R} \left| \Pr_P\!\left\{\frac1{\sqrt n}\sum_{i=1}^nG_{i,n}(P)\le x\right\}-\Phi(x) \right| =0.
\]
\end{lemma}

\begin{proof}
Suppose not. Then there are \(\varepsilon>0\), \(n_k\to\infty\) and \(P_k\in\mathcal P_{n_k}\) such that \(S_k:=n_k^{-1/2}\sum_{i\le n_k}G_{i,n_k}(P_k)\) satisfies \(\sup_x|\Pr_{P_k}(S_k\le x)-\Phi(x)|\ge\varepsilon\) for all \(k\). The summands \(n_k^{-1/2}G_{i,n_k}(P_k)\), \(i\le n_k\), are independent and centered, with variances summing to one and \(\sum_{i\le n_k}\E_{P_k}|n_k^{-1/2}G_{i,n_k}(P_k)|^{2+\varpi}\le Cn_k^{-\varpi/2}\to0\); by Lyapunov's central limit theorem \(S_k\) converges in distribution to \(N(0,1)\), and because \(\Phi\) is continuous the convergence of the distribution functions is uniform in \(x\) (P\'olya's theorem), a contradiction.
\end{proof}

\begin{proof}[Proof of~\Cref{theo:main}(iii)]
Set \(G_{i,n}(P)=\varphi_i(P)/\sigma(P)\). By~\eqref{eq:thy-dgp-if}, \(\E_PG_{i,n}(P)=0\) exactly, including under approximate sparsity, and condition~(6) gives the moment bound of~\Cref{lemm:uniform-clt}. Write \(S_n:=n^{-1/2}\sum_{i=1}^nG_{i,n}(P)\) and \(z:=z_{1-\alpha/2}\). By parts~(i) and~(ii) and \(\sigma^2(P)\ge c_\sigma\), \(\sqrt n(\widehat\theta-\thetao)/\sigma(P)=S_n+r_n(P)/\sigma(P)\) with \(r_n(P)/\sigma(P)=o_P(1)\), and \(|\widehat\sigma/\sigma(P)-1|\le|\widehat\sigma^2-\sigma^2(P)|/\sigma^2(P)=o_P(1)\). Fix \(\varepsilon\in(0,1/2)\) and let \(\mathcal E_n:=\{|r_n(P)|/\sigma(P)\le\varepsilon,\ |\widehat\sigma/\sigma(P)-1|\le\varepsilon\}\), so that \(\sup_P\Pr_P(\mathcal E_n^c)\to0\). On \(\mathcal E_n\), \(\{|S_n|\le z(1-\varepsilon)-\varepsilon\}\subseteq\{\thetao\in\widehat C_{1-\alpha}\}\subseteq\{|S_n|\le z(1+\varepsilon)+\varepsilon\}\). With \(d_n:=\sup_P\sup_x{|\Pr_P(S_n\le x)-\Phi(x)|}\to0\) (\Cref{lemm:uniform-clt}), for every \(P\in\Po\),
\[
2\Phi\{z(1-\varepsilon)-\varepsilon\}-1-2d_n-\Pr_P(\mathcal E_n^c) \le\Pr_P\{\thetao\in\widehat C_{1-\alpha}\} \le2\Phi\{z(1+\varepsilon)+\varepsilon\}-1+2d_n+\Pr_P(\mathcal E_n^c).
\]
Letting \(n\to\infty\) and then \(\varepsilon\downarrow0\) gives~\eqref{eq:thy-ci-validity}, by continuity of \(\Phi\).
\end{proof}

The argument is that of~\citet[Thms.~3.1--3.2]{chernozhukov2018double} for the linear score \(\widehat\psi_i(\theta)\), with nuisance \((\check\pi,\check\gamma)\); their theorems apply directly under the additional moment bound \(\sup_P\E_P|r_{g,i}|^{2q}\le C\) for some \(q>2\) (their Assumption~3.2(b) for the Jacobian score \(-Q_i\)), whereas the proof above uses only the fourth moments of~\Cref{def:dgpclass}. The rate conditions are evaluated at \(\thetao\) rather than at the root \(\bar\theta_n=\thetao+\Delta_{fg,n}/(J_0+\E_Pr_{g,i}^2)\) of the population moment, and \(\sqrt n|\bar\theta_n-\thetao|\to0\) by~\eqref{eq:thy-product-rate}.

\paragraph{Repeated partitions.} For a fixed number \(S\) of independent partitions into folds, the median of the \(S\) estimates and the median of \(\widehat\sigma_s^2+n(\widehat\theta_s-\widehat\theta^{\mathrm{med}})^2\) inherit the conclusions of~\Cref{theo:main}, as in~\citet[Cor.~3.3]{chernozhukov2018double}: each \(\widehat\theta_s\) satisfies~\eqref{eq:thy-uniform-expansion} with the same influence function, so the \(S\) estimates differ by \(o_P(n^{-1/2})\), and each \(\widehat\sigma_s^2\) is consistent.

\subsection{\texorpdfstring{Proof of~\Cref*{prop:plugin-inference}}{Proof of Proposition 5.4}}\label{app:plugin-proof}

Let \(\widetilde\Omega_\ell\) be the symmetric matrix that the variant uses in the held-out score of fold \(\ell\): \(\widetilde\Omega_\ell=\widehat\Omega^{\mathrm{full}}\) for the plug-in variant and \(\widetilde\Omega_\ell=\widehat\Omega_{-\ell}\) for the training-fold variant. Let \(\widehat N_v\) and \(\widehat J_v\) be the averages of \(N_i\) and \(Q_i\) with \(\widetilde\Omega_{\ell(i)}\) in place of \(\Omega_i\), \(\widehat\theta^{v}:=\widehat N_v/\widehat J_v\), and \(\Xi_\ell:=\bar\Omega_{(\ell)}-\widetilde\Omega_\ell\), where \(\bar\Omega_{(\ell)}\) is the held-out-fold average of~\eqref{eq:est-cm-plugin-identity}. The standard error of the variant defined in~\Cref{subsec:est-variants} is \(\widehat\sigma_v/\sqrt n\) with \(\widehat\sigma_v^2=\widehat J_v^{-2}\Pn\{\widehat\psi_i(\widehat\theta^{v})-\Pn\widehat\psi_i(\widehat\theta^{v})\}^2\); for the plug-in variant, the contributions \(\widehat\psi_i(\widehat\theta^{v})\) are the augmented contributions of~\Cref{subsec:est-variants}.

\emph{Identity.} Because \(\widehat N=\widehat N_v-\sum_\ell w_\ell\widehat\pi_{-\ell}'\Xi_\ell\widehat\gamma_{-\ell}\) and \(\widehat J=\widehat J_v-\sum_\ell w_\ell\widehat\gamma_{-\ell}'\Xi_\ell\widehat\gamma_{-\ell}\), and \(\widehat N_v-\widehat\theta^{v}\widehat J_v=0\), on \(\{\widehat J\ne0\}\)
\begin{equation}
\widehat\theta-\widehat\theta^{v}=-\widehat J^{-1}\sum_{\ell=1}^Lw_\ell\,\widehat\beta_\ell(\widehat\theta^{v})'\,\Xi_\ell\,\widehat\gamma_{-\ell},
\label{eq:app-variant-identity}
\end{equation}
with slopes only in the contraction; for the plug-in variant this is~\eqref{eq:est-cm-plugin-identity}.

\emph{The weights cancel the leading term.} Let \(E_n:=\sum_\ell w_\ell\Xi_\ell=\widehat\Omega^{\mathrm{full}}-\sum_\ell w_\ell\widetilde\Omega_\ell\). For the plug-in variant \(E_n=0\). For the training-fold variant, \(\sum_\ell w_\ell\widehat\Omega_{-\ell}=\sum_i c_i\Omega_i\) with \(c_i=\sum_{\ell\ne\ell(i)}n_\ell/\{n(n-n_\ell)\}\); since the fold sizes differ by at most one, \(n_\ell/(n-n_\ell)=1/(L-1)+O(n^{-1})\), so \(c_i=n^{-1}\{1+e_i\}\) with \(\max_i|e_i|\le C/n\), and \(E_n=-n^{-1}\sum_ie_i\Omega_i\). In both cases \(|u'E_nv|\le(C/n)\Pn|u'\Omega_iv|\) for all vectors \(u,v\), and in particular \(u'E_nv=O_P(n^{-1})\) for \(u,v\in\{\betao,\gammao,\pio\}\) by Markov's inequality, because \(\E_P|u'\Omega_iv|\le\|u'\Omega_iv\|_{\psi_1}\le C_{\psi}C^4\) by~\Cref{theo:cov-interface} and condition~(1).

\emph{Bounds.} Write \(\widehat\beta_\ell(\widehat\theta^{v})=\betao+\delta^s_{\beta,\ell}-(\widehat\theta^{v}-\thetao)\widehat\gamma_{-\ell}\) and \(\widehat\gamma_{-\ell}=\gammao+\delta^s_{\gamma,\ell}\), with slope errors \(\delta^s\). Then the sum in~\eqref{eq:app-variant-identity} equals \(\betao'E_n\gammao\) plus terms bounded by
\[
\max_\ell\bigl\{\|\delta^s_{\beta,\ell}\|_1\|\Xi_\ell\gammao\|_\infty+\|\delta^s_{\gamma,\ell}\|_1\|\Xi_\ell\betao\|_\infty+\|\delta^s_{\beta,\ell}\|_1\|\delta^s_{\gamma,\ell}\|_1\|\Xi_\ell\|_{\max}\bigr\}+|\widehat\theta^{v}-\thetao|\,\Bigl|\sum_\ell w_\ell\widehat\gamma_{-\ell}'\Xi_\ell\widehat\gamma_{-\ell}\Bigr| .
\]
Each \(\Xi_\ell\) is a difference of two of the averages \(\bar\Omega_{(\ell)}\), \(\widehat\Omega_{-\ell}\) and \(\widehat\Omega^{\mathrm{full}}\), each of which concentrates around \(\Omegao\) at the rate \(\sqrt{\log p/n}\) in the maximum norm and along the fixed vectors \(\betao\), \(\gammao\) and \(\pio\) (\Cref{prop:sparse-bilinear}, which requires \(\log p=o(n)\)). Since \(\sum_\ell w_\ell\gammao'\Xi_\ell\gammao=\gammao'E_n\gammao=O_P(n^{-1})\),
\begin{align*}
\Bigl|\sum_\ell w_\ell\widehat\gamma_{-\ell}'\Xi_\ell\widehat\gamma_{-\ell}\Bigr|
&\le O_P(n^{-1})+\max_\ell\bigl\{2\|\delta^s_{\gamma,\ell}\|_1\|\Xi_\ell\gammao\|_\infty+\|\delta^s_{\gamma,\ell}\|_1^2\|\Xi_\ell\|_{\max}\bigr\}\\
&=O_P\bigl(n^{-1}+\zeta_{g,n}\sqrt{\log p/n}\bigr)=o_P(1).
\end{align*}
The same argument with \(\widehat\pi_{-\ell}\) and \(\pio\) in place of one \(\widehat\gamma_{-\ell}\) and \(\gammao\), using \(\pio'E_n\gammao=O_P(n^{-1})\) and \(\|\Xi_\ell\pio\|_\infty=O_P(\sqrt{\log p/n})\), gives \(\widehat N_v-\widehat N=O_P(n^{-1}+(\zeta_{\pi,n}+\zeta_{g,n})\sqrt{\log p/n})=o_P(1)\), and \(\widehat N=\thetao\widehat J+\widehat M_n=O_P(1)\) by the proof of~\Cref{theo:main}(i), so \(\widehat N_v=O_P(1)\). Hence \(\widehat J_v=\widehat J+o_P(1)\), so \(\widehat J_v\ge c/4\) with probability approaching one, \(\widehat\theta^{v}-\thetao=O_P(1)\), and \(\sqrt n|\widehat\theta-\widehat\theta^{v}|=O_P({n^{-1/2}+\sqrt{\log p}\,\{\zeta_{\beta,n}+\zeta_{g,n}+\zeta_{\beta,n}\zeta_{g,n}\}})=o_P(1)\) by~\eqref{eq:thy-covariance-rate}. The variant therefore satisfies~\eqref{eq:thy-uniform-expansion}.

\emph{Variance.} \(\widehat\psi_i(\widehat\theta^{v})-\widehat\psi_i(\widehat\theta)=-(\widehat\theta^{v}-\widehat\theta)Q_i\), with \(\Pn Q_i^2=O_P(1)\), and \(\Pn\widehat\psi_i(\widehat\theta^{v})=-(\widehat\theta^{v}-\widehat\theta)\widehat J=o_P(n^{-1/2})\); with \(\widehat J_v-\widehat J=o_P(1)\), this gives \(\widehat\sigma_v^2-\widehat\sigma^2=o_P(1)\), and parts~(ii) and~(iii) follow from~\mbox{\Cref{theo:main}}.\qed

\subsection{Known covariance}\label{app:known-covariance}

\begin{corollary}[Known covariance]\label{coro:known-covariance}
Suppose that \(\Omegao\) is known and replaces \(\Omega_i\) in~\eqref{eq:est-NQ}, with first stages that may use \(\Omegao\) or the training-fold estimates \(\widehat\Omega_{-\ell}\), and let \(\widehat\sigma_{\Omegao}^2\) be~\eqref{eq:est-variance} computed from the resulting contributions. Suppose that~\Cref{def:dgpclass} holds with condition~(6) imposed on \(\varphi_i^{\Omegao}(P):=J_0(P)^{-1}\{\psii(\thetao,\betao,\gammao,\Omegao)-\E_P[\psii(\thetao,\betao,\gammao,\Omegao)]\}\) in place of \(\varphi_i(P)\), and that the first stages satisfy~\Cref{assu:first-stage-rates}. Then the conclusions of~\Cref{theo:main} hold with influence function \(\varphi_i^{\Omegao}\) and variance estimator \(\widehat\sigma_{\Omegao}^2\).
\end{corollary}

\begin{proof}
With \(\Omegao\) in place of \(\Omega_i\), the conditional means of~\Cref{lemm:cm-centering} are unchanged: with the fitted vectors held fixed, the products of the proxy residuals have conditional mean equal to that of the latent residuals plus \(\check\beta_\ell(\thetao)'\E[\xi^\ast_i\xi^{\ast\prime}_i]\check\gamma_\ell\), with \(\xi^\ast_i:=(0,\xi_i')'\), and the subtracted term \(\check\beta_\ell(\thetao)'\diag(0,\Omegao)\check\gamma_\ell\) cancels it exactly; and the perturbation bounds of~\Cref{lemm:cm-perturbation} hold for the corresponding differences, because \(\|\Omegao\|_{\mathrm{op}}\le C\). The proof of~\Cref{theo:main} then applies with \(\psi_i(\thetao,\betao,\gammao,\Omegao)\) in place of \(\psi^\times_{i,0}\), that is, with \(H_i\equiv0\); it uses the first stages only through~\Cref{assu:first-stage-rates}, which does not restrict their covariance input.
\end{proof}

\subsection{Measurement-specific intercepts}\label{app:source-intercepts}

Suppose that \(W_{ik}=\mu_k+X_i+\nu_{ik}\) for fixed vectors \(\mu_1,\ldots,\mu_K\), with \(\nu_{ik}\) as in~\Cref{def:model-repeated-measurement}. A common shift of all \(\mu_k\) is absorbed by the intercepts, so we normalize \(\sum_k\mu_k=0\); then \(Z_i=X_i+\xi_i\) is unchanged, and \(d_{ik}:=W_{ik}-Z_i=\mu_k+\nu_{ik}-\bar\nu_i\). The uncentered contributions~\eqref{eq:cov-bessel-row} have mean \(\Omegao+\{K(K-1)\}^{-1}\sum_k\mu_k\mu_k'\). For \(i\in I_\ell\), let \(\bar d_k^{(\ell)}:=n_\ell^{-1}\sum_{j\in I_\ell}d_{jk}\) and define the centered contribution
\begin{equation}
\Omega_i^{c}:=\frac{n_\ell}{n_\ell-1}\cdot\frac{1}{K(K-1)}\sum_{k=1}^K\bigl(d_{ik}-\bar d_k^{(\ell)}\bigr)\bigl(d_{ik}-\bar d_k^{(\ell)}\bigr)',
\label{eq:app-centered-contribution}
\end{equation}
and let the first stages use the training-sample analog with divisor \(|I_\ell^c|-1\). Let \(\Omega_i^{c,0}\) be~\eqref{eq:cov-bessel-row} computed from \(d_{ik}-\mu_k=\nu_{ik}-\bar\nu_i\), and \(H_i^c:=\Omega_i^{c,0}-\Omegao\).

\begin{corollary}[Measurement-specific intercepts]\label{coro:source-intercepts}
Suppose that \(W_{ik}=\mu_k+X_i+\nu_{ik}\) with \(\|\mu_k\|_2\le C\), that the conditions of~\Cref{theo:main} hold with condition~(6) imposed on~\eqref{eq:thy-dgp-if} with \(H_i^c\) in place of \(H_i\), and that \CORE{} is computed with the centered contributions~\eqref{eq:app-centered-contribution}. Then the conclusions of~\Cref{theo:main} hold with \(H_i^c\) in place of \(H_i\) in~\eqref{eq:thy-dgp-if}.
\end{corollary}

\begin{proof}
Because \(d_{ik}-\mu_k=\nu_{ik}-\bar\nu_i\), the estimator computed with \(\Omega_i^{c,0}\) is \CORE{} for measurements \(W_{ik}-\mu_k\) that satisfy~\Cref{def:model-repeated-measurement}, and for these measurements \(H_i=H_i^c\). \Cref{assu:first-stage-rates} concerns the first stages only; for CoCoLasso, centering the training-sample deviations subtracts outer products of averages that are \(O_P(\sqrt{\log p/n})\) in the maximum norm, so the concentration bounds of~\Cref{prop:sparse-bilinear} and the first-stage results of~\Cref{app:nuisance-proofs} are unaffected. It therefore suffices to show that replacing \(\Omega_i^{c,0}\) by \(\Omega_i^c\) changes \(\widehat N\) and \(\widehat J\) by \(o_P(n^{-1/2})\) and \(\widehat\sigma^2\) by \(o_P(1)\). Fix \(\ell\) and training-measurable vectors \(u,v\) with \(\|u\|_2+\|v\|_2=O_P(1)\), such as the fitted slopes. Expanding the centered cross-products,
\[
\mathbb P_{n,\ell}u'\Omega_i^cv-\mathbb P_{n,\ell}u'\Omega_i^{c,0}v=\frac{1}{n_\ell-1}\mathbb P_{n,\ell}u'\Omega_i^{c,0}v-\frac{n_\ell}{n_\ell-1}\cdot\frac{1}{K(K-1)}\sum_{k=1}^K\{u'(\bar d_k^{(\ell)}-\mu_k)\}\{v'(\bar d_k^{(\ell)}-\mu_k)\}.
\]
Conditionally on the training sample, \(u'(\bar d_k^{(\ell)}-\mu_k)\) is an average of \(n_\ell\) independent centered variables with variance at most \(C\|u\|_2^2\) by~\eqref{eq:thy-vector-subgaussian}, so it is \(O_P(n^{-1/2})\), and \(\mathbb P_{n,\ell}|u'\Omega_i^{c,0}v|=O_P(1)\) by~\Cref{theo:cov-interface}, applied conditionally on the training sample to the measurements \(W_{ik}-\mu_k\), and the conditional Markov inequality. Both terms are \(O_P(n^{-1})\), which gives the claim for \(\widehat N\) and \(\widehat J\). For the variance, write \(\tilde d_{ik}:=d_{ik}-\mu_k\) and \(\tilde d_k^{(\ell)}:=\bar d_k^{(\ell)}-\mu_k\); each \(u'\Omega_i^cv-u'\Omega_i^{c,0}v\) is bounded by \(C\{n^{-1}|u'\Omega_i^{c,0}v|+\sum_k(|u'\tilde d_{ik}|\,|v'\tilde d_k^{(\ell)}|+|v'\tilde d_{ik}|\,|u'\tilde d_k^{(\ell)}|+|u'\tilde d_k^{(\ell)}|\,|v'\tilde d_k^{(\ell)}|)\}\), whose empirical mean square is \(O_P(n^{-1})\).
\end{proof}
\section{First-Stage Learners}\label{app:nuisance-proofs}

\Cref{theo:main} requires only the prediction rates~\eqref{eq:thy-highlevel-pi}--\eqref{eq:thy-highlevel-gamma}, and~\Cref{prop:plugin-inference} adds the \(\ell_1\) rates and~\eqref{eq:thy-covariance-rate}.  This appendix gives sufficient conditions for the error bounds of the corrected Lasso and proves~\Cref{coro:cocolasso-core}.

\subsection{CoCoLasso as a first stage}\label{app:firststage-cl}

For \(R\in\{Y,D\}\), let \(b_{Y0}:=\pio\) and \(b_{D0}:=\gammao\), let \(s_R:=\|b_{R0}\|_0\), and write
\[
R_i=X_i'b_{R0}+r_{R,i}+\varepsilon_{R,i}, \qquad \E[\varepsilon_{R,i}\mid X_i]=0, \qquad \|r_R\|_{P,2}\le a_R,
\]
where \(r_Y=r_\pi\) and \(r_D=r_g\).  For a fixed constant \(c_0\ge2\), define the approximation-relaxed cone
\begin{equation*}
\mathcal C_R(c_0)
:=
\left\{\delta\in\R^{p_n}:
\|\delta_{\mathcal T_R^c}\|_1
\le
c_0\|\delta_{\mathcal T_R}\|_1
+
c_0\frac{a_R}{\lambda_R}\|X_i'\delta\|_{P,2}
\right\},
\end{equation*}
where \(\mathcal T_R=\supp(b_{R0})\) is the support of the target.

\begin{lemma}[Max-norm cost of a certified projection]\label{lemm:psd-projection-cost}
Fix $\lambda_{\mathrm{fl}}\ge0$ and a training fold, and let $\widehat\Sigma_Z$ denote the corrected Gram matrix. Suppose $\lambda_{\min}(\Sigma_X)\ge\lambda_{\mathrm{fl}}$. Let $A\succeq\lambda_{\mathrm{fl}} I$ satisfy
\begin{equation}
\|A-\widehat\Sigma_Z\|_{\max}\le\kappa\min_{B\succeq\lambda_{\mathrm{fl}} I}\|B-\widehat\Sigma_Z\|_{\max}+\bar\eta
\label{eq:thy-certified-projection}
\end{equation}
for some $\kappa\ge1$ and $\bar\eta\ge0$ (the exact projection is the case $\kappa=1$, $\bar\eta=0$). Then
\begin{equation*}
\|A-\widehat\Sigma_Z\|_{\max}\le\kappa\|\widehat\Sigma_Z-\Sigma_X\|_{\max}+\bar\eta,
\qquad
\|A-\Sigma_X\|_{\max}\le(1+\kappa)\|\widehat\Sigma_Z-\Sigma_X\|_{\max}+\bar\eta,
\end{equation*}
and, under the Gram-matrix bound~\eqref{eq:thy-gram-max} of~\Cref{prop:sparse-bilinear} with \(S=I_\ell^c\), every target vector satisfies
\begin{equation*}
\|(A-\widehat\Sigma_Z)b_{R0}\|_\infty
\le\|b_{R0}\|_1\bigl(\kappa\|\widehat\Sigma_Z-\Sigma_X\|_{\max}+\bar\eta\bigr)
=\Op\!\left(\|b_{R0}\|_1\sqrt{\frac{\log\bar p_n}{n}}\right)
\end{equation*}
whenever \(\bar\eta=\bar\eta_n=O(\sqrt{\log\bar p_n/n})\); this bound is \(\Op\{\sqrt{s_R\log\bar p_n/n}\}\) for an \(s_R\)-sparse target with \(\|b_{R0}\|_2\le C\).
\end{lemma}

\begin{proof}
Because $\lambda_{\min}(\Sigma_X)\ge\lambda_{\mathrm{fl}}$, $\Sigma_X$ is feasible, so the minimum in~\eqref{eq:thy-certified-projection} is at most $\|\Sigma_X-\widehat\Sigma_Z\|_{\max}$; this gives the first inequality, the triangle inequality gives the second, and $\|Mb\|_\infty\le\|M\|_{\max}\|b\|_1$ with the first inequality gives the last.  For the exact projection onto the positive-semidefinite cone ($\kappa=1$, $\bar\eta=0$, $\lambda_{\mathrm{fl}}=0$), the second inequality is the bound of~\citet[eq.~(2.3)]{datta2017cocolasso}.
\end{proof}

\begin{remark}[Certification and nonuniqueness]\label{rem:projection-certificate}
Condition~\eqref{eq:thy-certified-projection} can be verified in finite samples.  For every positive-semidefinite $Q$ with $\sum_{j,m}|Q_{jm}|\le1$ and every feasible $B$, $\langle Q,B-\lambda_{\mathrm{fl}} I\rangle\ge0$ and $|\langle Q,B-\widehat\Sigma_Z\rangle|\le\|B-\widehat\Sigma_Z\|_{\max}$, so $\langle Q,\lambda_{\mathrm{fl}} I-\widehat\Sigma_Z\rangle\le\min_{B\succeq\lambda_{\mathrm{fl}} I}\|B-\widehat\Sigma_Z\|_{\max}$; by conic duality this lower bound is attained at an optimal $Q$.  A computed $A\succeq\lambda_{\mathrm{fl}} I$ together with such a $Q$ satisfying $\|A-\widehat\Sigma_Z\|_{\max}\le\max\{\kappa\langle Q,\lambda_{\mathrm{fl}} I-\widehat\Sigma_Z\rangle,\ \langle Q,\lambda_{\mathrm{fl}} I-\widehat\Sigma_Z\rangle+\bar\eta\}$ therefore certifies~\eqref{eq:thy-certified-projection}.  The asymptotic statements concern sequences of estimators whose certificates use \(\bar\eta_n\le\sqrt{\log\bar p_n/n}\); \Cref{app:sim-numerics} gives the values used in the computations.  The max-norm projection is generally not unique; \Cref{lemm:psd-projection-cost} holds for every certified near-minimizer, and the computed estimator is defined jointly by the objective and the deterministic algorithm of~\Cref{app:sim-numerics}.  If $\lambda_{\min}(\Sigma_X)<\lambda_{\mathrm{fl}}$, comparing with $\Sigma_X+(\lambda_{\mathrm{fl}}-\lambda_{\min}(\Sigma_X))I$ adds $\kappa(\lambda_{\mathrm{fl}}-\lambda_{\min}(\Sigma_X))$ to the right-hand sides, which is not negligible for a fixed floor, so the floor is taken below the eigenvalue bound of condition~(5) of~\Cref{def:dgpclass}, \(\lambda_{\mathrm{fl}}<c_X\le\lambda_{\min}(\Sigma_X)\) (\Cref{subsec:est-nuisance}).
\end{remark}

\begin{proposition}[Sufficient conditions for corrected-Lasso error bounds]\label{prop:corrected-lasso-specialization}
Fix a training fold and let \(\widehat\Sigma_Z^+\) and \(\widehat c_{RZ}\) be the projected corrected Gram matrix and linear moment used in~\eqref{eq:est-cocolasso-objective}, where \(\widehat\Sigma_Z^+\) is any matrix satisfying~\eqref{eq:thy-certified-projection}, and let the penalty \(\lambda_R\) be deterministic.  Suppose that conditions~\textup{(1)}--\textup{(5)} of~\Cref{def:dgpclass} hold for every distribution in a class \(\Po\), not necessarily that of~\Cref{def:dgpclass}, and that, uniformly over \(P\in\Po\) and folds:

\begin{enumerate}[label=\textup{(\roman*)}]
\item the projected centered score
\begin{equation}
\mathcal S_R
:=
(\widehat\Sigma_Z^+-\Sigma_X)b_{R0}
-
\{\widehat c_{RZ}-\E_P[Z_iR_i]\}
\label{eq:thy-projected-score}
\end{equation}
satisfies \(\|\mathcal S_R\|_\infty\le\lambda_R/4\) with probability approaching one; and

\item there is a constant \(c_Q>0\) such that, with probability approaching one, every \(\delta\in\mathcal C_R(c_0)\) satisfies the restricted-eigenvalue condition
\begin{equation}
\delta'\widehat\Sigma_Z^+\delta
\ge
c_Q\|X_i'\delta\|_{P,2}^2.
\label{eq:thy-relaxed-compatibility}
\end{equation}
\end{enumerate}

Then the corrected-Lasso solution obeys
\begin{align}
\|X_i'(\widehat b_R-b_{R0})\|_{P,2}
&=
\Op\!\left(\sqrt{s_R}\lambda_R+a_R\right),
\label{eq:thy-cl-pred}\\
\|\widehat b_R-b_{R0}\|_1
&=
\Op\!\left(
s_R\lambda_R+\sqrt{s_R}\,a_R+\frac{a_R^2}{\lambda_R}
\right).
\label{eq:thy-cl-l1}
\end{align}
Let \(\rho_{R,n}\) and \(\zeta_{R,n}\) be the bounds in~\eqref{eq:thy-cl-pred} and~\eqref{eq:thy-cl-l1} with \(s_R\) and \(a_R\) replaced by deterministic bounds \(s_{R,n}\ge s_R\) and \(a_{R,n}\ge a_R\) (with \(\rho_{Y,n}=\rho_{\pi,n}\), \(\rho_{D,n}=\rho_{g,n}\), and likewise for \(\zeta\), \(s\) and \(a\)).  By the training-mean intercept bound of~\Cref{lem:affine-intercept}, the affine prediction errors are \(O_P(\rho_{R,n}+n^{-1/2}+\zeta_{R,n}\sqrt{\log p_n/n})\), which is \(O_P(\rho_{R,n})\) whenever \(\zeta_{R,n}\sqrt{\log p_n/n}=O(\rho_{R,n})\) and \(\rho_{R,n}\gtrsim n^{-1/2}\); the \(\ell_1\) rates of~\Cref{prop:plugin-inference} are \(\zeta_{R,n}\).
\end{proposition}

\begin{proof}[Proof of~\Cref{prop:corrected-lasso-specialization}]
The argument is the restricted-eigenvalue argument of~\citet{bickelritovtsybakov2009}, with the approximation error treated as in the Lasso bounds of~\citet[Thm.~1]{bellonichernozhukov2013least}; the new element is the projection term \((\widehat\Sigma_Z^+-\widehat\Sigma_Z)b_{R0}\) in the score~\eqref{eq:thy-projected-score}.  Set \(b_0:=b_{R0}\), \(\mathcal T:=\supp(b_0)\), \(\delta:=\widehat b-b_0\), \(\lambda:=\lambda_R\) and \(t:=\|X_i'\delta\|_{P,2}\).  All bounds below hold on the intersection \(\mathcal E_n\) of the events in parts~\textup{(i)} and~\textup{(ii)}, whose probability tends to one uniformly over \(P\in\Po\) and the fixed number of folds, so the stochastic-order conclusions follow from these deterministic bounds.

Comparing the objective~\eqref{eq:est-cocolasso-objective} at \(\widehat b\) and at \(b_0\) gives the basic inequality \(\frac12\delta'\widehat\Sigma_Z^+\delta\le-\{\widehat\Sigma_Z^+b_0-\widehat c_{RZ}\}'\delta+\lambda(\|b_0\|_1-\|b_0+\delta\|_1)\). Because \(\E_P[Z_iR_i]=\Sigma_Xb_0+v_R^{\mathrm{apx}}\) with \(v_R^{\mathrm{apx}}:=\E_P[X_ir_{R,i}]\), the gradient is \(\widehat\Sigma_Z^+b_0-\widehat c_{RZ}=\mathcal S_R-v_R^{\mathrm{apx}}\).  On \(\mathcal E_n\), \(|\mathcal S_R'\delta|\le(\lambda/4)\|\delta\|_1\); the coordinates of \(v_R^{\mathrm{apx}}\) need not vanish, so the approximation term is bounded through \(|(v_R^{\mathrm{apx}})'\delta|=|\E_P[r_{R,i}X_i'\delta]|\le a_Rt\).  With \(\|b_0\|_1-\|b_0+\delta\|_1\le\|\delta_{\mathcal T}\|_1-\|\delta_{\mathcal T^c}\|_1\),
\begin{equation}
\frac12\delta'\widehat\Sigma_Z^+\delta
+\frac{3\lambda}{4}\|\delta_{\mathcal T^c}\|_1
\le
\frac{5\lambda}{4}\|\delta_{\mathcal T}\|_1
+a_Rt.
\label{eq:app-cl-cone}
\end{equation}
Since \(\widehat\Sigma_Z^+\succeq0\), this gives \(\|\delta_{\mathcal T^c}\|_1\le\frac53\|\delta_{\mathcal T}\|_1+\frac{4a_R}{3\lambda}t\), so \(\delta\in\mathcal C_R(c_0)\) for every \(c_0\ge2\) before the restricted-eigenvalue condition is used.  By the Cauchy--Schwarz inequality and condition~(5), \(\|\delta_{\mathcal T}\|_1\le\sqrt{s_R}\|\delta\|_2\le c_X^{-1/2}\sqrt{s_R}\,t\).  Applying~\eqref{eq:thy-relaxed-compatibility} to the quadratic term in~\eqref{eq:app-cl-cone} and dropping the \(\ell_1\) term gives \(\frac{c_Q}{2}t^2\le\frac54c_X^{-1/2}\sqrt{s_R}\lambda t+a_Rt\), hence \(t\le C(\sqrt{s_R}\lambda+a_R)\), trivially if \(t=0\); this is~\eqref{eq:thy-cl-pred}.  The cone inequality then gives \(\|\delta\|_1\le\frac83\|\delta_{\mathcal T}\|_1+\frac{4a_R}{3\lambda}t\le C(\sqrt{s_R}+a_R/\lambda)t=O_P(s_R\lambda+\sqrt{s_R}\,a_R+a_R^2/\lambda)\), which is~\eqref{eq:thy-cl-l1}.  These are slope rates for centered regressions; by~\Cref{lem:affine-intercept}, the training-mean intercept adds to the affine prediction error \(O_P(n^{-1/2}+\zeta_{R,n}\sqrt{\log p_n/n})\), which is \(O_P(n^{-1/2})\) under~\eqref{eq:thy-covariance-rate}, and leaves~\eqref{eq:thy-cl-l1} unchanged.
\end{proof}

\begin{lemma}[Inexact penalized solutions]\label{lemm:kkt-perturbation}
Let $A\succeq0$, $\widehat c\in\R^{p_n}$, $\lambda>0$, and let $\widetilde b$ satisfy the optimality conditions of $\min_b\tfrac12 b'Ab-\widehat c'b+\lambda\|b\|_1$ up to $r\ge0$, that is, $\max_j\operatorname{dist}\bigl(-(A\widetilde b-\widehat c)_j,\;\lambda\,\partial|\widetilde b_j|\bigr)\le r$. Then $\widetilde b$ is an exact minimizer of $\tfrac12 b'Ab-(\widehat c+e)'b+\lambda\|b\|_1$ for some $e$ with $\|e\|_\infty\le r$. Consequently, in~\Cref{prop:corrected-lasso-specialization} an inexact solution with residual $r\le\lambda_R/8$ obeys the same conclusions provided $\|\mathcal S_R\|_\infty\le\lambda_R/8$ with probability approaching one.
\end{lemma}

\begin{proof}
For each $j$ choose $s_j\in\partial|\widetilde b_j|$ attaining the distance, and set $e_j:=(A\widetilde b-\widehat c)_j+\lambda s_j$; then $|e_j|\le r$ and $A\widetilde b-(\widehat c+e)+\lambda s=0$ with $s\in\partial\|\widetilde b\|_1$, the optimality condition of the perturbed convex problem.  In the proof of~\Cref{prop:corrected-lasso-specialization} the perturbation enters only through the empirical gradient, where it replaces $\mathcal S_R$ by $\mathcal S_R-e$.
\end{proof}

\begin{remark}[Restricted eigenvalues and the approximation-relaxed cone]\label{rem:relaxed-cone}
When \(\lambda_{\mathrm{fl}}>0\), as in the implemented estimator, condition~(ii) of~\Cref{prop:corrected-lasso-specialization} holds for every \(\delta\in\R^{p_n}\) with \(c_Q=\lambda_{\mathrm{fl}}/C\): \(\widehat\Sigma_Z^+\succeq\lambda_{\mathrm{fl}}I\) and \(\lambda_{\max}(\Sigma_X)\le C\) give \(\delta'\widehat\Sigma_Z^+\delta\ge\lambda_{\mathrm{fl}}\|\delta\|_2^2\ge(\lambda_{\mathrm{fl}}/C)\|X_i'\delta\|_{P,2}^2\).  Condition~(ii) is stated separately to permit constants that do not deteriorate with \(\lambda_{\mathrm{fl}}^{-1}\): with \(\lambda_{\mathrm{fl}}=10^{-4}\), the constants of~\eqref{eq:thy-cl-pred}--\eqref{eq:thy-cl-l1} obtained from \(c_Q=\lambda_{\mathrm{fl}}/C\) are of order \(10^4\).  Under exact sparsity, condition~(ii) holds with a constant free of the floor when \(s_{R,n}\sqrt{\log\bar p_n/n}\to0\) (\Cref{lemm:re-floor-free}), which the sparsity budgets of~\Cref{coro:cocolasso-core} imply.  Under approximate sparsity the relaxation term \(a_R/\lambda_R\) can grow with \(n\), and the same argument needs \((s_{R,n}+a_{R,n}^2/\lambda_R^2)\sqrt{\log\bar p_n/n}\to0\).
\end{remark}

\begin{lemma}[Restricted eigenvalues free of the floor]\label{lemm:re-floor-free}
Suppose conditions~\textup{(1)}--\textup{(5)} of~\Cref{def:dgpclass} hold, \(\log p_n=o(n)\), \(a_R=0\), and \(\widehat\Sigma_Z^+\) satisfies~\eqref{eq:thy-certified-projection} with a fixed \(\kappa\ge1\) and \(\bar\eta_n\le\sqrt{\log\bar p_n/n}\).  If \(s_{R,n}\sqrt{\log\bar p_n/n}\to0\) for a deterministic sequence \(s_{R,n}\ge\sup_{P\in\Po}s_R\), then, with probability approaching one uniformly over \(P\in\Po\), every \(\delta\) with \(\|\delta_{\mathcal T_R^c}\|_1\le c_0\|\delta_{\mathcal T_R}\|_1\) satisfies \(\delta'\widehat\Sigma_Z^+\delta\ge\{c_X/(2C)\}\|X_i'\delta\|_{P,2}^2\).
\end{lemma}

\begin{proof}
On the cone, \(\|\delta\|_1\le(1+c_0)\|\delta_{\mathcal T_R}\|_1\le(1+c_0)\sqrt{s_R}\|\delta\|_2\).  By~\Cref{lemm:psd-projection-cost} and~\eqref{eq:thy-gram-max} of~\Cref{prop:sparse-bilinear} with \(S=I_\ell^c\), \(\|\widehat\Sigma_Z^+-\Sigma_X\|_{\max}\le(1+\kappa)\|\widehat\Sigma_{Z,-\ell}-\Sigma_X\|_{\max}+\bar\eta_n=O_P(\sqrt{\log\bar p_n/n})\).  Hence, because \(s_R\le s_{R,n}\),
\[
\delta'\widehat\Sigma_Z^+\delta\ge\delta'\Sigma_X\delta-\|\delta\|_1^2\|\widehat\Sigma_Z^+-\Sigma_X\|_{\max}\ge\bigl\{c_X-(1+c_0)^2s_{R,n}\,O_P\bigl(\sqrt{\log\bar p_n/n}\bigr)\bigr\}\|\delta\|_2^2,
\]
which is at least \((c_X/2)\|\delta\|_2^2\) with probability approaching one uniformly over \(P\in\Po\), because \(s_{R,n}\sqrt{\log\bar p_n/n}\to0\); finally, \(\|X_i'\delta\|_{P,2}^2\le C\|\delta\|_2^2\).
\end{proof}

\begin{lemma}[Penalty level for the projected score]\label{lemm:penalty-level}
Suppose conditions~\textup{(1)}--\textup{(5)} of~\Cref{def:dgpclass} hold and \(\log p_n=o(n)\).  For a training fold and \(R\in\{Y,D\}\), let \(\widehat\Sigma_Z^+\) satisfy~\eqref{eq:thy-certified-projection} with a fixed \(\kappa\ge1\) and \(\bar\eta=\bar\eta_n\le\sqrt{\log\bar p_n/n}\), and let \(\mathcal S_R\) be the projected score~\eqref{eq:thy-projected-score}.  There is a constant \(C_\lambda\), depending only on \(\kappa\) and the constants of~\Cref{def:dgpclass}, such that, for every deterministic sequence \(s=s_n\) with \(s\ge\max\{1,s_R\}\) for every \(P\in\Po\), every deterministic penalty sequence with \(\lambda_R\ge C_\lambda\sqrt{s\log\bar p_n/n}\) satisfies
\[
\lim_{n\to\infty}\sup_{P\in\Po}\Pr_P\bigl(\|\mathcal S_R\|_\infty>\lambda_R/8\bigr)=0
\]
for every fold.  In particular, such penalties satisfy condition~\textup{(i)} of~\Cref{prop:corrected-lasso-specialization} and the score condition of~\Cref{lemm:kkt-perturbation}.
\end{lemma}

\begin{proof}
Write \(t_n:=\sqrt{\log\bar p_n/n}\), which tends to zero because \(\log p_n=o(n)\).  By~\eqref{eq:thy-projected-score} and~\Cref{lemm:psd-projection-cost}, for every realization at which \(\widehat\Sigma_Z^+\) is certified,
\begin{equation}
\|\mathcal S_R\|_\infty
\le\|b_{R0}\|_1\bigl\{(1+\kappa)\|\widehat\Sigma_{Z,-\ell}-\Sigma_X\|_{\max}+\bar\eta_n\bigr\}
+\|\widehat c_{RZ,-\ell}-\E_P[Z_iR_i]\|_\infty,
\label{eq:app-score-bound}
\end{equation}
where \(\|b_{R0}\|_1\le\sqrt{s_R}\|b_{R0}\|_2\le C\sqrt{s}\) by condition~(1) and \(s_R\le s\), with \(C\ge1\) without loss of generality.  It therefore suffices to find a constant \(M\) such that the events \(\{\|\widehat\Sigma_{Z,-\ell}-\Sigma_X\|_{\max}\le Mt_n\}\) and \(\{\|\widehat c_{RZ,-\ell}-\E_P[Z_iR_i]\|_\infty\le Mt_n\}\) have probability tending to one uniformly over \(P\in\Po\): on their intersection, \(\bar\eta_n\le t_n\) and \(s\ge1\) give \(\|\mathcal S_R\|_\infty\le\{C(2+\kappa)M+C\}\sqrt{s}\,t_n\), so \(C_\lambda:=8\{C(2+\kappa)M+C\}\) has the stated property.

Let \(m:=|I_\ell^c|\), so that \(m\ge n/3\) for \(n\ge3\) because \(L\ge2\) and the fold sizes differ by at most one.  Write \(\tilde R_i:=R_i-\E_PR_i\); replacing \(R_i\) by \(\tilde R_i\) changes neither the training-centered moment \(\widehat c_{RZ,-\ell}\) nor \(\E_P[Z_iR_i]\), because \(\E_PZ_i=0\).  By~\eqref{eq:model-covariance-correction} and \(\E_PZ_i=0\),
\begin{align*}
\widehat\Sigma_{Z,-\ell}-\Sigma_X
&=\{\mathbb P_{n,-\ell}(Z_iZ_i')-\E_P[Z_iZ_i']\}-\{\widehat\Omega_{-\ell}-\Omegao\}-\bar Z_{-\ell}\bar Z_{-\ell}',\\
\widehat c_{RZ,-\ell}-\E_P[Z_iR_i]
&=\{\mathbb P_{n,-\ell}(Z_i\tilde R_i)-\E_P[Z_i\tilde R_i]\}-\bar Z_{-\ell}\bar{\tilde R}_{-\ell}.
\end{align*}
Each entry of the bracketed terms is an average of \(m\) independent centered variables whose \(\psi_1\)-norms are bounded by a constant, by \dgpc{4} for \(Z_{ij}Z_{im}\), by~\Cref{lemm:app-cov-means} for \(Z_{ij}\tilde R_i\) and by~\Cref{theo:cov-interface} for \((H_i)_{jm}\).  Once \(n\) is large enough that \(Mt_n\le1\), \Cref{lemm:app-cov-bernstein} bounds the probability that a given entry exceeds \(Mt_n\) in absolute value by \(2\exp(-c_BmM^2t_n^2)\le2\bar p_n^{-c_BM^2/3}\), and a union bound over the at most \(3p_n^2\) entries gives at most \(6\bar p_n^{2-c_BM^2/3}\), which tends to zero uniformly over \(P\in\Po\) when \(c_BM^2>6\), because \(\bar p_n\ge n\to\infty\); the bound does not require \(p_n\to\infty\).  For the centering terms, each coordinate of \(\bar Z_{-\ell}\) is an average of \(m\) independent centered variables with \(\psi_2\)-norm at most \(2C\) by~\eqref{eq:thy-vector-subgaussian}, so the general Hoeffding inequality~\citep[Thm.~2.6.2]{vershynin2018high} and a union bound over \(p_n\) coordinates give \(\|\bar Z_{-\ell}\|_\infty\le Mt_n\) with probability tending to one uniformly, after enlarging \(M\) if necessary; then \(\|\bar Z_{-\ell}\bar Z_{-\ell}'\|_{\max}\le M^2t_n^2\le Mt_n\) for \(n\) large.  By~\Cref{lemm:app-cov-means}, \(\tilde R_i\) is centered with variance bounded uniformly, so \(|\bar{\tilde R}_{-\ell}|\le1\) with probability tending to one by Chebyshev's inequality, and then \(\|\bar Z_{-\ell}\bar{\tilde R}_{-\ell}\|_\infty\le Mt_n\).  Replacing \(M\) by \(3M\) absorbs these terms, which proves the claim.
\end{proof}

\begin{remark}[Penalty scale, sparsity cost and cross-validated penalties]\label{rem:penalty-scale}\label{rem:sparsity-cost}
Before projection, the fixed-direction score \((\widehat\Sigma_Z-\Sigma_X)b_{R0}-\{\widehat c_{RZ}-\E_P[Z_iR_i]\}\) contracts at the order \(\sqrt{\log p_n/n}\) by~\eqref{eq:thy-fixed-vector-contraction} and~\eqref{eq:thy-gram-fixed-vector} of~\Cref{prop:sparse-bilinear} with \(S=I_\ell^c\); subtracting the estimated covariance correction generates no sparsity cost.  The projection acts on \(b_{R0}\) only through the max-norm bound of~\Cref{lemm:psd-projection-cost}, so when \(\|b_{R0}\|_1\) is controlled only through \(\sqrt s\|b_{R0}\|_2\) and the projection is active, it is the source of the factor \(\sqrt s\) in the sufficient penalty of~\Cref{lemm:penalty-level} and of the budget of~\Cref{coro:cocolasso-core}(a).  The factor \(\log\bar p_n\), which equals \(\log p_n\) when \(p_n\ge n\), keeps the probability in condition~(i) of~\Cref{prop:corrected-lasso-specialization} tending to one even when \(p_n\) stays bounded.  \citet[Thm.~1]{datta2017cocolasso} require a penalty above a multiple of \(s\sqrt{\log p_n/n}\), which reflects the same mechanism when the coefficients are bounded in \(\ell_\infty\), so that \(\|b_{R0}\|_1\le s\|b_{R0}\|_\infty\); under our bound \(\|b_{R0}\|_2\le C\) the factor is \(\sqrt s\), and a penalty of the same order \(s\sqrt{\log p_n/n}\) appears in~\citet[Lem.~1]{lilima2021measurement}.  Bounded target \(\ell_1\) norms remove the factor (\Cref{coro:l1-bounded}), and so does an inactive projection, which \(p_n\le c_\ast n\) ensures with probability approaching one for the projections of~\Cref{lemm:inactive-projection} (\Cref{coro:inactive-projection}).  When the projection is active, some certified near-minimizer \(A\) with \(\kappa>1\) has \(\|(A-\widehat\Sigma_Z)b_{R0}\|_\infty\ge\frac{\kappa-1}{2}\,d\,\|b_{R0}\|_1\), where \(d>0\) is the max-norm distance of \(\widehat\Sigma_Z\) to \(\{B:B\succeq\lambda_{\mathrm{fl}}I\}\) (\Cref{rem:active-projection-tight}); whether the exact projection attains this order is not settled by these bounds.  Convexity itself is not the source of the cost: when the measurement errors are uncorrelated across coordinates, the targets are exactly sparse with \(s\le c(n/\log p)^{1/2}\), and the error variances are estimated at rate \((\log p/n)^{1/2}\) in sup-norm, the convex conic estimator of~\citet[Thm.~1 and Cor.~1]{bellonirosenbaumtsybakov2017conic} has \(\ell_q\) errors of order \(s^{1/q}(\log p/n)^{1/2}(1+\|b_{R0}\|_2)\), so the target enters through \(\|b_{R0}\|_2\) rather than \(\|b_{R0}\|_1\).  The distinction between \(\ell_2\)- and \(\ell_1\)-bounded coefficient classes matters for inference after regularization more generally: in a Gaussian linear model without measurement error and with many more controls than observations, the optimal linear bias-aware confidence interval for a scalar coefficient has half-length of order \((1+C\sqrt p)/\sqrt n\) when the \(p\) control coefficients \(b\) of the outcome equation satisfy \(\|b\|_2\le C\), against \((1+C\sqrt{\log p})/\sqrt n\) when \(\|b\|_1\le C\)~\citep[Sec.~4.3]{armstrongkolesarkwon2026biasaware}; that estimator and model differ from ours, so these results do not show that our \(\ell_2\) budget is necessary.

With deterministic penalties the corrected Lasso is the first stage covered by~\Crefinorder{coro:exact-sparsity,coro:l1-bounded,coro:inactive-projection}.  Penalties cross-validated over a grid whose smallest value is a fixed fraction of \(\|\widehat c_{RZ,-\ell}\|_\infty\) are not of these orders when \(\|\operatorname{Cov}_P(Z_i,R_i)\|_\infty\) is bounded away from zero (\Cref{app:cv-window}), so the feasible estimator is defined with the shrinking grid of~\Cref{rem:thy-cv-not-covered}; at every sample size in this paper, that grid coincides with the fixed-ratio grid.  Existing theory for the cross-validated Lasso~\citep{chetverikov2021cross} concerns the ordinary Lasso; \citet[Sec.~5]{datta2017cocolasso} propose calibrated cross-validation for CoCoLasso, and~\citet{dattazou2020note} prove the consistency of its leave-one-out version for a fixed number of covariates and a known error covariance.
\end{remark}

\begin{corollary}[Corrected-Lasso rates with \(\ell_2\)-bounded targets]\label{coro:exact-sparsity}\label{coro:approx-sparsity}
Let conditions~\textup{(1)}--\textup{(5)} of~\Cref{def:dgpclass} hold with \(c_X>\lambda_{\mathrm{fl}}>0\) and \(\log p_n=o(n)\), take \(s_{\pi,n},s_{g,n}\ge1\), and let the first stages be the corrected Lasso~\eqref{eq:est-cocolasso-objective} with projections certified as in~\eqref{eq:thy-certified-projection} with a fixed \(\kappa\ge1\) and \(\bar\eta_n\le\sqrt{\log\bar p_n/n}\), and the deterministic penalties
\begin{equation}
\lambda_Y=C_\lambda\sqrt{\frac{s_{\pi,n}\log\bar p_n}{n}},
\qquad
\lambda_D=C_\lambda\sqrt{\frac{s_{g,n}\log\bar p_n}{n}},
\label{eq:thy-cocolasso-penalty-scale}
\end{equation}
where \(C_\lambda\) is the constant of~\Cref{lemm:penalty-level}.  Define
\begin{gather*}
\rho_{\pi,n}=s_{\pi,n}\sqrt{\frac{\log\bar p_n}{n}}+a_{\pi,n},
\qquad
\rho_{g,n}=s_{g,n}\sqrt{\frac{\log\bar p_n}{n}}+a_{g,n},\\
\begin{aligned}
\zeta_{\pi,n}
&=
 s_{\pi,n}^{3/2}\sqrt{\frac{\log\bar p_n}{n}}
+\sqrt{s_{\pi,n}}a_{\pi,n}
+a_{\pi,n}^2\sqrt{\frac{n}{s_{\pi,n}\log\bar p_n}},\\
\zeta_{g,n}
&=
 s_{g,n}^{3/2}\sqrt{\frac{\log\bar p_n}{n}}
+\sqrt{s_{g,n}}a_{g,n}
+a_{g,n}^2\sqrt{\frac{n}{s_{g,n}\log\bar p_n}}.
\end{aligned}
\end{gather*}
Then, uniformly over \(P\) and folds, the slope prediction errors are \(O_P(\rho_{R,n})\) and the slope \(\ell_1\) errors \(O_P(\zeta_{R,n})\), and \(\zeta_{R,n}\sqrt{\log\bar p_n/n}\le(\sqrt{s_{R,n}}\sqrt{\log\bar p_n/n}+a_{R,n})\rho_{R,n}\), so the affine prediction errors are \(O_P(\rho_{R,n})\) whenever \(\rho_{R,n}\to0\).  Under weak-\(\ell_q\) coefficient decay, substituting the implied best-\(s\) approximation rates for \(a_{f,n}\), \(a_{g,n}\) and \(a_{\pi,n}\) gives sufficient conditions on the sparsity levels (\Cref{rem:polynomial-decay}).
\end{corollary}

\begin{proof}
The sparsity levels \(s_{\pi,n}\) and \(s_{g,n}\) of condition~(2) of~\Cref{def:dgpclass} are upper bounds on \(s_Y=\|\pio\|_0\) and \(s_D=\|\gammao\|_0\) for every \(P\).  \Cref{lemm:penalty-level}, applied with \(s=s_{\pi,n}\) for \(R=Y\) and with \(s=s_{g,n}\) for \(R=D\), gives condition~(i) of~\Cref{prop:corrected-lasso-specialization} at the penalties~\eqref{eq:thy-cocolasso-penalty-scale}, and condition~(ii) holds because \(\lambda_{\mathrm{fl}}>0\) (\Cref{rem:relaxed-cone}).  \Cref{prop:corrected-lasso-specialization} gives the slope bounds, because \(\sqrt{s_R}\lambda_R+a_R\le C_\lambda\rho_{R,n}\) and \(s_R\lambda_R+\sqrt{s_R}a_R+a_R^2/\lambda_R\le C\zeta_{R,n}\).  With \(t_n:=\sqrt{\log\bar p_n/n}\), \(\zeta_{R,n}t_n=s_{R,n}^{3/2}t_n^2+\sqrt{s_{R,n}}a_{R,n}t_n+a_{R,n}^2/\sqrt{s_{R,n}}\), while \((\sqrt{s_{R,n}}t_n+a_{R,n})\rho_{R,n}=s_{R,n}^{3/2}t_n^2+\sqrt{s_{R,n}}a_{R,n}t_n+a_{R,n}s_{R,n}t_n+a_{R,n}^2\), which is at least \(\zeta_{R,n}t_n\) because \(s_{R,n}\ge1\).  If \(\rho_{R,n}\to0\), then \(\sqrt{s_{R,n}}t_n+a_{R,n}\le2\rho_{R,n}\to0\) and \(n^{-1/2}\le(\log2)^{-1/2}t_n\le(\log2)^{-1/2}\rho_{R,n}\), so~\Cref{lem:affine-intercept} gives the affine rate.
\end{proof}

\subsection{Bounded target \texorpdfstring{\(\ell_1\)}{l1} norms}\label{app:l1-bounded}

The factor \(\sqrt s\), and with it the budget of~\Cref{coro:cocolasso-core}(a), enters through the bound \(\|b_{R0}\|_1\le\sqrt{s_R}\|b_{R0}\|_2\le C\sqrt s\) in the proof of~\Cref{lemm:penalty-level}; under the bound \(\max\{\|\pio\|_1,\|\gammao\|_1\}\le C_1\) the same argument needs no sparsity input.  \citet[Rem.~2]{lilima2020arxiv} note, without proof, that a bounded target \(\ell_1\) norm gives CoCoLasso the error bounds of the nonconvex and conic corrected estimators; the results below prove the claim for the rate conditions of~\Cref{assu:first-stage-rates}, with approximate sparsity and training-mean intercepts.  At fixed \(s\) the budgets of cases~(a) and~(b) of~\Cref{coro:cocolasso-core} both reduce to \(\log\bar p_n=o(\sqrt n)\).  From here on, \(t_n:=\sqrt{\log\bar p_n/n}\), and \(b_{R0}\), \(s_R\), \(a_R\) and \(r_{R,i}\) are as in~\Cref{app:firststage-cl}.

\begin{corollary}[Corrected-Lasso first stages with bounded target \(\ell_1\) norms]\label{coro:l1-bounded}
Fix a constant \(C_1<\infty\), the same for every \(n\), and let \(\mathcal P_{1,n}\) be the set of distributions that satisfy conditions~(1)--(6) of~\Cref{def:dgpclass} and \(\max\{\|\pio\|_1,\|\gammao\|_1\}\le C_1\).  Let the first stages be as in~\Cref{coro:exact-sparsity}, but with \(\lambda_Y=\lambda_D=C_\lambda^{(1)}\sqrt{\log\bar p_n/n}\), where \(C_\lambda^{(1)}\) is the constant of~\Cref{lemm:penalty-level-l1}, and let \((\rho_{\pi,n},\zeta_{\pi,n})\) and \((\rho_{g,n},\zeta_{g,n})\) be
\[
\rho=\sqrt{s\log\bar p_n/n}+a, \qquad \zeta=s\sqrt{\log\bar p_n/n}+\sqrt s\,a+a^2\sqrt{n/\log\bar p_n},
\]
evaluated at \((s,a)=(s_{\pi,n},a_{\pi,n})\) and at \((s,a)=(s_{g,n},a_{g,n})\), respectively, with \(s_{\pi,n},s_{g,n}\ge1\).  If \(\log p_n=o(n)\), then, uniformly over \(P\in\mathcal P_{1,n}\) and folds, the slope prediction errors are \(O_P(\rho_{R,n})\), the slope \(\ell_1\) errors are \(O_P(\zeta_{R,n})\), and \(\zeta_{R,n}\sqrt{\log\bar p_n/n}\le\rho_{R,n}^2\), so the affine prediction errors are \(O_P(\rho_{R,n})\) whenever \(\rho_{R,n}\to0\).  If the population linear-projection coefficients of \(f_0\) and \(g_0\) satisfy \(|b|_{(j)}\le Aj^{-\varsigma}\) with \(\varsigma>1\), the \(\ell_1\) bound holds for the best-\(s\) representations~\eqref{eq:model-best-s-beta}--\eqref{eq:model-best-s-gamma}, with \(C_1\) depending only on \(A\), \(\varsigma\) and the constants of~\Cref{def:dgpclass} (\Cref{lemm:l1-decay}).
\end{corollary}
The constants \(C_\lambda\) and \(C_\lambda^{(1)}\) depend on the constants of~\Cref{def:dgpclass}; in practice the penalties are chosen by cross-validation (\Cref{rem:thy-cv-not-covered}).

\begin{lemma}[Penalty level under bounded target \(\ell_1\) norms]\label{lemm:penalty-level-l1}
Let \(\mathcal P_{1,n}\) be as in~\Cref{coro:l1-bounded} \textup{(}condition~\textup{(6)} of~\Cref{def:dgpclass} is not used here\textup{)}, and suppose \(\log p_n=o(n)\).  For a training fold and \(R\in\{Y,D\}\), let \(\widehat\Sigma_Z^+\) satisfy~\eqref{eq:thy-certified-projection} with a fixed \(\kappa\ge1\) and \(\bar\eta=\bar\eta_n\le\sqrt{\log\bar p_n/n}\), and let \(\mathcal S_R\) be the projected score~\eqref{eq:thy-projected-score}.  There is a constant \(C_\lambda^{(1)}\), depending only on \(C_1\), \(\kappa\) and the constants of~\Cref{def:dgpclass}, such that every deterministic penalty sequence with \(\lambda_R\ge C_\lambda^{(1)}\sqrt{\log\bar p_n/n}\) satisfies
\[
\lim_{n\to\infty}\sup_{P\in\mathcal P_{1,n}}\Pr_P\bigl(\|\mathcal S_R\|_\infty>\lambda_R/8\bigr)=0
\]
for every fold.  In particular, such penalties satisfy condition~\textup{(i)} of~\Cref{prop:corrected-lasso-specialization} and the score condition of~\Cref{lemm:kkt-perturbation}; they depend neither on \(s_R\) nor on any sparsity bound.
\end{lemma}

\begin{proof}
The bound~\eqref{eq:app-score-bound} holds for every realization at which \(\widehat\Sigma_Z^+\) is certified; the proof of~\Cref{lemm:penalty-level} continues with \(\|b_{R0}\|_1\le C\sqrt s\), and here \(\|b_{R0}\|_1\le C_1\) instead.  The second paragraph of the proof of~\Cref{lemm:penalty-level} constructs a constant \(M\), depending only on the constants of~\Cref{def:dgpclass}, such that the events \(\mathcal E_{1,n}:=\{\|\widehat\Sigma_{Z,-\ell}-\Sigma_X\|_{\max}\le Mt_n\}\) and \(\mathcal E_{2,n}:=\{\|\widehat c_{RZ,-\ell}-\E_P[Z_iR_i]\|_\infty\le Mt_n\}\) have probability tending to one uniformly over the distributions that satisfy conditions~(1)--(5) and \(\log p_n=o(n)\); the construction involves neither \(s\), \(s_R\) nor \(b_{R0}\), so it applies a fortiori to \(\mathcal P_{1,n}\).  Take \(M\) after its final enlargement.  On \(\mathcal E_{1,n}\cap\mathcal E_{2,n}\), with \(\bar\eta_n\le t_n\) and \(C_1':=\max\{1,C_1\}\), \(\|\mathcal S_R\|_\infty\le C_1\{(1+\kappa)M+1\}t_n+Mt_n\le\{C_1'(2+\kappa)M+C_1'\}t_n\).  Hence \(C_\lambda^{(1)}:=8\{C_1'(2+\kappa)M+C_1'\}\) has the stated property: every \(\lambda_R\ge C_\lambda^{(1)}t_n\) gives \(\|\mathcal S_R\|_\infty\le\lambda_R/8\) on \(\mathcal E_{1,n}\cap\mathcal E_{2,n}\), and the number of folds is fixed.
\end{proof}

\begin{lemma}[Bounded \(\ell_1\) norms under polynomial decay]\label{lemm:l1-decay}
Let \(h_0\in\{f_0,g_0\}\), and let \(b^h:=\Sigma_X^{-1}\E_P[X_ih_0(X_i)]\) be its population linear-projection coefficient on all \(p_n\) coordinates, under the normalization \(\E X_i=0\) and \(\E h_0(X_i)=0\) of~\Cref{subsec:model-omega}; write \(e_h:=h_0(X_i)-X_i'b^h\).  Suppose that condition~\textup{(5)} of~\Cref{def:dgpclass} holds and that the ordered magnitudes satisfy \(|b^h|_{(j)}\le Aj^{-\varsigma}\) for all \(j\le p_n\), with fixed \(A>0\) and \(\varsigma>1\).  Then, for every \(s\in\{1,\ldots,p_n\}\) and every best-\(s\) coefficient vector \(b_s^\star\in\argmin_{\|b\|_0\le s}\E[\{h_0(X_i)-X_i'b\}^2]\), as in~\eqref{eq:model-best-s-beta}--\eqref{eq:model-best-s-gamma},
\begin{enumerate}[label=\textup{(\roman*)}]
\item \(\|b_s^\star\|_1\le A\{\varsigma/(\varsigma-1)+(C/c_X)^{3/2}(2\varsigma-1)^{-1/2}s^{1-\varsigma}\}\le A\{\varsigma/(\varsigma-1)+(C/c_X)^{3/2}(2\varsigma-1)^{-1/2}\}=:C_A\);
\item \(\|b_s^\star\|_2\le\|b^h\|_2+(C/c_X)^{3/2}(2\varsigma-1)^{-1/2}As^{1/2-\varsigma}\);
\item \(\E[\{h_0(X_i)-X_i'b_s^\star\}^2]\le\E[e_h^2]+CA^2s^{1-2\varsigma}/(2\varsigma-1)\).
\end{enumerate}
If \(f_0\) and \(g_0\) both satisfy these conditions, with the same \(A\) and \(\varsigma\), and \(\betao\) and \(\gammao\) are best-\(s\) vectors, then, under condition~\textup{(1)}, \(\max\{\|\pio\|_1,\|\gammao\|_1\}\le(1+C)C_A\) for every choice of the sparsity sequences.  The \(s\)-term truncation \(b^{(s)}\) of \(b^h\) satisfies \(\|b^{(s)}\|_1\le\|b^h\|_1\le A\varsigma/(\varsigma-1)\) directly.
\end{lemma}

\begin{proof}
Because \(\E[X_ie_h]=0\), every \(b\) satisfies \(\E[\{h_0(X_i)-X_i'b\}^2]=\E[e_h^2]+(b^h-b)'\Sigma_X(b^h-b)\).  The \(s\)-term truncation \(b^{(s)}\), which keeps the \(s\) largest coordinates of \(b^h\) in absolute value, is feasible, so the optimality of \(b_s^\star\) and condition~(5) give \((b^h-b_s^\star)'\Sigma_X(b^h-b_s^\star)\le(b^h-b^{(s)})'\Sigma_X(b^h-b^{(s)})\le C\sum_{j>s}|b^h|_{(j)}^2\le CA^2\sum_{j>s}j^{-2\varsigma}\le CA^2s^{1-2\varsigma}/(2\varsigma-1)\), which proves~(iii).  Let \(T:=\supp(b_s^\star)\), so that \(|T|\le s\); if \(T\) is empty, (i) and (ii) are trivial.  The left side of the last chain is at least \(c_X\|b^h-b_s^\star\|_2^2\ge c_X\|b^h_{T^c}\|_2^2\), because \(b_s^\star\) vanishes off \(T\), so \(\|b^h_{T^c}\|_2\le(C/c_X)^{1/2}As^{1/2-\varsigma}(2\varsigma-1)^{-1/2}\).  The vector \(b_s^\star\) also minimizes \(\E[\{h_0(X_i)-X_i'b\}^2]\) over the vectors supported in \(T\).  The block \(\Sigma_{TT}\) of \(\Sigma_X\) satisfies \(\Sigma_{TT}\succeq c_XI\) by condition~(5) and Cauchy interlacing, so this minimizer is unique and equals \(\Sigma_{TT}^{-1}\E[X_{iT}h_0(X_i)]\), where \(X_{iT}\) collects the coordinates in \(T\).  Since \(\E[X_{iT}h_0(X_i)]=\Sigma_{TT}b^h_T+\Sigma_{TT^c}b^h_{T^c}\), we have \(b^\star_{s,T}=b^h_T+\Sigma_{TT}^{-1}\Sigma_{TT^c}b^h_{T^c}\).  The vector \(\Sigma_{TT^c}b^h_{T^c}\) is the \(T\)-block of \(\Sigma_X\tilde v\), where \(\tilde v\) equals \(b^h\) on \(T^c\) and zero on \(T\), so its Euclidean norm is at most \(C\|b^h_{T^c}\|_2\); with \(\|\Sigma_{TT}^{-1}\|_{\mathrm{op}}\le1/c_X\), the correction term has Euclidean norm at most \((C/c_X)\|b^h_{T^c}\|_2\).  Consequently \(\|b_s^\star\|_1=\|b^\star_{s,T}\|_1\le\|b^h_T\|_1+\sqrt{|T|}\,(C/c_X)\|b^h_{T^c}\|_2\le\|b^h\|_1+(C/c_X)^{3/2}As^{1-\varsigma}(2\varsigma-1)^{-1/2}\), and \(\|b^h\|_1\le A\sum_{j\ge1}j^{-\varsigma}\le A\{1+\int_1^\infty x^{-\varsigma}\,dx\}=A\varsigma/(\varsigma-1)\), while \(s^{1-\varsigma}\le1\); this proves~(i).  Part~(ii) follows in the same way from \(\|b_s^\star\|_2\le\|b^h_T\|_2+(C/c_X)\|b^h_{T^c}\|_2\).  Finally, \(\|\pio\|_1\le\|\betao\|_1+|\thetao|\,\|\gammao\|_1\le(1+C)C_A\) because \(|\thetao|\le C\) by condition~(1), and \(\|b^{(s)}\|_1\le\|b^h\|_1\) because truncation removes coordinates.
\end{proof}

\begin{proof}[Proof of~\Cref{coro:l1-bounded}]
As in~\Cref{prop:corrected-lasso-specialization}, \(R=Y\) has target \(b_{Y0}=\pio\), sparsity \(s_Y\le s_{\pi,n}\) and approximation bound \(a_Y=a_{\pi,n}\), and \(R=D\) has \(\gammao\), \(s_D\le s_{g,n}\) and \(a_D=a_{g,n}\); write \(\rho_{Y,n}=\rho_{\pi,n}\), \(\rho_{D,n}=\rho_{g,n}\), and likewise for \(\zeta\), \(s\) and \(a\).

\emph{Rates.}  By~\Cref{lemm:penalty-level-l1}, the penalties \(\lambda_R=C_\lambda^{(1)}t_n\) satisfy condition~(i) of~\Cref{prop:corrected-lasso-specialization} and the score condition of~\Cref{lemm:kkt-perturbation} uniformly over \(\mathcal P_{1,n}\), and condition~(ii) holds because \(\lambda_{\mathrm{fl}}>0\) (\Cref{rem:relaxed-cone}).  \Cref{prop:corrected-lasso-specialization} then gives, uniformly over \(\mathcal P_{1,n}\) and folds, \(\|X_i'(\widehat b_R-b_{R0})\|_{P,2}=\Op(\sqrt{s_R}\lambda_R+a_R)=\Op(\rho_{R,n})\) and \(\|\widehat b_R-b_{R0}\|_1=\Op(s_R\lambda_R+\sqrt{s_R}\,a_R+a_R^2/\lambda_R)=\Op(\zeta_{R,n})\), because \(s_R\le s_{R,n}\), \(a_R\le a_{R,n}\) and \(a_R^2/\lambda_R=a_R^2\sqrt{n/\log\bar p_n}/C_\lambda^{(1)}\).

\emph{Intercepts.}  With \(t_n=\sqrt{\log\bar p_n/n}\), \(\zeta_{R,n}t_n=s_{R,n}t_n^2+\sqrt{s_{R,n}}a_{R,n}t_n+a_{R,n}^2\le(\sqrt{s_{R,n}}t_n+a_{R,n})^2=\rho_{R,n}^2\).  The training-mean intercepts add \(\Op(n^{-1/2}+\zeta_{R,n}\sqrt{\log p_n/n})\) to the affine prediction errors (\Cref{lem:affine-intercept}), which is \(\Op(\rho_{R,n})\) when \(\rho_{R,n}\to0\), because \(n^{-1/2}\le(\log2)^{-1/2}t_n\le(\log2)^{-1/2}\rho_{R,n}\).  With deterministic penalties the fold-\(\ell\) fit is a function of the training observations alone; the computed estimator is defined jointly by the objective and the deterministic algorithm (\Cref{rem:projection-certificate}).  The statement on polynomial decay is~\Cref{lemm:l1-decay}, with \(C_1=(1+C)C_A\).
\end{proof}

\begin{remark}[Polynomially decaying coefficients]\label{rem:polynomial-decay}
Suppose that, beyond heads of fixed size, the dictionary coefficients of \(f_0\) and \(g_0\) decay as \(|b_j|\asymp j^{-\varsigma}\) for some \(\varsigma>1/2\) (a weak-\(\ell_q\) decay with \(q=1/\varsigma\) in the sense of~\Cref{app:model-sparsity-details}), that \(p_n\ge2s_n\), and that condition~(5) holds.  The best-\(s_n\) approximation errors are then \(a_{f,n},a_{g,n}\asymp s_n^{1/2-\varsigma}\).  When these rates are substituted into the rates of~\Cref{coro:approx-sparsity} and the condition~\eqref{eq:thy-product-rate} with a common sparsity level \(s_n\), the cross term \(\sqrt n\,a_{g,n}\rho_{\beta,n}\) is at least of order \(s_n^{3/2-\varsigma}\sqrt{\log\bar p_n}\), and the approximation-derivative terms are at least of order \(\sqrt n\,s_n^{1-2\varsigma}\), because \(\rho_{\beta,n}\ge|\thetao|a_{g,n}\) and \(\rho_{g,n}\ge a_{g,n}\); these terms and the estimation term \(\sqrt n\,\rho_{\beta,n}\rho_{g,n}\) vanish together only if \(n^{1/(4\varsigma-2)}\ll s_n\ll n^{1/4}(\log\bar p_n)^{-1/2}\). Condition~\eqref{eq:thy-product-rate} can therefore hold for these rates along some sequence \(s_n\) only if \(\varsigma>3/2\).  With first stages whose prediction rates are of order \(\sqrt{s_n\log\bar p_n/n}\) plus the approximation error, which the corrected Lasso attains under bounded target \(\ell_1\) norms (cf.~\citealp[Rem.~2]{lilima2020arxiv}), the same calculation requires only \(\varsigma>1\); best-\(s\) representations have bounded \(\ell_1\) norms when \(\varsigma>1\) (\Cref{lemm:l1-decay}), and~\Cref{coro:l1-decay} shows that the class \(\varsigma>1\) then meets~\eqref{eq:thy-product-rate} for the rates of~\Cref{coro:l1-bounded} with deterministic penalties and suitable sparsity levels, for example when \(\log\bar p_n=O(\log n)\) and \(n^{1/(4\varsigma-2)}=o(p_n)\).  These thresholds come from first-stage rates; the generic condition~\eqref{eq:thy-product-rate}, and~\eqref{eq:thy-covariance-rate} for the variants of~\Cref{prop:plugin-inference}, restrict only those rates.
\end{remark}

\begin{corollary}[Polynomial decay under bounded target \texorpdfstring{\(\ell_1\)}{l1} norms]\label{coro:l1-decay}
Let conditions~\textup{(1)}--\textup{(6)} of~\Cref{def:dgpclass} hold, with linear nuisances \textup{(}\(e_f=e_g=0\) in the notation of~\Cref{lemm:l1-decay}\textup{)}, the best-\(s\) representations~\eqref{eq:model-best-s-beta}--\eqref{eq:model-best-s-gamma}, a common sparsity level \(s_{f,n}\asymp s_{g,n}\asymp s_{\pi,n}\asymp s_n\), and \(p_n\ge2\max\{s_{f,n},s_{g,n}\}\).  Suppose that the ordered coefficients of \(f_0\) and \(g_0\) satisfy \(|b|_{(j)}\le Aj^{-\varsigma}\) for all \(j\) and \(|b|_{(j)}\ge A^{-1}j^{-\varsigma}\) beyond heads of fixed size, for a fixed \(A>0\) and some \(\varsigma>1/2\), so that \(a_{f,n},a_{g,n}\asymp s_n^{1/2-\varsigma}\), and let the first stages be as in~\Cref{coro:l1-bounded}.
\begin{enumerate}[label=\textup{(\alph*)}]
\item If \(\varsigma>1\), the bound on the target \(\ell_1\) norms of~\Cref{coro:l1-bounded} holds by~\Cref{lemm:l1-decay}, and, for the rates of that corollary, condition~\eqref{eq:thy-product-rate} and, for the variants of~\Cref{prop:plugin-inference}, condition~\eqref{eq:thy-covariance-rate} hold along every sequence \(s_n\) with
\[
\max\bigl\{n^{1/(4\varsigma-2)},(\log\bar p_n)^{1/(2\varsigma-2)}\bigr\}\ll s_n\ll\frac{\sqrt n}{\log\bar p_n};
\]
such a sequence with \(2s_n\le p_n\) exists whenever both lower ends are \(o(\min\{p_n,\sqrt n/\log\bar p_n\})\), for example when \(\log\bar p_n=O(\log n)\) and \(n^{1/(4\varsigma-2)}=o(p_n)\).
\item If \(\varsigma\le1\), condition~\eqref{eq:thy-product-rate} fails for the rates of~\Cref{coro:l1-bounded} along every sequence \(s_n\), and the \(\ell_1\) norms of the population coefficient vectors \(b^h\) of \(f_0\) and \(g_0\) diverge as \(p_n\to\infty\), because \(|b|_{(j)}\ge A^{-1}j^{-\varsigma}\) beyond the head.
\end{enumerate}
These are sufficient conditions, not impossibility results for the estimator.
\end{corollary}

\begin{proof}
With \(e_f=e_g=0\), the argument for part~(iii) of~\Cref{lemm:l1-decay}, which uses only \(\varsigma>1/2\), gives \(a_{f,n},a_{g,n},a_{\pi,n}\lesssim s_n^{1/2-\varsigma}\); conversely, as in that proof, the best-\(s\) errors at \(s=s_{f,n}\) and \(s=s_{g,n}\) are at least \((c_X\sum_{s<j\le p_n}|b|_{(j)}^2)^{1/2}\ge cs_n^{1/2-\varsigma}\) by the lower bound on \(|b|_{(j)}\).

(a) By~\Cref{lemm:l1-decay}, the target \(\ell_1\) norms are at most \((1+C)C_A\) for every \(s_n\).  Every term of~\eqref{eq:thy-product-rate} and of~\eqref{eq:thy-covariance-rate} is bounded by a constant times \(T_1+T_2+T_3\), where \(T_1:=s_n\log\bar p_n/\sqrt n\), \(T_2:=s_n^{1-\varsigma}\sqrt{\log\bar p_n}\) and \(T_3:=\sqrt n\,s_n^{1-2\varsigma}\).  Indeed, \(\sqrt n\,a_{f,n}a_{g,n}\lesssim T_3\); \(\sqrt n\,a_{g,n}\rho_{\beta,n}\lesssim\sqrt n\,s_n^{1/2-\varsigma}\{\sqrt{s_n\log\bar p_n/n}+s_n^{1/2-\varsigma}\}=T_2+T_3\), and likewise for \(\sqrt n\,a_{f,n}\rho_{g,n}\); \(\sqrt n\,\rho_{\beta,n}\rho_{g,n}\lesssim T_1+2T_2+T_3\); and \(\sqrt{\log p_n}\,\zeta_{R,n}\lesssim\sqrt{\log\bar p_n}\{s_n\sqrt{\log\bar p_n/n}+s_n^{1-\varsigma}+s_n^{1-2\varsigma}\sqrt{n/\log\bar p_n}\}=T_1+T_2+T_3\).  Now \(T_1\to0\) if and only if \(s_n\ll\sqrt n/\log\bar p_n\), \(T_3\to0\) if and only if \(s_n\gg n^{1/(4\varsigma-2)}\), and, for \(\varsigma>1\), \(T_2\to0\) if and only if \(s_n\gg(\log\bar p_n)^{1/(2\varsigma-2)}\).  Let \(\underline s_n\) be the larger of the two lower ends and \(\bar s_n:=\min\{p_n/2,\sqrt n/\log\bar p_n\}\).  If \(\underline s_n=o(\bar s_n)\), then \(s_n:=\lceil(\underline s_n\bar s_n)^{1/2}\rceil\) lies in the window and satisfies \(2s_n\le p_n\) for large \(n\), because \(\underline s_n\to\infty\) and \((\underline s_n\bar s_n)^{1/2}\le s_n\le(\underline s_n/\bar s_n)^{1/2}\bar s_n+1\).  If \(\log\bar p_n=O(\log n)\) and \(n^{1/(4\varsigma-2)}=o(p_n)\), then \(\underline s_n=o(\bar s_n)\), because \(1/(4\varsigma-2)<1/2\) for \(\varsigma>1\) and \((\log\bar p_n)^{1/(2\varsigma-2)}=O\{(\log n)^{1/(2\varsigma-2)}\}=o(n^{1/(4\varsigma-2)})\).

(b) Since \(a_{g,n}\ge cs_n^{1/2-\varsigma}\) and \(\rho_{\beta,n}\ge\rho_{\pi,n}\ge\sqrt{s_{\pi,n}\log\bar p_n/n}\) with \(s_{\pi,n}\asymp s_n\), the cross term satisfies \(\sqrt n\,a_{g,n}\rho_{\beta,n}\ge c's_n^{1-\varsigma}\sqrt{\log\bar p_n}\), which does not tend to zero when \(\varsigma\le1\), because \(\log\bar p_n\ge\log n\to\infty\).  Finally, \(\sum_{j\le p_n}|b|_{(j)}\ge A^{-1}\sum_{j_0<j\le p_n}j^{-\varsigma}\), where \(j_0\) is the size of the head; the sum diverges as \(p_n\to\infty\) when \(\varsigma\le1\).
\end{proof}

\subsection{Inactive projection}\label{app:inactive-projection}

In the proof of~\Cref{lemm:penalty-level}, the factor \(\sqrt s\) can be confined to the projection increment \((\widehat\Sigma_Z^+-\widehat\Sigma_{Z,-\ell})b_{R0}\): the rest of \(\mathcal S_R\) is the unprojected score, which contracts at the order \(\sqrt{\log p_n/n}\) (\Cref{prop:sparse-bilinear} with \(S=I_\ell^c\)).  When \(p_n\) is at most a small fraction of \(n\), the corrected Gram matrix satisfies the floor with probability approaching one, and the increment then has sup norm at most \(\bar\eta_n\|b_{R0}\|_1\), which is \(O(t_n)\) when \(\bar\eta_n\le t_n/\sqrt{p_n}\) and zero when the projection leaves a feasible matrix unchanged (\Cref{lemm:inactive-projection}).

\begin{lemma}[Inactive projection when \(p_n\) is small relative to \(n\)]\label{lemm:inactive-projection}
Suppose conditions~\textup{(1)}--\textup{(5)} of~\Cref{def:dgpclass} hold, with \(C\ge1\) without loss of generality.  Let \(C_{\psi}=1\) be the constant of~\Cref{lemm:app-cov-product}, set \(B_1:=8C^2\), \(B_2:=8C_{\psi}C^3\) and \(B_3:=2C_{\psi}C/\sqrt{\log2}\), let \(c_r>0\) be the constant of~\Cref{lemm:app-cov-bernstein} for the bound \(B_r\) \((r=1,2,3)\), and set
\[
\varepsilon:=\min\Bigl\{1,\frac{c_X-\lambda_{\mathrm{fl}}}{10}\Bigr\}, \qquad c_\ast:=\frac{1}{\log 9}\min\Bigl\{\frac{c_1\varepsilon^2}{6},\frac{3c_3\varepsilon}{32}\Bigr\}.
\]
Suppose that \(p_n\le c_\ast n\) for all large \(n\); in particular, \(p_n/n\to0\) suffices.  For a training fold \(\ell\) and \(R\in\{Y,D\}\), let \(\widehat\Sigma_Z^+\) satisfy~\eqref{eq:thy-certified-projection} with a fixed \(\kappa\ge1\) and \(\bar\eta=\bar\eta_n\le t_n\), as in~\Cref{lemm:penalty-level}, and suppose in addition that either \(\bar\eta_n\le t_n/\sqrt{p_n}\), or \(\widehat\Sigma_Z^+=\widehat\Sigma_{Z,-\ell}\) whenever \(\widehat\Sigma_{Z,-\ell}\succeq\lambda_{\mathrm{fl}}I\).  The algorithm of~\Cref{app:sim-numerics} has the second property up to rounding: it returns \(\widehat\Sigma_{Z,-\ell}\) itself when the computed smallest eigenvalue of \(\widehat\Sigma_{Z,-\ell}\) is at least \(\lambda_{\mathrm{fl}}\), and on the event \(\mathcal F_n\) of part~(i) the exact smallest eigenvalue exceeds \(\lambda_{\mathrm{fl}}\) by at least \((c_X-\lambda_{\mathrm{fl}})/2\).  Let \(\inf_P\) and \(\sup_P\) run over the distributions that satisfy conditions~\textup{(1)}--\textup{(5)} with the given \(p_n\).
\begin{enumerate}[label=\textup{(\roman*)}]
\item The event \(\mathcal F_n:=\{\lambda_{\min}(\widehat\Sigma_{Z,-\ell})\ge(c_X+\lambda_{\mathrm{fl}})/2\}\) satisfies \(\lim_{n\to\infty}\inf_P\Pr_P(\mathcal F_n)=1\).  On \(\mathcal F_n\), \(\widehat\Sigma_{Z,-\ell}\succeq\lambda_{\mathrm{fl}}I\) and \(\|\widehat\Sigma_Z^+-\widehat\Sigma_{Z,-\ell}\|_{\max}\le\bar\eta_n\), with \(\widehat\Sigma_Z^+=\widehat\Sigma_{Z,-\ell}\) in the second case.
\item There is a constant \(C_\lambda^{(2)}\), depending only on the constants of~\Cref{def:dgpclass}, such that every deterministic penalty sequence with \(\lambda_R\ge C_\lambda^{(2)}t_n\) satisfies
\[
\lim_{n\to\infty}\sup_P\Pr_P\bigl(\|\mathcal S_R\|_\infty>\lambda_R/8\bigr)=0
\]
for every fold, where \(\mathcal S_R\) is the projected score~\eqref{eq:thy-projected-score}.  Such penalties satisfy condition~\textup{(i)} of~\Cref{prop:corrected-lasso-specialization} and the score condition of~\Cref{lemm:kkt-perturbation}; they depend neither on \(s_R\) nor on \(\|b_{R0}\|_1\).
\end{enumerate}
\end{lemma}

\begin{proof}
Write \(m:=|I_\ell^c|\ge n/3\), \(\widehat\Sigma:=\widehat\Sigma_{Z,-\ell}\) and \(b_0:=b_{R0}\), so that \(\|b_0\|_2\le C\) by condition~(1).  For a scalar \(G\), \(\|G^2\|_{\psi_1}=\|G\|_{\psi_2}^2\) by the definitions of the two norms, and \(\|G-\E G\|_{\psi_1}\le2\|G\|_{\psi_1}\) by~\Cref{lemm:app-cov-product}.  By~\eqref{eq:thy-vector-subgaussian}, every unit vector \(x\) satisfies \(\|x'X_i\|_{\psi_2}\le C\) and \(\|x'\nu_{ik}\|_{\psi_2}\le C\), hence \(\|x'Z_i\|_{\psi_2}\le2C\) and \(\|x'(\nu_{ik}-\bar\nu_i)\|_{\psi_2}\le2C\).

\emph{Step 1 (nets).}  Let \(\mathcal N\) be a maximal \(1/4\)-separated subset of the unit sphere \(S^{p_n-1}\).  Every unit vector lies within \(1/4\) of \(\mathcal N\), and the balls of radius \(1/8\) centered at the points of \(\mathcal N\) are disjoint and contained in the ball of radius \(9/8\), so \(|\mathcal N|\le9^{p_n}\).  For a symmetric matrix \(A\), take a unit \(x\) with \(|x'Ax|=\|A\|_{\mathrm{op}}\) and \(y\in\mathcal N\) with \(\|x-y\|_2\le1/4\); then \(|x'Ax-y'Ay|\le|(x-y)'Ax|+|y'A(x-y)|\le\|A\|_{\mathrm{op}}/2\), so \(\|A\|_{\mathrm{op}}\le2\max_{y\in\mathcal N}|y'Ay|\).  For a vector \(z\ne0\), the same choice with \(x=z/\|z\|_2\) gives \(y'z\ge\frac34\|z\|_2\), so \(\|z\|_2\le\frac43\max_{y\in\mathcal N}|y'z|\).

\emph{Step 2 (decomposition).}  As in the proof of~\Cref{lemm:penalty-level}, \(\widehat\Sigma-\Sigma_X=\bar Q^{(1)}-\bar Q^{(2)}-\bar Z_{-\ell}\bar Z_{-\ell}'\), with the symmetric matrices \(\bar Q^{(1)}:=\mathbb P_{n,-\ell}(Z_iZ_i')-\E_P[Z_iZ_i']\) and \(\bar Q^{(2)}:=\widehat\Omega_{-\ell}-\Omegao=\mathbb P_{n,-\ell}H_i\).

\emph{Step 3 (quadratic forms).}  Fix a unit vector \(x\).  The variables \(x'(Z_iZ_i'-\E_PZ_iZ_i')x=(x'Z_i)^2-\E_P(x'Z_i)^2\) and \(x'H_ix=\{K(K-1)\}^{-1}\sum_{k=1}^K\{x'(\nu_{ik}-\bar\nu_i)\}^2-\E_P[x'\Omega_ix]\) are centered with \(\psi_1\)-norm at most \(B_1\); for the second, the sum has \(\psi_1\)-norm at most \(K(2C)^2/\{K(K-1)\}\le4C^2\). The observations of the training fold are independent, so~\Cref{lemm:app-cov-bernstein} with \(t=\varepsilon\le1\) gives \(\Pr_P(|x'\bar Q^{(r)}x|>\varepsilon)\le2\exp(-c_1m\varepsilon^2)\le2\exp(-c_1n\varepsilon^2/3)\) for \(r=1,2\).  By Step~1 and a union bound over \(\mathcal N\), once \(p_n\le c_\ast n\),
\[
\Pr_P\bigl(\|\bar Q^{(r)}\|_{\mathrm{op}}>2\varepsilon\bigr) \le2\exp\bigl(p_n\log9-c_1n\varepsilon^2/3\bigr) \le2\exp(-c_1n\varepsilon^2/6).
\]

\emph{Step 4 (training mean).}  For a unit vector \(x\), \(x'Z_i\) is centered and, by~\Cref{lemm:app-cov-product} applied with the constant \(G_2=1\), whose \(\psi_2\)-norm is \((\log2)^{-1/2}\), has \(\psi_1\)-norm at most \(B_3\). \Cref{lemm:app-cov-bernstein} with \(u:=\frac34\varepsilon^{1/2}\le1\) gives \(\Pr_P(|x'\bar Z_{-\ell}|>u)\le2\exp(-c_3mu^2)\le2\exp(-3c_3n\varepsilon/16)\).  Since \((\frac43u)^2=\varepsilon\), Step~1 and a union bound give, once \(p_n\le c_\ast n\),
\[
\Pr_P\bigl(\|\bar Z_{-\ell}\|_2^2>\varepsilon\bigr) \le\Pr_P\Bigl(\max_{y\in\mathcal N}|y'\bar Z_{-\ell}|>u\Bigr) \le2e^{p_n\log9-3c_3n\varepsilon/16} \le2e^{-3c_3n\varepsilon/32}.
\]

\emph{Step 5 (part (i)).}  By Steps~2--4, with probability at least \(1-4e^{-c_1n\varepsilon^2/6}-2e^{-3c_3n\varepsilon/32}\), which tends to one uniformly in \(P\), \(\|\widehat\Sigma-\Sigma_X\|_{\mathrm{op}}\le2\varepsilon+2\varepsilon+\varepsilon\le(c_X-\lambda_{\mathrm{fl}})/2\). By Weyl's inequality and condition~(5), \(\lambda_{\min}(\widehat\Sigma)\ge(c_X+\lambda_{\mathrm{fl}})/2\) on this event, which is therefore contained in \(\mathcal F_n\).  On \(\mathcal F_n\), \(\widehat\Sigma\succeq\lambda_{\mathrm{fl}}I\) is itself feasible, so the minimum in~\eqref{eq:thy-certified-projection} is zero and \(\|\widehat\Sigma_Z^+-\widehat\Sigma\|_{\max}\le\bar\eta_n\); in the second case \(\widehat\Sigma_Z^+=\widehat\Sigma\).

\emph{Step 6 (part (ii)).}  For every realization,
\[
\mathcal S_R=(\widehat\Sigma_Z^+-\widehat\Sigma)b_0+\bar Q^{(1)}b_0-\bar Q^{(2)}b_0-\bar Z_{-\ell}(\bar Z_{-\ell}'b_0) -\{\widehat c_{RZ,-\ell}-\E_P[Z_iR_i]\}.
\]
(a) On \(\mathcal F_n\), \(\|(\widehat\Sigma_Z^+-\widehat\Sigma)b_0\|_\infty\le\bar\eta_n\|b_0\|_1\le\bar\eta_n\sqrt{p_n}\|b_0\|_2\le Ct_n\) in the first case, and the term is zero in the second. (b) The \(j\)th coordinates of \(\bar Q^{(1)}b_0\) and \(\bar Q^{(2)}b_0\) are training averages of the centered variables \(Z_{ij}Z_i'b_0-\E_P[Z_{ij}Z_i'b_0]\) and \(e_j'H_ib_0=\{K(K-1)\}^{-1}\sum_{k=1}^K\{e_j'(\nu_{ik}-\bar\nu_i)\}\{b_0'(\nu_{ik}-\bar\nu_i)\}-\E_P[e_j'\Omega_ib_0]\), which have \(\psi_1\)-norm at most \(B_2\) by~\Cref{lemm:app-cov-product}, because the factors have \(\psi_2\)-norms at most \(2C\) and \(2C\|b_0\|_2\le2C^2\).  Choose \(M\ge1\), at least the constant of the proof of~\Cref{lemm:penalty-level-l1}, with \(c_2M^2>3\) and \(c_3M^2>3\).  For \(n\) so large that \(Mt_n\le1\), \Cref{lemm:app-cov-bernstein}, \(m\ge n/3\) and a union bound over \(2p_n\le2\bar p_n\) coordinates bound the probability that \(\|\bar Q^{(1)}b_0\|_\infty\) or \(\|\bar Q^{(2)}b_0\|_\infty\) exceeds \(Mt_n\) by \(4\bar p_n^{\,1-c_2M^2/3}\to0\), because \(\bar p_n\ge n\). (c) Each \(\bar Z_{-\ell,j}\) is an average of centered variables with \(\psi_1\)-norm at most \(B_3\), and the same argument gives \(\Pr_P(\|\bar Z_{-\ell}\|_\infty>Mt_n)\le2\bar p_n^{\,1-c_3M^2/3}\to0\).  Also \(\E_P(\bar Z_{-\ell}'b_0)^2=b_0'(\Sigma_X+\Omegao)b_0/m\le6C^3/n\), so \(|\bar Z_{-\ell}'b_0|\le1\) with probability tending to one by Chebyshev's inequality, and then \(\|\bar Z_{-\ell}(\bar Z_{-\ell}'b_0)\|_\infty\le Mt_n\). (d) \(\|\widehat c_{RZ,-\ell}-\E_P[Z_iR_i]\|_\infty\le Mt_n\) on the event \(\mathcal E_{2,n}\) of the proof of~\Cref{lemm:penalty-level-l1}, whose probability tends to one uniformly. On the intersection of \(\mathcal F_n\) with the events in (b)--(d), \(\|\mathcal S_R\|_\infty\le(4M+C)t_n\), so \(C_\lambda^{(2)}:=8(4M+C)\) has the stated property.  All bounds are uniform in \(P\), \(t_n\to0\) because \(\log\bar p_n\le\log n+\log\max\{1,c_\ast\}\), and the number of folds is fixed.
\end{proof}

\begin{corollary}[CoCoLasso first stages with an inactive projection]\label{coro:inactive-projection}
Let \(c_\ast\) and \(C_\lambda^{(2)}\) be the constants of~\Cref{lemm:inactive-projection}, let \(p_n\le c_\ast n\) for all large \(n\), and let \(\mathcal P_{\circ,n}\) be the set of distributions that satisfy conditions~\textup{(1)}--\textup{(5)} of~\Cref{def:dgpclass}. Let the first stages be the corrected Lasso~\eqref{eq:est-cocolasso-objective} with \(\lambda_{\mathrm{fl}}>0\), projections as in~\Cref{lemm:inactive-projection}, and \(\lambda_Y=\lambda_D=C_\lambda^{(2)}\sqrt{\log\bar p_n/n}\).  With the rates of~\Cref{coro:l1-bounded}, the conclusions of that corollary on the slope, \(\ell_1\) and affine prediction errors hold uniformly over \(\mathcal P_{\circ,n}\), with no bound on \(\|\pio\|_1\) or \(\|\gammao\|_1\) beyond condition~\textup{(1)}.
\end{corollary}
When \(p_n\ge|I_\ell^c|\), the projection is always active (\Cref{rem:active-projection-tight}) and~\Crefinorder{coro:exact-sparsity,coro:l1-bounded} apply instead; \Cref{coro:window-cv} covers penalties selected by any rule, cross-validation included, from a shrinking grid.

\begin{proof}[Proof of~\Cref{coro:inactive-projection}]
The proof of \Cref{coro:l1-bounded} uses the bound on the target \(\ell_1\) norms only in \Cref{lemm:penalty-level-l1}, so that proof applies verbatim with \(\mathcal P_{\circ,n}\) in place of \(\mathcal P_{1,n}\) and with part~(ii) of~\Cref{lemm:inactive-projection} in place of~\Cref{lemm:penalty-level-l1}; this part of the lemma gives condition~(i) of~\Cref{prop:corrected-lasso-specialization} and the score condition of~\Cref{lemm:kkt-perturbation} uniformly over \(\mathcal P_{\circ,n}\).  Condition~(ii) holds because \(\lambda_{\mathrm{fl}}>0\) (\Cref{rem:relaxed-cone}); in the second case of~\Cref{lemm:inactive-projection}, on \(\mathcal F_n\), it holds even with \(c_Q=(c_X+\lambda_{\mathrm{fl}})/(2C)\), because \(\delta'\widehat\Sigma_{Z,-\ell}\delta\ge\frac{c_X+\lambda_{\mathrm{fl}}}2\|\delta\|_2^2\) and \(\|X_i'\delta\|_{P,2}^2\le C\|\delta\|_2^2\), so the constants do not deteriorate with \(\lambda_{\mathrm{fl}}^{-1}\).  \Cref{prop:corrected-lasso-specialization} then gives the prediction rate \(\Op(\sqrt{s_R}\lambda_R+a_R)=\Op(\rho_{R,n})\) and the \(\ell_1\) rate \(\Op(s_R\lambda_R+\sqrt{s_R}a_R+a_R^2/\lambda_R)=\Op(\zeta_{R,n})\), since \(a_R^2/\lambda_R=a_R^2\sqrt{n/\log\bar p_n}/C_\lambda^{(2)}\); the intercepts are handled as in the proof of~\Cref{coro:l1-bounded}, through~\Cref{lem:affine-intercept}.  Because \(p_n\le c_\ast n\), \(\log\bar p_n\le\log n+\log\max\{1,c_\ast\}\le\{1+\log\max\{1,c_\ast\}/\log2\}\log n\), so the rates may be written with \(\log n\) in place of \(\log\bar p_n\).
\end{proof}

\begin{remark}[The active projection]\label{rem:active-projection-tight}
Let \(d:=\min_{B\succeq\lambda_{\mathrm{fl}}I}\|B-\widehat\Sigma_{Z,-\ell}\|_{\max}\) be the max-norm distance of the corrected Gram matrix to the floored set, with \(\lambda_{\mathrm{fl}}>0\).  If \(p_n\ge|I_\ell^c|\), then \(d>0\) in every sample: the matrix \(\mathbb P_{n,-\ell}\{(Z_i-\bar Z_{-\ell})(Z_i-\bar Z_{-\ell})'\}\) has rank at most \(|I_\ell^c|-1<p_n\), so a unit vector \(x\) in its null space has \(x'\widehat\Sigma_{Z,-\ell}x=-x'\widehat\Omega_{-\ell}x\le0<\lambda_{\mathrm{fl}}\), because \(\widehat\Omega_{-\ell}\) is an average of the positive-semidefinite \(\Omega_i\).  Now let \(\kappa>1\) and \(b\neq0\).  Every matrix \(A\) satisfying~\eqref{eq:thy-certified-projection} has \(\|(A-\widehat\Sigma_{Z,-\ell})b\|_\infty\le(\kappa d+\bar\eta)\|b\|_1\), by \(\|Mb\|_\infty\le\|M\|_{\max}\|b\|_1\), and some matrix satisfying it with \(\bar\eta=0\) has \(\|(A-\widehat\Sigma_{Z,-\ell})b\|_\infty\ge\frac{\kappa-1}{2}\,d\,\|b\|_1\).  For the latter, let \(A^\ast\) attain the minimum, which exists because the floored set is closed and the objective is coercive, let \(v:=\operatorname{sign}(b)\), and set \(A':=A^\ast+(\kappa-1)d\,vv'\).  Then \(A'\succeq A^\ast\succeq\lambda_{\mathrm{fl}}I\) and \(\|A'-\widehat\Sigma_{Z,-\ell}\|_{\max}\le d+(\kappa-1)d\|vv'\|_{\max}=\kappa d\), so both matrices satisfy~\eqref{eq:thy-certified-projection} with \(\bar\eta=0\); since \((A'-A^\ast)b=(\kappa-1)d\,\|b\|_1v\) and \(\|v\|_\infty=1\), the triangle inequality shows that one of them has \(\|(A-\widehat\Sigma_{Z,-\ell})b\|_\infty\ge\frac12\|(A'-A^\ast)b\|_\infty=\frac{\kappa-1}{2}d\|b\|_1\). Over the certified near-minimizers with \(\kappa>1\) and \(\bar\eta=0\), the worst-case projection increment at \(b_{R0}\) is therefore of the exact order \(d\,\|b_{R0}\|_1\); a tolerance \(\bar\eta>0\) adds up to \(\bar\eta\|b_{R0}\|_1\), which \(\widehat\Sigma_{Z,-\ell}+\bar\eta\,vv'\), with \(v=\operatorname{sign}(b_{R0})\), attains when \(d=0\).  An improvement on~\Cref{lemm:penalty-level} valid for all of them requires \(d\,\|b_{R0}\|_1=\Op(t_n)\), that is, \(d=\Op(t_n/\sqrt{s_R})\) when \(\|b_{R0}\|_1\asymp\sqrt{s_R}\), whereas~\Cref{lemm:psd-projection-cost} uses only \(d\le\|\widehat\Sigma_{Z,-\ell}-\Sigma_X\|_{\max}=\Op(t_n)\).  \Cref{lemm:inactive-projection} gives \(d=0\) with probability tending to one when \(p_n\le c_\ast n\), and its additional condition on the projections, \(\bar\eta_n\le t_n/\sqrt{p_n}\) or the return of a feasible \(\widehat\Sigma_{Z,-\ell}\) unchanged, bounds the remaining term \(\bar\eta_n\|b_{R0}\|_1\) by \(Ct_n\) or removes it.
\end{remark}

\begin{example}[Distributions that meet every condition]\label{exam:fs-nonempty}
Let \(K=2\), \(\thetao=1\), \(f_0(x)=x'\betao\) and \(g_0(x)=x'\gammao\), and let \(X_i\sim N(0,I_{p_n})\), \(U_i\sim N(0,1)\), \(V_i\sim N(0,1)\) and \(\nu_{i1},\nu_{i2}\sim N(0,\sigma_\nu^2I_{p_n})\) be independent.  Conditions~(1)--(5) of~\Cref{def:dgpclass} hold with constants that depend only on \(\sigma_\nu\) and on a bound for \(\|\betao\|_2+\|\gammao\|_2\): \(\E[U_i\mid D_i,X_i]=0\), \(\E[V_i\mid X_i]=0\), \(J_0=1\), \(\Sigma_X=I\) (so \(c_X=1\) exceeds any floor \(\lambda_{\mathrm{fl}}<1\)), \(\Omegao=\frac{\sigma_\nu^2}{2}I\), the \(\psi_2\)-norms are those of Gaussian variables, and the \(\psi_1\)-norms of condition~(4) follow from~\Cref{lemm:app-cov-product}, with \(\operatorname{Var}(Y_i)=\|\pio\|_2^2+2\).  For condition~(6), \(\xi_i=\bar\nu_i\) and \(\tilde\nu_i:=(\nu_{i1}-\nu_{i2})/2\) are independent \(N(0,\frac{\sigma_\nu^2}{2}I)\) vectors with \(\Omega_i=\tilde\nu_i\tilde\nu_i'\), so that, by~\eqref{eq:score-corrected} and~\eqref{eq:thy-dgp-if}, \(J_0\varphi_i=U_iV_i-U_i\xi_i'\gammao-V_i\xi_i'\betao+(\xi_i'\betao)(\xi_i'\gammao)-(\tilde\nu_i'\betao)(\tilde\nu_i'\gammao)\), a sum of five uncorrelated terms; hence \(\sigma^2(P)=1+\frac{\sigma_\nu^2}{2}(\|\betao\|_2^2+\|\gammao\|_2^2)+\frac{\sigma_\nu^4}{2}\{\|\betao\|_2^2\|\gammao\|_2^2+(\betao'\gammao)^2\}\in[1,C_\sigma]\), and \(\E\varphi_i^4\) is bounded.  With \(p_n=2n\), \(\betao=e_1\) and \(\gammao=e_2\), cases~(a), (b) and~(d) of~\Cref{coro:cocolasso-core} apply, because \(s_{\pi,n}=2\), \(s_{g,n}=1\), \(s_{g,n}(s_{\pi,n}+s_{g,n})\log\bar p_n=3\log(2n)=o(\sqrt n)\), \(\upsilon_n^2(s_{\pi,n}+s_{g,n})\log\bar p_n=3\upsilon_n^2\log(2n)=o(\sqrt n)\), \(\|\pio\|_1=2\) and \(\|\gammao\|_1=1\), and, because \(\Sigma_X=I\), \(\|\operatorname{Cov}(Z_i,Y_i)\|_\infty=\|\pio\|_\infty=1\) and \(\|\operatorname{Cov}(Z_i,D_i)\|_\infty=\|\gammao\|_\infty=1\), so case~(d) holds with \(c_{ZR}=1\). \Cref{prop:plugin-inference} also applies, because \((s_{\pi,n}^{3/2}+s_{g,n}^{3/2})\log\bar p_n=o(\sqrt n)\).  With \(p_n=\lceil n^{9/10}\rceil\), \(s_n=\lceil n^{2/5}\rceil\), \(\gammao=s_n^{-1/2}\sum_{j\le s_n}e_j\) and \(\betao=s_n^{-1/2}\sum_{s_n<j\le2s_n}e_j\), case~(c) applies, because \(p_n/n\to0\) and \((s_{\pi,n}+s_{g,n})\log n=3s_n\log n=o(\sqrt n)\), whereas \(\|\pio\|_1=2\sqrt{s_n}\to\infty\) and \(s_{g,n}(s_{\pi,n}+s_{g,n})\log\bar p_n=3s_n^2\log n\ge3n^{4/5}\log n\) is not \(o(\sqrt n)\), so for these representations neither the \(\ell_1\) bound of case~(b) nor the budget of case~(a) holds.
\end{example}

\subsection{Penalties selected from a data-dependent grid}\label{app:cv-window}

Over a penalty grid whose smallest value is a fixed fraction of \(\|\widehat c_{RZ,-\ell}\|_\infty\), such as the grid of~\Cref{app:sim-numerics} without the factor \(\min\{1,100\,\upsilon_mt_m\}\), the first stage is inconsistent, whatever penalty is selected, when \(\|\operatorname{Cov}_P(Z_i,R_i)\|_\infty\) is bounded away from zero (\Cref{prop:grid-floor}); over a grid whose largest value eventually shrinks with the training-sample size, such as the shrinking grid of~\Cref{rem:thy-cv-not-covered}, penalties selected by any rule, cross-validation included, are covered asymptotically under the conditions of~\Cref{coro:window-cv}.

\begin{proposition}[Fixed-ratio penalty grids]\label{prop:grid-floor}
Let conditions~\textup{(1)}--\textup{(5)} of~\Cref{def:dgpclass} hold with \(\log p_n=o(n)\), fix \(R\in\{Y,D\}\) and \(c_{ZR}>0\), and let \(\mathcal P_n^R(c_{ZR})\) be the set of distributions that satisfy these conditions and \(\|\operatorname{Cov}_P(Z_i,R_i)\|_\infty\ge c_{ZR}\).  For a training fold \(\ell\), let \(\widehat\Sigma_Z^+\) satisfy~\eqref{eq:thy-certified-projection} with \(\lambda_{\mathrm{fl}}\in[0,c_X]\), a fixed \(\kappa\ge1\) and \(\bar\eta=\bar\eta_n\le t_n\), and let \(\widehat b_R\) satisfy the optimality conditions of~\eqref{eq:est-cocolasso-objective} at a penalty \(\widehat\lambda_R>0\) with \(\widehat\lambda_R\ge u_0\|\widehat c_{RZ,-\ell}\|_\infty\), up to a residual at most \(\widehat\lambda_R/8\) (\Cref{lemm:kkt-perturbation}), where \(u_0\in(0,1]\) is fixed and \(\widehat\lambda_R\) may depend on the data in any way.  There are \(\epsilon_0>0\), depending only on \(c_{ZR}\), \(u_0\) and the constants of~\Cref{def:dgpclass}, an integer \(n_0\), depending in addition on \(\kappa\) and the sequences \((a_{R,n})\) and \((t_n)\), and events \(\mathcal E_n\), which depend neither on \(\widehat\lambda_R\) nor on \(\widehat b_R\), such that \(\lim_{n\to\infty}\inf_{P\in\mathcal P_n^R(c_{ZR})}\Pr_P(\mathcal E_n)=1\) and, for every \(n\ge n_0\) and \(P\in\mathcal P_n^R(c_{ZR})\),
\[
\|X_i'(\widehat b_R-b_{R0})\|_{P,2}>\epsilon_0\qquad\text{on }\mathcal E_n .
\]
Consequently, if the penalty is selected, by any rule, from a grid whose values are all at least \(u_0\|\widehat c_{RZ,-\ell}\|_\infty\), such as the grid of~\Cref{app:sim-numerics} without the factor \(\min\{1,100\,\upsilon_mt_m\}\) \textup{(}\(u_0=0.01\)\textup{)}, then along every sequence of distributions in \(\mathcal P_n^R(c_{ZR})\) the affine first-stage error of~\Cref{assu:first-stage-rates} exceeds \(\epsilon_0\) with probability approaching one, and~\Cref{assu:first-stage-rates} fails, for every rate \(\rho_{R,n}\to0\), over every class of distributions that contains such a sequence.
\end{proposition}

\begin{proof}
Suppress the fold index and write \(\widehat c:=\widehat c_{RZ,-\ell}\), \(c:=\E_P[Z_iR_i]\), which equals \(\operatorname{Cov}_P(Z_i,R_i)\) because \(\E_PZ_i=0\) (\Cref{subsec:model-omega}), \(v:=c-\Sigma_Xb_{R0}=\E_P[X_ir_{R,i}]\), \(\delta:=\widehat b_R-b_{R0}\) and \(\widehat\lambda:=\widehat\lambda_R\).  By condition~(5), \(\|v\|_2=\sup_{\|u\|_2=1}\E_P[r_{R,i}X_i'u]\le C^{1/2}a_R\), and by condition~(1), \(\|b_{R0}\|_2\le C\); hence \(\|c\|_2\le C_c:=C^2+C^{1/2}\sup_na_{R,n}\), which is finite because \(a_{R,n}\to0\) by condition~(2).  The second paragraph of the proof of~\Cref{lemm:penalty-level} constructs a constant \(M\), depending only on the constants of~\Cref{def:dgpclass}, such that the events \(\{\|\widehat\Sigma_{Z,-\ell}-\Sigma_X\|_{\max}\le Mt_n\}\) and \(\{\|\widehat c-c\|_\infty\le Mt_n\}\) have probability tending to one uniformly over the distributions that satisfy conditions~(1)--(5) and \(\log p_n=o(n)\); let \(\mathcal E_n\) be their intersection.  On \(\mathcal E_n\), \Cref{lemm:psd-projection-cost} and \(\bar\eta_n\le t_n\) give \(\|\widehat\Sigma_Z^+-\Sigma_X\|_{\max}\le M't_n\) with \(M':=(1+\kappa)M+1\).

Put \(\lambda_{\mathrm{lo}}:=u_0c_{ZR}/2\), \(K_1:=4C_c/(3\lambda_{\mathrm{lo}})\), \(\epsilon_0:=\min\{c_X^{1/2}c_{ZR}/(4C),\,\lambda_{\mathrm{lo}}/(4C^{1/2})\}\) and \(C_b:=C+c_X^{-1/2}\epsilon_0\), and let \(n_0\) be such that, for every \(n\ge n_0\),
\[
Mt_n\le\frac{\lambda_{\mathrm{lo}}}8,\qquad C^{1/2}a_{R,n}\le\min\Bigl\{\frac{\lambda_{\mathrm{lo}}}4,\frac{c_{ZR}}2\Bigr\},\qquad Mt_nK_1+M't_nK_1^2C_b\le\frac{\lambda_{\mathrm{lo}}}4 .
\]
Fix \(n\ge n_0\) and \(P\in\mathcal P_n^R(c_{ZR})\), and work on \(\mathcal E_n\).  The bound \(Mt_n\le\lambda_{\mathrm{lo}}/8\le c_{ZR}/2\) gives \(\|\widehat c\|_\infty\ge\|c\|_\infty-Mt_n\ge c_{ZR}/2\), so \(\widehat\lambda\ge\lambda_{\mathrm{lo}}\).  By~\Cref{lemm:kkt-perturbation}, \(\widehat b_R\) exactly minimizes \(\tfrac12b'\widehat\Sigma_Z^+b-(\widehat c+e)'b+\widehat\lambda\|b\|_1\) for some \(e\) with \(\|e\|_\infty\le\widehat\lambda/8\).

\emph{Step 1 (\(\ell_1\) and \(\ell_2\) norms of the fit).}  The perturbed objective at \(\widehat b_R\) is at most its value \(0\) at \(b=0\), and \(\widehat\Sigma_Z^+\succeq0\), so
\[
\widehat\lambda\|\widehat b_R\|_1\le(\widehat c+e)'\widehat b_R\le\|c\|_2\|\widehat b_R\|_2+\bigl(Mt_n+\widehat\lambda/8\bigr)\|\widehat b_R\|_1\le C_c\|\widehat b_R\|_2+\frac{\widehat\lambda}4\|\widehat b_R\|_1,
\]
because \(Mt_n\le\lambda_{\mathrm{lo}}/8\le\widehat\lambda/8\).  Hence \(\|\widehat b_R\|_1\le4C_c\|\widehat b_R\|_2/(3\widehat\lambda)\le K_1\|\widehat b_R\|_2\).

\emph{Step 2 (optimality identity).}  Every subgradient \(u\) of \(\|\cdot\|_1\) at \(\widehat b_R\) satisfies \(\widehat b_R'u=\|\widehat b_R\|_1\), so the optimality conditions of the perturbed problem give \(\widehat\lambda\|\widehat b_R\|_1=\widehat b_R'(\widehat c+e-\widehat\Sigma_Z^+\widehat b_R)\), and \(c=\Sigma_Xb_{R0}+v\) gives \(\widehat c+e-\widehat\Sigma_Z^+\widehat b_R=v-\Sigma_X\delta+(\widehat c-c)+e-(\widehat\Sigma_Z^+-\Sigma_X)\widehat b_R\).  Now \(|\widehat b_R'v|=|\E_P[r_{R,i}X_i'\widehat b_R]|\le a_R\|X_i'\widehat b_R\|_{P,2}\le C^{1/2}a_R\|\widehat b_R\|_2\); by the Cauchy--Schwarz inequality in the \(\Sigma_X\) inner product, \(|\widehat b_R'\Sigma_X\delta|\le C^{1/2}\|\widehat b_R\|_2\|X_i'\delta\|_{P,2}\); \(|\widehat b_R'(\widehat c-c)|\le Mt_nK_1\|\widehat b_R\|_2\) and \(|\widehat b_R'(\widehat\Sigma_Z^+-\Sigma_X)\widehat b_R|\le M't_nK_1^2\|\widehat b_R\|_2^2\) by Step~1; and \(|\widehat b_R'e|\le(\widehat\lambda/8)\|\widehat b_R\|_1\).  Since \(\|\widehat b_R\|_2\le\|\widehat b_R\|_1\) and \(\widehat\lambda\ge\lambda_{\mathrm{lo}}\),
\[
\frac78\lambda_{\mathrm{lo}}\|\widehat b_R\|_2\le\frac78\widehat\lambda\|\widehat b_R\|_1 \le\Bigl\{C^{1/2}\bigl(a_R+\|X_i'\delta\|_{P,2}\bigr)+Mt_nK_1+M't_nK_1^2\|\widehat b_R\|_2\Bigr\}\|\widehat b_R\|_2 .
\]

\emph{Step 3 (contradiction).}  By condition~(5), \(\|X_i'b_{R0}\|_{P,2}\ge c_X^{1/2}\|b_{R0}\|_2\ge c_X^{1/2}\|\Sigma_Xb_{R0}\|_\infty/C\ge c_X^{1/2}(\|c\|_\infty-\|v\|_2)/C\ge c_X^{1/2}(c_{ZR}-C^{1/2}a_R)/C\ge c_X^{1/2}c_{ZR}/(2C)\).  Suppose that \(\|X_i'\delta\|_{P,2}\le\epsilon_0\).  Then \(\|X_i'\widehat b_R\|_{P,2}\ge c_X^{1/2}c_{ZR}/(4C)>0\), so \(\widehat b_R\ne0\), and \(\|\widehat b_R\|_2\le\|b_{R0}\|_2+c_X^{-1/2}\|X_i'\delta\|_{P,2}\le C_b\).  Dividing the last display by \(\|\widehat b_R\|_2\) and using \(C^{1/2}\epsilon_0\le\lambda_{\mathrm{lo}}/4\) gives \(\frac78\lambda_{\mathrm{lo}}\le C^{1/2}a_R+\frac14\lambda_{\mathrm{lo}}+Mt_nK_1+M't_nK_1^2C_b\le\frac34\lambda_{\mathrm{lo}}\), which is false because \(\lambda_{\mathrm{lo}}>0\).  Hence \(\|X_i'\delta\|_{P,2}>\epsilon_0\) on \(\mathcal E_n\).

For the consequence, every penalty selected from such a grid satisfies the hypothesis.  Because \(\E_PX_i=0\), every constant \(\mu\) satisfies \(\|\mu+X_i'\delta\|_{P,2}^2=\mu^2+\|X_i'\delta\|_{P,2}^2\), so the affine error of~\Cref{assu:first-stage-rates}, whatever the fitted intercept, exceeds \(\epsilon_0\) on \(\mathcal E_n\).  Along every sequence of distributions in \(\mathcal P_n^R(c_{ZR})\) this happens with probability at least \(\Pr_P(\mathcal E_n)\to1\), which is incompatible with an \(\Op(\rho_{R,n})\) bound for any \(\rho_{R,n}\to0\).
\end{proof}

The proposition uses neither sparsity nor a rate for the selected penalty.  The obstruction disappears when every candidate penalty is of a slowly inflated rate order.

\begin{corollary}[First-stage bounds uniformly over a penalty window]\label{coro:penalty-window}
Let the distributions \(P\) range over a class that satisfies conditions~\textup{(1)}--\textup{(5)} of~\Cref{def:dgpclass}, fix a training fold and \(R\in\{Y,D\}\), and let \(\widehat\Sigma_Z^+\succeq\lambda_{\mathrm{fl}}I\), with \(\lambda_{\mathrm{fl}}>0\), satisfy~\eqref{eq:thy-certified-projection}.  Let \(\lambda_{R,n}>0\) be deterministic with \(\lim_{n\to\infty}\sup_P\Pr_P(\|\mathcal S_R\|_\infty>\lambda_{R,n}/8)=0\), let \(C_u\ge1\) be a constant, and put \(\mathcal E_n:=\{\|\mathcal S_R\|_\infty\le\lambda_{R,n}/8\}\).  On \(\mathcal E_n\), simultaneously for every \(\lambda\in[\lambda_{R,n},C_u\lambda_{R,n}]\) and every \(\widetilde b\) that satisfies the optimality conditions of~\eqref{eq:est-cocolasso-objective} at penalty \(\lambda\) up to a residual at most \(\lambda/8\) (\Cref{lemm:kkt-perturbation}), the error \(\delta:=\widetilde b-b_{R0}\) satisfies
\[
\|X_i'\delta\|_{P,2}\le B_{R,n}:=\frac{2C}{\lambda_{\mathrm{fl}}}\Bigl(\frac54c_X^{-1/2}C_u\sqrt{s_R}\,\lambda_{R,n}+a_R\Bigr), \qquad \|\delta\|_1\le\Bigl(\frac83c_X^{-1/2}\sqrt{s_R}+\frac{4a_R}{3\lambda_{R,n}}\Bigr)B_{R,n}.
\]
Now let \(\widehat\lambda_R\) be a function of the training observations \(\{i\notin I_\ell\}\) and of an auxiliary randomization independent of the data, with \(\lim_{n\to\infty}\sup_P\Pr_P(\widehat\lambda_R\notin[\lambda_{R,n},C_u\lambda_{R,n}])=0\).  Then the fit computed at \(\widehat\lambda_R\) with residual at most \(\widehat\lambda_R/8\) satisfies~\eqref{eq:thy-cl-pred}--\eqref{eq:thy-cl-l1} with \(\lambda_R\) replaced by \(\lambda_{R,n}\), uniformly over \(P\) and folds, and it is a function of the training observations and the auxiliary randomization, as~\Cref{assu:first-stage-rates} permits.  In particular, \Crefinorder{coro:exact-sparsity,coro:l1-bounded} remain true, with the same rates up to constant factors, when their deterministic penalties \(\lambda_R\) are replaced by such \(\widehat\lambda_R\) with \(\lambda_{R,n}:=\lambda_R\).
\end{corollary}

\begin{proof}
Work on \(\mathcal E_n\) and fix \(\lambda\in[\lambda_{R,n},C_u\lambda_{R,n}]\) and \(\widetilde b\) as stated.  By~\Cref{lemm:kkt-perturbation}, \(\widetilde b\) exactly minimizes \(\tfrac12b'\widehat\Sigma_Z^+b-(\widehat c_{RZ}+e)'b+\lambda\|b\|_1\) for some \(e\) with \(\|e\|_\infty\le\lambda/8\), so the proof of~\Cref{prop:corrected-lasso-specialization} applies with \(\mathcal S_R-e\) in place of \(\mathcal S_R\).  Since \(\|\mathcal S_R-e\|_\infty\le\lambda_{R,n}/8+\lambda/8\le\lambda/4\), \eqref{eq:app-cl-cone} holds with this \(\lambda\), \(t:=\|X_i'\delta\|_{P,2}\) and \(\mathcal T=\supp(b_{R0})\).  With \(\delta'\widehat\Sigma_Z^+\delta\ge(\lambda_{\mathrm{fl}}/C)t^2\) for every \(\delta\) (\Cref{rem:relaxed-cone}) and \(\|\delta_{\mathcal T}\|_1\le c_X^{-1/2}\sqrt{s_R}\,t\), dropping the \(\ell_1\) term gives \(t\le(2C/\lambda_{\mathrm{fl}})\{\tfrac54c_X^{-1/2}\sqrt{s_R}\,\lambda+a_R\}\le B_{R,n}\), because \(\lambda\le C_u\lambda_{R,n}\), and dropping the quadratic term gives \(\|\delta_{\mathcal T^c}\|_1\le\tfrac53\|\delta_{\mathcal T}\|_1+\tfrac{4a_R}{3\lambda}t\), hence \(\|\delta\|_1\le\tfrac83c_X^{-1/2}\sqrt{s_R}\,t+\tfrac{4a_R}{3\lambda_{R,n}}t\), because \(\lambda\ge\lambda_{R,n}\).  The event \(\mathcal E_n\) depends neither on \(\lambda\) nor on \(\widetilde b\), so both bounds hold simultaneously.

For the consequence, on \(\mathcal E_n\cap\{\widehat\lambda_R\in[\lambda_{R,n},C_u\lambda_{R,n}]\}\), whose probability tends to one uniformly, the fit at \(\widehat\lambda_R\) is one of the vectors \(\widetilde b\) above, so \(\|X_i'\delta\|_{P,2}=\Op(\sqrt{s_R}\lambda_{R,n}+a_R)\) and \(\|\delta\|_1=\Op(s_R\lambda_{R,n}+\sqrt{s_R}\,a_R+a_R^2/\lambda_{R,n})\), uniformly over \(P\) and the fixed number of folds.  The fit is computed from the training observations at the penalty \(\widehat\lambda_R\) by the deterministic algorithm of~\Cref{app:sim-numerics}, so it is a function of the training observations and the auxiliary randomization, and~\Cref{lem:affine-intercept}, which bounds the training-mean intercept pathwise in \(\|\delta\|_1\), applies to it.  In the proofs of~\Crefinorder{coro:exact-sparsity,coro:l1-bounded}, the deterministic penalty enters only through condition~(i) of~\Cref{prop:corrected-lasso-specialization}, supplied by~\Cref{lemm:penalty-level} or~\Cref{lemm:penalty-level-l1} at \(\lambda_{R,n}=\lambda_R\), and through the two rate bounds, which hold here with constants multiplied by \(C_u\); the measurability just stated replaces the sentence on deterministic penalties in the proof of~\Cref{coro:l1-bounded}.
\end{proof}

\begin{corollary}[Penalties selected from a shrinking grid]\label{coro:window-cv}
Let \(\upsilon:[1,\infty)\to[1,\infty)\) be nondecreasing, with \(\upsilon(x)\to\infty\) as \(x\to\infty\) and \(\upsilon(x)\le C_\upsilon\upsilon(x/3)\) for \(x\ge3\) and a constant \(C_\upsilon\ge1\), and write \(\upsilon_n:=\upsilon(n)\).  Fix \(G\ge2\), \(u_0\in(0,1)\) and \(c_{\mathrm{top}}>0\).  For fold \(\ell\) and \(R\in\{Y,D\}\), with \(m:=|I_\ell^c|\) and \(t_m:=\{\log\max(p_n,m)/m\}^{1/2}\), let
\[
\lambda_{\mathrm{sh}}:=\|\widehat c_{RZ,-\ell}\|_\infty\min\{1,\,c_{\mathrm{top}}\upsilon(m)t_m\}, \qquad \mathcal L_{R,-\ell}:=\bigl\{\lambda_{\mathrm{sh}}\,u_0^{(g-1)/(G-1)}:\ g=1,\ldots,G\bigr\},
\]
and let \(\widehat\lambda_R\in\mathcal L_{R,-\ell}\) be selected by any rule that is a function of the training observations and of an auxiliary randomization independent of the data, for example the calibrated cross-validation rule of~\Cref{app:sim-numerics}.  Let the first stages be the corrected Lasso~\eqref{eq:est-cocolasso-objective} fitted on the training fold at \(\widehat\lambda_R\), with \(\lambda_{\mathrm{fl}}>0\), projections certified as in~\eqref{eq:thy-certified-projection} with a fixed \(\kappa\ge1\) and \(\bar\eta=\bar\eta_n\le t_n\), and fits accepted with residual at most \(\widehat\lambda_R/8\).  Fix \(c_{ZR}>0\) and let \(\mathcal P^{c_{ZR}}_{1,n}\) be the set of distributions in \(\mathcal P_{1,n}\) (\Cref{coro:l1-bounded}) with \(\|\operatorname{Cov}_P(Z_i,Y_i)\|_\infty\ge c_{ZR}\) and \(\|\operatorname{Cov}_P(Z_i,D_i)\|_\infty\ge c_{ZR}\).  Let \((\rho_{\pi,n},\zeta_{\pi,n})\) and \((\rho_{g,n},\zeta_{g,n})\) be
\[
\rho=\upsilon_n\sqrt{s\log\bar p_n/n}+a, \qquad \zeta=\upsilon_ns\sqrt{\log\bar p_n/n}+\sqrt s\,a+\upsilon_n^{-1}a^2\sqrt{n/\log\bar p_n},
\]
evaluated at \((s,a)=(s_{\pi,n},a_{\pi,n})\) and at \((s,a)=(s_{g,n},a_{g,n})\), respectively, with \(s_{\pi,n},s_{g,n}\ge1\).  If \(\upsilon_n\sqrt{\log\bar p_n/n}\to0\), then, uniformly over \(P\in\mathcal P^{c_{ZR}}_{1,n}\) and folds, the slope prediction errors are \(O_P(\rho_{R,n})\), the slope \(\ell_1\) errors are \(O_P(\zeta_{R,n})\), and the affine prediction errors are \(O_P(\rho_{R,n})\) whenever \(\rho_{R,n}\to0\); the selected penalties and the fits are functions of the training observations and the auxiliary randomization.  The function \(\upsilon(x)=\max\{1,\log\log x\}\), with \(\upsilon(x)=1\) for \(x\le e\), satisfies the conditions above with \(C_\upsilon=3/2\); with it, \(G=30\), \(u_0=0.01\) and \(c_{\mathrm{top}}=100\), \(\mathcal L_{R,-\ell}\) is the shrinking grid of~\Cref{rem:thy-cv-not-covered} and~\Cref{app:sim-numerics}, which coincides with the fixed-ratio grid of~\Cref{prop:grid-floor} with \(u_0=0.01\) whenever \(100\,\upsilon(m)t_m\ge1\).
\end{corollary}

\begin{proof}
Write \(c:=\operatorname{Cov}_P(Z_i,R_i)\), \(\upsilon_m:=\upsilon(m)\) and \(\bar C_{ZR}:=CC_1+C^{1/2}\sup_na_{R,n}\).  By assumption \(\upsilon_nt_n\to0\), so in particular \(\log p_n=o(n)\); in the applications of the corollary this follows from~\eqref{eq:thy-product-rate}, because \(s_{\pi,n},s_{g,n}\ge1\) gives \(\rho_{\beta,n}\ge\rho_{\pi,n}\ge\upsilon_nt_n\ge t_n\) and \(\rho_{g,n}\ge t_n\), and \(\sqrt n\,t_n\ge(\log2)^{1/2}\), so that \(\sqrt n\,\rho_{\beta,n}\rho_{g,n}\to0\) forces \(\rho_{\pi,n}\to0\) and \(\rho_{g,n}\to0\).

Since \(L\ge2\) and the fold sizes differ by at most one, \(n/3\le m\le n\) for \(n\ge3\).  For \(n\ge9\), \(t_n\le t_m\le\sqrt3\,t_n\): the upper bound holds because \(\max(p_n,m)\le\bar p_n\) and \(m\ge n/3\); for the lower bound, \(t_m^2\ge\log p_n/n=t_n^2\) if \(p_n\ge n\), and otherwise \(t_m^2\ge\log m/m\ge\log n/n=t_n^2\), because \(x\mapsto\log x/x\) decreases on \([e,\infty)\) and \(3\le m\le n\).  The condition \(\upsilon(x)\le C_\upsilon\upsilon(x/3)\) and monotonicity give \(\upsilon_n/C_\upsilon\le\upsilon(n/3)\le\upsilon_m\le\upsilon_n\).  On \(\mathcal P_{1,n}\), \(\|c\|_\infty\le\|\Sigma_Xb_{R0}\|_\infty+\|\E_P[X_ir_{R,i}]\|_\infty\le CC_1+C^{1/2}a_R\le\bar C_{ZR}\), because the entries of \(\Sigma_X\) are bounded by \(\lambda_{\max}(\Sigma_X)\le C\) and \(\|b_{R0}\|_1\le C_1\).  By the proof of~\Cref{lemm:penalty-level}, there is a constant \(M\) such that \(\mathcal E_{c,n}:=\{\|\widehat c_{RZ,-\ell}-c\|_\infty\le Mt_n\}\) has probability tending to one uniformly over \(\mathcal P_{1,n}\).  Take \(n\ge9\) so large that \(Mt_n\le c_{ZR}/2\) and \(\sqrt3\,c_{\mathrm{top}}\upsilon_nt_n<1\).  For \(P\in\mathcal P^{c_{ZR}}_{1,n}\), on \(\mathcal E_{c,n}\), \(\|\widehat c_{RZ,-\ell}\|_\infty\in[c_{ZR}/2,2\bar C_{ZR}]\), because \(c_{ZR}\le\|c\|_\infty\le\bar C_{ZR}\), and \(c_{\mathrm{top}}\upsilon_mt_m\le\sqrt3\,c_{\mathrm{top}}\upsilon_nt_n<1\), so \(\lambda_{\mathrm{sh}}=c_{\mathrm{top}}\upsilon_mt_m\|\widehat c_{RZ,-\ell}\|_\infty\).  Every element of \(\mathcal L_{R,-\ell}\), and hence \(\widehat\lambda_R\), then lies in \([u_0\lambda_{\mathrm{sh}},\lambda_{\mathrm{sh}}]\subset[\lambda_{R,n},C_u\lambda_{R,n}]\), where \(\lambda_{R,n}:=u_0c_{\mathrm{top}}c_{ZR}\upsilon_nt_n/(2C_\upsilon)\) and \(C_u:=4\sqrt3\,C_\upsilon\bar C_{ZR}/(u_0c_{ZR})\ge1\), because \(u_0\lambda_{\mathrm{sh}}\ge u_0c_{\mathrm{top}}(\upsilon_n/C_\upsilon)t_n(c_{ZR}/2)\) and \(\lambda_{\mathrm{sh}}\le c_{\mathrm{top}}\upsilon_n\sqrt3\,t_n\,2\bar C_{ZR}\).  Since \(\upsilon_n\to\infty\), \(\lambda_{R,n}\ge C^{(1)}_\lambda t_n\) for every \(n\) beyond a threshold that depends only on the constants and on \((\upsilon_n)\), so~\Cref{lemm:penalty-level-l1} gives \(\lim_{n\to\infty}\sup_{P\in\mathcal P_{1,n}}\Pr_P(\|\mathcal S_R\|_\infty>\lambda_{R,n}/8)=0\), and \(\Pr_P(\widehat\lambda_R\notin[\lambda_{R,n},C_u\lambda_{R,n}])\le\Pr_P(\mathcal E_{c,n}^c)\to0\) uniformly over \(\mathcal P^{c_{ZR}}_{1,n}\).  \Cref{coro:penalty-window}, with \(P\) ranging over \(\mathcal P^{c_{ZR}}_{1,n}\), gives, uniformly over \(P\) and folds, \(\|X_i'(\widehat b_R-b_{R0})\|_{P,2}=\Op(\upsilon_n\sqrt{s_R}\,t_n+a_R)=\Op(\rho_{R,n})\) and \(\|\widehat b_R-b_{R0}\|_1=\Op(\upsilon_ns_Rt_n+\sqrt{s_R}\,a_R+a_R^2/(\upsilon_nt_n))=\Op(\zeta_{R,n})\), because \(s_R\le s_{R,n}\) and \(a_R\le a_{R,n}\).

For the intercepts, if \(\rho_{R,n}\to0\), then \(\sqrt{s_{R,n}}\,t_n\le\rho_{R,n}\le1\) and \(a_{R,n}\le\rho_{R,n}\le1\) for large \(n\), and \(\zeta_{R,n}t_n=\upsilon_ns_{R,n}t_n^2+\sqrt{s_{R,n}}\,a_{R,n}t_n+\upsilon_n^{-1}a_{R,n}^2\le3\rho_{R,n}\), because \(\upsilon_ns_{R,n}t_n^2\le\upsilon_n\sqrt{s_{R,n}}t_n\cdot\sqrt{s_{R,n}}t_n\le\rho_{R,n}\).  The intercept term of~\Cref{lem:affine-intercept} is therefore \(\Op(\rho_{R,n})\), with \(n^{-1/2}\le(\log2)^{-1/2}\rho_{R,n}\) because \(\upsilon_n\ge1\).

Finally, let \(\upsilon(x)=\max\{1,\log\log x\}\).  If \(3\le x\le3e^e\), then \(\upsilon(x/3)=1\) and \(\upsilon(x)\le\log(\log3+e)<3/2\).  If \(x>3e^e\), then \(\log(x/3)>e\) and \(\log\log x-\log\log(x/3)=\log\{1+\log3/\log(x/3)\}\le\log3/e\), so \(\upsilon(x)/\upsilon(x/3)\le1+\log3/e<3/2\).  With \(G=30\), \(u_0=0.01\) and \(c_{\mathrm{top}}=100\), \(\mathcal L_{R,-\ell}\) consists of the thirty values \(\lambda_{\mathrm{sh}}0.01^{(g-1)/29}\) of the shrinking grid of~\Cref{rem:thy-cv-not-covered} and~\Cref{app:sim-numerics}, and if \(100\,\upsilon(m)t_m\ge1\), then \(\lambda_{\mathrm{sh}}=\|\widehat c_{RZ,-\ell}\|_\infty=\lambda_{\mathrm{top}}\), so that \(\mathcal L_{R,-\ell}\) is the fixed-ratio grid of~\Cref{prop:grid-floor} with \(u_0=0.01\).
\end{proof}

The condition \(\|\operatorname{Cov}_P(Z_i,R_i)\|_\infty\ge c_{ZR}\) keeps the smallest candidate above \(C^{(1)}_\lambda t_n\); without it, the whole grid can fall below the level that condition~(i) of~\Cref{prop:corrected-lasso-specialization} requires.

\subsection{Training-mean intercepts}\label{app:affine-intercepts}

\begin{lemma}[Training intercept error]\label{lem:affine-intercept}
Consider the training-centered corrected-Lasso fit for $R\in\{Y,D\}$, with slope error $\|\widehat b_R-b_{R0}\|_1=O_P(\zeta_{R,n})$; let $\widehat\mu_R$ denote its training-mean intercept (\Cref{sec:estimator}) and $\mu_{R0}$ the population intercept after population centering.  Under conditions~\textup{(1)}--\textup{(4)} of~\Cref{def:dgpclass},
\[
 |\widehat\mu_R-\mu_{R0}|=O_P\left(n^{-1/2}+\zeta_{R,n}\sqrt{\log p_n/n}\right).
\]
Thus the intercept error is $O_P(n^{-1/2})$ when $\zeta_{R,n}\sqrt{\log p_n}=O(1)$, as under~\eqref{eq:thy-covariance-rate}, and of the order of the prediction rate when $\zeta_{R,n}\sqrt{\log p_n/n}=O(\rho_{R,n})$ and $\rho_{R,n}\gtrsim n^{-1/2}$.  The slope prediction bounds extend to affine predictions by adding this term.
\end{lemma}
\begin{proof}
With training means denoted by bars, $\widehat\mu_R=\bar R-\bar Z'\widehat b_R$, so that, after the population mean of the approximation residual is absorbed into $\mu_{R0}$, $\widehat\mu_R-\mu_{R0}=\overline{R-\mu_{R0}-Z'b_{R0}}-\bar Z'(\widehat b_R-b_{R0})$.  The first term has mean zero and variance bounded uniformly by a constant divided by the training-sample size, hence is $O_P(n^{-1/2})$.  The second is bounded pathwise by $\|\bar Z\|_\infty\|\widehat b_R-b_{R0}\|_1$, which requires no independence between the fitted slopes and the training means, and $\|\bar Z\|_\infty=O_P(\sqrt{\log p_n/n})$ by vector sub-Gaussianity and a union bound over coordinates, uniformly over the fixed number of folds.
\end{proof}

\subsection{\texorpdfstring{Proof of~\Cref{coro:cocolasso-core}}{Proof of Corollary 5.6}}\label{app:cocolasso-core-proof}

In each case the stated class satisfies conditions~(1)--(6) of~\Cref{def:dgpclass}, and~\eqref{eq:thy-product-rate} holds for the stated rates by assumption; this condition implies \(\rho_{\pi,n},\rho_{g,n}\to0\) and \(\log p_n=o(n)\), because \(s_{\pi,n},s_{g,n}\ge1\) gives \(\rho_{\beta,n}\ge\rho_{\pi,n}\ge\sqrt{\log\bar p_n/n}\) and \(\rho_{g,n}\ge\sqrt{\log\bar p_n/n}\), and \(\sqrt n\sqrt{\log\bar p_n/n}\ge(\log2)^{1/2}\).  The domination conditions \(a_{\pi,n}\le\rho_{\pi,n}\) and \(a_{g,n}\le\rho_{g,n}\) hold by the form of the rates.  It remains to verify~\Cref{assu:first-stage-rates}.

\emph{Case~(a).} \Cref{coro:exact-sparsity} gives affine prediction errors \(O_P(\rho_{R,n})\).  With deterministic penalties the fits are functions of the training observations.  Under exact sparsity, \(\Delta_{fg,n}=0\), the approximation terms vanish, and \(\sqrt n\,\rho_{\beta,n}\rho_{g,n}\le(1+C)s_{g,n}(s_{\pi,n}+s_{g,n})\log\bar p_n/\sqrt n\), because \(\rho_{\beta,n}\le(s_{\pi,n}+Cs_{g,n})\sqrt{\log\bar p_n/n}\).

\emph{Case~(b).} \Cref{coro:l1-bounded} gives affine prediction errors \(O_P(\rho_{R,n})\).  Under exact sparsity, \(\rho_{\pi,n}\rho_{g,n}=\sqrt{s_{\pi,n}s_{g,n}}\log\bar p_n/n\le(s_{\pi,n}+s_{g,n})\log\bar p_n/(2n)\) and \(\rho_{g,n}^2\le(s_{\pi,n}+s_{g,n})\log\bar p_n/n\), so \(\sqrt n\,\rho_{\beta,n}\rho_{g,n}\le(\tfrac12+C)(s_{\pi,n}+s_{g,n})\log\bar p_n/\sqrt n\).

\emph{Case~(c).} The projections are assumed, in addition, to satisfy \(\bar\eta_n\le\sqrt{\log\bar p_n/(np_n)}\) or to return \(\widehat\Sigma_{Z,-\ell}\) unchanged whenever \(\widehat\Sigma_{Z,-\ell}\succeq\lambda_{\mathrm{fl}}I\), so they are projections as in~\Cref{lemm:inactive-projection}, and~\Cref{coro:inactive-projection} gives the rates of case~(b) with \(\log n\) in place of \(\log\bar p_n\); the same calculation applies.

\emph{Case~(d).} \Cref{coro:window-cv} gives affine prediction errors \(O_P(\rho_{R,n})\) for penalties that are functions of the training observations and an auxiliary randomization, as~\Cref{assu:first-stage-rates} permits.  Under exact sparsity the calculation of case~(b) gives \(\sqrt n\,\rho_{\beta,n}\rho_{g,n}\le(\tfrac12+C)\upsilon_n^2(s_{\pi,n}+s_{g,n})\log\bar p_n/\sqrt n\).

In every case~\Cref{theo:main} applies.  The same corollaries give the \(\ell_1\) rates \(\zeta_{R,n}\), so the variants are covered by~\Cref{prop:plugin-inference} when, in addition,~\eqref{eq:thy-covariance-rate} holds for them; under exact sparsity in case~(a) this is \((s_{\pi,n}^{3/2}+s_{g,n}^{3/2})\log\bar p_n=o(\sqrt n)\), and in cases~(b)--(d) it is implied by the stated budgets.\qed

\subsection{The Loh--Wainwright corrected Lasso}\label{app:fs-lw}\label{subsec:fs-lw}

The corrected Lasso of~\citet{loh2012high} keeps the corrected Gram matrix unprojected and controls the possibly nonconvex objective with an \(\ell_1\) constraint. With the same \(\widehat\Sigma_{Z,-\ell}\) and \(\widehat c_{RZ,-\ell}\), it computes a stationary point of
\begin{equation}
\min_{b\in\R^{p_n}:\ \|b\|_1\le B_{R,-\ell}}
\Bigl\{
\tfrac12 b'\widehat\Sigma_{Z,-\ell}b-\widehat c_{RZ,-\ell}'b+\lambda_R\|b\|_1
\Bigr\},
\label{eq:fs-lw-objective}
\end{equation}
the \(\ell_1\)-constrained penalized program~(2.7) of~\citet{loh2012high} and program~(12) of~\citet{lohwainwright2015regularized}, with the radius \(B_{R,-\ell}\) set to twice the \(\ell_1\) norm of the cross-validated ordinary Lasso fitted to the training proxy (\Cref{app:sim-firststage}).  The penalty is chosen by the cross-validation rule of CoCoLasso, applied to this learner's own fits, so the two learners differ only in the training objective.  When \(\widehat\Sigma_{Z,-\ell}\) is indefinite, stationary points need not be unique; \citet[Cor.~1]{lohwainwright2015regularized} bound the error of every stationary point when the penalty lies between a constant multiple of \(\sqrt{\log p/n}\) and a constant divided by the radius, the radius makes the target feasible and \(n\) is at least a constant times \(\max\{B_{R,-\ell}^2,s_R\}\log p\), for a corrected Gram matrix whose error covariance is known or estimated from replicates~\citep[Sec.~3.2.1]{loh2012high}.  We do not verify these conditions for cross-validated penalties and data-dependent radii, and~\Cref{app:sim-firststage} reports this learner only as a numerical check of the two implementations: when the projection of \(\widehat\Sigma_{Z,-\ell}\) is inactive, the program~\eqref{eq:fs-lw-objective} at a given penalty has the CoCoLasso solution unless the constraint \(\|b\|_1\le B_{R,-\ell}\) binds.
\section{Simulation Details}\label{app:simulation-supplement}

This appendix gives additional details on the designs, procedures and results of~\Cref{sec:simulations}.

\subsection{The canonical sequence}\label{app:sim-canonical-extra}

\paragraph{The first-stage remainder.} By~\Cref{lemm:cm-centering}, with linear nuisances the conditional mean of the held-out score of fold \(\ell\) at \(\thetao\) is \(\mathrm{Rem}_\ell:=\E[(T_i'\delta_{\beta,\ell})(T_i'\delta_{\gamma,\ell})]\), with \(T_i=(1,X_i')'\) and the errors \(\delta_{\beta,\ell}\) and \(\delta_{\gamma,\ell}\) of the fold-\(\ell\) affine coefficients defined in~\Cref{app:nuisance-and-inference-proofs}. The stored fold fits, the true coefficient vectors and the covariance \(\Sigma_X\) of \(X_i\) determine it exactly: \(\mathrm{Rem}_\ell=\delta^{0}_{\beta,\ell}\delta^{0}_{\gamma,\ell}+(\delta^s_{\beta,\ell})'\Sigma_X\delta^s_{\gamma,\ell}\), with intercept errors \(\delta^0\) and slope errors \(\delta^s\). The bias of \CORE{} implied by the first stages is \(b_{\mathrm{Rem}}:=\widehat J^{-1}\sum_\ell(n_\ell/n)\mathrm{Rem}_\ell\). \Cref{tab:app-remainder} compares it with the Monte Carlo bias. In the approximately sparse and growing-support designs the true vectors are the full coefficient vectors, so the same formula applies.

\begin{table}[!htbp]\centering\tablesize
\caption{The bias of \CORE{} and the first-stage remainder.}\label{tab:app-remainder}
{\setlength{\tabcolsep}{5pt}\begin{tabular}{lrrrrr}
\toprule
$(n,p)$ & Bias & Mean $b_{\mathrm{Rem}}$ & Median $b_{\mathrm{Rem}}$ & Coverage & Recentered coverage \\
\midrule
\multicolumn{6}{l}{\textit{Canonical sequence}}\\
$(500,100)$ & 0.036 & 0.047 & 0.044 & 0.928 & 0.946 \\
$(1000,200)$ & 0.028 & 0.031 & 0.029 & 0.937 & 0.950 \\
$(2000,500)$ & 0.019 & 0.021 & 0.019 & 0.937 & 0.955 \\
$(4000,1000)$ & 0.015 & 0.013 & 0.013 & 0.936 & 0.953 \\
\addlinespace
\multicolumn{6}{l}{\textit{Growing exact sparsity}}\\
$(500,100)$ & 0.093 & 0.095 & 0.092 & 0.880 & 0.938 \\
$(1000,200)$ & 0.105 & 0.106 & 0.107 & 0.856 & 0.950 \\
$(2000,500)$ & 0.124 & 0.128 & 0.129 & 0.682 & 0.924 \\
$(4000,1000)$ & 0.110 & 0.109 & 0.109 & 0.554 & 0.958 \\
\bottomrule
\end{tabular}}
\par\smallskip\parbox{0.95\linewidth}{\footnotesize \(b_{\mathrm{Rem}}\): bias implied by the first-stage remainder, computed on every replication from the stored fits; Recentered coverage: coverage of the interval \(\widehat\theta-b_{\mathrm{Rem}}\pm z_{0.975}\widehat\sigma/\sqrt n\), an infeasible diagnostic. Canonical rows: 1{,}000 replications; growing-support rows: 500.}
\end{table}

\paragraph{Variants.} \Cref{tab:app-variants} reports the plug-in and training-fold variants of~\Cref{subsec:est-variants} on the fits of \CORE{}. The plug-in variant with its augmented standard error and \CORE{} are close in every design; with the standard error that ignores covariance estimation, the plug-in variant covers less often in the canonical sequence.

\begin{table}[!htbp]\centering\tablesize
\caption{\CORE{} and its variants on the same fits.}\label{tab:app-variants}
{\setlength{\tabcolsep}{4pt}\begin{tabular}{llrrrrr}
\toprule
Design & Procedure & Bias & SD & SE/SD & Coverage & Length \\
\midrule
Canonical $(500,100)$ & \CORE{} & 0.036 & 0.183 & 0.96 & 0.928 & 0.693 \\
 & Plug-in variant & 0.033 & 0.184 & 0.96 & 0.928 & 0.694 \\
 & \quad SE ignoring covariance estimation &  &  & 0.94 & 0.922 & 0.675 \\
\addlinespace
Canonical $(1000,200)$ & \CORE{} & 0.028 & 0.129 & 0.99 & 0.937 & 0.500 \\
 & Plug-in variant & 0.027 & 0.129 & 0.99 & 0.939 & 0.501 \\
 & \quad SE ignoring covariance estimation &  &  & 0.96 & 0.931 & 0.486 \\
\addlinespace
Canonical $(2000,500)$ & \CORE{} & 0.019 & 0.093 & 0.98 & 0.937 & 0.357 \\
 & Plug-in variant & 0.018 & 0.093 & 0.98 & 0.938 & 0.357 \\
 & \quad SE ignoring covariance estimation &  &  & 0.95 & 0.929 & 0.346 \\
\addlinespace
Canonical $(4000,1000)$ & \CORE{} & 0.015 & 0.066 & 0.99 & 0.936 & 0.255 \\
 & Plug-in variant & 0.015 & 0.066 & 0.99 & 0.940 & 0.255 \\
 & \quad SE ignoring covariance estimation &  &  & 0.96 & 0.932 & 0.246 \\
\addlinespace
Design 1 & \CORE{} & 0.028 & 0.134 & 0.95 & 0.934 & 0.497 \\
 & Plug-in variant & 0.025 & 0.134 & 0.95 & 0.936 & 0.498 \\
 & Training-fold variant & 0.025 & 0.134 & 0.95 & 0.935 & 0.498 \\
\addlinespace
Design 2 & \CORE{} & 0.037 & 0.155 & 0.93 & 0.920 & 0.563 \\
 & Plug-in variant & 0.032 & 0.156 & 0.93 & 0.921 & 0.565 \\
 & Training-fold variant & 0.031 & 0.156 & 0.92 & 0.922 & 0.565 \\
\addlinespace
Design 3 & \CORE{} & 0.043 & 0.159 & 0.95 & 0.918 & 0.591 \\
 & Plug-in variant & 0.036 & 0.160 & 0.94 & 0.920 & 0.594 \\
 & Training-fold variant & 0.034 & 0.161 & 0.94 & 0.918 & 0.595 \\
\addlinespace
Design 4 & \CORE{} & 0.037 & 0.141 & 0.93 & 0.918 & 0.515 \\
 & Plug-in variant & 0.035 & 0.142 & 0.93 & 0.921 & 0.516 \\
 & Training-fold variant & 0.035 & 0.142 & 0.93 & 0.921 & 0.516 \\
\addlinespace
Design 5 & \CORE{} & 0.034 & 0.133 & 0.96 & 0.935 & 0.501 \\
 & Plug-in variant & 0.031 & 0.134 & 0.96 & 0.937 & 0.502 \\
 & Training-fold variant & 0.031 & 0.134 & 0.96 & 0.936 & 0.502 \\
\addlinespace
Design 6 & \CORE{} & 0.017 & 0.094 & 0.96 & 0.932 & 0.355 \\
 & Plug-in variant & 0.016 & 0.094 & 0.96 & 0.934 & 0.355 \\
 & Training-fold variant & 0.016 & 0.094 & 0.96 & 0.934 & 0.355 \\
\addlinespace
Design 7 & \CORE{} & 0.027 & 0.113 & 0.96 & 0.930 & 0.426 \\
 & Plug-in variant & 0.022 & 0.114 & 0.96 & 0.934 & 0.427 \\
 & Training-fold variant & 0.021 & 0.114 & 0.95 & 0.936 & 0.428 \\
\bottomrule
\end{tabular}}
\par\smallskip\parbox{0.95\linewidth}{\footnotesize Canonical designs as in~\Cref{tab:sim-canonical}; designs 1--7 as in~\Cref{tab:sim-hetero}. The plug-in and training-fold variants use the standard errors of~\Cref{subsec:est-variants} (\Cref{prop:plugin-inference}), which include the covariance-estimation term, except in the indented rows labeled ``SE ignoring covariance estimation'' under the plug-in variant, which use the plug-in estimate with the standard error of~\Cref{subsec:est-variants} without the covariance-estimation term.}
\end{table}

\paragraph{Corrected least squares.} Classical corrected least squares on all measurements is computed on the draws of \CORE{}, with the unit-clustered standard error of~\Cref{rem:stacked-iv} and with a delete-group jackknife standard error (20 groups, critical values of \(t_{19}\)). In the four canonical designs it is biased by \ensuremath{-}0.360 to \ensuremath{-}0.231, its standard deviation and RMSE are 1.61--1.92 and 2.09--5.46 times those of \CORE{}, and its intervals cover at 0.154--0.871 with the clustered standard error and at 0.500--0.968 with the much longer jackknife intervals.

\subsection{Approximately sparse and growing-support designs}\label{app:sim-partc}

\paragraph{Designs.} The sixteen designs change only the coordinates of $\betao$ and $\gammao$ outside each vector's canonical support. In the decay families all of these are nonzero: the $j$th of them, at coordinate $m$, has magnitude $0.3\,j^{-\varsigma}$ in $\betao$ and $0.15\,j^{-\varsigma}$ in $\gammao$ and sign $(-1)^m$. The growing-support designs have entries $0.3(-1)^m$ in $\betao$ and $0.15(-1)^m$ in $\gammao$ at the remaining coordinates $m$ of $\{1,\ldots,\lceil3\log p\rceil\}$. With \(s=\lceil3\log p\rceil\), \(\|\pio\|_1=0.45s+6.35\), \(\|\gammao\|_1=0.15s+1.95\), \(\|\pio\|_2^2=0.2025s+15.8025\) and \(\|\gammao\|_2^2=0.0225s+1.9325\); in the canonical and decay designs these norms are at most \(8.9\) in \(\ell_1\) and \(4.2\) in \(\ell_2\) for every \(p\). No fixed \(C\) in condition~(1) of~\Cref{def:dgpclass}, or \(C_1\) in~\Cref{coro:cocolasso-core}(b) and~(d), bounds them as \(p\to\infty\), whereas the sparsity budgets of~\Cref{coro:cocolasso-core} are \(o(\sqrt n)\) when \(s_\pi=s_g=\lceil3\log p\rceil\) and \(p\le n\), so the sequence leaves the classes of~\Cref{sec:theory} through its coefficient norms, though slowly: the sum bounded in condition~(1) is 10.85--11.14 along the sequence and 10.47 in the canonical sequence. The nuisances are linear, so the target is $\thetao$ (\Cref{rem:projection-target}).

\paragraph{Penalties.} In the approximately sparse and growing-support designs, the median cross-validated penalty of a design is 2.73--3.89 times \(\sqrt{\log\bar p_n/n}\) for the outcome equation and 0.74--1.20 times \(\sqrt{\log\bar p_n/n}\) for the treatment equation, and none of the 180{,}000 penalties selected by \CORE{} in the canonical, approximately sparse, growing-support and heterogeneous-error designs lies at an end of the grid.

\paragraph{First stages.} Let \(r_\pi\) and \(r_\gamma\) be the root mean squared errors of the cross-fitted slope predictions on the latent controls, \(r_\pi^2:=n^{-1}\sum_\ell\sum_{i\in I_\ell}\{X_i'(\widehat\pi_{-\ell}-\pio)\}^2\) and analogously for \(r_\gamma\); because \(\rho_{\beta,n}\ge\rho_{\pi,n}\), the last term of~\eqref{eq:thy-product-rate} is at least \(\sqrt n\,\rho_{\pi,n}\rho_{g,n}\), whose empirical counterpart is \(\sqrt n\,r_\pi r_\gamma\). In the growing-support designs, the means of \(r_\pi\) and \(r_\gamma\) fall only from 1.13 to 0.96 and from 0.38 to 0.31 along the sequence, and the mean of \(\sqrt n\,r_\pi r_\gamma\) rises from 9.6 to 19.0; in the canonical sequence the three means fall from 0.82 to 0.36, from 0.22 to 0.10 and from 4.1 to 2.2. In the simulated designs, neither the sparsity budgets nor the coefficient norms account for the rise: along the sequence, \(s_g(s_\pi+s_g)\log\bar p_n/\sqrt n\) stays between 109 and 123, \((s_\pi+s_g)\log\bar p_n/\sqrt n\) falls from 7.8 to 5.5, the \(\ell_2\) norms of \(\pio\) and \(\gammao\) rise only from 4.32 to 4.48 and from 1.50 to 1.55, and their \(\ell_1\) norms grow from 12.65 to 15.80 and from 4.05 to 5.10. The level of the cross-validated penalties accounts for most of it. On the coordinates \(j\ge7\), where only added coefficients enter, \(\pi_{0j}=0.45(-1)^j\) and \(\gamma_{0j}=0.15(-1)^j\); because the signs alternate and the controls are positively correlated, \((\Sigma_X\pio)_j\) and \((\Sigma_X\gammao)_j\) are only a quarter to a half of \(\pi_{0j}\) and \(\gamma_{0j}\) for \(12\le j\le\lceil3\log p\rceil-2\). The Lasso on the population moments, which minimizes \(\tfrac12b'\Sigma_Xb-(\Sigma_Xb_0)'b+\lambda\|b\|_1\) for a target \(b_0\) and is the population analog of both CoCoLasso and the Lasso on the latent controls, therefore leaves these coefficients largely unfitted: at the median cross-validated penalties of each design, its coefficients on the coordinates \(j\ge7\) average less than 3\% of the true ones at the three smaller sample sizes and 19\% (outcome) and 34\% (treatment) at \((4000,1000)\), and its remainder implies a bias of 1.10--1.22 times the mean \(b_{\mathrm{Rem}}\) of~\Cref{tab:app-remainder}. The first-stage errors are therefore mostly the regularization bias of the cross-validated penalties on many coefficients of fixed size, which becomes small only at larger sample sizes. Apart from this remainder, \CORE{} behaves as the theory predicts: SE/SD is 0.91--1.03, and the infeasible intervals recentered by \(b_{\mathrm{Rem}}\) cover at 0.924--0.958.

\paragraph{Error-free references.} ORACLE-L and ORACLE-DML (as in~\Cref{fig:sim-sparsity}; ORACLE-DML uses the fits of ORACLE-L, the uncorrected score and the conventional standard error) are computed on the replications of the growing-support designs. The two references share their first-stage slopes, and hence the slope part of Rem, in every replication. Their biases, 0.060--0.090, are close to the median of $b^{s}_{\mathrm{Rem}}$, the bias implied by the slope part of the quadratic remainder, and the interval recentered by $b^{s}_{\mathrm{Rem}}$ covers more often in every design (\Cref{tab:app-refset-grow-remainder}). \mbox{ORACLE-L} covers more often than \mbox{ORACLE-DML} mainly because its standard deviation is 1.24--1.34 times larger: its score uses the measurements. ORACLE-L evaluates the score of the plug-in variant of \CORE{} (\Cref{subsec:est-variants}), which uses \(\Pn\Omega_i\) in place of \(\Omega_i\), with that variant's standard error, on the same draws and folds as the variant; the two therefore differ only in their first stages (in~\Cref{fig:app-firststage-canonical}, ORACLE-L instead uses the score of \CORE{}). Their standard deviations differ by at most 4\%, and the bias of the plug-in variant, 0.090--0.122, is 1.19--1.82 times that of ORACLE-L (for \CORE{}, 0.093--0.124 and 1.24--1.85 times): learning the first stages from noisy measurements adds to the second-order term.

\begin{table}[!htbp]\centering\tablesize
\caption{Growing-support designs: the error-free references and their quadratic remainder.}\label{tab:app-refset-grow-remainder}\label{tab:app-refset-grow}
{\setlength{\tabcolsep}{3pt}\begin{tabular}{llrrrrrr}
\toprule
$(n,p)$ & Procedure & Share Rem $>0$ & Median $b^{s}_{\mathrm{Rem}}$ & Bias & SD & Coverage & Recentered coverage \\
\midrule
$(500,100)$ & ORACLE-L & 97.4\% & 0.071 & 0.075 & 0.183 & 0.918 & 0.930 \\
 & ORACLE-DML & 97.4\% & 0.070 & 0.077 & 0.138 & 0.886 & 0.934 \\
$(1000,200)$ & ORACLE-L & 99.0\% & 0.080 & 0.080 & 0.116 & 0.912 & 0.958 \\
 & ORACLE-DML & 99.0\% & 0.080 & 0.082 & 0.093 & 0.848 & 0.946 \\
$(2000,500)$ & ORACLE-L & 100.0\% & 0.087 & 0.084 & 0.091 & 0.814 & 0.916 \\
 & ORACLE-DML & 100.0\% & 0.088 & 0.090 & 0.068 & 0.716 & 0.936 \\
$(4000,1000)$ & ORACLE-L & 100.0\% & 0.057 & 0.060 & 0.059 & 0.862 & 0.964 \\
 & ORACLE-DML & 100.0\% & 0.057 & 0.063 & 0.046 & 0.732 & 0.942 \\
\bottomrule
\end{tabular}}
\par\smallskip\parbox{0.95\linewidth}{\footnotesize Rem: the slope part \(\sum_\ell(n_\ell/n)(\delta^s_{\beta,\ell})'\Sigma_X\delta^s_{\gamma,\ell}\) of the first-stage remainder of~\Cref{lemm:cm-centering}, computed from the stored fold slopes and the true coefficients on each of the 500 replications; \(b^{s}_{\mathrm{Rem}}\): the bias implied by Rem; Recentered coverage: coverage of the interval recentered by \(b^{s}_{\mathrm{Rem}}\).}
\end{table}

\subsection{Heterogeneous-error designs}\label{app:sim-hetero}

The seven designs of~\Cref{subsec:sim-hetero} use the generator of the canonical sequence with the latent design \(\Sigma_X=\Sigma_{0.5}\), \(V_i\sim N(0,1)\), \(U_i\sim N(0,9)\), \(\thetao=1\) and coefficients \(\betao\) with entries \((3,1.5,2)\) and \(\gammao\) with entries \((0.8,1,0.6)\), both on coordinates \(\{1,3,5\}\). The error of measurement \(k\) is \(\nu_{ik}=s_k^{1/2}\tau_i\,\epsilon_{ik}\), where \(\epsilon_{ik}=L_Ae_{ik}\), with \(L_A\) the lower-triangular Cholesky factor of \(\Sigma_A=\tfrac14\Sigma_{0.5}\), and the \(e_{ik}\) are independent across \(i\), \(k\) and coordinates, standard normal in designs~1, 2, 5 and~6 and standardized \(t_5\) in designs~3, 4 and~7. The source scales are \(s=(1,1)\) in designs~1, 5 and~6, \(s=(1,2)\) in designs~2, 3 and~7, and \(s=(1,1.5,2.5)\) with \(K=3\) in design~4. In designs~3, 4 and~7 the scale \(\tau_i^2\) is proportional to \(1+\min\{3,|X_{i1}|/m_1\}^2\) and normalized to have sample mean one, where \(m_1\) is the sample root mean square of \(X_{i1}\); in the other designs \(\tau_i=1\). The errors are independent of \((U_i,V_i)\) and, given \(X_1,\ldots,X_n\), of each other across measurements, so~\Cref{def:model-repeated-measurement} holds (the two sample normalizations of \(\tau_i\) make the units exchangeable rather than independent), with \(\Omegao=K^{-2}\sum_ks_k\Sigma_A\), which, as in the canonical sequence, is proportional to \(\Sigma_X\). In design~5, \(f_0(x)=x'\betao+0.3\,\Phi(x_1)x_2\), so that \(\betao\) is the coefficient of the linear part; because \(\E[X_i\Phi(X_{i1})X_{i2}]=\Sigma_Xe_2/2\), the population projection coefficient of \(f_0\) on \(X_i\) is \(\betao+0.15e_2\). The seed of replication \(r\) in design \(d\) is \(7{,}000{,}000+10{,}000d+r\), and the folds use the seed plus seven. None of the 6{,}000 replications of the seven designs produced a numerical failure.

\subsection{First-stage learners}\label{app:sim-firststage}

\paragraph{The Loh--Wainwright corrected Lasso (LW).} We take the stationary point of~\eqref{eq:fs-lw-objective} reached by a deterministic homotopy from \(b=0\) along CoCoLasso's penalty grid, with the validation Gram matrices projected at CoCoLasso's tolerances. The radius, one per outer fold and nuisance, is $c_R=2$ times the $\ell_1$ norm of the ordinary Lasso fitted to the training proxy with the same grid and inner split (at the smallest penalty if that fit is zero). This is the largest radius in the grids over which~\citet[p.~820]{sorensenfrigessithoresen2015measurement} and~\citet[p.~2411]{datta2017cocolasso} cross-validate the radius of the program without a penalty; here the radius is fixed and the penalty is cross-validated. A fit is accepted if it is feasible and its stationarity residual meets the acceptance bound of~\Cref{app:sim-numerics}.  Along the canonical sequence, all 1{,}000 replications are used at the three smaller designs and the first 250 at \((4000,1000)\).

\paragraph{Results.} Along the canonical sequence (\Cref{fig:app-firststage-canonical}) the correction of the first stages matters with the corrected score as well: the coverage of the ordinary Lasso with the corrected score falls short of that of \CORE{} by 0.083 at \((500,100)\) and by 0.384 at \((4000,1000)\), whereas ORACLE-L covers more often than \CORE{} by 0.010 to 0.020. At a given penalty, the Loh--Wainwright program differs from CoCoLasso only when its \(\ell_1\) constraint binds or the projection of the training Gram matrix is active, and the two learners give the same \CORE{} estimate in 972, 996 and 998 of 1{,}000 replications at the three smaller designs and 249 of 250 at $(4000,1000)$; the Loh--Wainwright learner covers at 0.912--0.937. The comparison is therefore a check of the two implementations rather than of the sensitivity of \CORE{} to the first-stage learner.

\begin{figure}[t]
\centering
\includegraphics[width=\linewidth]{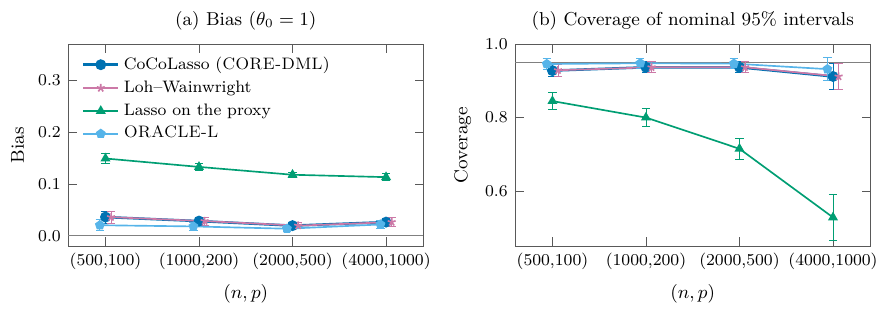}
\caption{First-stage learners with the cross-measurement score along the canonical sequence: all 1{,}000 replications at the three smaller designs and the first 250 at \((4000,1000)\), on which \CORE{} covers at 0.912. (a)~Bias (\(\thetao=1\)) and (b)~coverage of nominal \(95\%\) intervals, with \(\pm1.96\) Monte Carlo standard-error bars. Lasso on the proxy: the ordinary Lasso on the uncorrected proxy Gram matrix. ORACLE-L is infeasible.}
\label{fig:app-firststage-canonical}
\end{figure}

\subsection{Comparisons with other procedures}\label{app:sim-comparison-suite}

\paragraph{Single orderings of the cross-measurement moment.} The single ordering \((k,h)\) forms the residuals of \(Y\) and \(D\) with measurements \(k\) and~\(h\), respectively, from the fits, folds and intercepts of \CORE{}, whose population limit it therefore shares (\Cref{app:measurement-one-ordering}); with two measurements \CORE{} is the average of the two orderings. In the canonical sequence, on 1{,}000 replications per design, a single ordering covers at 0.932--0.947, with standard deviations 1.15--1.23 times and intervals 1.18--1.19 times those of \CORE{}. At the true nuisance parameters, the symmetrized score has 0.69 times the variance of one ordering (\Cref{app:measurement-one-ordering}).

\subsection{Numerical implementation}\label{app:sim-numerics}

\Cref{alg:coredml} summarizes the computation of the feasible \CORE{} estimator of~\Cref{def:core-dml}.  The contractions \(u'\Omega_iv\) are computed from the deviations, \(\{K(K-1)\}^{-1}\sum_k\{u'(W_{ik}-Z_i)\}\{v'(W_{ik}-Z_i)\}\), so no \(p\times p\) matrix is formed per unit.

\begin{algorithm}[!htbp]
\caption{Feasible \CORE{} with the CoCoLasso first stage.}\label{alg:coredml}
\begin{algorithmic}[1]
\Require Observations $\{Y_i,D_i,W_{i1},\ldots,W_{iK}\}_{i=1}^n$; number of folds $L$; nominal level $1-\alpha$; projection floor $\lambda_{\mathrm{fl}}$.
\Procedure{\CORE{}}{$Y,D,W,L,\alpha,\lambda_{\mathrm{fl}}$}
  \State Form $Z_i=K^{-1}\sum_k W_{ik}$ and the within-unit deviations $W_{ik}-Z_i$, which determine the contributions $\Omega_i$ of~\eqref{eq:cov-bessel-row}.
  \State Independently of the observations, partition $\{1,\ldots,n\}$ into folds $I_1,\ldots,I_L$ whose sizes differ by at most one.
  \For{$\ell=1,\ldots,L$}
    \State Compute $\widehat\Omega_{-\ell}=\mathbb P_{n,-\ell}\Omega_i$, $\widehat\Sigma_{Z,-\ell}$ and $\widehat c_{RZ,-\ell}$, $R\in\{Y,D\}$, on $I_\ell^c$.
    \State Replace each corrected Gram matrix used below (inner training, inner validation and full training) by its certified projection~\eqref{eq:est-psd-projection} (\Cref{app:sim-numerics}).
    \State For $R\in\{Y,D\}$, select $\lambda_R$ by four-fold calibrated cross-validation within $I_\ell^c$ over the shrinking grid of thirty log-spaced values from $\lambda_{\mathrm{sh}}$ down to $\lambda_{\mathrm{sh}}/100$ (\Cref{rem:thy-cv-not-covered}; at the sample sizes reported here $\lambda_{\mathrm{sh}}=\lambda_{\mathrm{top}}=\|\widehat c_{RZ,-\ell}\|_\infty$), with criterion~\eqref{eq:est-calibrated-cv-loss}, in which both inner corrected Gram matrices subtract $\widehat\Omega_{-\ell}$ and the inner validation observations are centered at the inner-training means, and with ties going to the larger penalty.
    \State Fit $\widehat\pi_{-\ell}$ and $\widehat\gamma_{-\ell}$ on $I_\ell^c$ by~\eqref{eq:est-cocolasso-objective} at the selected penalties, and form the training-mean intercepts.
    \State For $i\in I_\ell$, form the residuals $\widetilde Y_i,\widetilde D_i$ and the contractions $\widehat\pi_{-\ell}'\Omega_i\widehat\gamma_{-\ell}$ and $\widehat\gamma_{-\ell}'\Omega_i\widehat\gamma_{-\ell}$, which give $N_i$ and $Q_i$ of~\eqref{eq:est-NQ}.
  \EndFor
  \State Compute $\widehat\theta=\Pn N_i/\Pn Q_i$, $\widehat\psi_i=N_i-\widehat\theta Q_i$ and $\widehat\sigma^2=(\Pn Q_i)^{-2}\Pn\widehat\psi_i^2$, with $\widehat\theta=\widehat\sigma=0$ if $\Pn Q_i=0$ (\Cref{def:core-generic}).
  \State \Return $(\widehat\theta,\widehat\sigma,\widehat C_{1-\alpha})$ with $\widehat C_{1-\alpha}$ from~\eqref{eq:est-ci}.
\EndProcedure
\end{algorithmic}
\end{algorithm}

The projection~\eqref{eq:est-psd-projection} uses the floor \(\lambda_{\mathrm{fl}}=10^{-4}\) and is computed by the two-block alternating direction method of multipliers with eigenvalue clipping of~\citet[App.~A, Algorithm~1]{datta2017cocolasso}, initialized at the Frobenius projection, with the over-relaxation and residual balancing of~\citet{boyd2011distributed} and relaxation parameter 1.6. A computed matrix \(A\succeq\lambda_{\mathrm{fl}}I\) is accepted when its distance to the corrected Gram matrix \(G\) exceeds the dual lower bound \(\langle Q,\lambda_{\mathrm{fl}}I-G\rangle\) of~\Cref{rem:projection-certificate} by at most \(\max\{\bar\delta\|A-G\|_{\max},10^{-10}\}\), which certifies~\eqref{eq:thy-certified-projection} with \(\kappa=1/(1-\bar\delta)\) and \(\bar\eta=10^{-10}\); at every sample size used here, \(10^{-10}\le\sqrt{\log\bar p_n/n}\). If the computed smallest eigenvalue of \(G\) is at least \(\lambda_{\mathrm{fl}}\), the projection returns \(G\) itself. The relative tolerance is \(\bar\delta=0.05\) in the canonical sequence and the heterogeneous-error designs, and \(\bar\delta=0.01\) in the approximately sparse and growing-support designs and the applications; in the first three replications of designs~1 and~3 of~\Cref{subsec:sim-hetero}, estimates and standard errors computed at the two tolerances agree to eight decimals. Penalized quadratic programs are solved by warm-started coordinate descent with active-set polishing; a fit is accepted when its residual in the Karush--Kuhn--Tucker conditions is at most \(\max\{10^{-9}\Lambda_{\mathrm{num}},t_{\mathrm{num}}\}\), where \(\Lambda_{\mathrm{num}}=\max(1,\|\widehat c\|_\infty,\lambda)\) and \(t_{\mathrm{num}}=64\varepsilon_{\mathrm{mach}}\max_i(\sum_j|A_{ij}||b_j|+|\widehat c_i|)\), far below the residual \(\lambda/8\) that~\Cref{lemm:kkt-perturbation} allows. A fit that fails these checks is recorded as failed. Because the floor is absolute, the estimator is invariant to the location but not to the scale of the controls, which are standardized in the online labor market application, rescaled by fixed constants in the Caltech application (\Cref{app:cal-estimation}) and have unit variances in the simulations.

For each first stage, four-fold calibrated cross-validation within the outer training sample selects among the thirty log-spaced penalty values \(\lambda_{\mathrm{sh}}0.01^{(g-1)/29}\), \(g=1,\ldots,30\), of the shrinking grid of~\Cref{rem:thy-cv-not-covered}, where \(\lambda_{\mathrm{sh}}=\lambda_{\mathrm{top}}\min\{1,100\upsilon_mt_m\}\), \(\lambda_{\mathrm{top}}=\|\widehat c_{RZ,-\ell}\|_\infty\), \(t_m=\sqrt{\log\max(p_n,m)/m}\), \(\upsilon_m=\max\{1,\log\log m\}\) and \(m=|I_\ell^c|\); at the sample sizes of~\Cref{sec:simulations,sec:application}, \(100\upsilon_mt_m\ge1\), so \(\lambda_{\mathrm{sh}}=\lambda_{\mathrm{top}}\); otherwise the function \texttt{core\_dml} of the replication package applies the factor. The inner folds are drawn separately for \(Y\) and \(D\). In each inner split the observations of both inner folds are centered at the means of the inner training fold, and both corrected Gram matrices subtract the outer-training estimate \(\widehat\Omega_{-\ell}\): a fit \(b\) on the inner training fold uses the certified projection of \(\mathbb P_{\mathrm{tr}}\{(Z_i-\bar Z_{\mathrm{tr}})(Z_i-\bar Z_{\mathrm{tr}})'\}-\widehat\Omega_{-\ell}\), and \(\widehat\Sigma^+_{\mathrm{val}}\) is the certified projection of \(\mathbb P_{\mathrm{val}}\{(Z_i-\bar Z_{\mathrm{tr}})(Z_i-\bar Z_{\mathrm{tr}})'\}-\widehat\Omega_{-\ell}\), with \(\widehat c_{RZ,\mathrm{val}}=\mathbb P_{\mathrm{val}}\{(Z_i-\bar Z_{\mathrm{tr}})(R_i-\bar R_{\mathrm{tr}})\}\). The corrected validation loss of \(b\) is
\begin{equation}
 b'\widehat\Sigma^+_{\mathrm{val}}b-2\widehat c_{RZ,\mathrm{val}}'b,
 \label{eq:est-calibrated-cv-loss}
\end{equation}
summed over the four inner folds, with ties resolved in favor of the larger penalty.
\section{Empirical Applications: Additional Details}\label{app:application-supplement}

\subsection{The Caltech Cohort Study}\label{app:cal}

\paragraph{Data and sample.}\phantomsection\label{app:cal-data} The archive of~\citet{gillensnowbergyariv2019data}, the replication files of~\citet{gillensnowbergyariv2019} (GSY), contains one student-level data file, \texttt{GillenSnowbergYarivME.dta}, with 783 students. The outcome is \texttt{sumsCompete2015}, the choice of the tournament for the second round of the spring 2015 addition task, and \(D_i\) is \texttt{male}. The first-round score is \texttt{sumsCorrectPR2015}, the number of correct sums in the first round; despite the suffix of its name, that round was paid as a tournament. The stated rank is \texttt{sumsRank2015} and enters as indicators for ranks 2, 3 and~4. The repeated measurements are \texttt{riskyUrn20MaxValue2014} and \texttt{riskyUrn30MaxValue2014} and their counterparts for the ambiguous and compound urns. Nine students lack at least one of the six fall valuations and none lacks a spring variable, which leaves 774 students.

\paragraph{Coding of the valuations.}\phantomsection\label{app:cal-coding} For each urn the archive records the smallest sure amount, on a grid of 10 tokens, that the student preferred to the lottery, from 10 to 100 tokens for the 20-ball urns and from 10 to 150 tokens for the 30-ball urns, so the switching point lies within the 10 tokens below the recorded value. We code both lists at the midpoints of these intervals (\Cref{subsec:app-caltech}). The code of GSY subtracts 10 tokens from the 20-ball value, a choice its comments describe as making the risk-neutral value the modal response, and codes the 30-ball value as we do. This choice raises the normalized 30-ball valuations by 5 tokens relative to the 20-ball valuations: under their coding the mean differences are 4.65--4.78 tokens and equality of the means is rejected (\(\chi^2(3)=336\)), whereas under the midpoint coding the differences are below half a token in absolute value (\Cref{tab:gsymeas}). The two codings differ by a constant shift of the 20-ball measurement, which the centered contrasts of~\Cref{rem:source-intercepts} remove exactly: with centered contrasts the two codings give identical estimates, and without centering \CORE{} gives 12.74 points on the primary partition under either coding (the two estimates differ by less than 0.001 points).

\paragraph{Estimation.}\phantomsection\label{app:cal-estimation} The first stages are CoCoLasso fits on the seven controls. For penalization, the valuations are measured in tens of tokens, the first-round score in units of five sums, and the rank indicators are multiplied by two. The recorded controls take the same value in both measurements, so their rows and columns of every \(\Omega_i\) vanish (\Cref{rem:mixed-controls}). Penalties are selected by four-fold cross-validation over 30 values with the loss of~\Cref{app:sim-numerics}, and the projection uses the relative tolerance \(\bar\delta=0.01\). On the primary partition no selected penalty lies at an end of the grid, the projection is inactive in every training fold, and the smallest eigenvalue of the corrected Gram matrices of the training folds is at least 0.148. Partition \(s=1,\dots,100\) assigns the students to five folds at random with seed \(20261120+s-1\); the first is the primary partition. Following~\citet[Sec.~3.4]{chernozhukov2018double}, the reported estimate is the median \(\widehat\theta^{\mathrm{med}}\) of the partition estimates \(\widehat\theta_s\), and its standard error is \(\{\operatorname{median}_s[\widehat{\mathrm{se}}_s^{\,2}+(\widehat\theta_s-\widehat\theta^{\mathrm{med}})^2]\}^{1/2}\), where \(\widehat{\mathrm{se}}_s\) is the standard error of partition~\(s\). On the primary partition \CORE{} is 12.74 points with standard error 2.99. Corrected least squares on all measurements (the last row of~\Cref{tab:gsymain}) is the stacked estimator of~\Cref{rem:stacked-iv}, with the standard error clustered by student.

\paragraph{The first stages.} On the primary partition, the fold-averaged slopes imply that, holding the other controls fixed, the fitted probability of being male differs by at most 1.51 percentage points, and the fitted entry probability by at most 0.53 points, between students whose valuation of one lottery differs by 10 tokens. The averaged valuations have correlations of 0.012 to 0.026 in absolute value with gender; given the first-round score and the rank indicators, their partial correlations with gender and with entry are at most 0.024 in absolute value.

\paragraph{Tests of the observable restrictions.}\phantomsection\label{app:cal-tests} Let \(\Delta_i:=(W_{i2}-W_{i1})/2\) collect the stake contrasts of the three valuations. Under~\Cref{def:model-repeated-measurement}, \(\Delta_i\) is a difference of errors with conditional mean zero given \((X_i,D_i,Y_i)\), so it has mean zero and is uncorrelated with \(D_i\) and \(Y_i\). The correlation test stacks the six sample covariances of \(\Delta_i\) with \(D_i\) and with \(Y_i\), estimates their joint variance from the sample covariance of the products of the centered variables and compares the Wald statistic with the \(\chi^2(6)\) distribution. The test of equal means compares \(n\bar\Delta'\widehat V_\Delta^{-1}\bar\Delta\), with \(\widehat V_\Delta\) the sample covariance matrix of \(\Delta_i\), with the \(\chi^2(3)\) distribution. For the confidence ratings the contrasts are the differences of consecutive ratings, which gives four covariances and two means; for the fall--spring pairs they are the three spring-minus-fall differences. Let \(\widehat c_{kh}\) be the sample covariance of ratings \(k\) and \(h\) (\(\widehat c_{12}=0.705\), \(\widehat c_{13}=0.693\) and \(\widehat c_{23}=0.708\)). The test of equal covariances for the confidence ratings stacks \(\widehat c_{12}-\widehat c_{13}\) and \(\widehat c_{12}-\widehat c_{23}\), estimates their joint variance from the sample covariance of the products of the centered ratings and compares the Wald statistic, which does not depend on the pair taken as reference, with the \(\chi^2(2)\) distribution; in a one-factor model with three indicators, it is a test of equal loadings. It uses covariances rather than the raw moments of~\Cref{rem:cov-weights-identification} because the rating means differ. With the scale of the first rating normalized to one, the implied scales of the second and third ratings are \(\widehat c_{23}/\widehat c_{13}\) and \(\widehat c_{23}/\widehat c_{12}\) (\Cref{rem:source-intercepts}), with delta-method standard errors. An error component common to the three ratings adds the same amount to each covariance: it is not detected, and it moves the implied scales toward one.

\paragraph{Errors shared across stakes.}\phantomsection\label{app:cal-sensitivity} \Cref{tab:gsymult} multiplies every covariance contribution by \(m\) in the first stages, the score and the variance. If the errors of the two stakes share a component with covariance \(\bar C=\bar\chi(\Sigma_{\nu,1}+\Sigma_{\nu,2})/2\), then \(\E_P\Omega_i=(\Sigma_{\nu,1}+\Sigma_{\nu,2}-2\bar C)/4\) and the covariance of the averaged error is \((\Sigma_{\nu,1}+\Sigma_{\nu,2}+2\bar C)/4=m\,\E_P\Omega_i\), with \(m=(1+\bar\chi)/(1-\bar\chi)\) (\Cref{app:sim-crossdep}); the cross-stake correlation of the errors of each control is then at least \(\bar\chi\), with equality when the two stakes have equal error covariances. Multipliers above 2.31 make the implied covariance matrix of the latent controls, the sample covariance of the controls minus \(m\) times the covariance estimate, indefinite, so at the point estimates values of \(\bar\chi\) above 0.40, and hence, when the two stakes have equal error covariances, cross-stake correlations above 0.40, are inconsistent with the data. The threshold is itself estimated: a bootstrap over students (2{,}000 resamples) gives a 95\% interval of 1.71 to 2.91 for the multiplier and 0.26 to 0.49 for \(\bar\chi\). Up to \(m=2\), the gap rises from 13.41 to 13.82 points.

\begin{table}[!htbp]\centering\tablesize
\caption{Sensitivity of the Caltech estimate to errors shared across stakes.}\label{tab:gsymult}
\begin{tabular*}{\linewidth}{@{\extracolsep{\fill}}rrrcc@{}}
\toprule
Multiplier $m$ & Shared fraction $\bar\chi$ & Gap & Range across partitions & Implied latent covariance p.d. \\
\midrule
1.00 & 0.00 & 13.41 & [12.51, 14.46] & Yes \\
1.22 & 0.10 & 13.46 & [12.55, 14.51] & Yes \\
1.67 & 0.25 & 13.65 & [12.54, 14.72] & Yes \\
2.00 & 0.33 & 13.82 & [12.94, 15.44] & Yes \\
2.50 & 0.43 & 14.22 & [12.52, 15.43] & No \\
3.00 & 0.50 & 14.37 & [6.57, 15.71] & No \\
\bottomrule
\end{tabular*}
\par\smallskip\parbox{\linewidth}{\footnotesize Gaps in percentage points: medians and ranges over the 100 partitions of~\Cref{tab:gsymain}. \(\bar\chi=(m-1)/(m+1)\) is the cross-stake error correlation that the multiplier represents when the two stakes have equal error covariances, and a lower bound on that correlation otherwise. Last column: whether the sample covariance of the controls minus \(m\) times the covariance estimate is positive definite, that is, whether the implied covariance matrix of the latent controls is positive definite at the estimates.}
\end{table}

\paragraph{Alternative repeated controls.}\phantomsection\label{app:cal-alternatives} The confidence ratings are \texttt{guess}\(k\)\texttt{Confidence2015}, \(k=1,2,3\): students saw three pictures of jars of jelly beans, guessed the number in each and rated their confidence in each guess on a six-point scale~\citep{gillensnowbergyariv2019}. We treat the three ratings as measurements of one latent control with a common scale and rating-specific intercepts (\Cref{rem:source-intercepts}); the sample has 783 students. The fall--spring design uses the second risky-project allocation (\texttt{secondRiskyProjectAllocation}, in units of 20 tokens for penalization), the average of the two fair-urn valuations as shares of their prizes (in units of 0.1) and self-rated willingness to take risks (\texttt{qualitativeRisk}), each measured in fall 2014 and spring 2015, for 775 students. Both designs include the first-round score and stated rank, with the same scaling as the main design.

\subsection{The online labor market}\label{app:mturk}

\paragraph{Controls and measurements.}\phantomsection\label{app:emp-sample} The sample is the task sample labeled \texttt{ipeirotis} in the replication archive of~\citet{dubejacobsnaidusuri2020data}; the retained tasks were first recorded between 2014 and 2016. The dictionary uses only clean-data information available at posting, not the generated measurements or realized completion rates, hit counts and timing gaps. It consists of signed hashed unigram and bigram counts of the title, description, keywords and qualification text, scaled to unit Euclidean norm within each field (48, 112, 48 and 48 bins; 256 coordinates), shares from five-topic models of the first three fields (15), scores for six task categories from the same fields (18) and 16 exactly measured controls (the logarithm of one plus the time allotted, the numbers of characters and words of the four fields, and the year and the sine and cosine of the month, day of the week and hour at which the task was first recorded); all coordinates are standardized. The shares of each topic model sum to one for 100.0\%, 95.5\% and 99.7\% of tasks (title, description and keyword models), so the 305-coordinate dictionary has one exact and two near linear dependences; dropping one share per model raises the smallest eigenvalue of the sample covariance matrix to \ensuremath{8.0\times10^{-4}}, above the floor \(\lambda_{\mathrm{fl}}=10^{-4}\). The choice of the dropped share changes the penalized fits but leaves the linear span unchanged for the title model and nearly so for the other two. With \(\sigma_\varrho^2=(1-\varrho)/\varrho\) the error variance of a single measurement, the average of \(K\) measurements has reliability \(1/(1+\sigma_\varrho^2/K)\), from \(0.947\) to \(0.400\) in the primary designs, a range that includes reliabilities reported in earnings-validation studies~\citep{bound2001measurement,abowdstinson2013estimating}. In the AR\((0.5)\) design the errors of the 289 error-prone coordinates of the full dictionary have covariance \(\sigma_\varrho^2\) times the AR\((0.5)\) correlation matrix in column order.

\paragraph{Benchmark and procedures.}\phantomsection\label{app:emp-procedures} The benchmark uses one task per requester and a dictionary and learner specific to this exercise, so it is not comparable with the estimates of~\citet{dubejacobsnaidusuri2020}. \NAIVEDML{} also passes its Gram matrices through the floored projection. All procedures and draws share five outer folds and the tuning rule of~\Cref{app:sim-numerics}.

\paragraph{Excluded draws and the plug-in variant.}\phantomsection\label{app:emp-complete} Ten of the 1{,}400 draws, at most five in any design, stopped with an error in the symmetric eigenvalue computation and are excluded. On the remaining draws the plug-in variant of~\Cref{subsec:est-variants}, with the standard error defined there (\Cref{prop:plugin-inference}), differs from \CORE{} by at most 0.19 benchmark standard errors.

\paragraph{Corrected least squares.} We also compute least squares on the clean controls, \(\widehat\theta_{\mathrm{LS}}=\ensuremath{-}0.399\) with standard error 0.013, and, on all 1{,}400 draws, corrected least squares on all measurements with the clustered standard error of~\Cref{rem:stacked-iv} and a delete-group jackknife standard error (20 groups, critical values of \(t_{19}\)). Corrected least squares fails our positive-definiteness safeguard if the corrected Gram matrix of the controls has smallest eigenvalue below \(10^{-4}\). It fails in every draw with \(\varrho\le0.60\), where that matrix is indefinite, and in 2 of the AR\((0.5)\) draws. Where it passes, its mean displacement from \(\widehat\theta_{\mathrm{LS}}\) is between \ensuremath{-}0.08 and 0.00 times the standard error of \(\widehat\theta_{\mathrm{LS}}\), and its intervals contain \(\widehat\theta_{\mathrm{LS}}\) in every draw with either standard error; the jackknife fails in 5 more draws.

\subsection{\texorpdfstring{The procedure of~\citet{belloni2017confidence}}{The procedure of Belloni, Chernozhukov and Kaul}}\label{app:emp-bck}

In the six designs whose errors are independent across coordinates, the error covariance is diagonal, as~\citet{belloni2017confidence} (BCK) assume. On every draw of these designs on which \CORE{} could be computed (1{,}191 of the 1{,}200; \Cref{app:emp-complete}) we compute their Algorithm~2 for the coefficient of~\(D\), with regressors \(z=(D,Z')'\), estimated error covariance \(\widehat\Gamma=\diag\{0,\diag(\widehat\Omega^{\mathrm{full}})\}\) and no cross-fitting. As in their simulations, Steps~1 and~2 use the conic estimator of~\citet*{bellonirosenbaumtsybakov2017conic}, which, for a response~\(R\), solves
\begin{equation*}
\min_{b,\,t\ge0}\bigl\{\|b\|_1+\lambda_c t\bigr\}
\quad\text{subject to}\quad
\bigl\|\widehat c_{Rz}-\widehat Ab\bigr\|_\infty\le\mu t+\tau_R,
\quad
\|b\|_2\le t,
\end{equation*}
where \(\widehat c_{Rz}\) is the sample covariance of the regressors and \(R\), and \(\widehat A\) is the sample covariance matrix of the regressors minus their estimated error covariance (eq.~(9) of~\citealp{bellonirosenbaumtsybakov2017conic}, with an unrestricted parameter set). Step~1 has response~\(Y\), regressors~\(z\) and error covariance~\(\widehat\Gamma\); Step~2 has response~\(D\), regressors~\(Z\) and error covariance \(\diag(\widehat\Omega^{\mathrm{full}})\); Step~3 is their closed form. As in their simulations, coefficients below \(10^{-7}\) in absolute value are set to zero; solutions are accepted at a relative duality gap of at most \(10^{-8}\) (SCS, \citealp{odonoghue2016scs}, with ECOS, \citealp{domahidi2013ecos}, as the fallback). The primary tuning is that of their simulations, \(\tau_R=\mu=\{\log((p_n+1)/0.05)/n\}^{1/2}\) in both steps, with \(\lambda_c=1\), a value their paper does not state and one of the three used in the simulations of~\citet*{bellonirosenbaumtsybakov2017conic}. The secondary tuning keeps~\(\mu\) and sets \(\tau_R=\mathrm{sd}(R)\,\mu\), with \(\mathrm{sd}(R)\) the sample standard deviation of the response. Their simulations take the covariance as known; with an estimated covariance, the asymptotic variance is that of their representation~(3.1) (Corollary~3), which adds the covariance-estimation term, and we use its plug-in estimate. The refitted variant replaces each first step by corrected least squares on its selected support and fails if, in either step, the submatrix of \(\widehat A\) on that support admits no Cholesky factorization or has a reciprocal condition number below \(10^{-10}\); their Condition~B is stated to include refitted first steps, but their paper does not define a refit, and we do not establish its rates. With the covariance known, our implementation without refitting rejects at 0.054 and 0.044 in the cells (\(n=350\), \(p=300\), measurement-error standard deviation 0.5) of their Tables~1 and~3 in which they report 0.066 and 0.054, within Monte Carlo bands of about 0.03 (500 replications).

\Cref{tab:app-mturk-bck} compares \CORE{} and the BCK procedure with the clean least-squares estimate, because the BCK procedure, like least squares, uses the full sample without cross-fitting. On the clean controls, \CORE{} with the covariance set to zero is the clean DML benchmark, 0.45 standard errors of \(\widehat\theta_{\mathrm{LS}}\) from \(\widehat\theta_{\mathrm{LS}}\), and the BCK procedure with the primary tuning is 1.11 such standard errors away without refitting and 0.62 with it; these clean-data gaps reflect the regularization of each procedure, not measurement error. Under measurement error, \CORE{} and the refitted BCK procedure with the primary tuning stay within about half a standard error of their clean-data values at reliabilities of 0.60 and above. Measured from each procedure's own clean-data value, that is, after subtracting the clean-controls rows of~\Cref{tab:app-mturk-bck}, the refitted BCK procedure with the primary tuning is less displaced than \CORE{} at \(\varrho\le0.40\) and more displaced at \(\varrho\ge0.80\). At lower reliabilities the mean displacement of \CORE{} grows, whereas that of the refitted BCK procedure does not, although its dispersion grows and the refit fails in 17 of the 200 draws at \(\varrho=0.25\) with the primary tuning (19 draws in all designs) and in 34 draws in all designs with the secondary tuning. Without refitting, the procedure with the primary tuning is likewise less displaced than \CORE{} from its own clean-data value at \(\varrho\le0.60\) and more displaced at \(\varrho\ge0.80\), and with the secondary tuning both versions are less displaced than \CORE{} in every design; on the clean controls, however, the procedure with the secondary tuning and without refitting is 1.89 standard errors of \(\widehat\theta_{\mathrm{LS}}\) from \(\widehat\theta_{\mathrm{LS}}\). The design with errors correlated across coordinates, which BCK's Conditions~A(v) and~D exclude, is not included.

\begin{table}[!htbp]\centering\tablesize
\caption{\CORE{} and the procedure of~\citet{belloni2017confidence} in the online labor market.}\label{tab:app-mturk-bck}
{\setlength{\tabcolsep}{3.5pt}\setlength{\defaultaddspace}{0.2em}\begin{tabular}{@{}llrrrrr@{}}
\toprule
 & & & Mean & Mean abs. & RMS & \\
Design & Procedure & Draws & displ. & displ. & displ. & Incl. \\
\midrule
Clean controls & \CORE{} & & 0.45 & & & Yes \\
 & BCK, primary tuning, no refit & & 1.11 & & & Yes \\
 & BCK, primary tuning, refitted & & 0.62 & & & Yes \\
 & BCK, secondary tuning, no refit & & 1.89 & & & No \\
 & BCK, secondary tuning, refitted & & 0.63 & & & Yes \\
\addlinespace
$\varrho=0.90$, $K=2$ & \CORE{} & 200 & 0.48 & 0.48 & 0.49 & 1.000 \\
 & BCK, primary tuning, no refit & 200 & 1.15 & 1.15 & 1.15 & 1.000 \\
 & BCK, primary tuning, refitted & 200 & 0.47 & 0.47 & 0.49 & 1.000 \\
 & BCK, secondary tuning, no refit & 200 & 1.90 & 1.90 & 1.90 & 0.430 \\
 & BCK, secondary tuning, refitted & 200 & 0.63 & 0.63 & 0.65 & 1.000 \\
\addlinespace
$\varrho=0.80$, $K=2$ & \CORE{} & 198 & 0.52 & 0.52 & 0.54 & 1.000 \\
 & BCK, primary tuning, no refit & 198 & 1.20 & 1.20 & 1.20 & 1.000 \\
 & BCK, primary tuning, refitted & 198 & 0.43 & 0.43 & 0.46 & 1.000 \\
 & BCK, secondary tuning, no refit & 198 & 1.92 & 1.92 & 1.92 & 0.399 \\
 & BCK, secondary tuning, refitted & 198 & 0.61 & 0.61 & 0.63 & 1.000 \\
\addlinespace
$\varrho=0.60$, $K=2$ & \CORE{} & 195 & 0.65 & 0.65 & 0.69 & 1.000 \\
 & BCK, primary tuning, no refit & 195 & 1.26 & 1.26 & 1.27 & 1.000 \\
 & BCK, primary tuning, refitted & 195 & 0.44 & 0.46 & 0.52 & 1.000 \\
 & BCK, secondary tuning, no refit & 195 & 1.91 & 1.91 & 1.91 & 0.477 \\
 & BCK, secondary tuning, refitted & 195 & 0.56 & 0.56 & 0.62 & 1.000 \\
\addlinespace
$\varrho=0.40$, $K=2$ & \CORE{} & 199 & 0.83 & 0.84 & 0.91 & 1.000 \\
 & BCK, primary tuning, no refit & 199 & 1.31 & 1.31 & 1.34 & 0.985 \\
 & BCK, primary tuning, refitted & 197 & 0.40 & 0.46 & 0.58 & 1.000 \\
 & BCK, secondary tuning, no refit & 199 & 1.85 & 1.85 & 1.86 & 0.673 \\
 & BCK, secondary tuning, refitted & 196 & 0.49 & 0.65 & 0.99 & 0.990 \\
\addlinespace
$\varrho=0.25$, $K=2$ & \CORE{} & 200 & 1.26 & 1.26 & 1.36 & 0.970 \\
 & BCK, primary tuning, no refit & 200 & 1.44 & 1.44 & 1.50 & 0.920 \\
 & BCK, primary tuning, refitted & 183 & 0.43 & 0.75 & 0.95 & 0.989 \\
 & BCK, secondary tuning, no refit & 200 & 1.85 & 1.85 & 1.89 & 0.635 \\
 & BCK, secondary tuning, refitted & 169 & 0.68 & 1.09 & 2.05 & 0.976 \\
\addlinespace
$\varrho=0.80$, $K=3$ & \CORE{} & 199 & 0.49 & 0.49 & 0.50 & 1.000 \\
 & BCK, primary tuning, no refit & 199 & 1.16 & 1.16 & 1.17 & 1.000 \\
 & BCK, primary tuning, refitted & 199 & 0.44 & 0.44 & 0.47 & 1.000 \\
 & BCK, secondary tuning, no refit & 199 & 1.90 & 1.90 & 1.90 & 0.432 \\
 & BCK, secondary tuning, refitted & 199 & 0.61 & 0.61 & 0.63 & 1.000 \\
\bottomrule
\end{tabular}}
\par\smallskip\parbox{0.97\linewidth}{\footnotesize Columns as in~\Cref{tab:app-main}, with displacements measured from the clean least-squares estimate \(\widehat\theta_{\mathrm{LS}}=\ensuremath{-}0.399\), in units of its standard error 0.013, and positive when toward zero. BCK:~\citet{belloni2017confidence}. Incl.: share of nominal 95\% intervals that contain \(\widehat\theta_{\mathrm{LS}}\) (on the clean controls, whether the interval contains it). Primary tuning: the tuning of the simulations of~\citet{belloni2017confidence}; secondary tuning: \(\tau_R=\mathrm{sd}(R)\mu\). Refitted: corrected least squares on the selected supports. Draws: draws on which the procedure passed its numerical checks, among those on which \CORE{} could be computed; on the clean controls, \CORE{} is DML with the covariance set to zero.}
\end{table}
\endgroup

\end{document}